\documentclass[preprintnumbers,floatfix,aps]{revtex4}
\usepackage{amsmath}
\usepackage{amsfonts}
\usepackage{amssymb}
\usepackage{physymb}
\usepackage{graphicx}
\usepackage{epsfig}

\usepackage{dcolumn}
\usepackage{bm}
\usepackage{fancybox}
\usepackage{multirow}
\usepackage{float}
\usepackage{hyperref}    
\usepackage{epstopdf}

\usepackage[utf8]{inputenc}

\newcommand{\Punkte}{0}

{\noindent {\bf Exercise {#1}.} #2 \vspace{0.2cm} \\  
\renewcommand{\Punkte}{#3}}%
{\mbox{\hspace{3ex}} \hfill {\bf \Punkte~\mbox{Points}}\bigskip }

\newenvironment{Exercise*}[2]%
{\noindent {\bf Exercise {#1}*.} #2 \vspace{0.2cm} \\ 
renewcommand{\Punkte}{#2}}%
{\mbox{\hspace{2ex}} \hfill {\bf \Punkte~\mbox{Points}}\bigskip }

{\noindent {\bf Exercise {#1}.}\vspace{0.2cm} #2}%

\newcounter{enum1}

\newcounter{enuma}

\begin{document}

\title{{\bf Discrete Solitons and Vortices in Anisotropic Hexagonal and Honeycomb Lattices}}
\author{Q.E. Hoq}
\affiliation{ Department of Mathematics,
Western New England University, Springfield, MA, 01119, USA}

\author{P.G. Kevrekidis}
\email{kevrekid@math.umass.edu}
\affiliation{Department of Mathematics and Statistics, University of Massachusetts,
Amherst, Massachusetts 01003-4515 USA}

\affiliation{Center for Nonlinear Studies and Theoretical Division, Los Alamos
National Laboratory, Los Alamos, NM 87544}

\author{A.R. Bishop}
\affiliation{Center for Nonlinear Studies and Theoretical Division, Los Alamos
National Laboratory, Los Alamos, NM 87544}

\begin{abstract} 
In the present work, 
we consider the self-focusing
discrete nonlinear Schr{\"o}dinger equation on hexagonal 
and honeycomb lattice geometries. 
Our emphasis is on the study of the effects of anisotropy, motivated
by the tunability afforded in recent optical and atomic physics 
experiments. We find that important classes of solutions, such
as the so-called discrete vortices, undergo destabilizing bifurcations
as the relevant anisotropy control parameter is varied. We quantify
these bifurcations by means of explicit analytical calculations
of the solutions, as well as of their spectral linearization
eigenvalues. Finally, we corroborate the relevant stability picture
through direct numerical computations. In the latter, we observe the
prototypical manifestation of these instabilities to be the 
spontaneous rearrangement of the solution, for larger
values of the coupling, into localized waveforms
typically centered over fewer sites than the original unstable structure.
For weak coupling, the instability
appears to result in a robust breathing of the relevant
waveforms.
\end{abstract}

\maketitle

\section{Introduction}

In both optical media~\cite{segev} and atomic systems, such 
as Bose-Einstein condensates (BECs)~\cite{morsch}, in the past 
two decades there has been
a tremendous amount of effort focused on understanding the
implications of periodic lattices. In the former case, both the
realms of optical waveguides~\cite{christo} 
and of photorefractive crystals~\cite{segev2}
have played crucial roles towards the analysis and
experimental realization of states such as discrete solitons,
and vortices, as well as of more complex waveforms, including
ring structures, necklaces, gap solitons and many others.
In atomic BECs, on the other hand, the emphasis has
not only been on corresponding matter waves~\cite{konotop},
but also on quantum phenomena beyond the realm of mean-field
models~\cite{bloch}. 

In recent years, the emphasis has somewhat shifted 
from the consideration of the more customary square lattices
to the examination
of lattices of hexagonal or honeycomb form. There, a source
of emphasis has again been localization and 
self-trapping in the form of solitonic and vortical
structures~\cite{Law1,Terhalle,apl_zhang}, but also other aspects
have been studied including, e.g., Bloch states~\cite{zhig_bloch}.
A significant fraction of the focus has been on the emulation
by these optical systems of ``photonic graphene'', leading to numerous 
remarkable features, including the creation, destruction
and experimental observation of topologically protected, so-called,
edge states~\cite{recht1,recht2}, and also the emergence
of pseudospin and angular momentum~\cite{zhig3}. In the atomic
realm too, considerably tunable and flexible optical lattices
of both a hexagonal and honeycomb form have been produced
for single~\cite{klaus1} and multi-species~\cite{klaus2}
experiments. While much of the interest in this context lies
within quantum mechanical transitions, such as the superfluid-insulator
transition~\cite{bloch}, the atoms can, very controllably, be considered in the
superfluid regime where a mean-field description paralleling
the optical one is suitable. As an aside, it is relevant
to mention that more complex lattice structures including
e.g. Kagom{\'e} lattices are also a subject of ongoing
consideration~\cite{kag1,kag2} and
are within the realm of experimental possibility
in both settings.

At the mathematical level, there exists a prototypical model that
combines the suitable lattice geometry, the discreteness
and the nonlinearity. As a result, it  captures the principal features
of the experimental observations, at least as regards the emerging coherent
structures. This model is the so-called discrete
nonlinear Schr{\"o}dinger (DNLS) equation, which has been a subject of
intense theoretical and numerical investigation~\cite{dnls}.
Our aim in the present work is to utilize this DNLS model in 
order to capture the impact of anisotropy on the hexagonal
and honeycomb lattices. This is in part motivated by the
studies in optical photorefractive systems such as the
work of~\cite{Terhalle} where both unstretched and stretched
lattices were used and in both cases the coupling was
anisotropic (varying in one direction between $20 \%$ and $80 \%$ of the
coupling in the other directions). Such a systematic
study is also motivated by the atomic realm of, e.g.,~\cite{klaus1},
where the full control of the optical beam intensities, wavenumbers 
and phases
that create the lattice trapping the atoms can straightforwardly 
be used to produce different types of lattices (e.g. both
hexagonal and honeycomb) and different anisotropies.


Our aim here is to provide a systematic analysis of the different
types of solutions that are possible in the anisotropic system.
Starting from the isotropic two-dimensional limit, we
vary the strength of the interaction along a particular direction.
Progressively this leads from a two-dimensional configuration, e.g. in 
the honeycomb case, to
an uncoupled set of quasi-one-dimensional configurations. As a 
result, we can appreciate that numerous states among those
that exist in the two-dimensional (2d) setting should 
disappear at a suitable critical point
as we approach the 1d regime. For instance, the discrete
vortices belong to this category, as there are no solutions with
nontrivial vorticity in one-dimensional DNLS lattices~\cite{dnls,Pelinovsky}.
Here, we intend to provide a quantification of the relevant
solutions, as well as to provide a road map for their dynamical
destabilization by evaluating their dominant linearization
eigenvalues. Both of these steps are performed analytically
(to leading order) permitting a complete characterization of the bifurcation
events/destabilization or disappearance of different branches of solutions.
This is done for the prototypical unit cell of each lattice i.e.,
for a triangular cell within the hexagonal lattice and a hexagonal
cell within the honeycomb lattice, although it can be straightforwardly
generalized to other cases. Once the existence, stability and bifurcations
of the relevant solutions are determined, then their potential
instabilities (and spectral properties) are also explored numerically.
Finally, these findings are corroborated by direct numerical
computations illustrating the tendency of the (unstable) 
dynamics towards (typically) fewer sites than the original structure.
In the case of the 3-site cell in the hexagonal lattice, we observe
a tendency of the dynamics towards the 
ground (single-site) state of the model for stronger couplings, or
towards robust breathing excitations in the case of weaker couplings.
In the case of the 6-site cell of the honeycomb lattice, even 
for stronger couplings, multi-site excitations (of different
types -- see details below) were typically found to
persist over the evolution scales of dynamical propagation considered
here.

Our presentation is structured as follows. In section II, we
present our systematic analytical findings regarding the existence
and stability of solutions for each of the lattices in their respective unit
cells. Then in section III, we present the corresponding numerical findings,
as well as examine, the fate of dynamically unstable solutions. 
Finally, in section IV, we summarize our results and present
some challenges for future studies.

\section{Theoretical Analysis}

To study the two geometries of interest, we consider the following 
discrete nonlinear Schr\"{o}dinger equation

\begin{equation} \label{eq:DNLS}
i\frac{du_{m,n}}{dz} = -\varepsilon \Delta_2 u_{m,n} - |u_{m,n}|^2 u_{m,n} 
\end{equation}
with the two-dimensional anisotropic discrete Laplacian
\begin{equation} \label{eq:Laplacian}
\Delta_2 u_{m,n} = \sum_{m',n'\in N_1} u_{m',n'} + \sum_{m'',n''\in N_2} \delta u_{m'',n''} - \frac{|N|}{3}(2+ \delta) u_{m,n}.
\end{equation}
Here, the constant $\varepsilon$ denotes the strength of the linear 
coupling between nearest neighbor sites in the isotropic case 
and the anisotropy is controlled by the parameter $\delta$ 
$(0 \leq \delta \leq 1)$. Since both of the grids of interest
are rotationally 
invariant, hence there is a freedom in selecting 
the direction of the anisotropy. 
A value of
$\delta = 1$ yields the isotropic lattice with a uniform nearest 
neighbor coupling of $\varepsilon$, whereas $\delta=0$ completely 
decouples sites in the direction parallel to a particular lattice
direction, chosen without loss of generality.
Physically, the field $u_{m,n}$ represents (in our optical
example) the envelope of the 
electric field in the waveguide, while in that case $z$ is the propagation 
coordinate. For BECs, the field represents the atomic wavefunction
in the corresponding well of the optical lattice, while $z$
in that realization is replaced by the time $t$.
The summations are over disjoint subsets, $N_1$ and 
$N_2$, of the set $N=N_1 \cup N_2$ of nearest neighbors where 
$|N|=6$ for the hexagonal lattice and $|N|=3$ for the honeycomb 
lattice. The set $N_2$ is the set of nearest neighbors joined to 
$u_{m,n}$ by the (anisotropic) coupling, $\delta\varepsilon$, 
while the remaining nearest neighbors belong to the set $N_1$ 
and have a coupling of $\varepsilon$ to $u_{m,n}$. Note that 
in the case $\delta=0$, the hexagonal grid becomes the usual 
rectangular grid, while the honeycomb grid becomes a parallel 
set of one-dimensional grids with (in both geometries) a nearest 
neighbor coupling of $\varepsilon$.

We are interested in stationary solutions of the form 
$u_{m,n}=\exp(i \Lambda z)v_{m,n}$, where $\Lambda$ is the propagation 
constant in optics or the chemical potential in BECs. Then, $v_{m,n}$ 
satisfies the steady-state equation
\begin{equation} \label{eq:steadystate}
\Lambda v_{m,n} = \varepsilon \Delta_2 v_{m,n} + |v_{m,n}|^2 v_{m,n}.
\end{equation}
In the anticontinuum (AC) limit of uncoupled sites~\cite{macaub}, 
(i.e. when $\varepsilon \rightarrow 0$), the solutions of 
Eq.~\eqref{eq:steadystate} are 
$v_{m,n} = 0$ and $v_{m,n} = \sqrt{\Lambda}\exp(i \theta_{m,n})$. 
Thus, at the AC limit, explicit solutions of the form 
$u_k=\sqrt{\Lambda}\exp(i \theta_k)\exp(i \Lambda z)$ can be found 
over contours $M$ of the uncoupled lattice points for arbitrary 
$\theta_k \in [0,2\pi)$, where the nodes are indexed by $k$. For the 
current work, in the hexagonal lattice, $k$ will index the sites 
along a three-site one-dimensional closed contour ($|M|=3$), while 
in the honeycomb lattice, $k$ will index the sites along a six-site 
one-dimensional closed contour ($|M|=6$), i.e., we will
consider the principal cells of the respective
lattices. Without loss of generality, 
we set $\Lambda=1$. 

Following an analysis similar to that of~\cite{Pelinovsky, Pelinovsky2},
we then find that the necessary (leading order) 
conditions for solutions over a discrete contour to 
persist for $\varepsilon > 0$ are given by 
\begin{equation} \label{eq:necessary}
F_k = \delta_{k,k-1} \sin(\theta_k - \theta_{k-1}) + \delta_{k,k+1} \sin(\theta_k - \theta_{k+1}) = 0,
\end{equation}
where we have the periodic condition 
$\theta_{k+|M|}=\theta_k$ for $k=1,...,|M|$, 
and the coefficients $\delta_{k,k-1}$ and $\delta_{k,k+1}$, provide 
the lattice anisotropy as defined by:

\begin{equation}
\delta_{k,l}=
\begin{cases}
\delta,       &\text{if the segment of $M$ joining adjacent nodes $k$ and $l$ is parallel to the 
anisotropic direction}\\
1,            &\text{if the segment of $M$ joining adjacent nodes $k$ and $l$ is not parallel to the 
anisotropic direction.}
\end{cases}
\end{equation}

We will study the behavior of some solutions of the variety 
described above in both the hexagonal and honeycomb lattices 
when the ``background" coupling, $\varepsilon$, is fixed at 
a small value and the anisotropy is switched on (i.e. $\delta < 1)$. 
We will see that the anisotropy may stabilize or destabilize
some solutions via, typically, pitchfork (i.e.,
symmetry breaking) bifurcations. Theoretically we 
will find that in the weak coupling limit, 
this transition from stability to instability, 
or vice-versa, occurs when the solution collides with another 
solution at $\delta = 0.5$ which then persists as
the lattice becomes more anisotropic, until there is a complete 
decoupling in the prescribed direction. 

Again, adapting the results of~\cite{Pelinovsky, Pelinovsky2}
(see also the exposition of~\cite{dnls}),
the stability of lattice excitations can be determined 
(to leading order) from 
the eigenvalues $\gamma_k$ of the $|M|\times|M|$ Jacobian 
matrix of the form
\begin{equation} \label{Jacobian}
\partial F_k/\partial \theta_l = 
\begin{cases}
\delta_{k,k-1} \cos(\theta_k - \theta_{k-1}) + \delta_{k,k+1} \cos(\theta_k - \theta_{k+1}),    &\text{$k=l$}\\
-\delta_{k,l} \cos(\theta_k - \theta_l),                                                         &\text{$k=l\pm 1$}\\
0,                                                                                              &\text{$|l-k|\geq 2$.}
\end{cases}
\end{equation}
When the excited nodes in the lattice are adjacent, the 
(near zero) stability eigenvalues $\lambda_k$ of the full problem 
are given by $\lambda_k = \pm\sqrt{2\gamma_k \varepsilon}$ \cite{Pelinovsky, Pelinovsky2}.
We now examine some explicit examples of this general
theoretical formulation of the anisotropic DNLS problem.

\subsection{Hexagonal Lattice}
In this subsection, we will analytically track the effects of 
anisotropy on the stability of various three-site configurations 
in the hexagonal lattice and make predictions about what the 
original configurations in the isotropic lattice transform into. 
The notation $[a, b, c]$ is employed to describe the three-site 
contour with phases $a$, $b$ and $c$ at the nodes (while a corresponding 
notation will be used for six-site contours). The contours we 
discuss here are: 
(1) $[0, 2\pi/3, 4\pi/3]$ (charge one vortex), (2) $[0, \pi, 0]$, 
and (3) $[0, 0, 0]$. These are the principal (up to trivial
transformations of phase) 3-site solutions of the hexagonal
lattice cell.

(1) We begin with the three-site single charged vortex
($\theta_1=0, |\Delta \theta| = \frac{2\pi}{3}$). From 
Eq.~\eqref{eq:necessary} with $|M|=3$ and with the anisotropy 
lying between the sites with phases $\theta_1$ and $\theta_3$, 
we obtain the following relationship 
between the phases: 
\begin{eqnarray} \label{eq:3siteequations}
\begin{aligned}
\theta_3 &=& \theta_1 + 2\arccos \left(-\frac{1}{2 \delta} \right)~~~~~0.5 \leq \delta \leq 1 \\
\theta_2 &=& \frac{\theta_1 + \theta_3}{2}  ~~~~~~~~~~~~~~~~~~~~~~~(\text{mod}~ 2\pi).
\end{aligned}
\end{eqnarray}
Using this phase profile in  Eq.~\eqref{Jacobian}, the Jacobian 
matrix becomes 
\begin{equation} \label{eq:3Jacobian}
\mathbf{J} =
\begin{pmatrix}
\cos \left( \frac{\theta_1 - \theta_3}{2} \right) + \delta \cos \left( \theta_1 - \theta_3 \right)   &   
-\cos \left( \frac{\theta_1 - \theta_3}{2} \right)   &   -\delta \cos \left( \theta_1 - \theta_3 \right)\\
-\cos \left( \frac{\theta_1 - \theta_3}{2} \right)   &   2\cos \left( \frac{\theta_1 - \theta_3}{2} \right)   &
-\cos \left( \frac{\theta_1 - \theta_3}{2} \right)\\
-\delta \cos \left( \theta_1 - \theta_3 \right)   &   -\cos \left( \frac{\theta_1 - \theta_3}{2} \right)   &
\cos \left( \frac{\theta_1 - \theta_3}{2} \right) + \delta \cos \left( \theta_1 - \theta_3 \right)
\end{pmatrix}.
\end{equation}
Thus, for $0.5 \leq \delta \leq 1$ the eigenvalues $\gamma_k$ are  
found to be 
$\gamma_1 = 0$, 
$\gamma_2 = \cos \left( \frac{\theta_1 - \theta_3}{2} \right) + 2\delta \cos \left( \theta_1 - \theta_3 \right)$ and, 
$\gamma_3 = 3\cos \left( \frac{\theta_1 - \theta_3}{2} \right)$. 
In particular, for the hexagonal three-site charge-one vortex 
(which satifies Eq.~\eqref{eq:3siteequations}) the stability 
eigenvalues for $0.5 \leq \delta \leq 1$ are found to be 
$\lambda_1 = \{0, 0\}$, 
$\lambda_2 = \pm \sqrt{\frac{(4\delta^2 - 1)\varepsilon}{\delta}}i$, 
and $\lambda_3 = \pm \sqrt{\frac{3\varepsilon}{\delta}}i$.
For this range of values of $\delta$, all of the quantities in the radicals
within the eigenvalue pairs remain nonnegative and, for sufficiently 
small $\varepsilon$, the pair $\lambda_3$ will not collide with the 
continuous spectrum. Hence, this vortex remains stable throughout 
this interval of anisotropy. Note that the only eigenvalue pair that moves 
along the imaginary axis towards the origin of the spectral 
plane and thus has the potential to bring about instability 
when $0 \leq \delta < 0.5$ is 
$\lambda_2 = \pm \sqrt{\frac{(4\delta^2 - 1)\varepsilon}{\delta}}i$.  
As the above solution of Eq.~\eqref{eq:3siteequations}
cannot be continued below $\delta=0.5$, additional analysis 
is needed to reveal the outcome for a further 
increase in anisotropy, i.e. for $0 \leq \delta < 0.5$.

From Eq.~\eqref{eq:3siteequations}, we can also determine the theoretical 
predictions for the changes in the relative phases (mod $2\pi$)
as a function of $\delta$ for $0.5 \leq \delta \leq 1$:
\begin{eqnarray}  \label{eq:3siterelphase}
|\theta_2 - \theta_1| = |\theta_3 - \theta_2| = \frac{1}{2}|\theta_1 - \theta_3| = \arccos \left(-\frac{1}{2 \delta} \right).
\end{eqnarray}
In the isotropic case (i.e. when $\delta = 1$) for all the 
relative phases we have $|\Delta \theta| = \frac{2\pi}{3}$. 
As the anisotropy increases, reaching $\delta \rightarrow 0.5$, we 
find that $|\theta_2 - \theta_1| \rightarrow \pi$, 
$|\theta_3 - \theta_2| \rightarrow \pi$, and $|\theta_1 - \theta_3| \rightarrow 0$.
Thus, at this critical point, the single-charge vortex
 $[0, \frac{2\pi}{3}, \frac{4\pi}{3}]$, 
merges with the configuration
$[\theta_1, \theta_1+\pi, \theta_1+2\pi] = [\theta_1, \theta_1+\pi, \theta_1] (\text{mod}~ 2\pi)$ 
i.e., with the $[0, \pi, 0]$ state. 
Exploring the latter state and its stability for
$0 \leq \delta \leq 0.5$ (or, in fact, for any $\delta$ since
the solution persists $\forall \delta$), the corresponding
Jacobian of Eq.~\eqref{Jacobian} 
assumes the form
\begin{equation} \label{eq:3Jacobian_leq_0_5}
\mathbf{J} =
\begin{pmatrix}
-1+\delta          &     1          &          -\delta   \\
1                  &    -2          &              1     \\
-\delta            &     1          &         -1+\delta  \\
\end{pmatrix}.
\end{equation}
The associated stability eigenvalues become
$\lambda_1 = \{0, 0\}$,
$\lambda_2 = \pm \sqrt{6\varepsilon}i$, and
$\lambda_3 = \pm \sqrt{2(1 - 2\delta)\varepsilon}i$.
Of particular note is the eigenvalue pair $\lambda_3$ which
remains imaginary for $0 \leq \delta <0.5$. Thus, this solution
is stable for $0 \leq \delta \leq 0.5$, but it destabilizes
through a pitchfork bifurcation, giving rise to the discrete vortex
(inheriting its stability) for $\delta > 0.5$.

(2) Given our analysis of the  $[0, \pi, 0]$ state above in 
connection with the vortex bifurcation, we will not discuss 
further here the case with anisotropy between sites with 
phase $0$. 

For the $[0, \pi, 0]$ state with anisotropy between the  
site with phase $\pi$ and one of the $0$ phase sites,
the Jacobian matrix is
\begin{equation} \label{eq:3Jacobian_0p0}
\mathbf{J} =
\begin{pmatrix}
0          &         1          &        -1      \\
1          &     -1-\delta      &       \delta   \\
-1         &       \delta       &       1-\delta \\
\end{pmatrix}
\end{equation}
from which we find the stability eigenvalues to be
$\lambda_1 = \{0, 0\}$ and
$\lambda_2 = \pm \sqrt{2(-\delta + \sqrt{\delta^2 + 3})\varepsilon}$,
$\lambda_3 = \pm \sqrt{2(\delta + \sqrt{\delta^2 + 3})\varepsilon}i$.
The eigenvalue pair $\lambda_2$ remains real valued throughout
the change of $\delta$ and thus we see that this solution remains
unstable. 

(3) In the isotropic lattice the stability eigenvalues of 
the hexagonal $[0, 0, 0]$ configuration are found to be 
$\lambda_1 = \{0, 0\}$,
$\lambda_2 = \pm \sqrt{6\varepsilon}$, and
$\lambda_3 = \pm \sqrt{6\varepsilon}$.
Not surprisingly, due to the adjacent in-phase sites,
this is unstable. In the case of $\delta \neq 1$, this 
instability persists. From our above analysis, the 
relevant eigenvalues can be directly found to be 
$\lambda_1 = \{0, 0\}$,
$\lambda_2 = \pm \sqrt{6\varepsilon}$, and
$\lambda_3 = \pm \sqrt{2 (2 \delta + 1) \varepsilon}$.
Despite the significant dependence of $\lambda_2$ on $\delta$, 
we see that this configuration is indeed generically
expected to be unstable.

\subsection{Honeycomb Lattice}
We now analytically examine the stability 
and transformation of solutions in the anisotropic honeycomb 
lattice and its six-site cell 
($|M|=6$). Similar to what was done for the hexagonal 
lattice, we will deduce relationships between the phases for a 
few prototypical configurations in the honeycomb lattice. 
In each solution, the anisotropy 
lies between the sites with phases $\theta_3$ and $\theta_4$ 
and also between sites with phases $\theta_6$ 
and $\theta_1$. The structures we will present here assume
the following form in the isotropic limit of $\delta=1$: 
(1) $[0, \frac{\pi}{3}, \frac{2\pi}{3}, \pi, \frac{4\pi}{3}, \frac{5\pi}{3}]$
(charge 1 vortex),
(2) $[0, \frac{2\pi}{3}, \frac{4\pi}{3}, 2\pi, \frac{8\pi}{3}, \frac{10\pi}{3}]$
(charge 2 vortex),
(3) $[0, \pi, 0, \pi, 0, \pi]$
(4) $[0, 0, 0, 0, 0, 0]$. 
While additional configurations are possible here (in particular all
combinations of $0$ and $\pi$ phase are possible within the
6 sites), these configurations are the most interesting
ones and will provide a basic understanding of the stability
properties of the anisotropic system.

(1) The honeycomb six-site charge 1 vortex 
($\theta_1=0, |\Delta \theta| = \frac{\pi}{3}$) satisfies
the following phase relationships which can be deduced 
from Eq.\eqref{eq:necessary}:
\begin{eqnarray} \label{eq:6siteequations1}
\begin{aligned}
\theta_3 &=& \theta_1 + 2\arccos\left(\frac{1}{2 \delta}\right)\hspace{85pt} 0.5 \leq \delta \leq 1 \\ 
\theta_2 &=& \frac{\theta_1 + \theta_3}{2} \hspace{185pt} \\
\theta_4 &=& \theta_1 + \pi \hspace{190pt} \\
\theta_5 &=& \frac{\theta_1 + \theta_3}{2} + \pi \hspace{165pt} \\
\theta_6 &=& \theta_3 + \pi \hspace{145pt} (\text{mod}~ 2\pi).
\end{aligned}
\end{eqnarray}
Using these results with the Jacobian matrix of Eq.~\eqref{Jacobian}, 
we find that the stability eigenvalues of the charge $1$ vortex are 
$\lambda_1 = \{0, 0\}$, 
$\lambda_2 = \pm \sqrt{\frac{\varepsilon}{\delta}}$, 
$\lambda_3 = \pm \sqrt{\frac{(4\delta^2 - 1)\varepsilon}{\delta}}$, 
$\lambda_4 = \pm \sqrt{\frac{3\varepsilon}{\delta}}$, and \\
$\lambda_{5, 6} = \pm \sqrt{\left( \frac{1+4\delta^2}{2\delta} \pm \sqrt{-4 + \frac{21}{8\delta^2} 
+ 9\cos(\alpha(\delta)\beta(\delta)) + 4\delta^2 \cos(4\alpha(\delta)\beta(\delta)) 
+ \left(\frac{16\delta^2 - 9}{2}\right) \cos^2(\alpha(\delta)) 
- 8\delta^2 \cos^4(\alpha(\delta))}\right) \varepsilon}$
where $\alpha(\delta) = \theta_1 + \arccos\left(\frac{1}{2\delta}\right)$, 
$\beta(\delta) = \arccos\left(\frac{1}{2\delta}\right)$.
It is clear that at least the eigenvalue pairs $\lambda_1$, 
$\lambda_2$, and $\lambda_3$ all remain real-valued as 
$\delta \rightarrow 0.5$ and hence the charge 1 vortex 
is unstable on the entire interval $0.5 \leq \delta \leq 1$.  

The changes in the relative phases (mod $2\pi$) as a function 
of $\delta$ for $0.5 \leq \delta \leq 1$ are given by:
\begin{eqnarray*}
|\theta_2-\theta_1| = |\theta_3-\theta_2| = |\theta_5-\theta_4| = |\theta_6-\theta_5| = \arccos \left(\frac{1}{2 \delta} \right) \\
|\theta_4-\theta_3| = \pi - 2\arccos \left(\frac{1}{2 \delta} \right) \\
|\theta_1-\theta_6| = \pi + 2\arccos \left(\frac{1}{2 \delta} \right).
\end{eqnarray*}
In the isotropic case ($\delta=1$), all the relative 
phases satisfy $|\Delta \theta| = \frac{\pi}{3}$. But as $\delta \rightarrow 0.5$, 
we find that $|\theta_{i+1}-\theta_i| \rightarrow 0$ for 
$i = 1, 2, 4, 5$, while $|\theta_4-\theta_3| \rightarrow \pi$ and,
$|\theta_1-\theta_6| \rightarrow \pi$. Also as $\delta \rightarrow 0.5$,
from Eq.\eqref{eq:6siteequations1} we theoretically predict that 
the charge 1 vortex in the isotropic lattice merges at with
$[\theta_1, \theta_1, \theta_1, \theta_1+\pi, \theta_1+\pi, \theta_1+\pi]$
at $\delta = 0.5$.
As was shown in more detail with the hexagonal three-site 
charge $1$ vortex, the stability eigenvalues for the interval
$0 \leq \delta \leq 0.5$ can be obtained and appear more simply as 
$\lambda_1 = \{0, 0\}$,  
$\lambda_2 = \pm \sqrt{2\varepsilon}$, 
$\lambda_3 = \pm \sqrt{2(1-2\delta)\varepsilon}$, 
$\lambda_4 = \pm \sqrt{6\varepsilon}$ and,
$\lambda_{5, 6} = \pm \sqrt{(3 - 2\delta \pm \sqrt{4\delta^2 + 4\delta +9})\varepsilon}$.
Thus, we have instability throughout the change in
the anisotropy. More generally, the established ``rule of thumb''
for self-focusing nonlinearities is that whenever sites of the
same phase are adjacent to each other, the configuration will
inherit an instability associated with a real eigenvalue pair.
Hence, it is natural to expect, based on the structural form
of the configuration $[\theta_1, \theta_1, \theta_1, 
\theta_1+\pi, \theta_1+\pi, \theta_1+\pi]$, that it will be unstable
for all values of $\delta$. Nevertheless, one of its real pairs
for $\delta < 0.5$ will become imaginary for $\delta > 0.5$, giving
rise to an unstable daughter state, namely the single charge
vortex solution.

(2) The honeycomb six-site charge-2 vortex 
($\theta_1=0, \Delta \theta = \frac{2\pi}{3}$ for $\delta=1$) satisfies
\begin{eqnarray} \label{eq:6siteequations2}
\begin{aligned}
\theta_3 &=& \theta_1 + 2\arccos\left(-\frac{1}{2 \delta}\right)  \hspace{110pt}     0.5 \leq \delta \leq 1 \\
\theta_2 &=& \frac{\theta_1 + \theta_3}{2} \hspace{220pt} \\
\theta_4 &=& \theta_1 + 2 \pi \hspace{220pt} \\
\theta_5 &=& \frac{\theta_1 + \theta_3}{2} + 2 \pi \hspace{195pt} \\
\theta_6 &=& \theta_3 + 2 \pi \hspace{30pt} \hspace{145pt} (\text{mod}~ 4\pi)
\end{aligned}
\end{eqnarray}
and the stability eigenvalues for the charge $2$ vortex are 
$\lambda_1 = \{0, 0\}$, 
$\lambda_2 = \pm \sqrt{\frac{\varepsilon}{\delta}}i$, 
$\lambda_3 = \pm \sqrt{\frac{(4\delta^2 - 1)\varepsilon}{\delta}}i$, 
$\lambda_4 = \pm \sqrt{\frac{3\varepsilon}{\delta}}i$, and \\
$\lambda_{5, 6} = \pm \sqrt{\left( \frac{1+4\delta^2}{2\delta} \pm \sqrt{-4 + \frac{21}{8\delta^2} 
+ 9\cos(\alpha(-\delta)\beta(-\delta)) + 4\delta^2 \cos(4\alpha(-\delta)\beta(-\delta)) 
+ \left(\frac{16\delta^2 - 9}{2}\right) \cos^2(\alpha(-\delta)) 
- 8\delta^2 \cos^4(\alpha(-\delta))}\right) \varepsilon}i$,
where as before $\alpha(\delta) = \theta_1 + \arccos\left(\frac{1}{2\delta}\right)$, 
$\beta(\delta) = \arccos\left(\frac{1}{2\delta}\right)$.

The changes in the relative phases as a function 
of $\delta$ for $0.5 \leq \delta \leq 1$ are given by:
\begin{eqnarray*}
|\theta_2-\theta_1| = |\theta_3-\theta_2| = |\theta_5-\theta_4| = |\theta_6-\theta_5| = \arccos \left(-\frac{1}{2 \delta} \right) \\
|\theta_4-\theta_3| = 2\pi - 2\arccos \left(-\frac{1}{2 \delta} \right) \\
|\theta_1-\theta_6| = 2\pi + 2\arccos \left(-\frac{1}{2 \delta} \right).
\end{eqnarray*}
In the isotropic honeycomb lattice ($\delta=1$), all the relative 
phases satisfy $|\Delta \theta| = \frac{2\pi}{3}$. But as $\delta \rightarrow 0.5$, 
we find that $|\theta_{i+1}-\theta_i| \rightarrow \pi$ for 
$i = 1, 2, 4, 5$, while $|\theta_4-\theta_3| \rightarrow 0$ and,
$|\theta_1-\theta_6| \rightarrow 0$. Also, as $\delta \rightarrow 0.5$,
from Eq.~\eqref{eq:6siteequations2} we theoretically predict that 
the charge 2 vortex in the isotropic lattice merges at $\delta = 0.5$ with
$[\theta_1, \theta_1+\pi, \theta_1+2\pi, \theta_1+2\pi, \theta_1+3\pi, \theta_1+4\pi] = 
[\theta_1, \theta_1+\pi, \theta_1, \theta_1, \theta_1+\pi, \theta_1]~~~(\text{mod}~ 2\pi)$.
The stability eigenvalues of the latter state for 
$0 \leq \delta \leq 0.5$ are
$\lambda_1 = \{0, 0 \}$,
$\lambda_2 = \pm \sqrt{2\varepsilon}i$,
$\lambda_3 = \pm \sqrt{2(1-2\delta)\varepsilon}i$,
$\lambda_4 = \pm \sqrt{6\varepsilon}i$,
$\lambda_5 = \pm \sqrt{(-3 + 2\delta - \sqrt{4\delta^2+4\delta+9})\varepsilon} = \pm \sqrt{(3-2\delta+\sqrt{4\delta^2+4\delta+9})\varepsilon}i$,
$\lambda_6 = \pm \sqrt{(-3 + 2\delta + \sqrt{4\delta^2+4\delta+9})\varepsilon}$.
Note that $\lambda_6$ is real for $0 \leq \delta \leq 0.5$ 
and so it is expected that the stable vortex of the isotropic
limit with $\delta=1$ becomes unstable
for some $\delta$ in $0.5 < \delta < 1$. As regards
the state $[\theta_1, \theta_1+\pi, \theta_1, \theta_1, \theta_1+\pi, \theta_1]~~~(\text{mod}~ 2\pi)$, the above analysis predicts that it also undergoes
a supercritical pitchfork bifurcation giving rise to the charge
2 vortex which inherits its (in)stability properties and is eventually
full stabilized for a larger value of $\delta$ in $0.5 < \delta < 1$.
In fact, using the expression for $\lambda_6$ given above for
the charge-2 vortex, we find that the relevant critical point
is $\delta=0.716$, which we will compare with our numerical computations
in the following section.

(3) In a similar way to the analysis done above, the 
stability eigenvalues for the six-site configuration 
of alternating sites $[0, \pi, 0, \pi, 0, \pi]$ are 
found to be
$\lambda_1 = \{0, 0\}$,
$\lambda_2 = \pm \sqrt{6\varepsilon}i$,
$\lambda_3 = \pm \sqrt{2\varepsilon}i$,
$\lambda_4 = \pm \sqrt{2(2\delta + 1)\varepsilon}i$
$\lambda_{5,6} = \pm \sqrt{(3 + 2\delta \pm \sqrt{4\delta^2 - 4\delta + 9})\varepsilon}i$.
The imaginary eigenvalues indicate that this configuration 
is stable both in the isotropic and anisotropic lattices.

(4) Finally the six-site configuration of in-phase sites 
$[0, 0, 0, 0, 0, 0]$ has the stability eigenvalues 
$\lambda_1 = \{0, 0\}$,
$\lambda_2 = \pm \sqrt{6\varepsilon}$,
$\lambda_3 = \pm \sqrt{2\varepsilon}$,
$\lambda_4 = \pm \sqrt{2(2\delta + 1)\varepsilon}$
$\lambda_{5,6} = \pm \sqrt{(3 + 2\delta \pm \sqrt{4\delta^2 - 4\delta + 9})\varepsilon}$.
Given the presence of real eigenvalues, this configuration 
is unstable. Similarly to what we saw previously for the 3-site
configuration in the hexagonal lattice, the presence of
adjacent in-phase sites is detrimental to the stability
of this configuration for arbitrary values of $\delta$. Hence,
no stabilization of the relevant state is anticipated, independently
of the particular value of the anisotropy parameter $\delta$.

\section{Numerical Results}

In this section we present our numerical findings for the 
various configurations in the hexagonal and honeycomb lattices 
and compare these results with the theoretical findings from 
the previous section. In all cases, we use a Newton-Raphson
fixed point iteration to identify the full numerical
solution over the two-dimensional lattices. This process is
initiated  
at the AC limit and continued to a small coupling while 
maintaining $\delta = 1$ to yield the isotropic case. In most 
of the examples we consider, the Newton-Raphson interation 
is continued to $\varepsilon = 0.01$, but a few cases will 
also be examined where the fixed point iteration 
is continued to a higher value of $\varepsilon$. 
The anisotropy (as described in a previous section) is introduced 
by letting $\delta$ deviate from the isotropic
unity value and performing a continuation in decreasing values
of the parameter towards $\delta \rightarrow 0$. 
We present figures that show each configuration, along with 
its phase portrait and spectral plane at some $\delta$ 
before and after the relevant bifurcation points, comparing
the latter with our theoretical predictions. 
It is generally found that for values of the coupling, $\varepsilon$, 
near the AC limit the theoretical predictions match the 
numerical results very well.

The theoretical predictions of the linearization eigenvalues 
will be compared to the numerical results for the linear stability 
of the stationary solution $v_{m,n}\exp(iz)$ by using the
ansatz
\begin{equation}
u_{m,n} = e^{iz} \left[ v_{m,n} + \gamma(a_{m,n} e^{\lambda z} + b_{m,n}^* e^{\lambda^* z}) \right].
\end{equation}
The eigenvalue problem that follows is then solved for the
eigenvalues $\lambda$ and eigenvectors $(a_{m,n}, b_{m,n})^T$.
The asterisk denotes complex conjugate while T denotes transpose.

\subsection{Hexagonal Lattice} 
\begin{figure}[tbh]
\centering
\includegraphics[width=5cm,height=4cm]{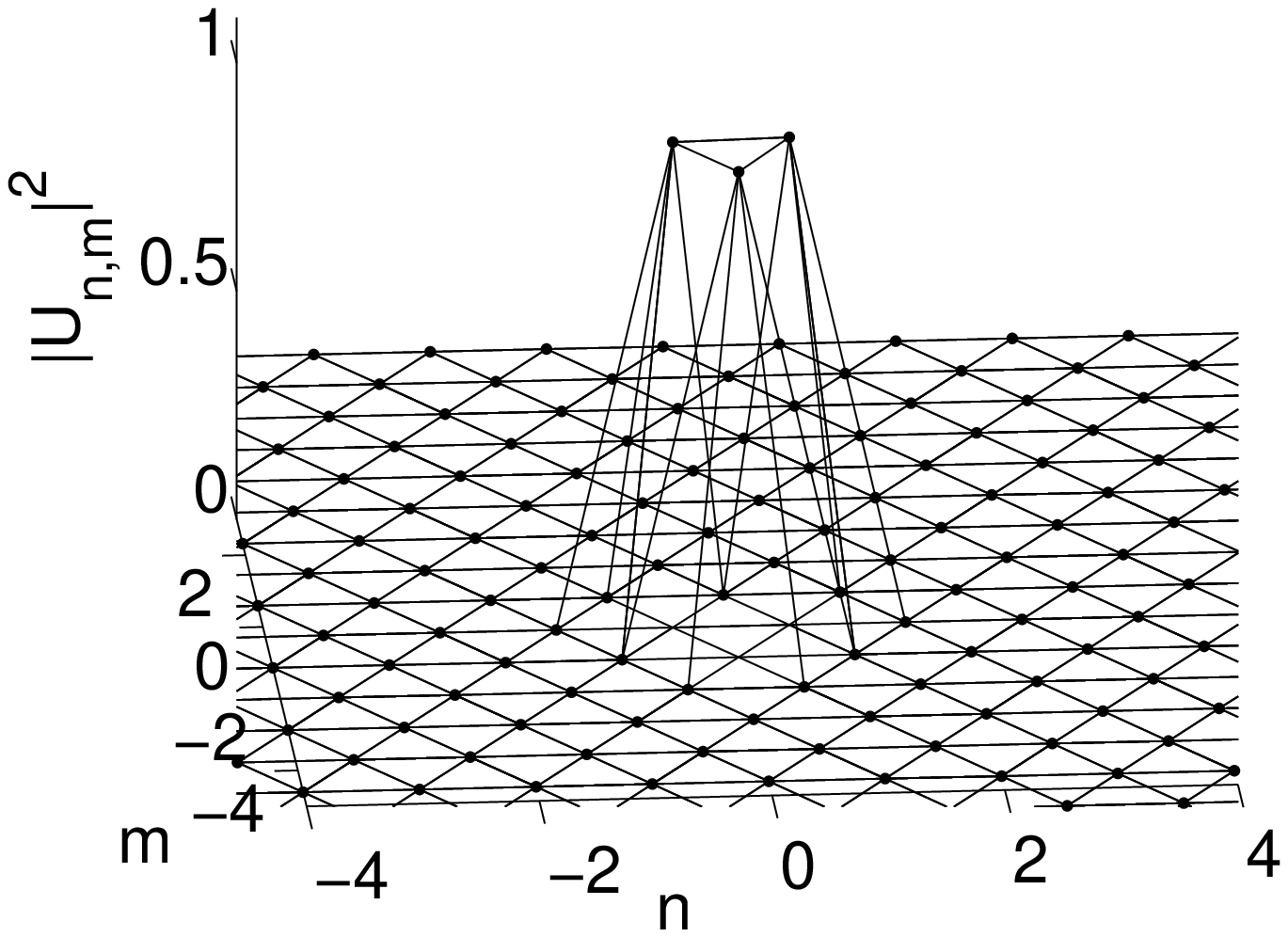}   
\includegraphics[width=5cm,height=4cm]{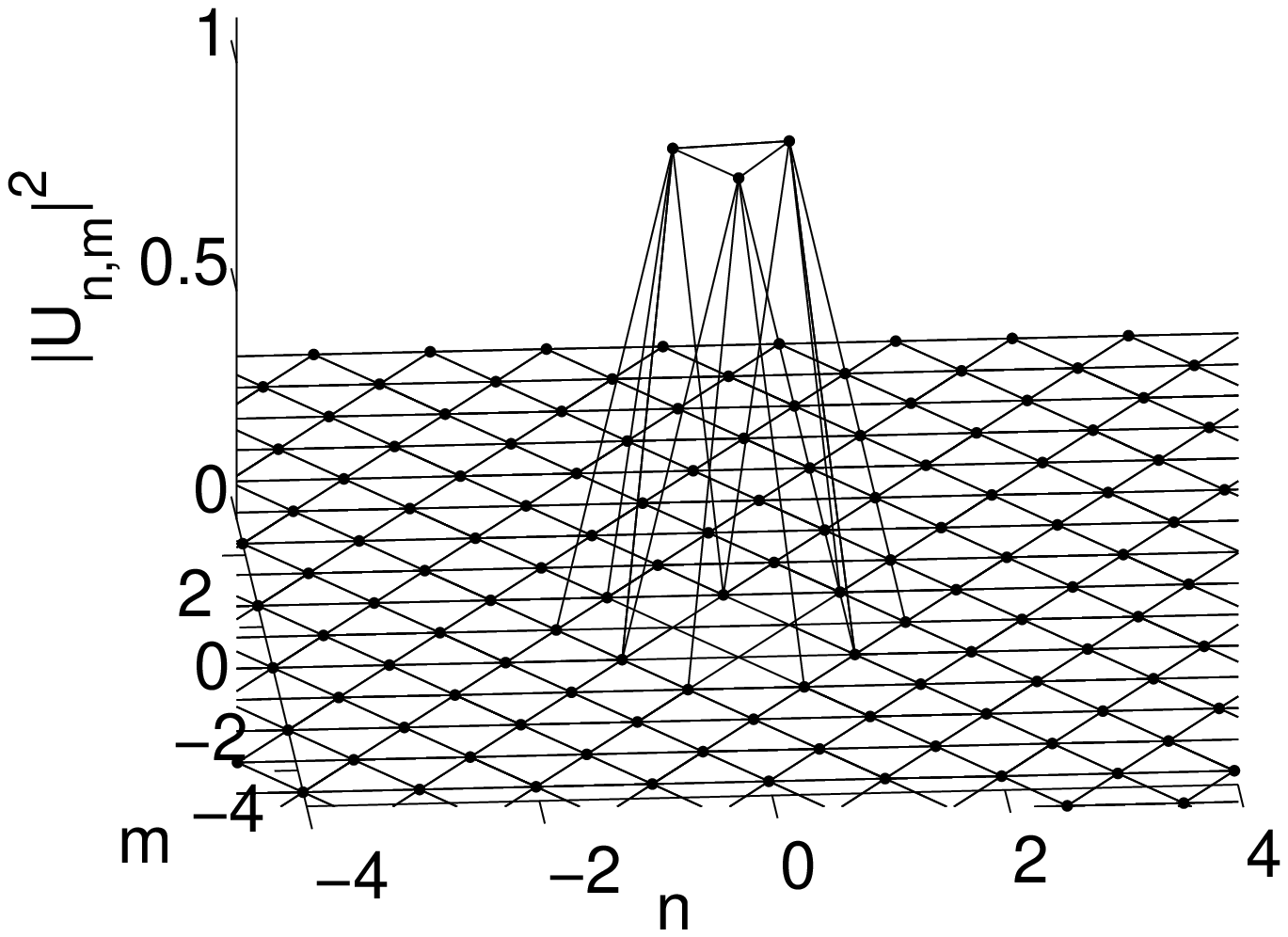}\\
\includegraphics[width=5cm,height=4cm]{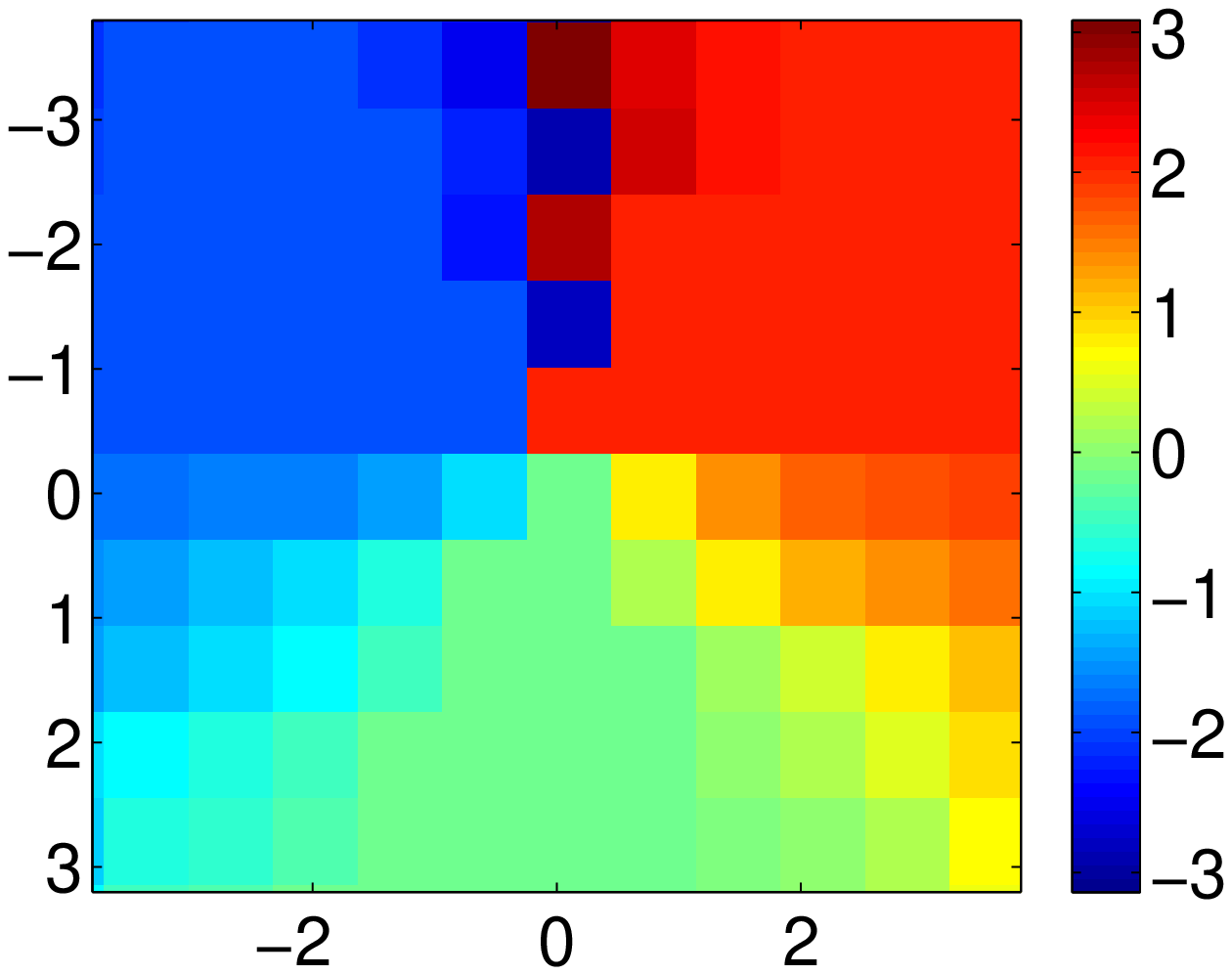}
\includegraphics[width=5cm,height=4cm]{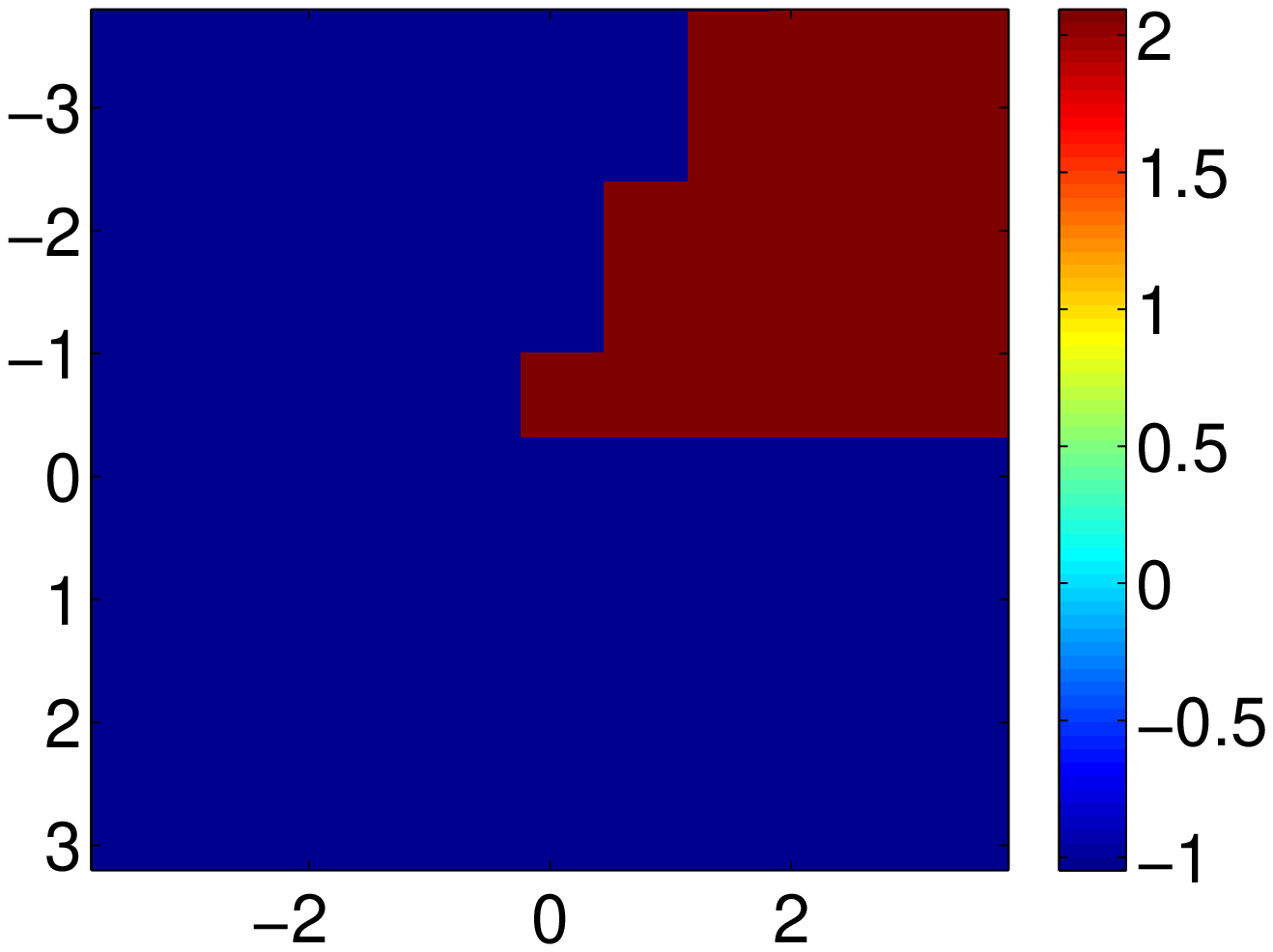}\\
\includegraphics[width=5cm,height=4cm]{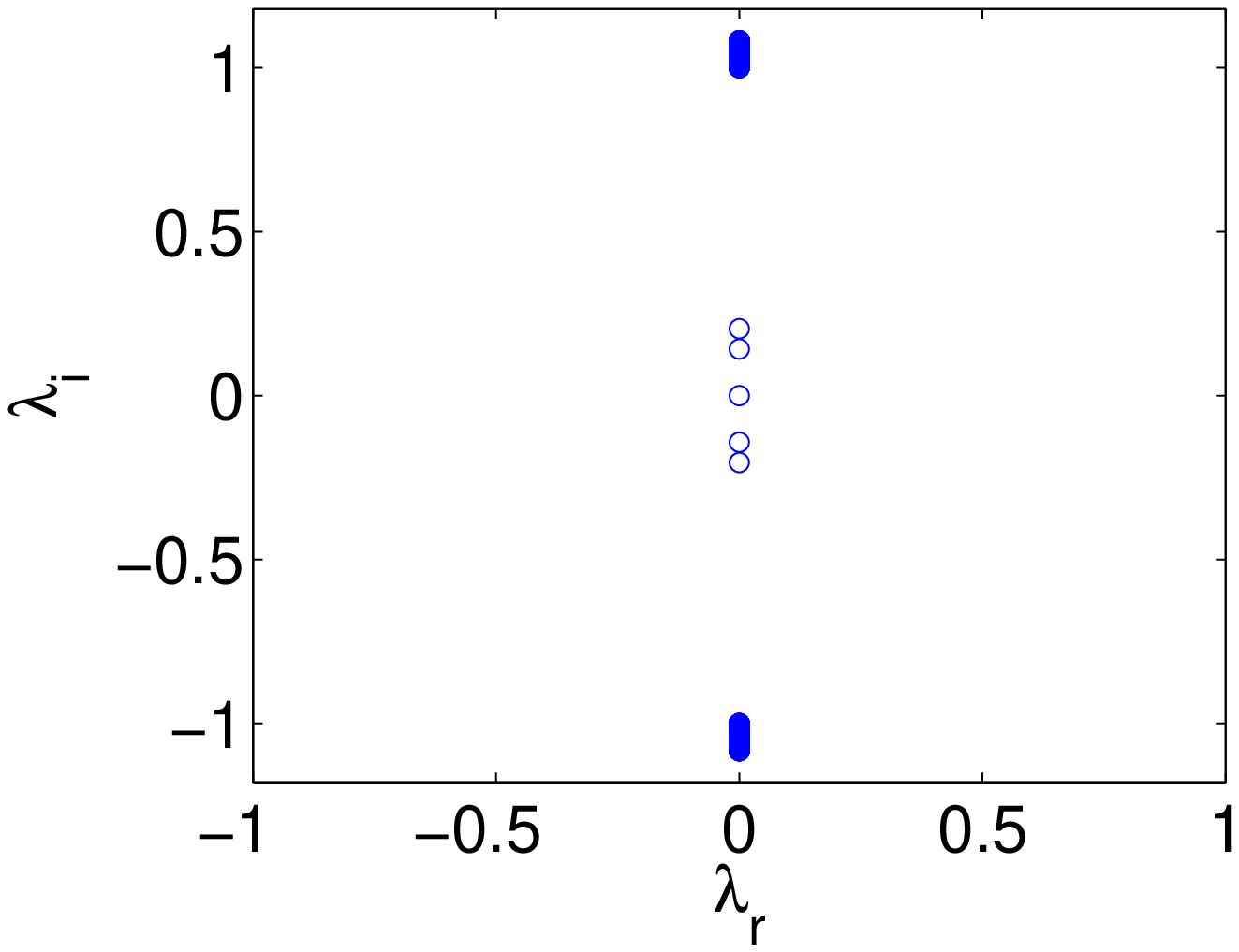}
\includegraphics[width=5cm,height=4cm]{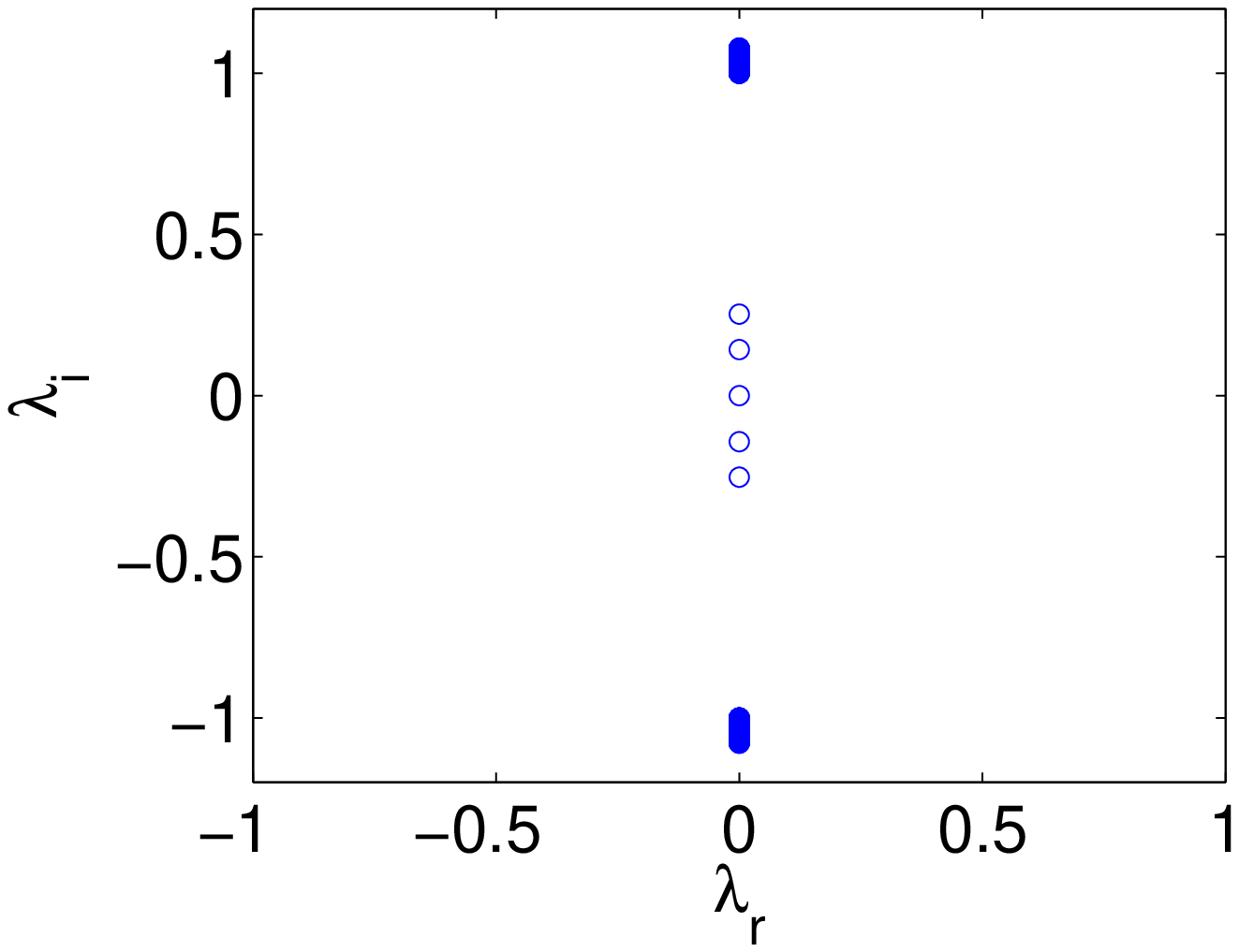}\\
\includegraphics[width=5cm,height=4cm]{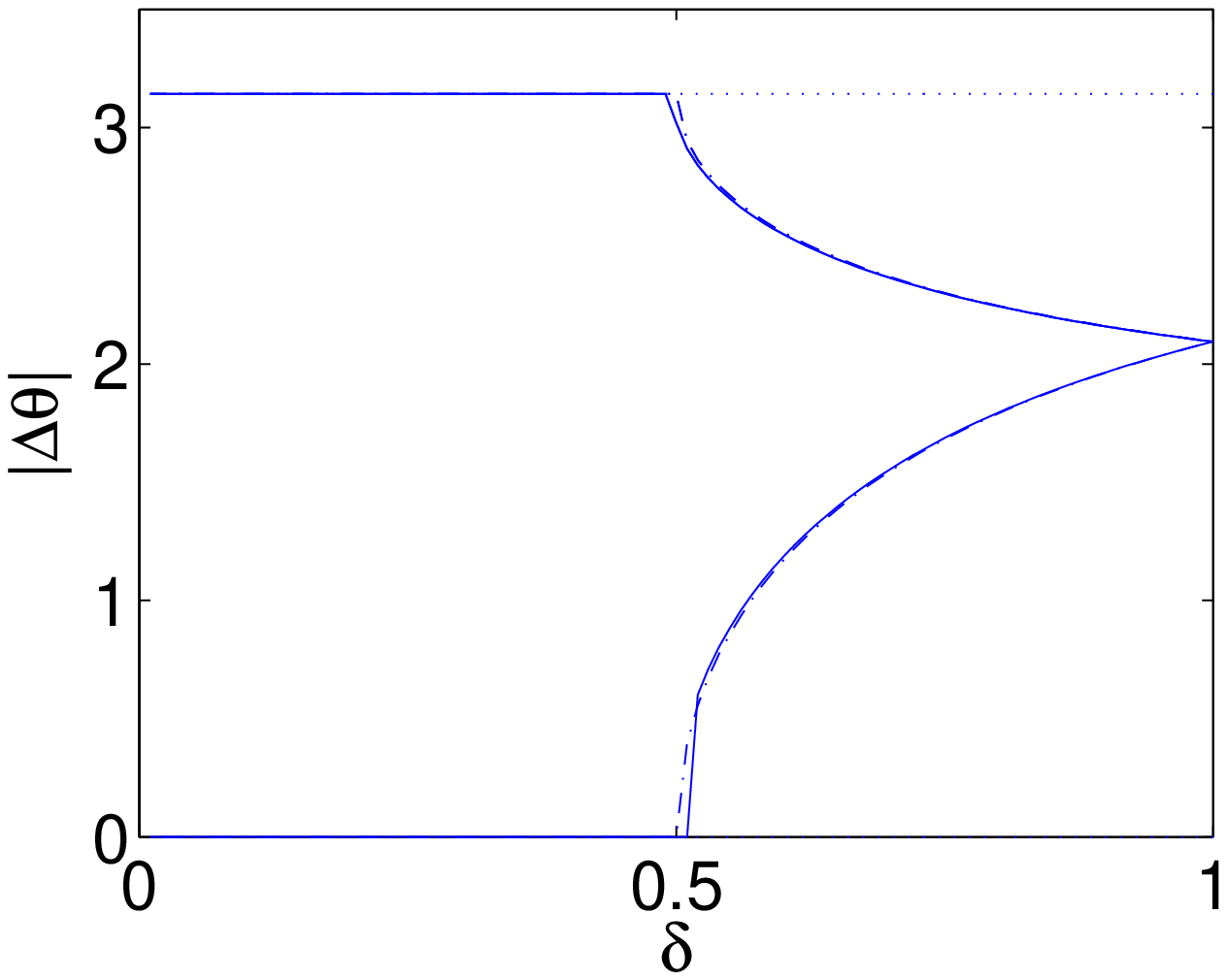}
\includegraphics[width=5cm,height=4cm]{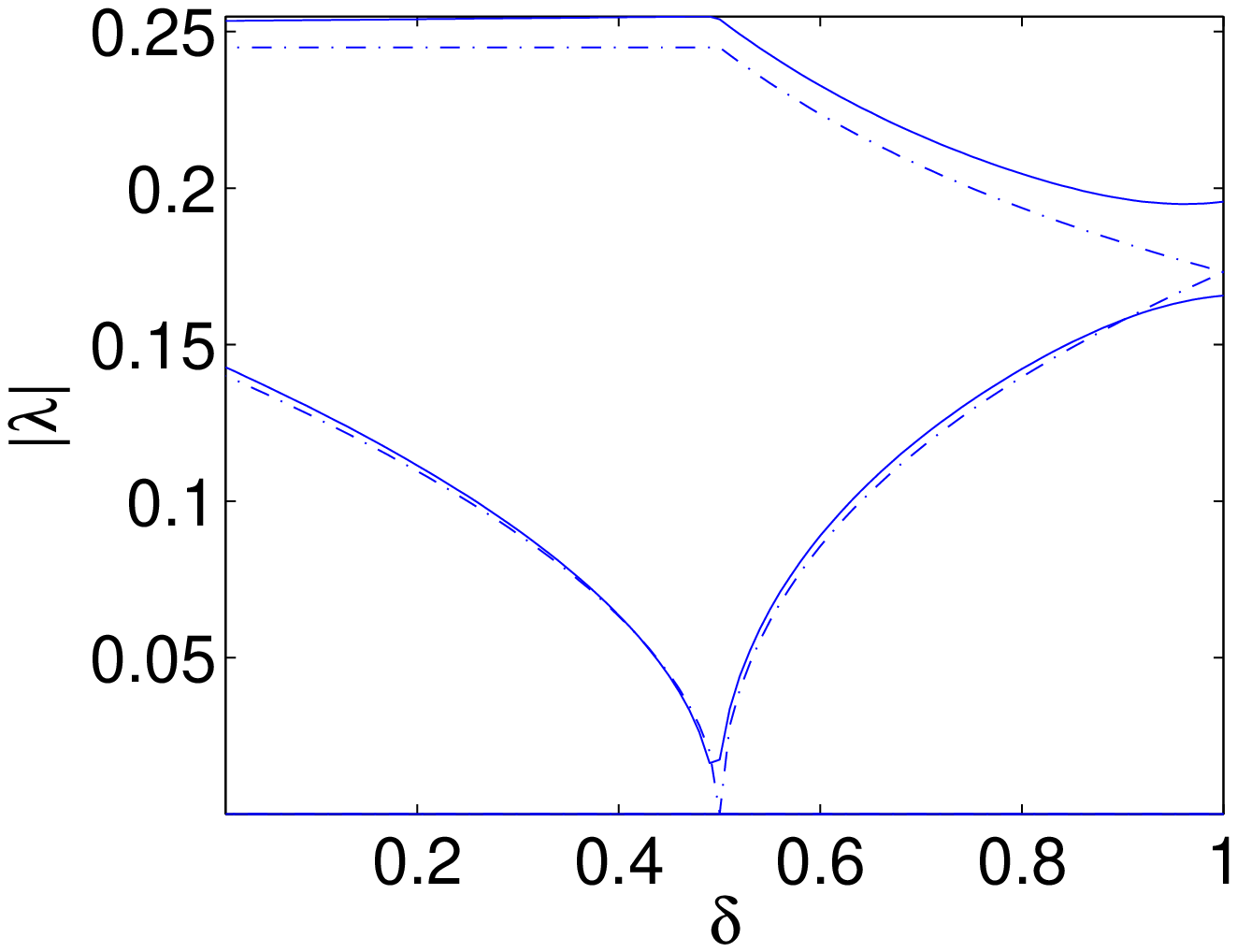}
\vspace{-0.4cm}
\caption{Hexagonal three-site charge $1$ vortex i.e. the hexagonal 
$[0, 2\pi/3, 4\pi/3]$ configuration (at $\varepsilon = 0.01$). The 
anisotropy here is activated between the sites with (initial) phases $0$ 
and $4\pi/3$ when $\delta=1$. The top row displays the modulus 
squared of the configuration corresponding to anisotropic parameter 
$\delta=0.8$ (left panel) and $\delta=0$ (right panel). The second 
row shows the phase portraits and the third row shows the spectral 
plane for the same values of the anisotropy, $\delta=0.8$ (left), 
and $\delta=0$ (right). In the fourth row, the left panel shows the 
comparison between the theoretical (dash-dot lines) and numerical 
(solid lines) changes in the relative phases. The charge 1 vortex 
collides at $\delta=0.5$ with the stable 
(for lower values of $\delta$)
hexagonal $[-\pi/3, 2\pi/3, -\pi/3]$ 
configuration. When $\delta=0$ this is equivalent to the 
$[-\pi/3, 2\pi/3, -\pi/3]$ configuration, i.e., effectively
a $[0, \pi, 0]$ configuration. The fourth row right panel 
shows a theoretical (dash-dot lines) versus numerical (solid lines) 
comparison of the stability eigenvalues for $0 \leq \delta \leq 1$.}
\label{fig:hex3site_C1}
\end{figure}

(1) We start with the hexagonal lattice and begin by presenting 
the results of the continuation for the three site charge $1$ vortex with 
coupling $\varepsilon = 0.01$ (see Fig.~\ref{fig:hex3site_C1}). 
Initially, at $\delta = 1$, the configuration has phases 
$\theta_1 = 0, \theta_2 = 2\pi/3, \theta_3 = 4\pi/3$,
with the anisotropy to be activated between the sites with 
phases $\theta_1$ and $\theta_3$. The top row of
Fig.~\ref{fig:hex3site_C1} shows the square modulus of the field
($|u_{n,m}|^2$) for the
vortex at $\delta = 0.8$ (left panel) and at $\delta = 0$ 
(right panel), while the second and third rows show, 
respectively, the corresponding phase portraits and spectral 
planes. The left panel of the fourth row shows the change 
in the relative phases ($|\Delta \theta|$) and the right 
panel traces the linear stability eigenvalues with the change 
in anisotropy. In both images the theory (dash-dot lines) 
compares extremely well with the numerical (solid lines) 
results, over the entire interval of continuation of the
anisotropy parameter $\delta$.
In the case of $\delta \rightarrow 0$ it is evident that
the configuration has changed its character into
a $[0, \pi, 0]$ configuration, as theoretically predicted.
The relative phase and eigenvalue predictions of the bottom row
indeed confirm
that the pitchfork 
bifurcation takes place at $\delta=0.5$ 
and leads to a collision with the configuration $[-\pi/3, 2\pi/3, -\pi/3]$ 
at $\delta=0.5$ (which as theoretically predicted has the 
form $[\theta_1, \theta_1+\pi, \theta_1]$, i.e., up to a
trivial phase is a $[0, \pi, 0]$ configuration). 
We note in passing that we have also examined
the relevant continuation and bifurcation for other values
of $\varepsilon$ (such as $\varepsilon=0.05$), finding
similar qualitative results (although quantitative details, such
as the critical value of $\delta$ do change).



\begin{figure}[tbh]
\centering
\includegraphics[width=5cm,height=4cm]{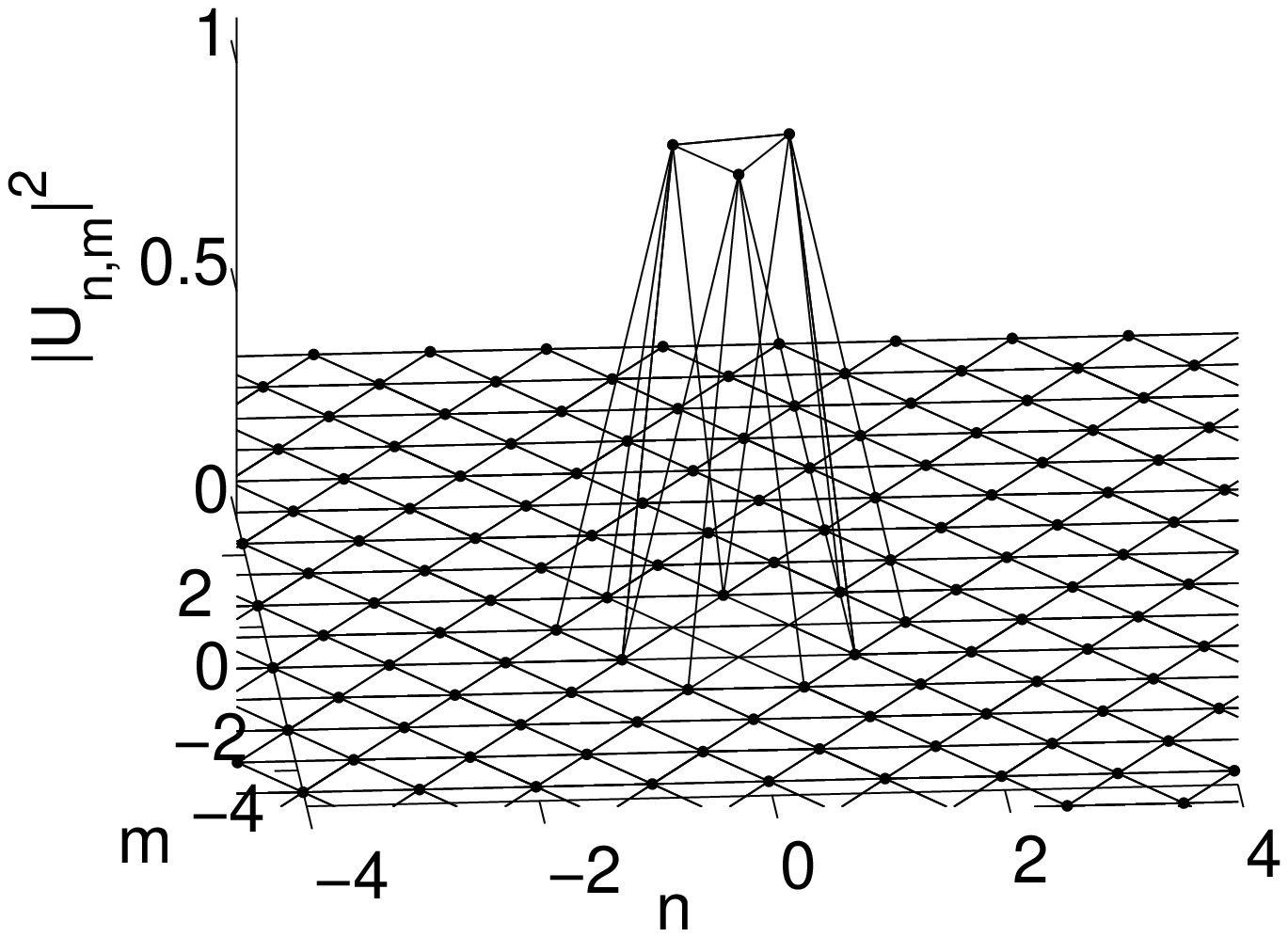}
\includegraphics[width=5cm,height=4cm]{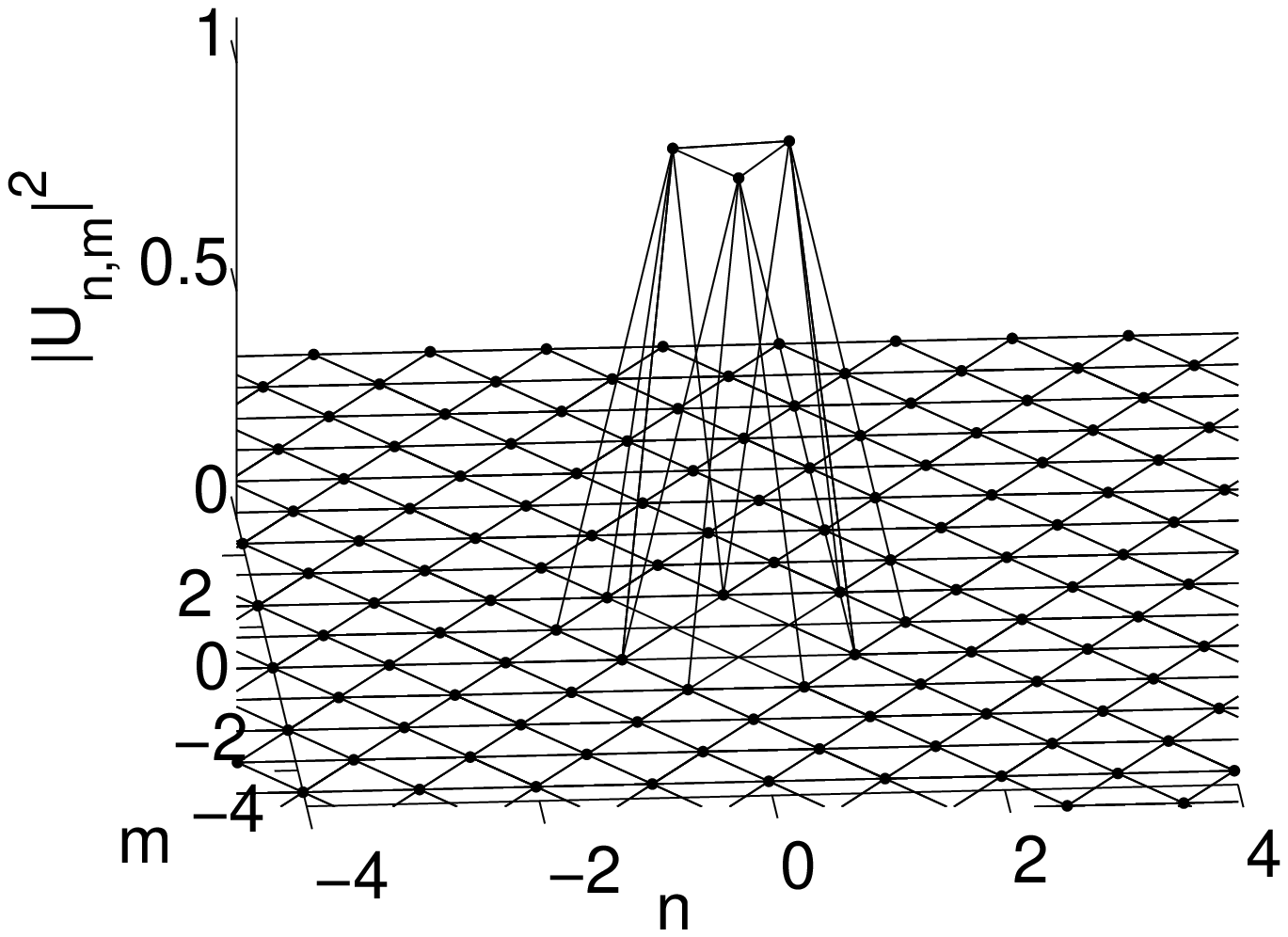}\\
\includegraphics[width=5cm,height=4cm]{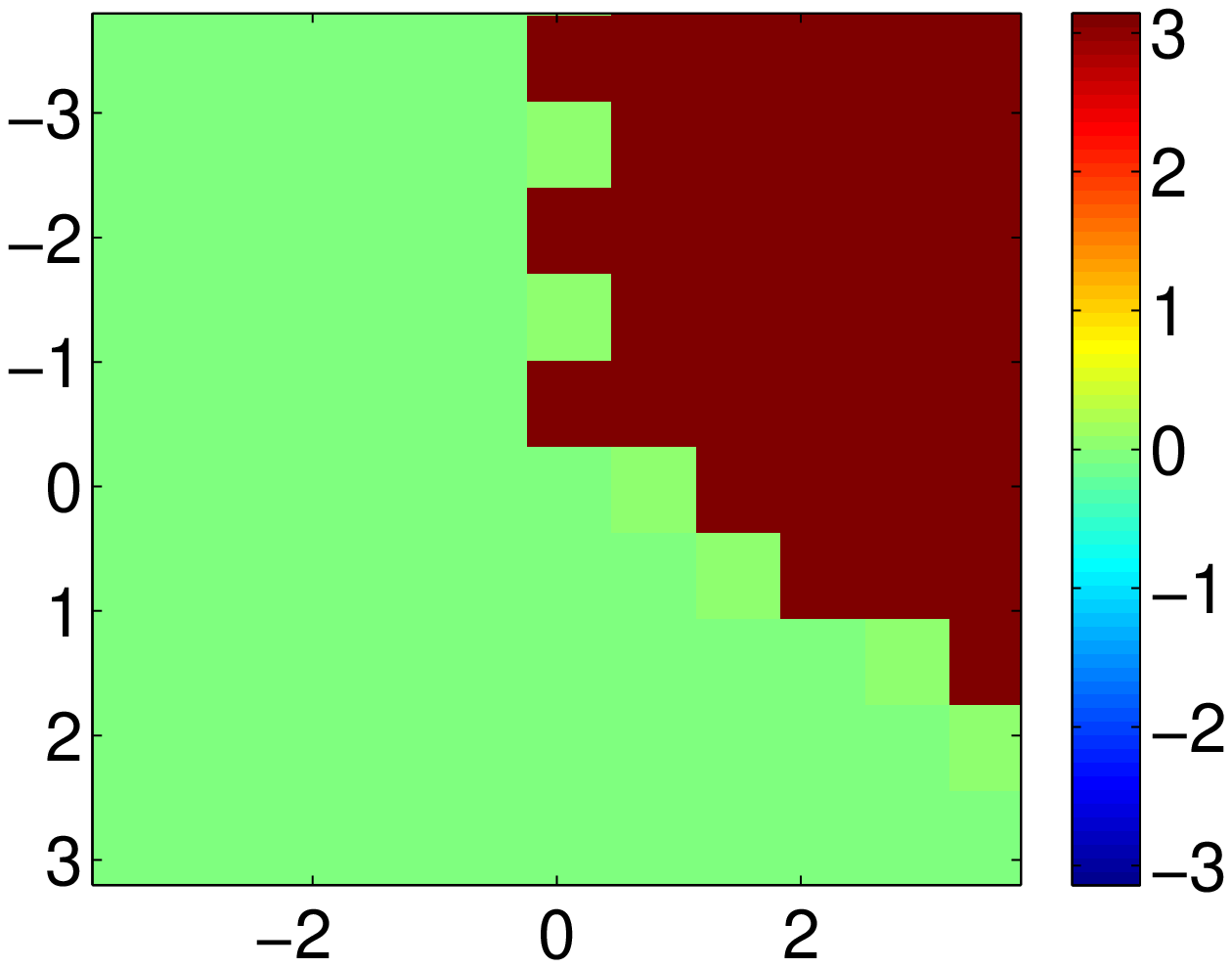}
\includegraphics[width=5cm,height=4cm]{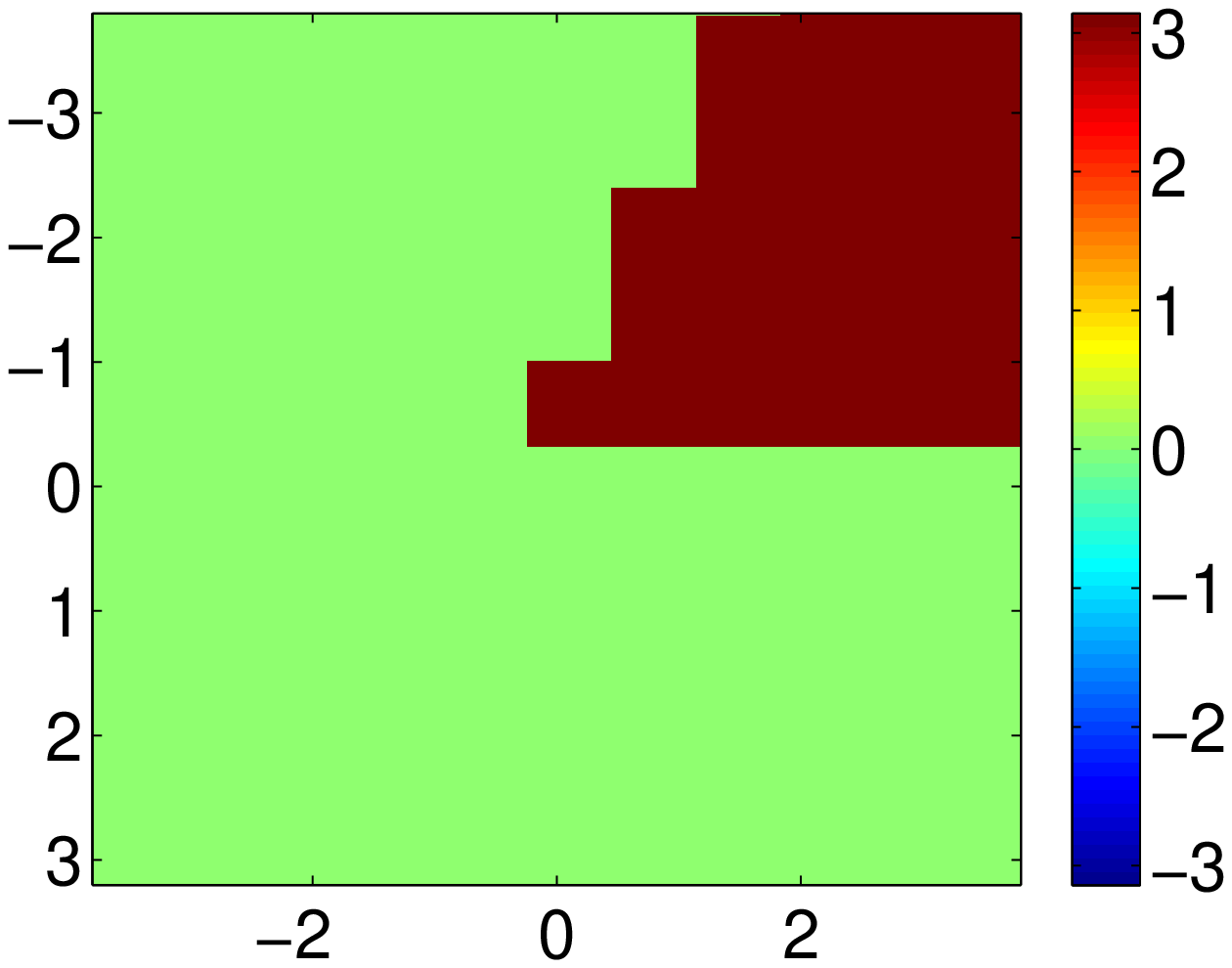}\\
\includegraphics[width=5cm,height=4cm]{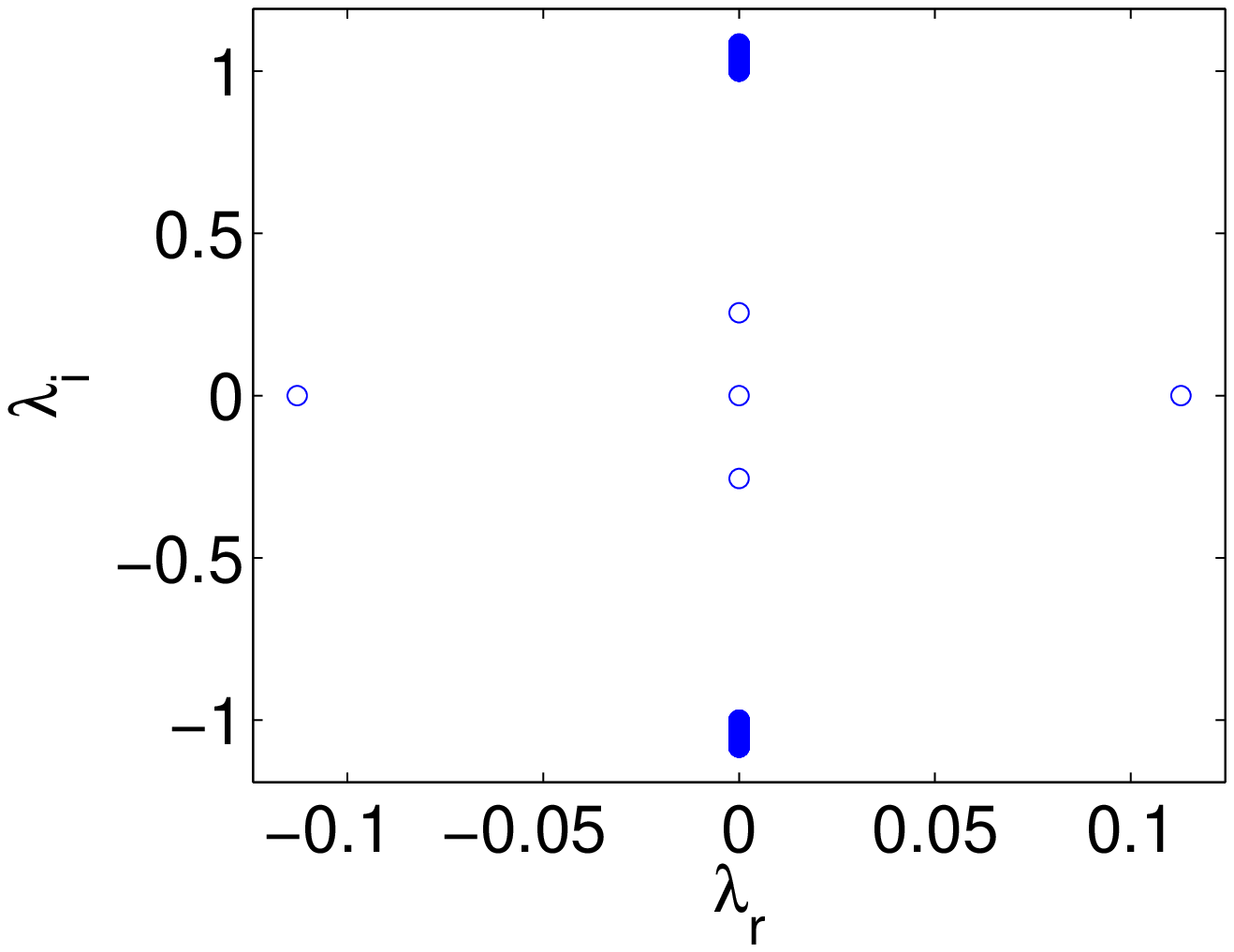}
\includegraphics[width=5cm,height=4cm]{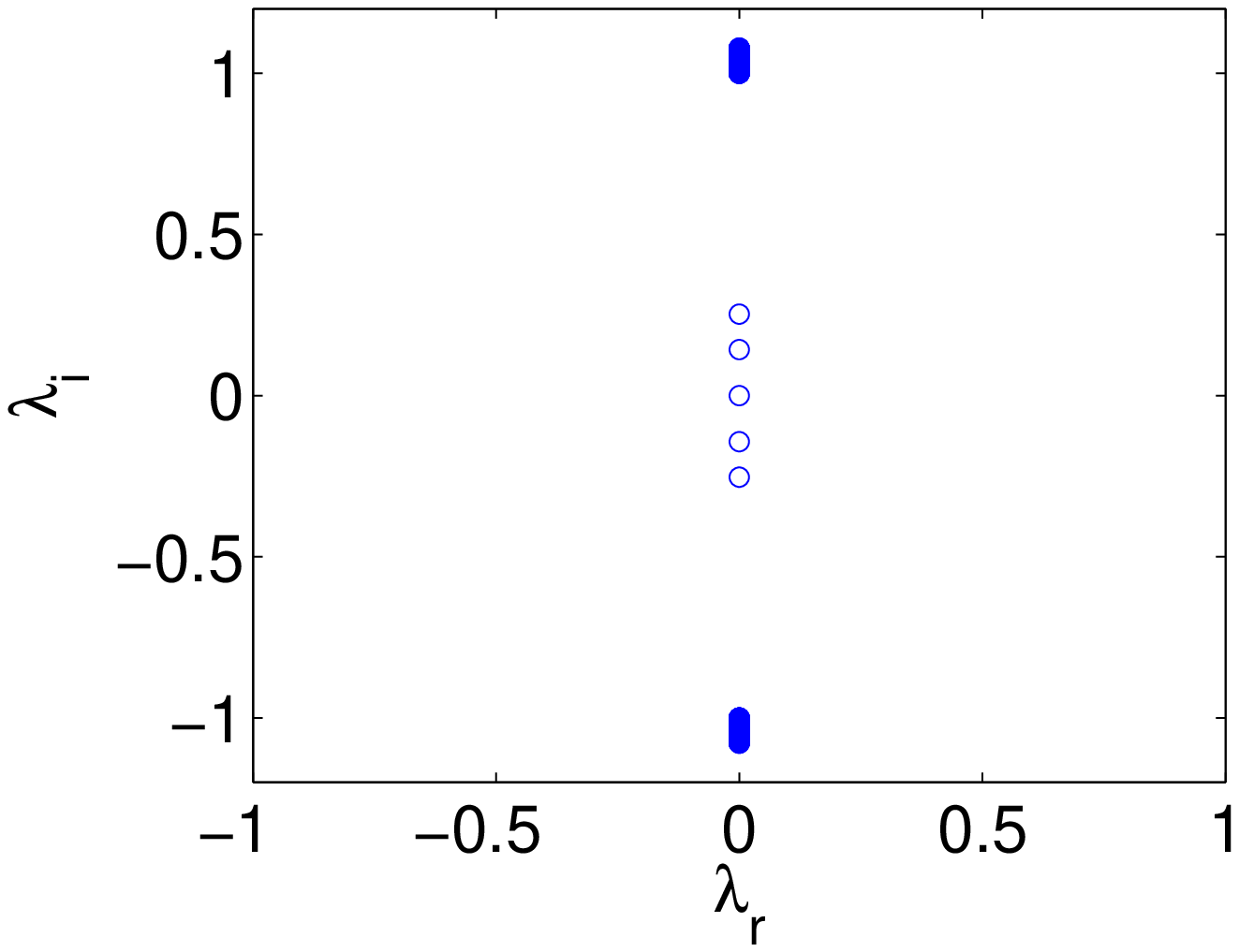}\\
\vspace{-0.4cm}
\caption{Hexagonal three-site $[0, \pi, 0]$ configuration 
($\varepsilon = 0.01$).
The top row displays the modulus squared of the field 
at $\delta=0.8$ (left column) 
and $\delta=0$ (right column). The second row shows the corresponding 
phase portraits while the third row displays the respective spectral planes.
In this case, the anisotropy is invoked between the two nodes with phase $0$.
Hence, as discussed in the text a stabilization is observed for
$\delta < 0.5$.} 
\label{fig:hex3site1_0_pi_0}
\end{figure}

\begin{figure}[tbh]
\centering
\includegraphics[width=5cm,height=4cm]{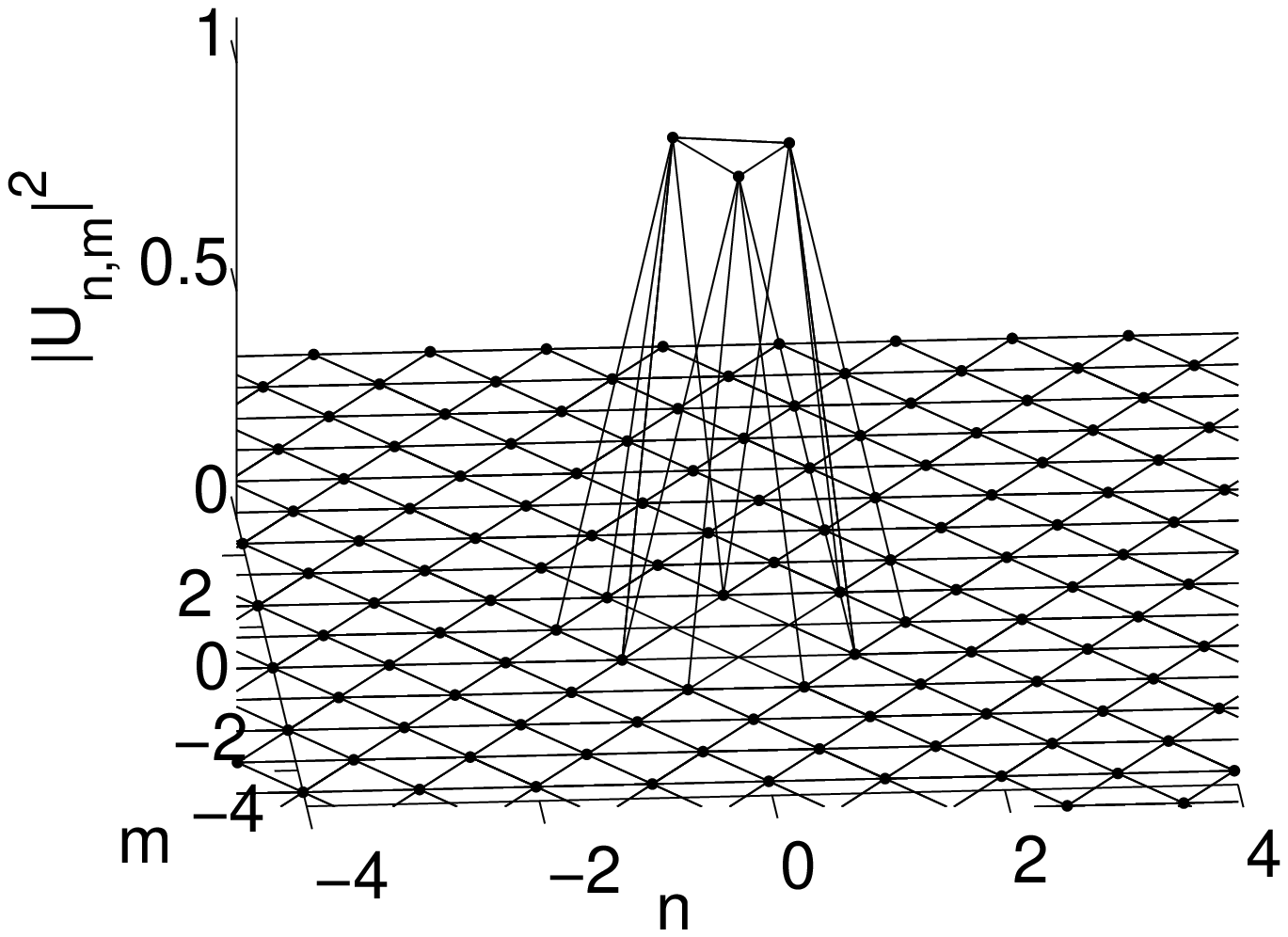}
\includegraphics[width=5cm,height=4cm]{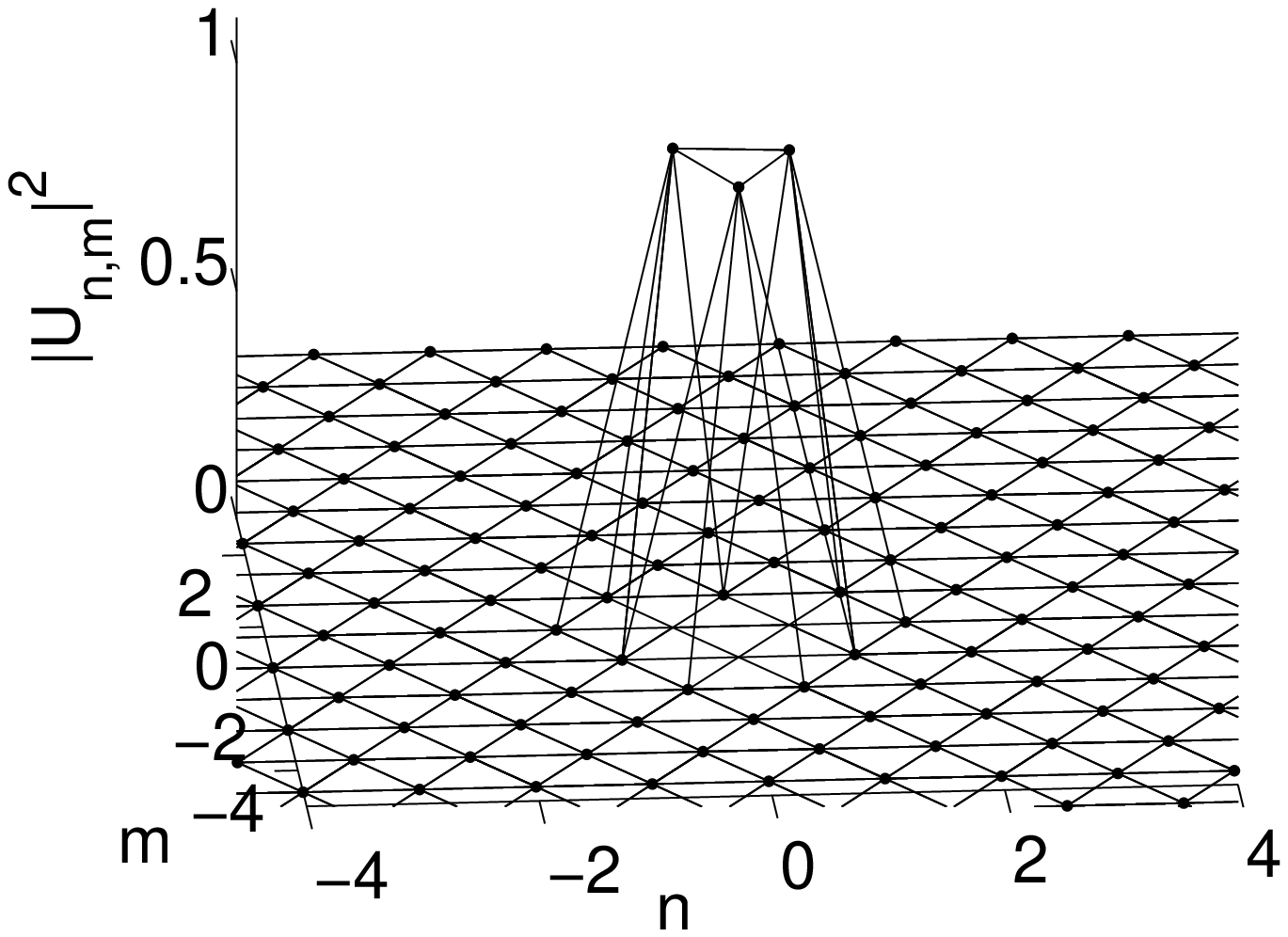}\\
\includegraphics[width=5cm,height=4cm]{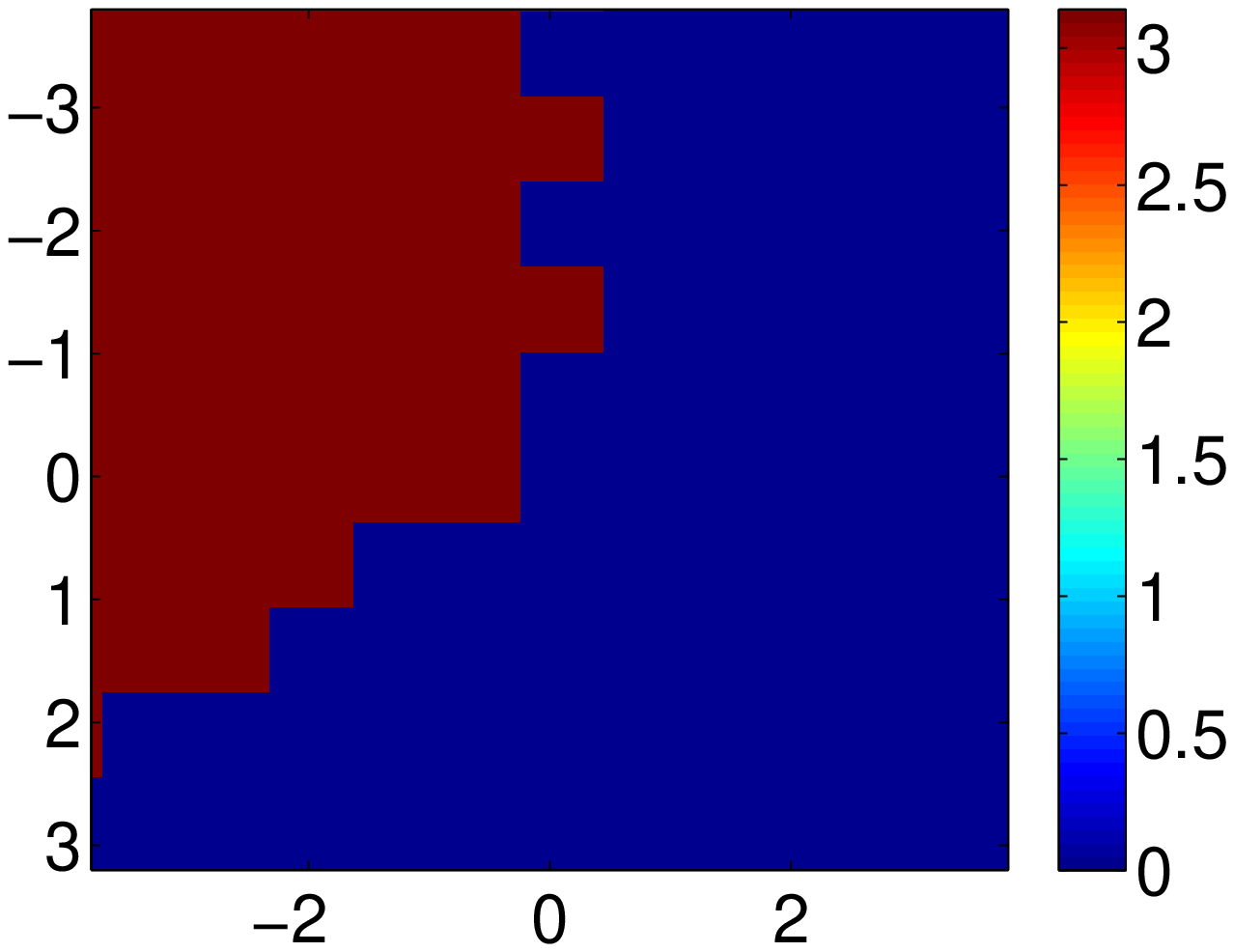}
\includegraphics[width=5cm,height=4cm]{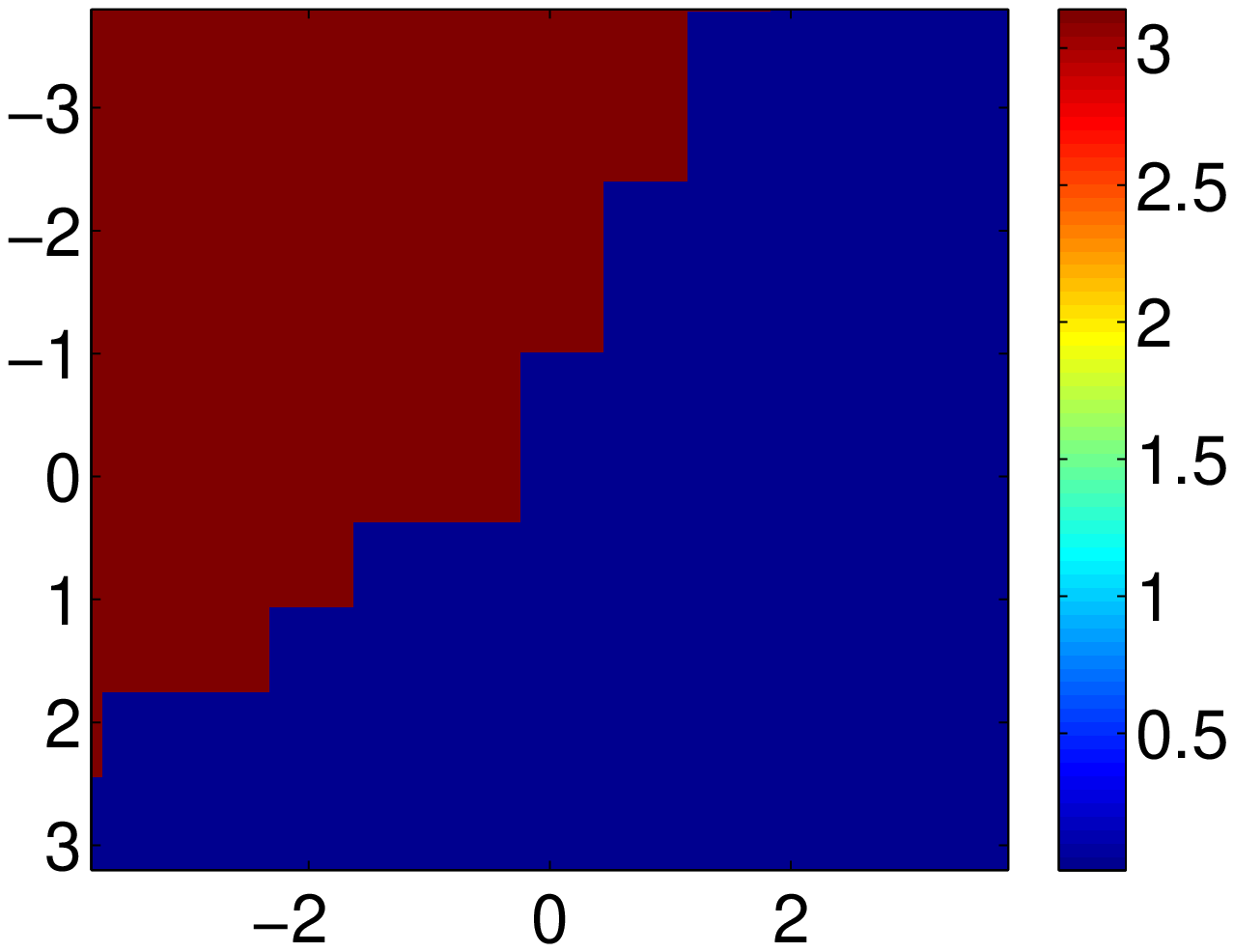}\\
\includegraphics[width=5cm,height=4cm]{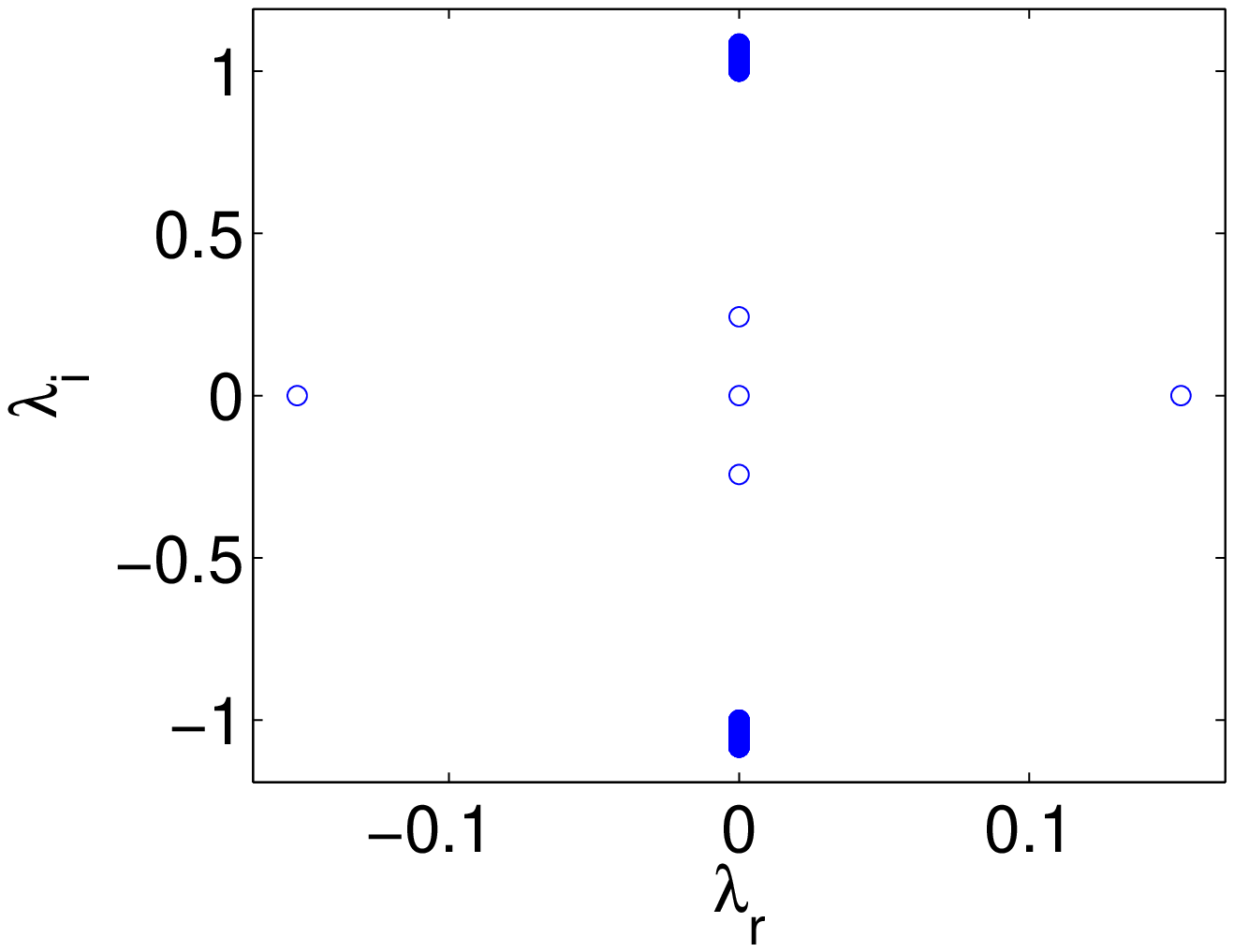}
\includegraphics[width=5cm,height=4cm]{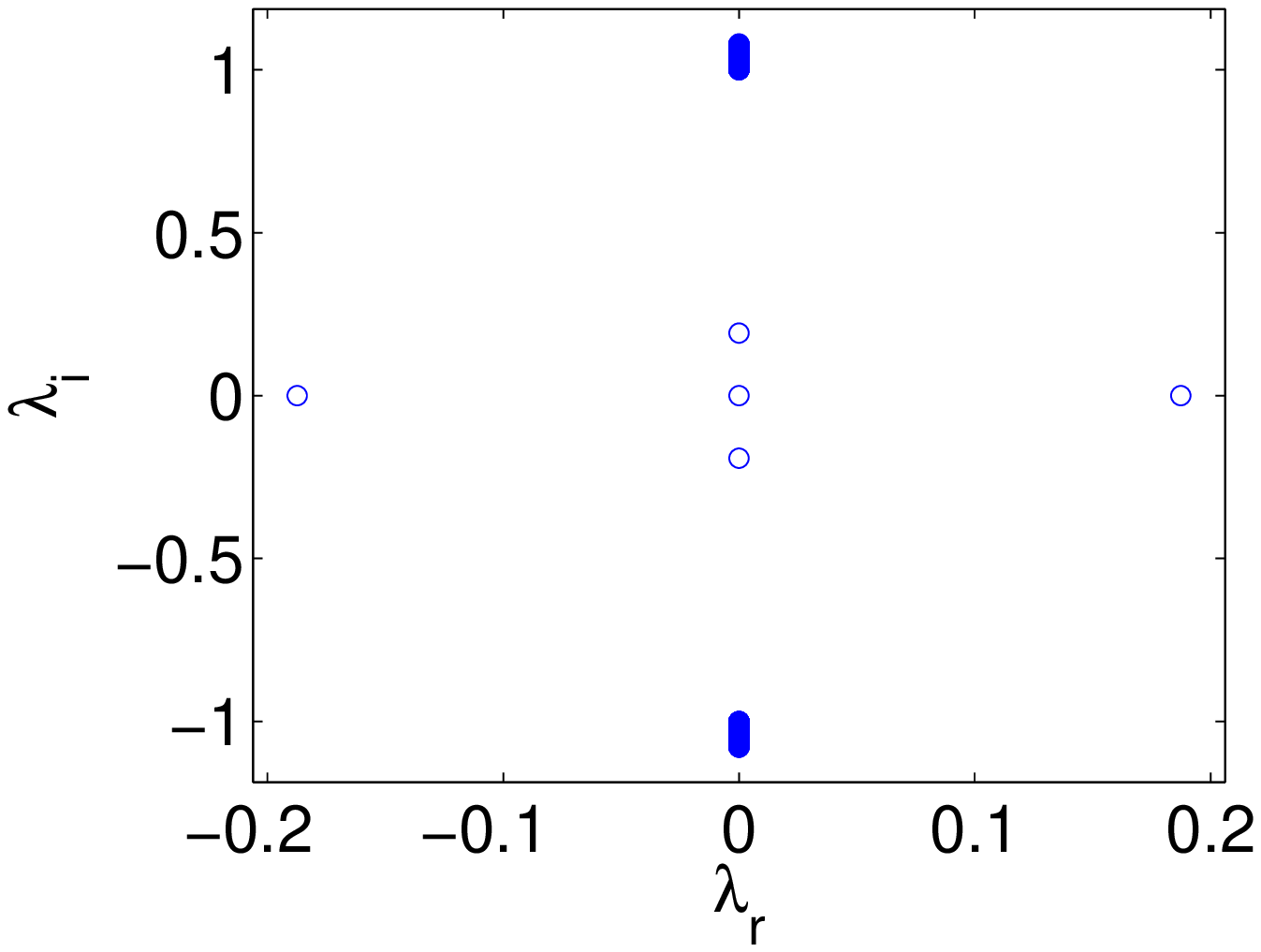}\\
\vspace{-0.4cm}
\caption{Hexagonal three-site $[0, \pi, 0]$ configuration ($\varepsilon = 0.01$). 
The top row displays the modulus squared of the field
with $\delta=0.8$ (left column) 
and $\delta=0$ (right column). The anisotropy is invoked between the node 
with phase $\pi$ and one of the nodes with phase $0$. At $\delta=0$ 
the result is equivalent to an effective $[0, 0, \pi]$ configuration,
along a line. Hence, in this case the instability is preserved for
all values of $\delta$.}
\label{fig:hex3site2_0_pi_0}
\end{figure}

(2) The hexagonal $[0, \pi, 0]$ configuration can be seen in 
Figs.~\ref{fig:hex3site1_0_pi_0} and \ref{fig:hex3site2_0_pi_0}. 
In the isotropic lattice of $\varepsilon = 0.01$ this solution 
is unstable. However stability can be achieved if the anisotropy 
is activated between the the two nodes with phase $0$, while any 
other placement maintains the instability. 
This is because if the nodes with the same phase are
connected at the $\delta \rightarrow 0$ limit (where we can
think of the three nodes as effectively being on a straight
line), then, as discussed above, the instability due to a real
eigenvalue pair will be maintained.
In the case where 
a stabilization effect is observed (in Fig.~\ref{fig:hex3site1_0_pi_0}) 
$[0, \pi, 0]$,  as $\delta$ decreases, a weakened 
bond between the $0$ phase nodes results and eventually
an effective $[0, \pi, 0]$ state along a line is effectively obtained which
is well known to be stable, as a one-dimensional
configuration, for small $\varepsilon$~\cite{Pelinovsky}. 
The change of stability occurs 
at $\delta = 0.49$, in excellent agreement with the theoretical
prediction of $\delta=0.5$. On the other hand, an anisotropic
weakening of the bond 
between the site with phase $\pi$ and either of the $0$ phase 
sites does not bring about stability and eventually 
just yields an unstable waveform (along an effective line)  
$[0, 0, \pi]$, which has been demonstrated to be unstable
in 1d settings~\cite{Pelinovsky}. 

\begin{figure}[tbh]
\centering
\includegraphics[width=5cm,height=4cm]{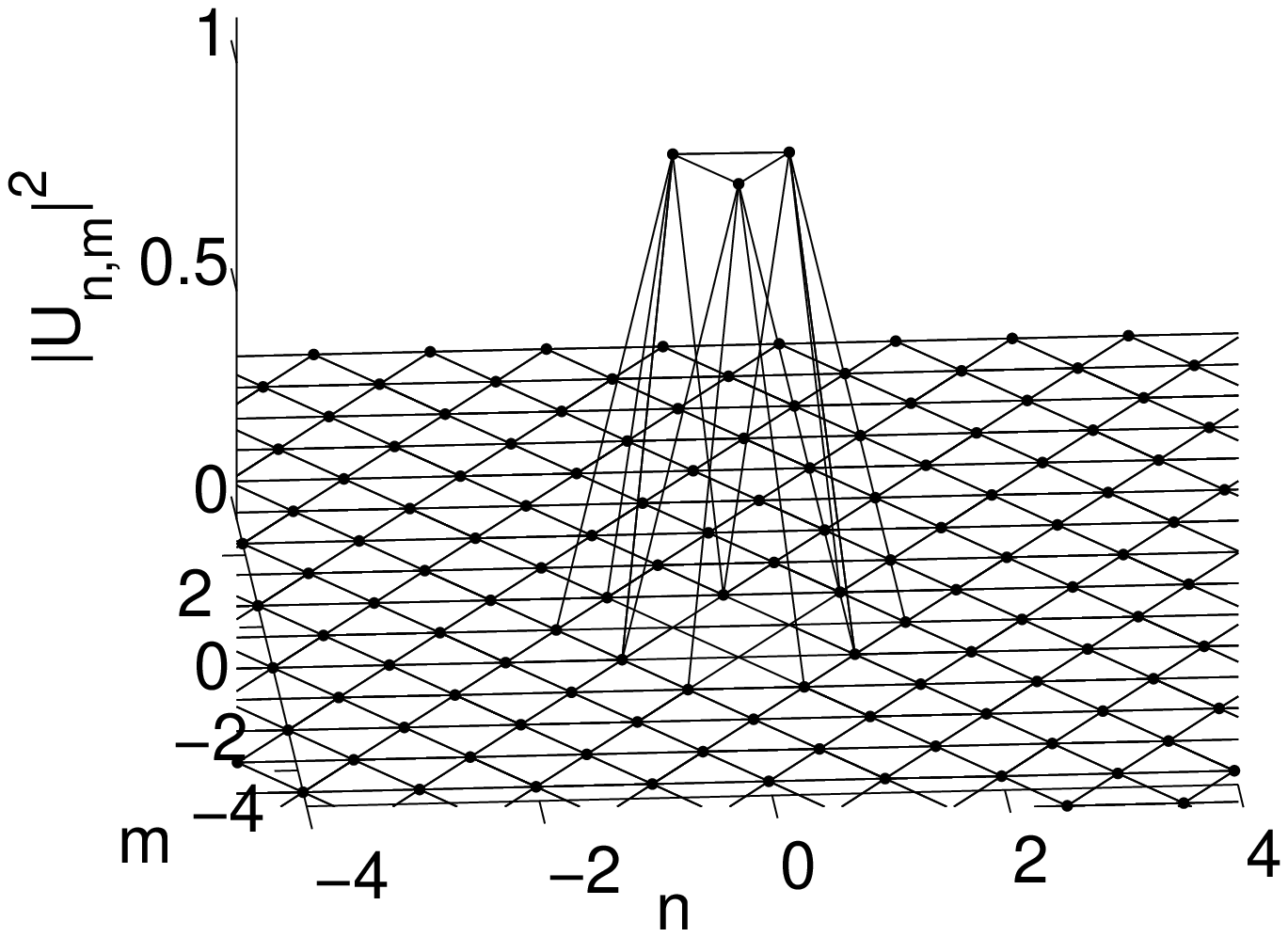}
\includegraphics[width=5cm,height=4cm]{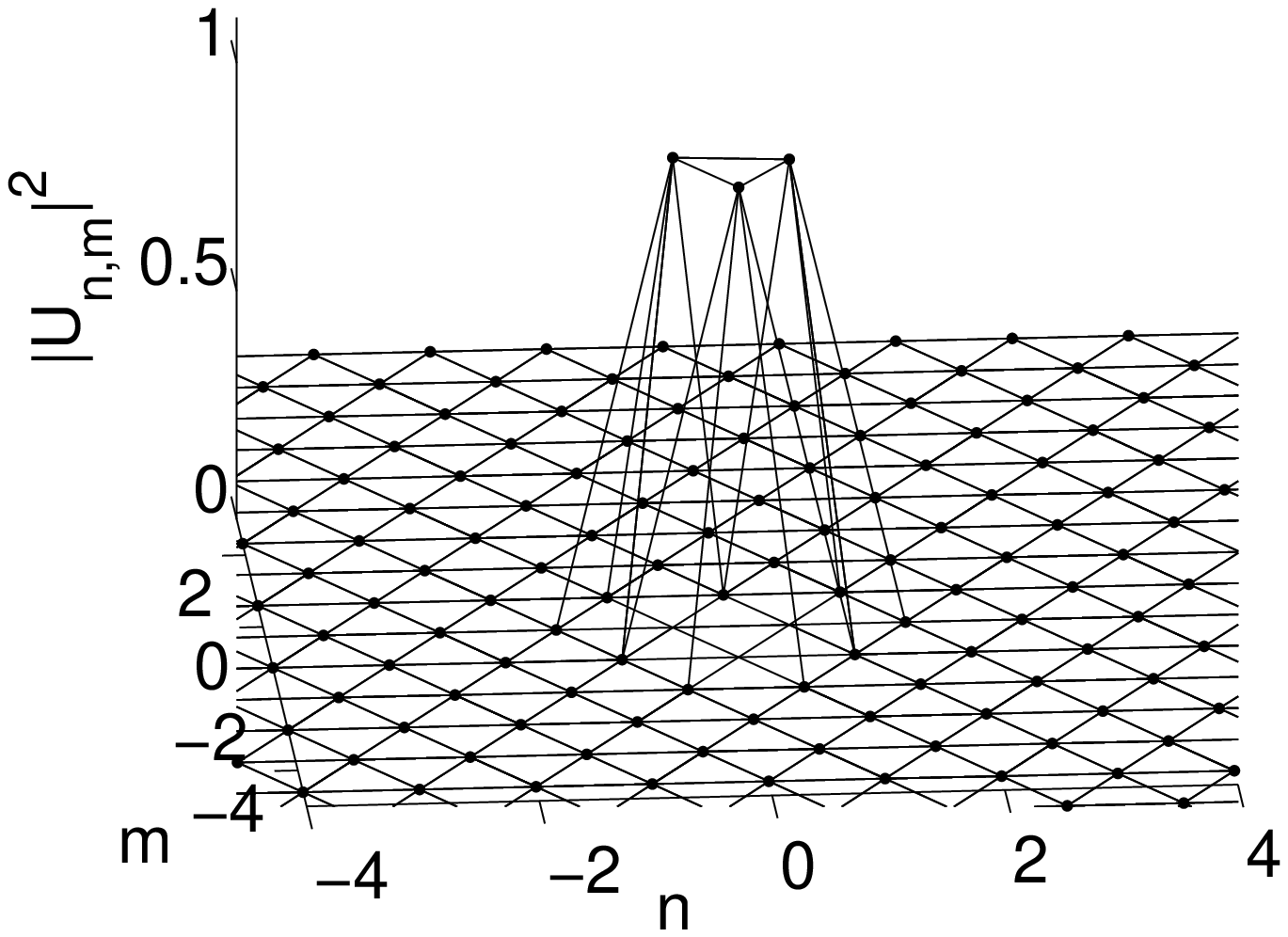}\\
\includegraphics[width=5cm,height=4cm]{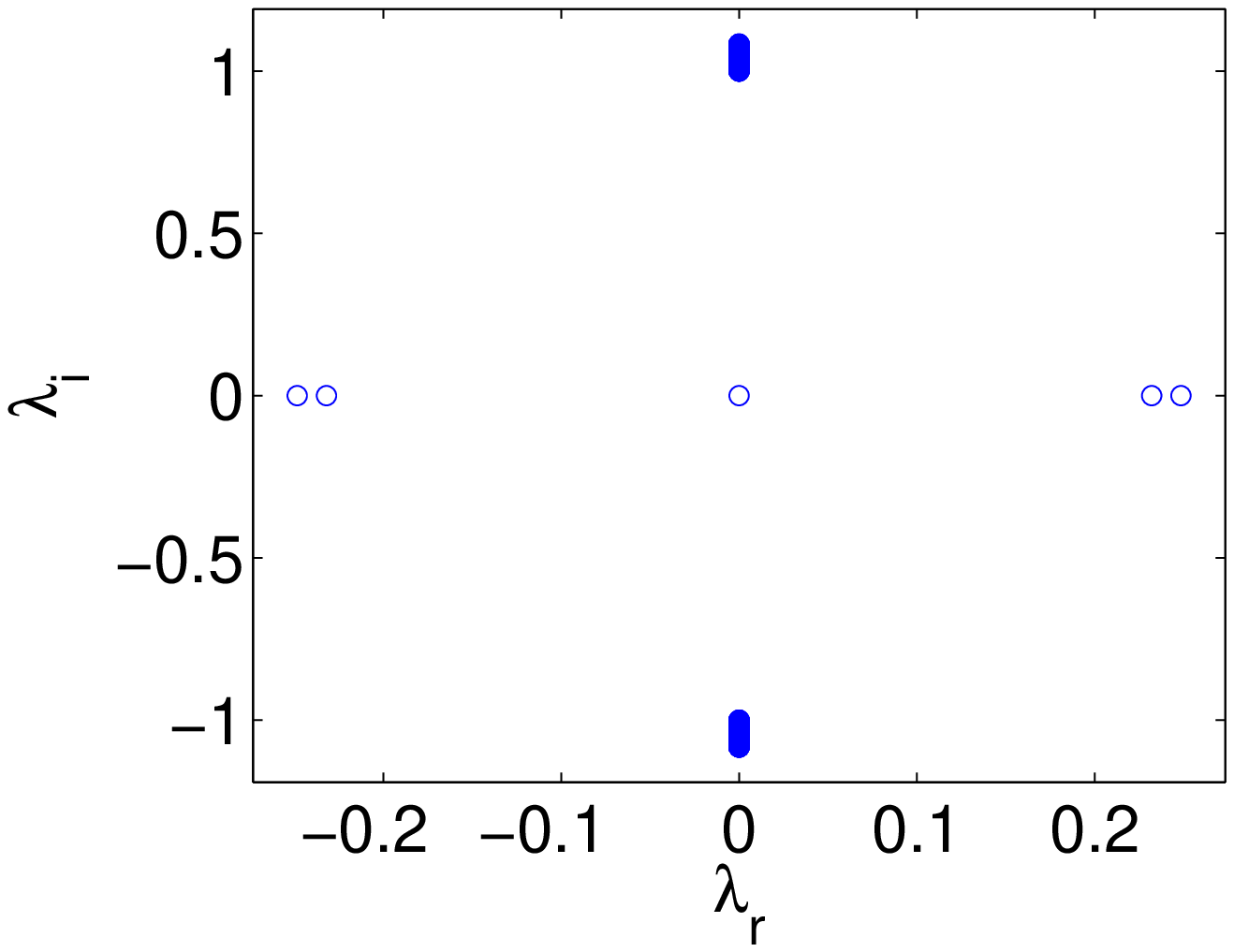}
\includegraphics[width=5cm,height=4cm]{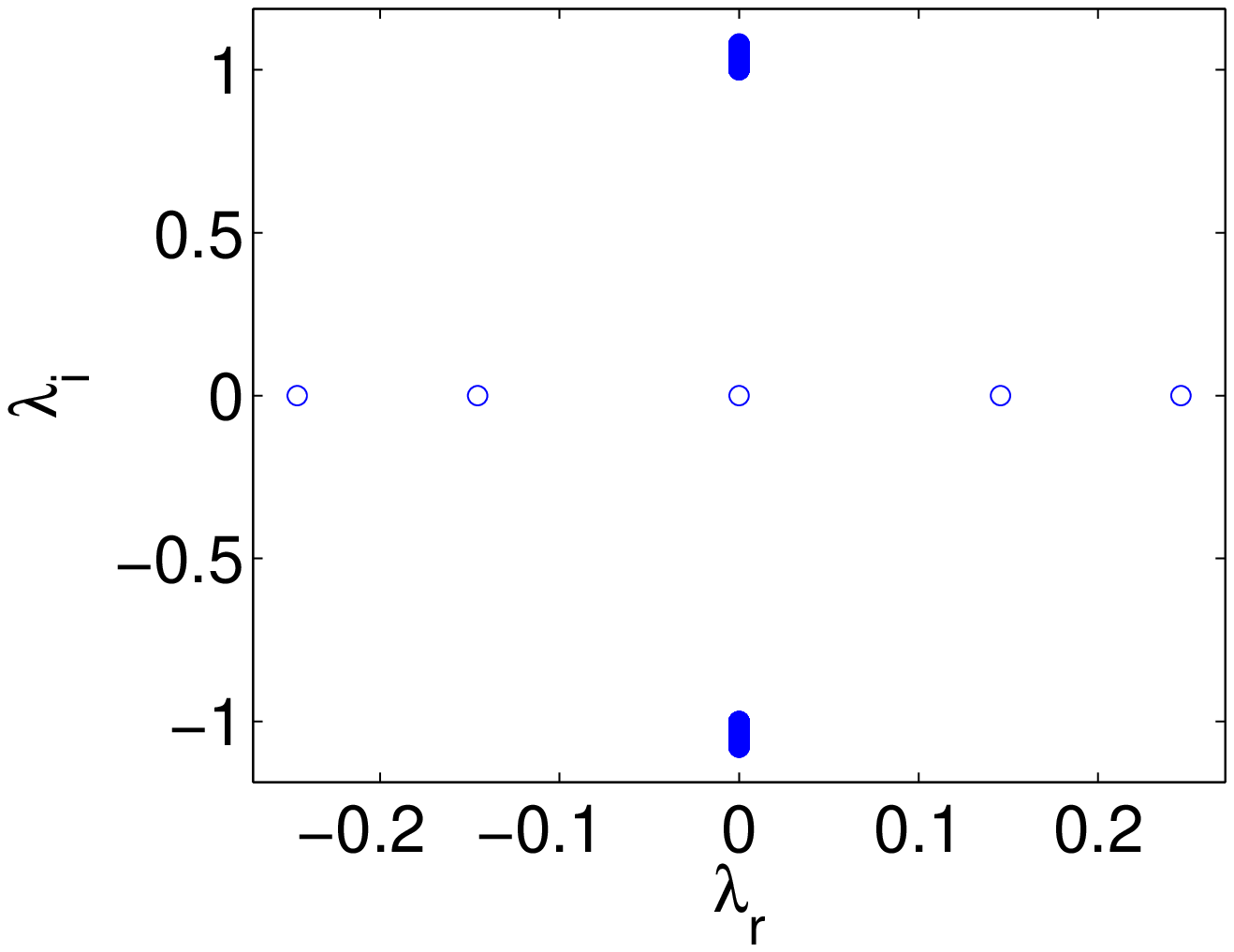}
\vspace{-0.4cm}
\caption{Hexagonal three-site $[0, 0, 0]$ configuration ($\varepsilon = 0.01$). 
The top row displays the modulus squared of the field
with anisotropy $0.8$ (left column) and $0$ (right column). 
The configuration is found to be unstable for all values of $\delta$,
as also predicted theoretically.}
\label{fig:hex3site_0_0_0}
\end{figure}

(3) The hexagonal $[0, 0, 0]$ configuration (Fig.~\ref{fig:hex3site_0_0_0}) 
is, not surprisingly, unstable in the isotropic lattice 
($\varepsilon = 0.01$) due to the adjacent in-phase sites and,
in full accordance with our theoretical predictions,  
 this remains unstable as $\delta \rightarrow 0$.

\subsection{Honeycomb Lattice}

\begin{figure}[tbh]
\centering
\includegraphics[width=5cm,height=4cm]{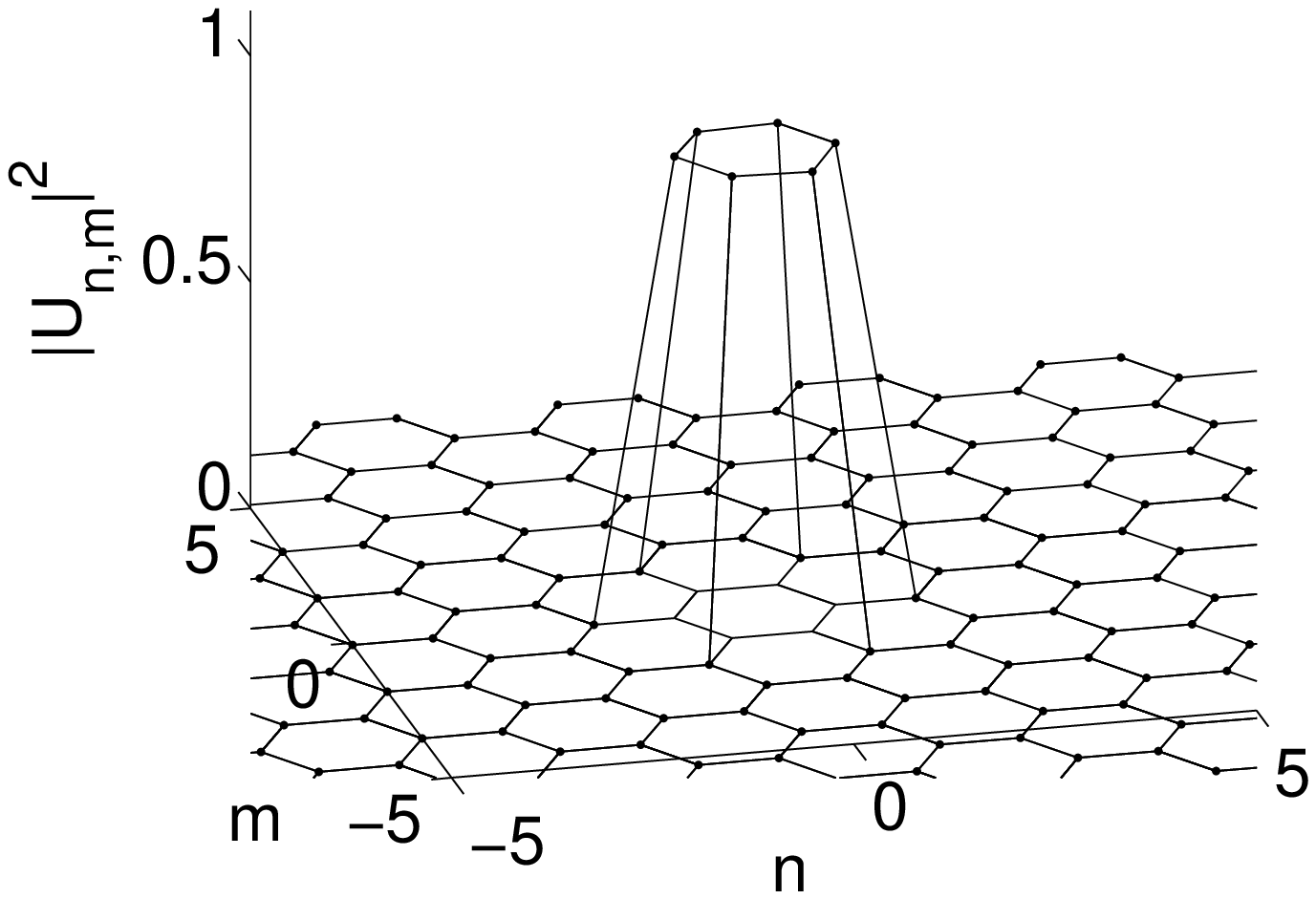}
\includegraphics[width=5cm,height=4cm]{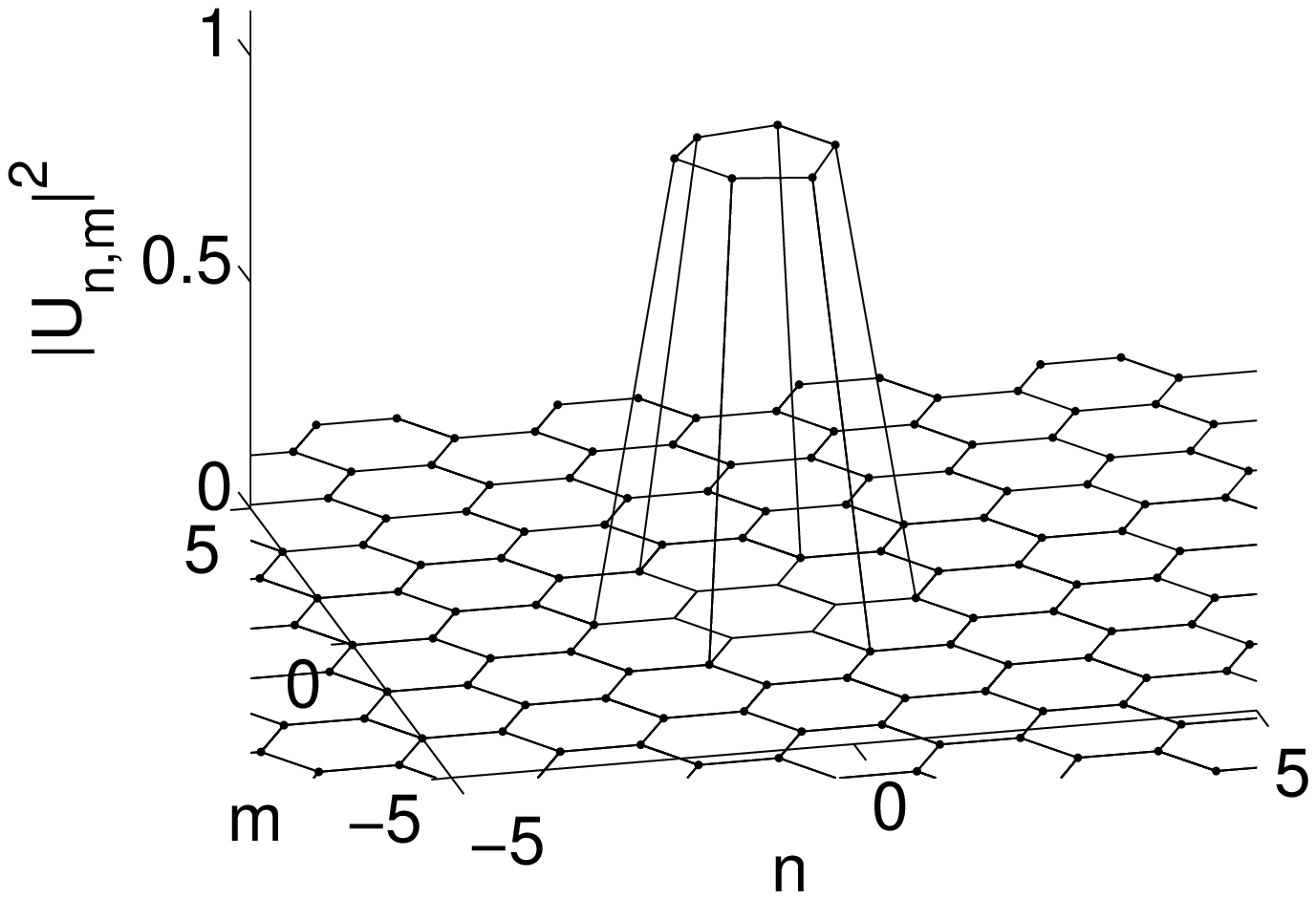}\\
\includegraphics[width=5cm,height=4cm]{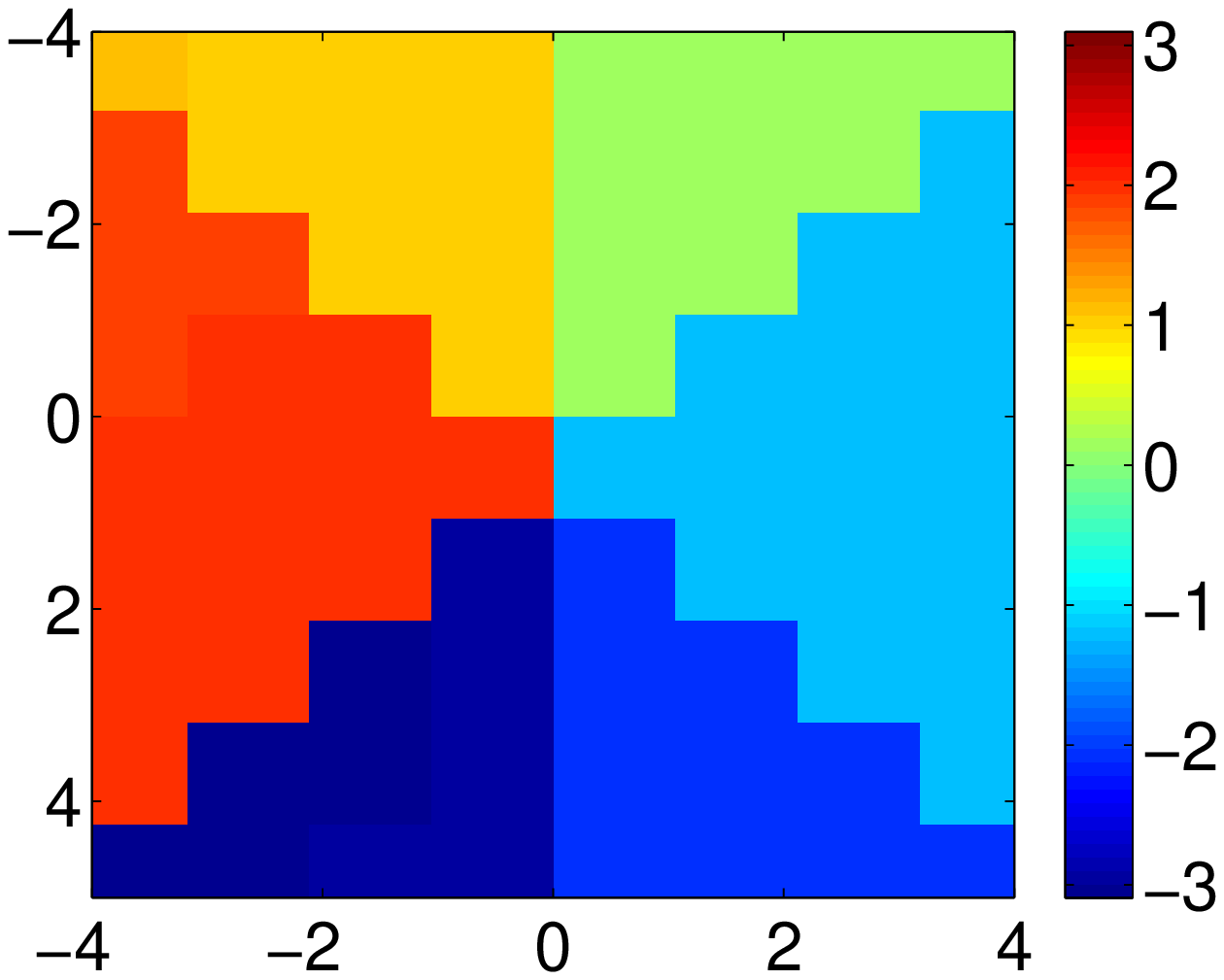}
\includegraphics[width=5cm,height=4cm]{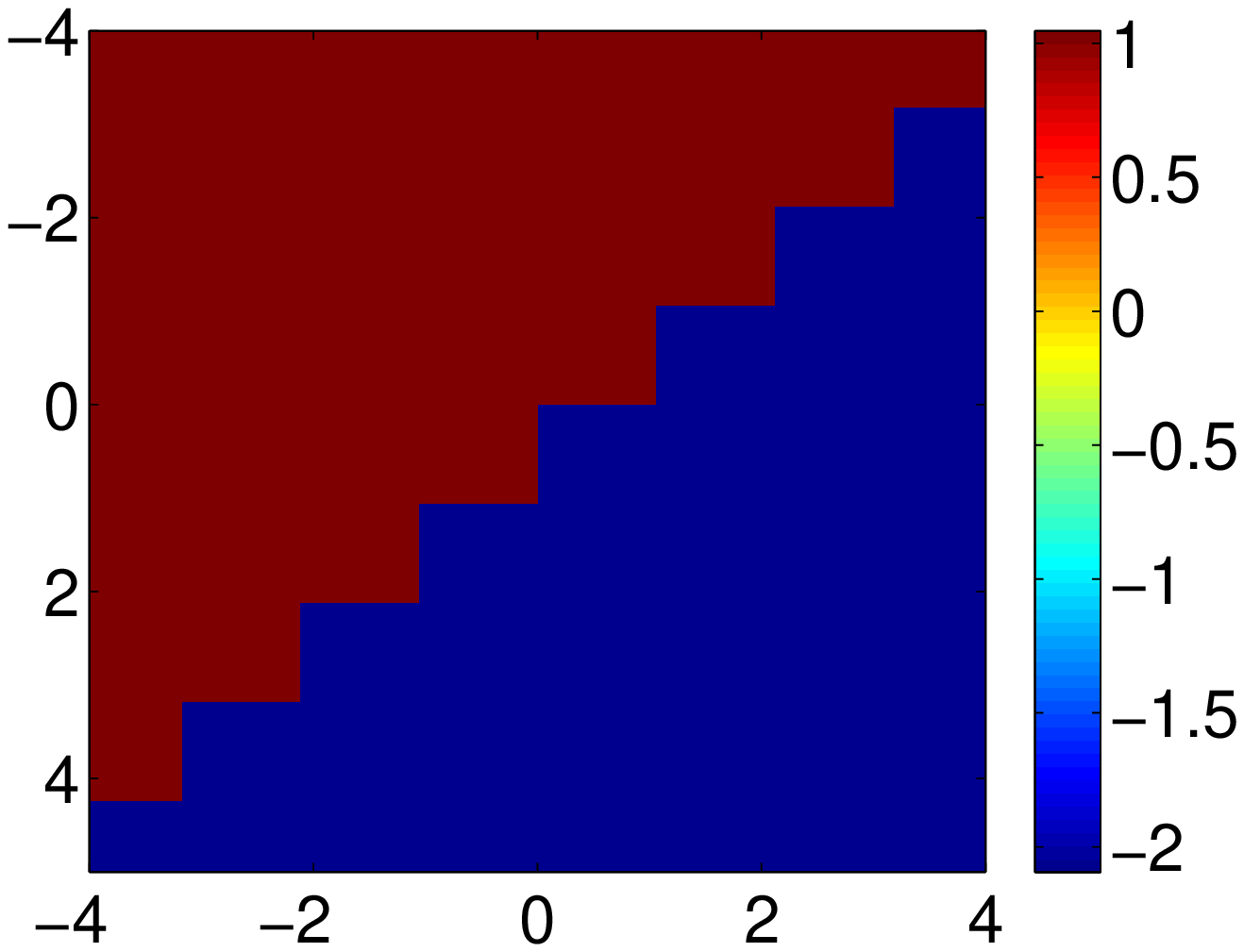}\\
\includegraphics[width=5cm,height=4cm]{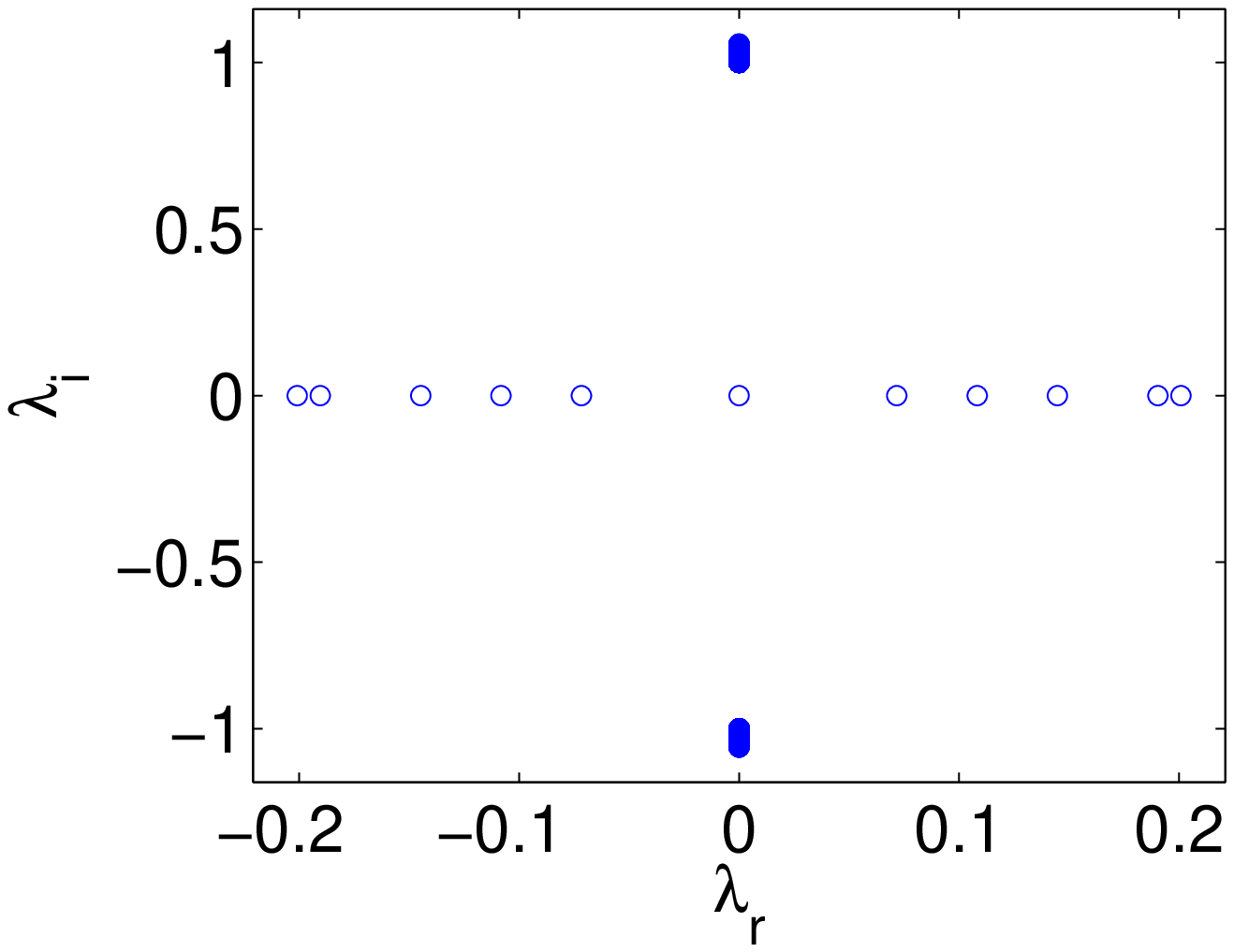}
\includegraphics[width=5cm,height=4cm]{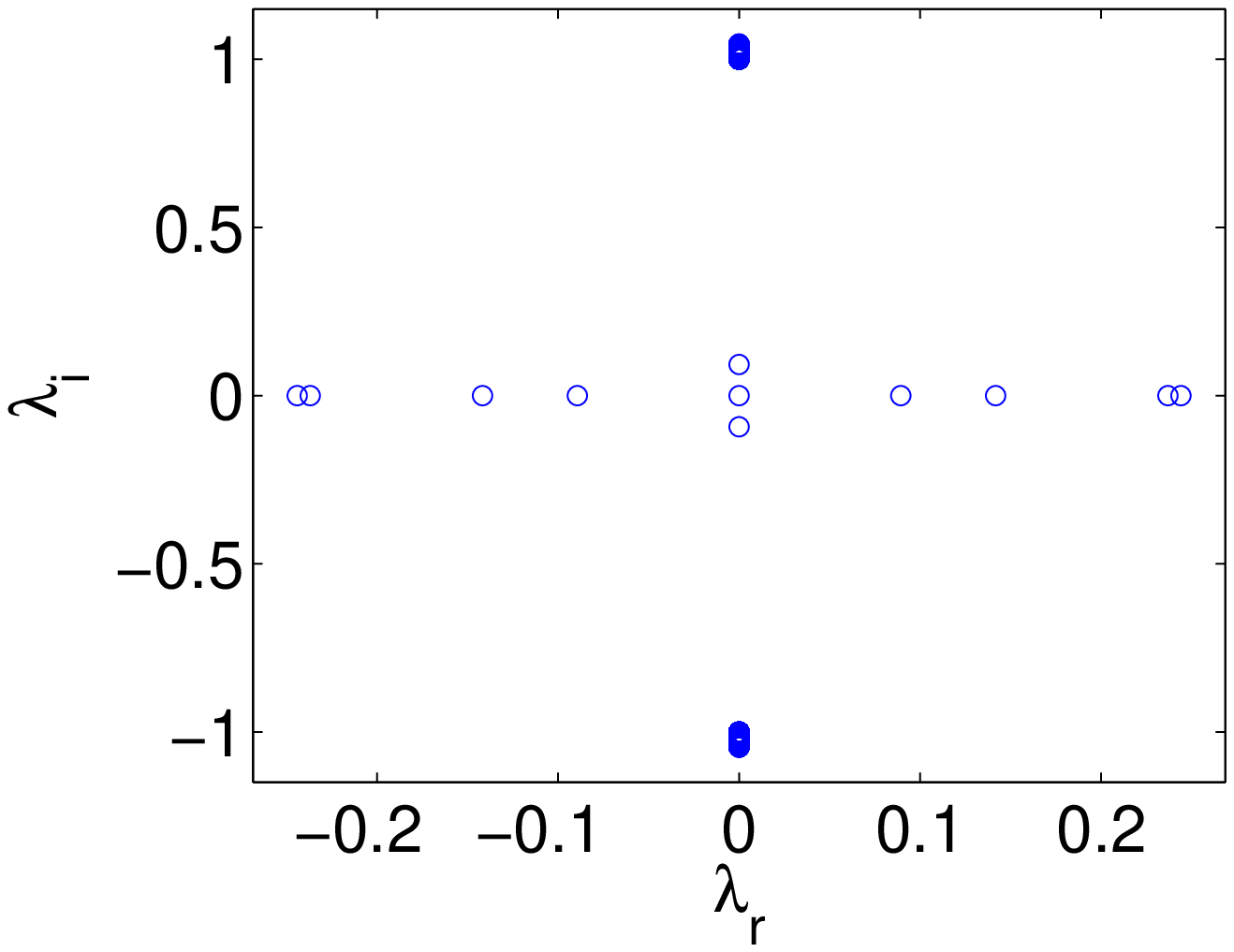}\\
\includegraphics[width=5cm,height=4cm]{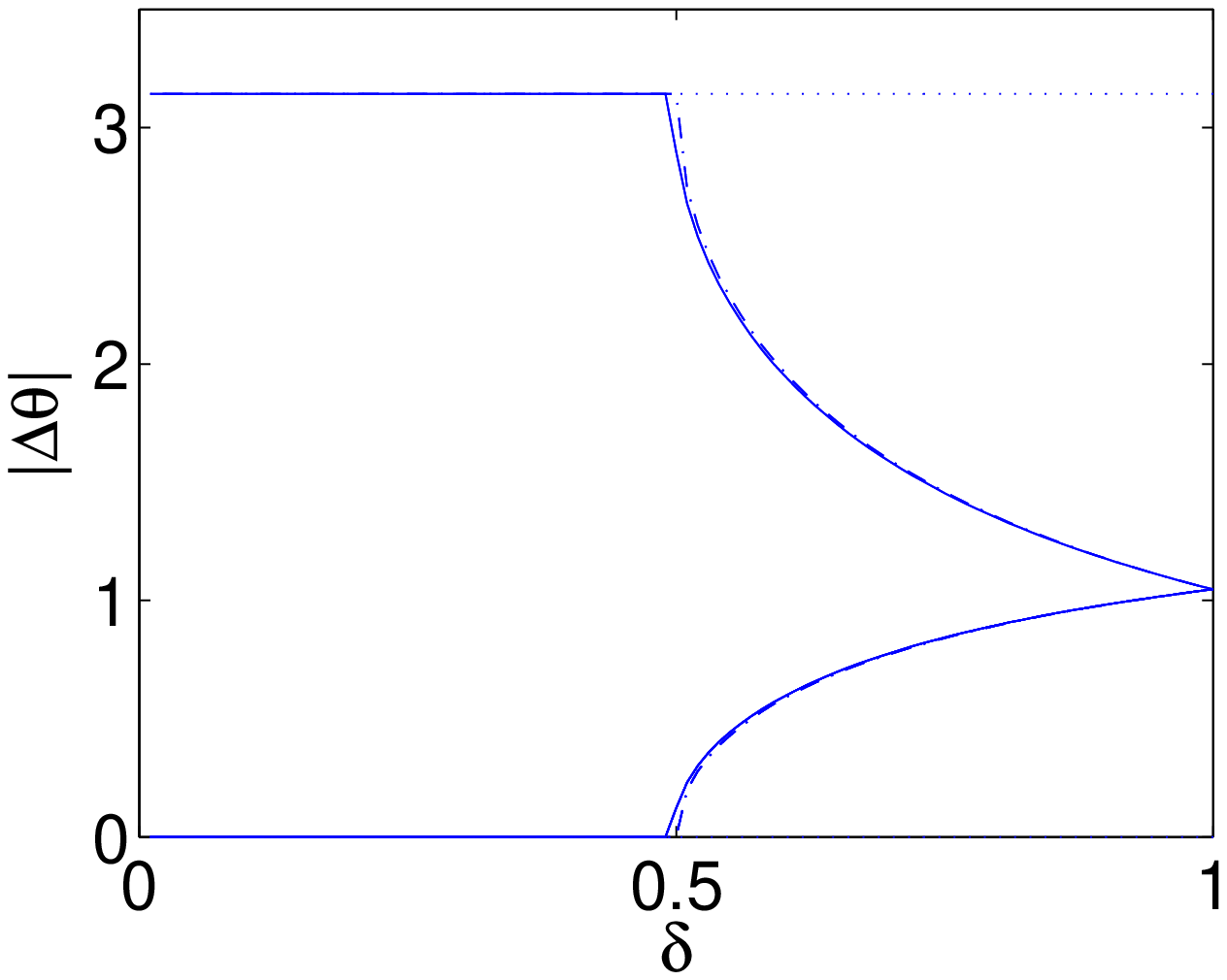}
\includegraphics[width=5cm,height=4cm]{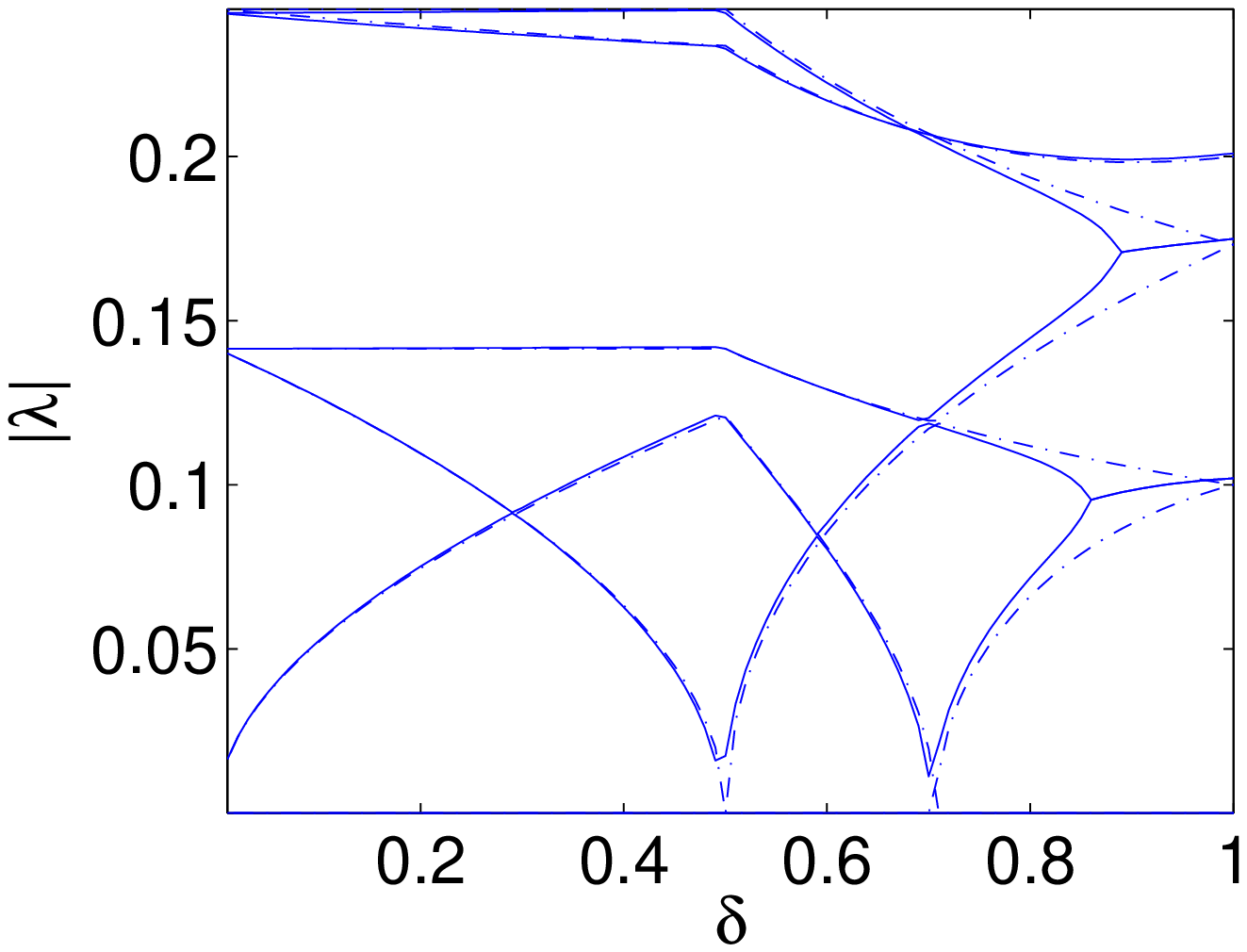}
\vspace{-0.4cm}
\caption{Honeycomb six-site charge $1$ vortex i.e. the 
honeycomb $[0, \pi/3, 2\pi/3, \pi, 4\pi/3, 5\pi/3]$ 
configuration at the isotropic limit ($\varepsilon = 0.01$). The top row 
displays the modulus squared of the 
configuration corresponding to anisotropic parameter $\delta=0.8$ (left panel) 
and $\delta=0.3$ (right panel). The second row shows the phase portraits 
and the 
third row shows the spectral plane for the same values of the anisotropy, 
$\delta=0.8$ (left), 
and $\delta=0.3$ (right). In the last row the left panel shows the 
comparison between 
the theoretical (dash-dot lines) and numerical (solid lines) changes in the 
relative 
phases. The vortex collides at $\delta=0.5$ with the 
$[\pi/3, \pi/3, \pi/3, -2\pi/3, -2\pi/3, -2\pi/3]$ 
configuration. At $\delta=0$, for the present form of anisotropy,
this is equivalent to the two configurations
along a line, namely 
$[\pi/3, \pi/3, \pi/3]$ and $[-2\pi/3, -2\pi/3, -2\pi/3]$. The bottom right 
panel shows 
the comparison of the theoretical (dash-dot lines) versus numerical (solid lines) 
linear stability eigenvalues for $0 \leq \delta \leq 1$.}
\label{fig:hon6site_C1}
\end{figure}

We now discuss the effects of anisotropy in the honeycomb 
lattice. For the case of the six-site charge $1$ vortex
(see Fig.~\ref{fig:hon6site_C1}), we again see a very good
comparison between the numerical results and the theory.  
The unstable charge $1$ vortex remains unstable throughout
the anisotropic variation of $\delta$ in the interval 
$[0,1)$. At about $\delta = 0.70$ one 
of the real eigenvalue pairs $\lambda_{5, 6}$ collides with the 
origin of the spectral plane and becomes purely imaginary. 
At $\delta = 0.5$ the
pair which was theoretically predicted to be 
$\lambda_3 = \pm \sqrt{\frac{(4\delta^2 - 1)\varepsilon}{\delta}}$
collides with the pair $\lambda_1$, giving rise to the
bifurcation that was theoretically predicted to arise
at this critical point. 
Indeed this bifurcation transforms 
the vortex for $\delta < 0.5$ into the unstable 
$[\pi/3, \pi/3, \pi/3, 2\pi/3, 2\pi/3, 2\pi/3]$ honeycomb
configuration (i.e., a $[0, 0, 0, \pi, \pi, \pi]$ state up
to a trivial phase shift, as discussed in section II).

\begin{figure}[tbh]
\centering
\includegraphics[width=5cm,height=4cm]{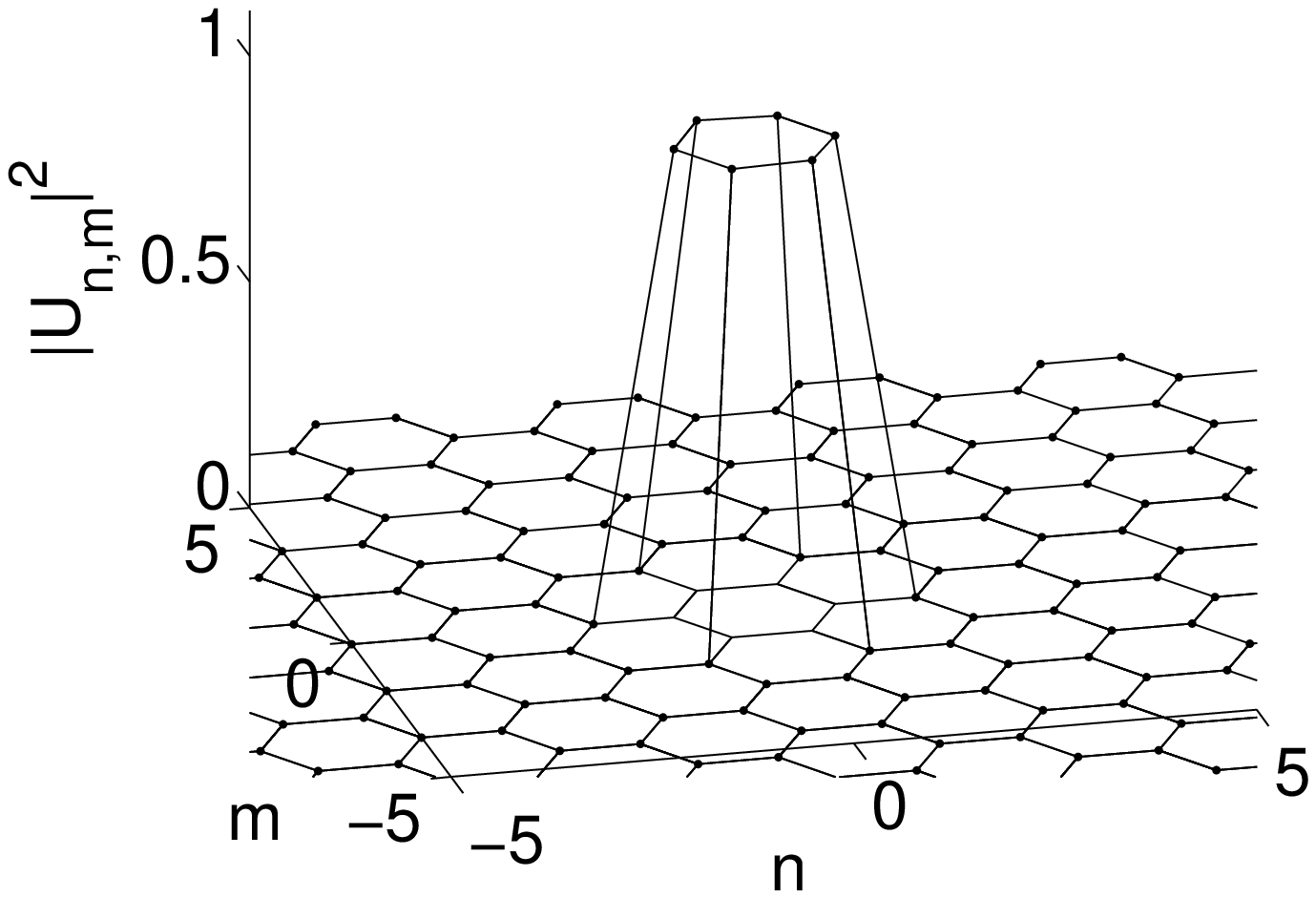}
\includegraphics[width=5cm,height=4cm]{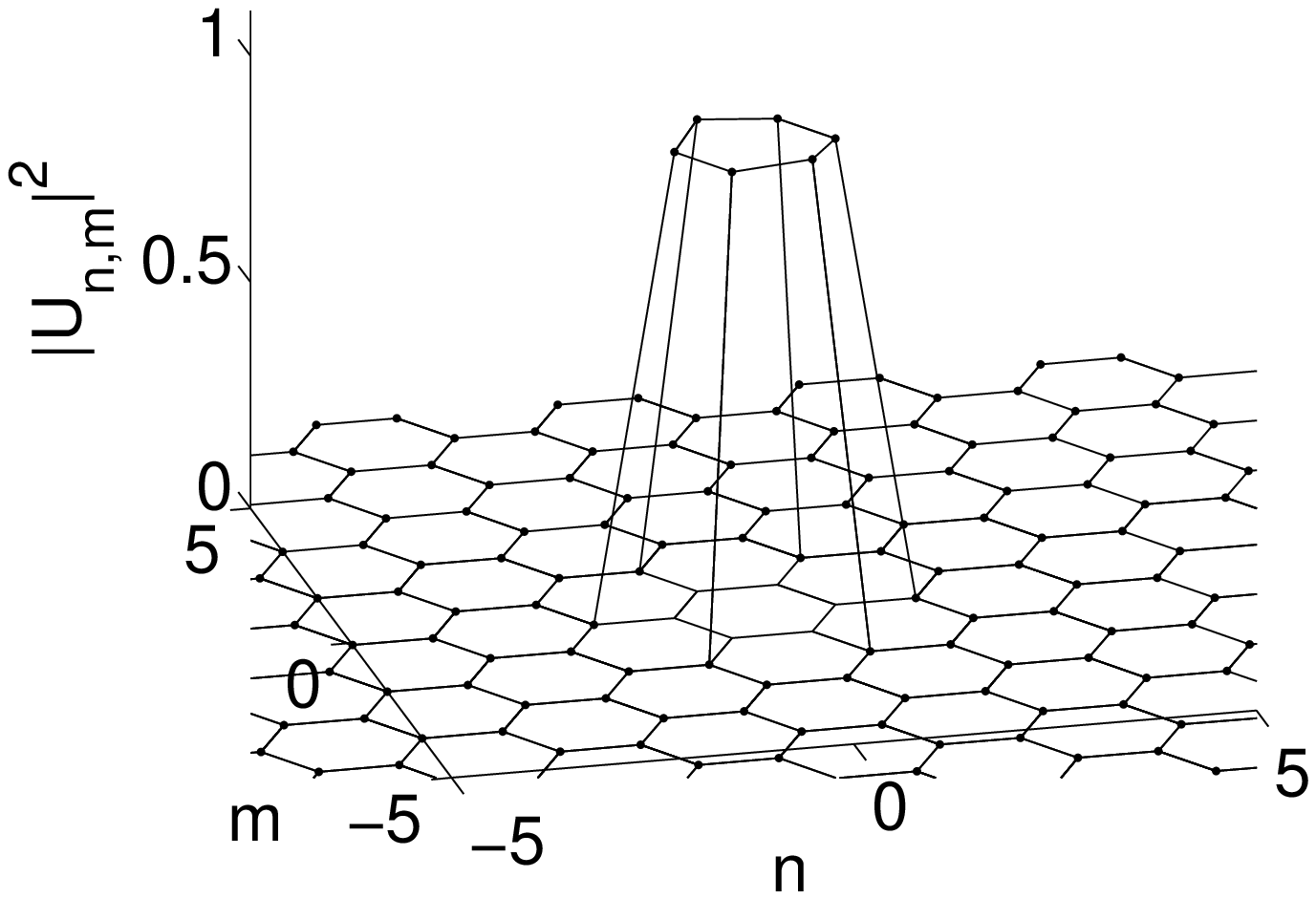}\\
\includegraphics[width=5cm,height=4cm]{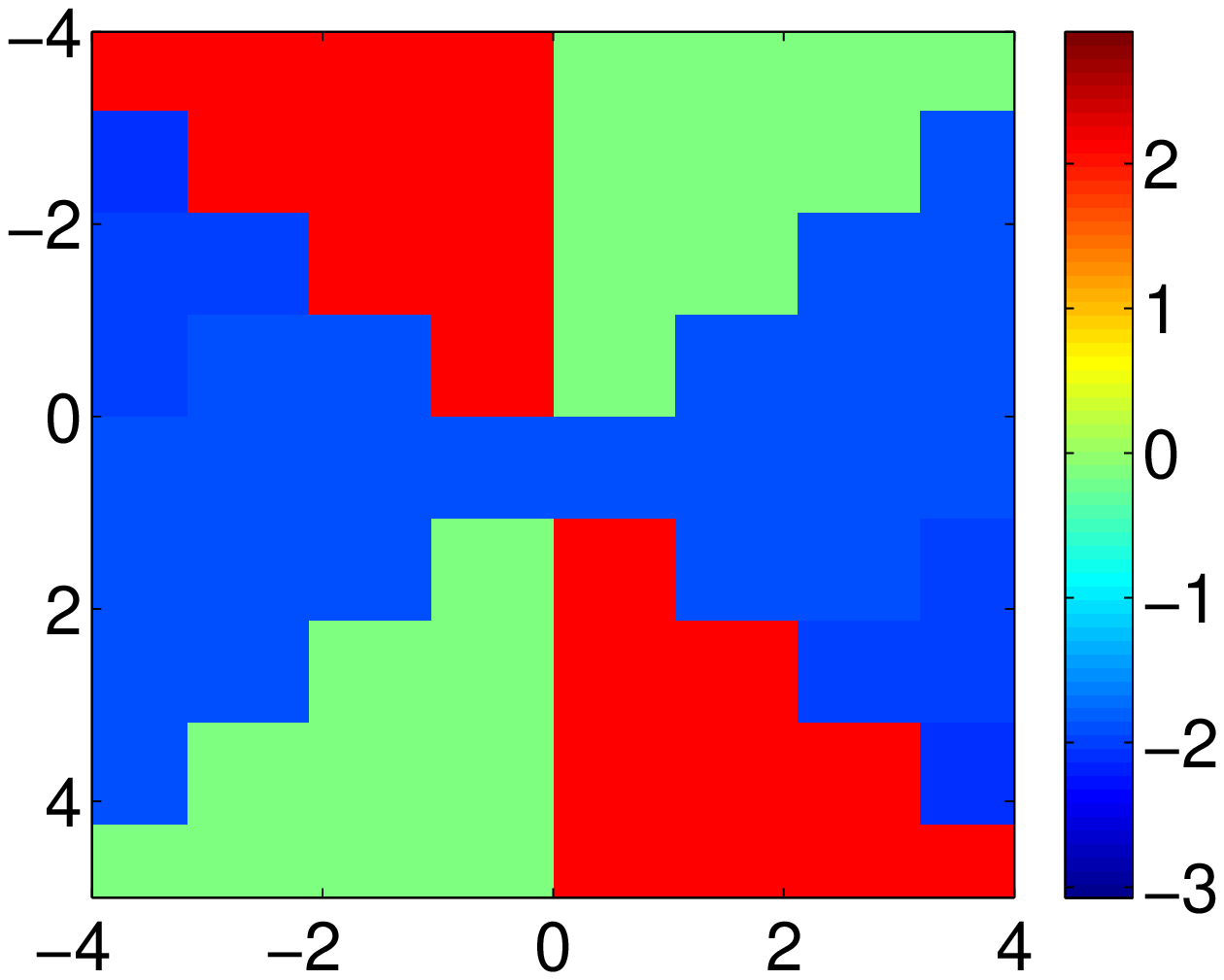}
\includegraphics[width=5cm,height=4cm]{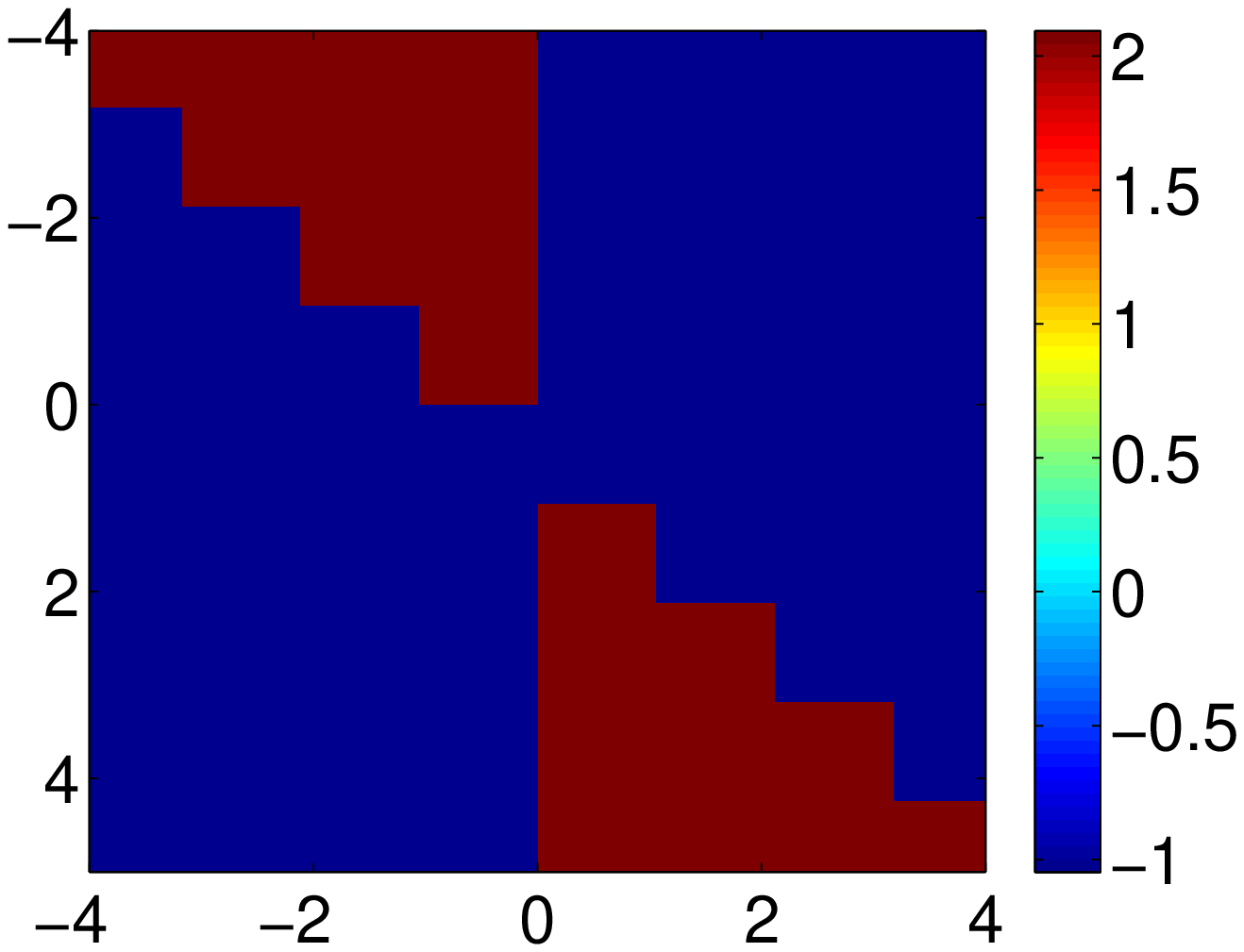}\\
\includegraphics[width=5cm,height=4cm]{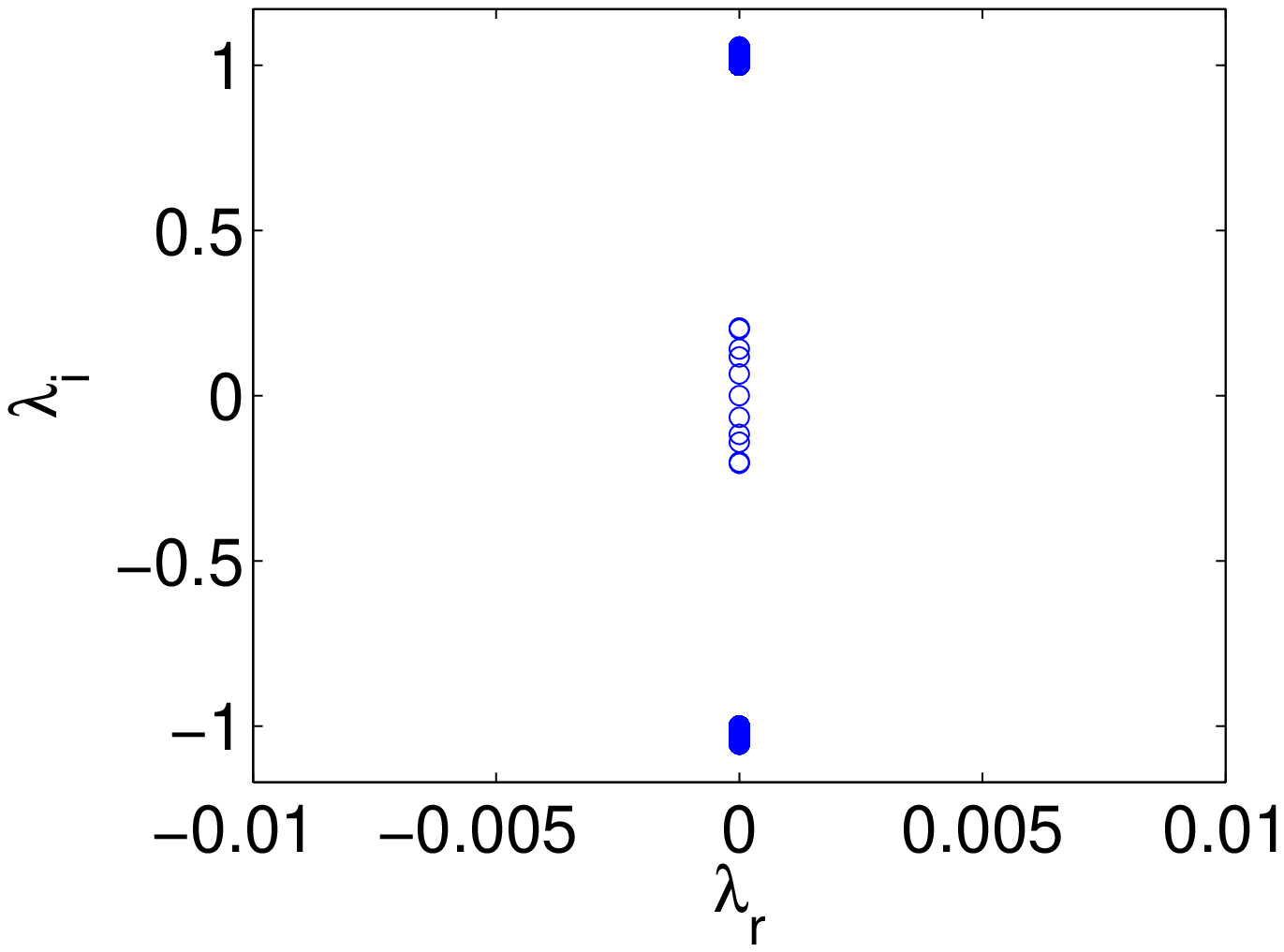}
\includegraphics[width=5cm,height=4cm]{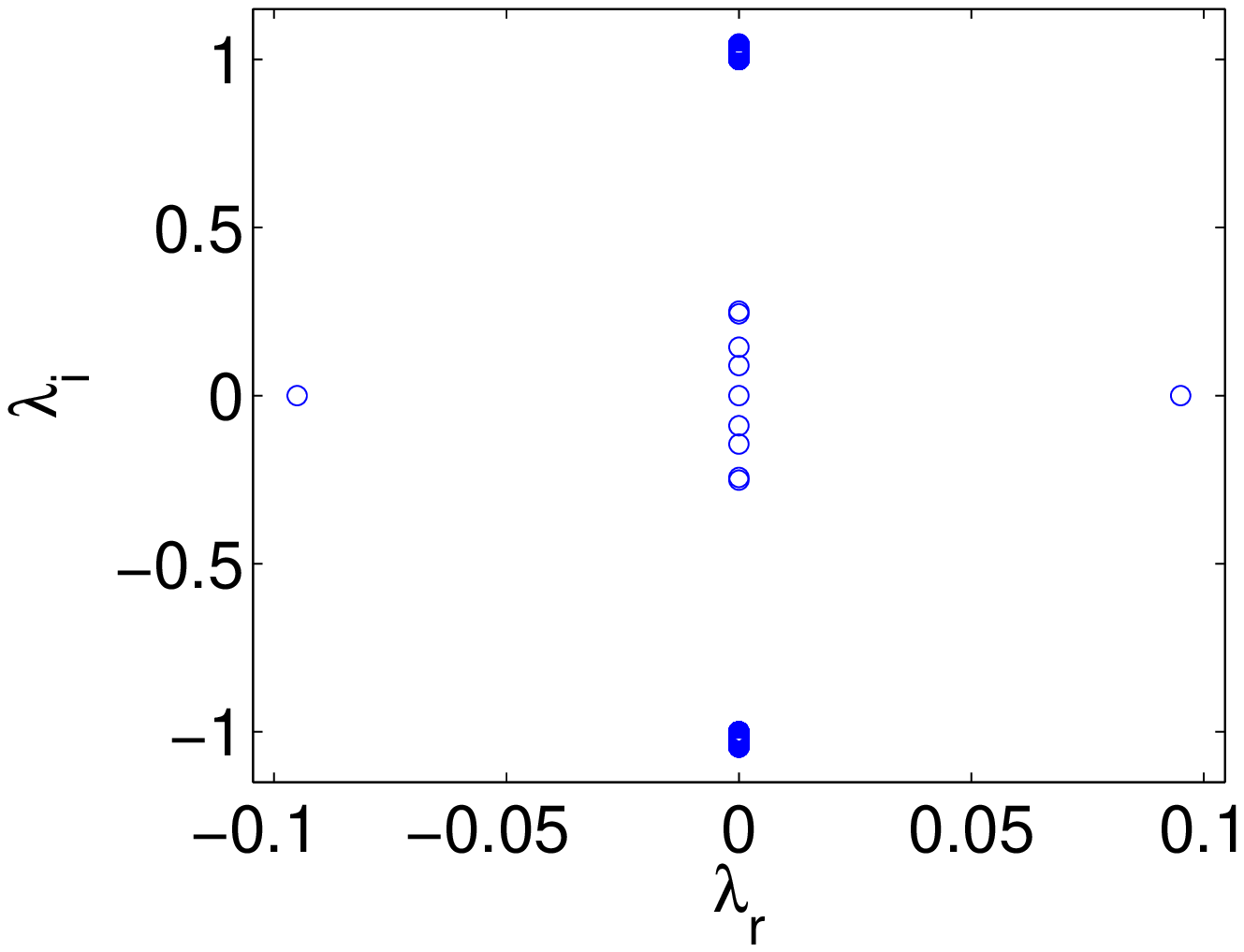}\\
\includegraphics[width=5cm,height=4cm]{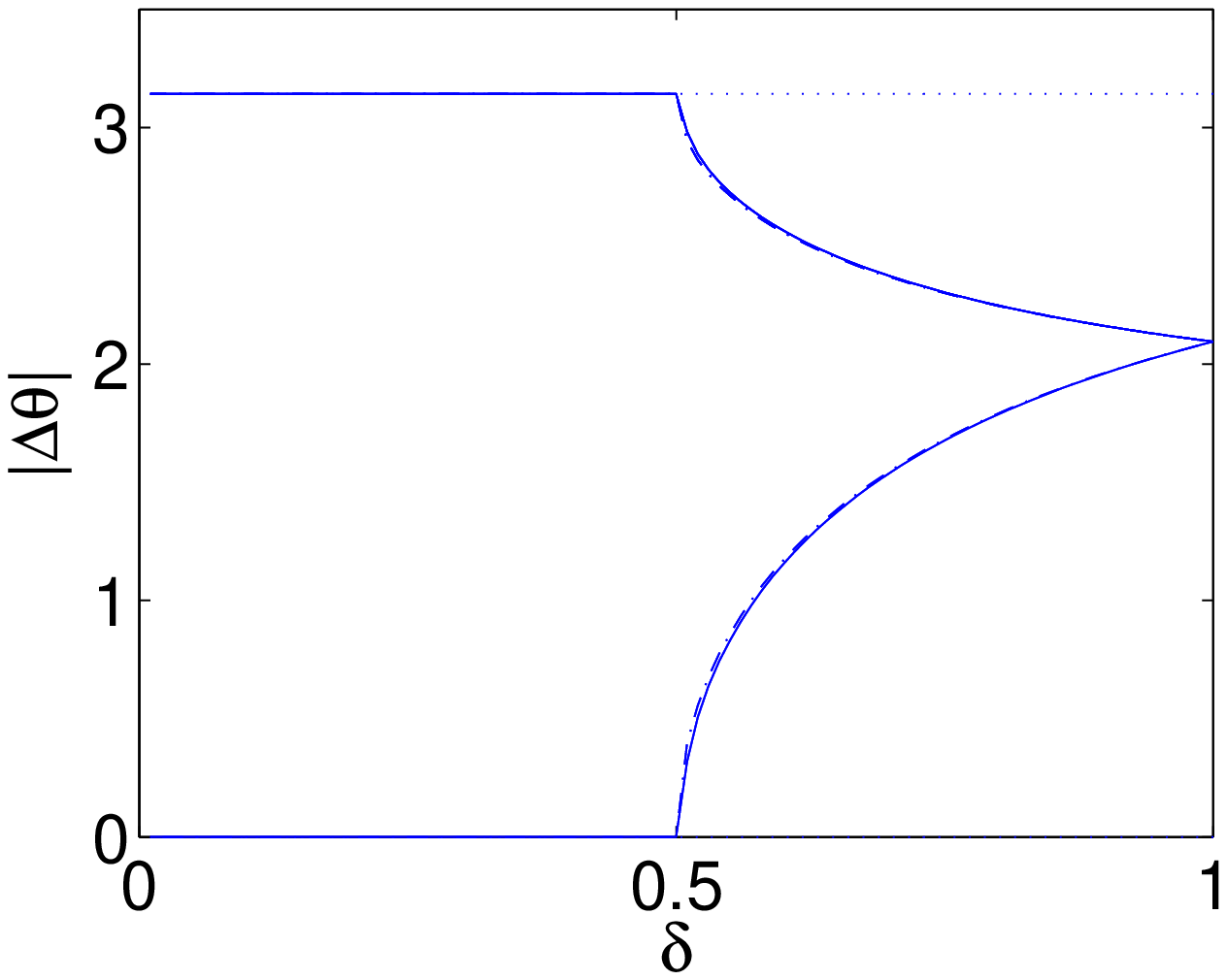}
\includegraphics[width=5cm,height=4cm]{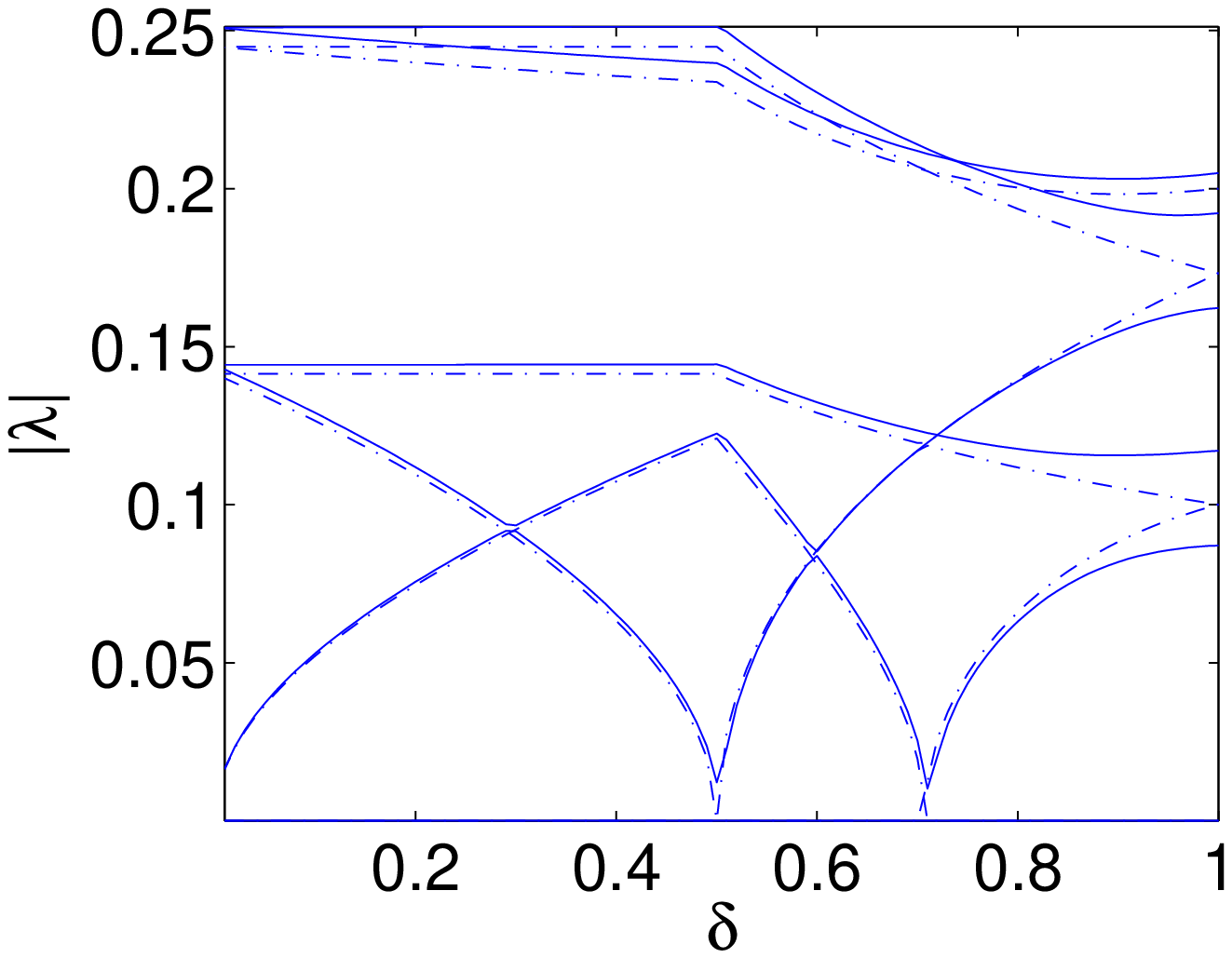}
\vspace{-0.4cm}
\caption{Honeycomb six-site charge $2$ vortex i.e. the 
honeycomb $[0, 2\pi/3, 4\pi/3, 2\pi, 8\pi/3, 10\pi/3]$ 
configuration ($\varepsilon = 0.01$). The top row displays the modulus 
squared of the 
configuration corresponding to anisotropic parameter $\delta=0.8$ (left panel) and 
$\delta=0.3$ (right panel). The second row shows the phase portraits and the third 
row shows the spectral plane for the same values of the anisotropy, $\delta=0.8$ (left), 
and $\delta=0.3$ (right). In the last row the left panel shows the change in the 
relative phases. The charge 2 vortex collides at $\delta=0.5$ with the 
$[-\pi/3, 2\pi/3, -\pi/3, -\pi/3, 2\pi/3, -\pi/3]$ configuration. 
The bottom right panel shows the theoretical (dash-dot lines) versus numerical 
(solid lines) comparisons of the linear stability eigenvalues for 
$0 \leq \delta \leq 1$.}
\label{fig:hon6site_C2}
\end{figure}

The stable honeycomb charge $2$ vortex (shown in Fig.~\ref{fig:hon6site_C2}) 
becomes unstable at approximately $\delta = 0.70$, in very good
agreement with the theoretical prediction of
$\delta=0.716$; cf. the relevant discussion in section II. The instability 
arises due to a pair of eigenvalues from $\lambda_{5, 6}$ becoming 
real-valued. Subsequently, as theoretically predicted, for
$\delta=0.5$, a pitchfork bifurcation eventually transforms
the vortex into the state 
$[-\pi/3, 2\pi/3, -\pi/3, -\pi/3, 2\pi/3, -\pi/3]$,
(i.e., a $[0, \pi, 0, 0, \pi, 0]$ state up to a trivial
phase shift). The latter configuration has the same
stability characteristics for $\delta < 0.5$ that
the vortex state possesses in the vicinity of $\delta > 0.5$. Therefore,
it possesses a single real eigenvalue pair as confirmed
in the right panels of Fig.~\ref{fig:hon6site_C2}.
Naturally, for $\delta > 0.5$, this $[0, \pi, 0, 0, \pi, 0]$
persists but acquires a second real pair as a result of 
the supercritical pitchfork bifurcation. 

\begin{figure}[tbh]
\centering
\includegraphics[width=5cm,height=4cm]{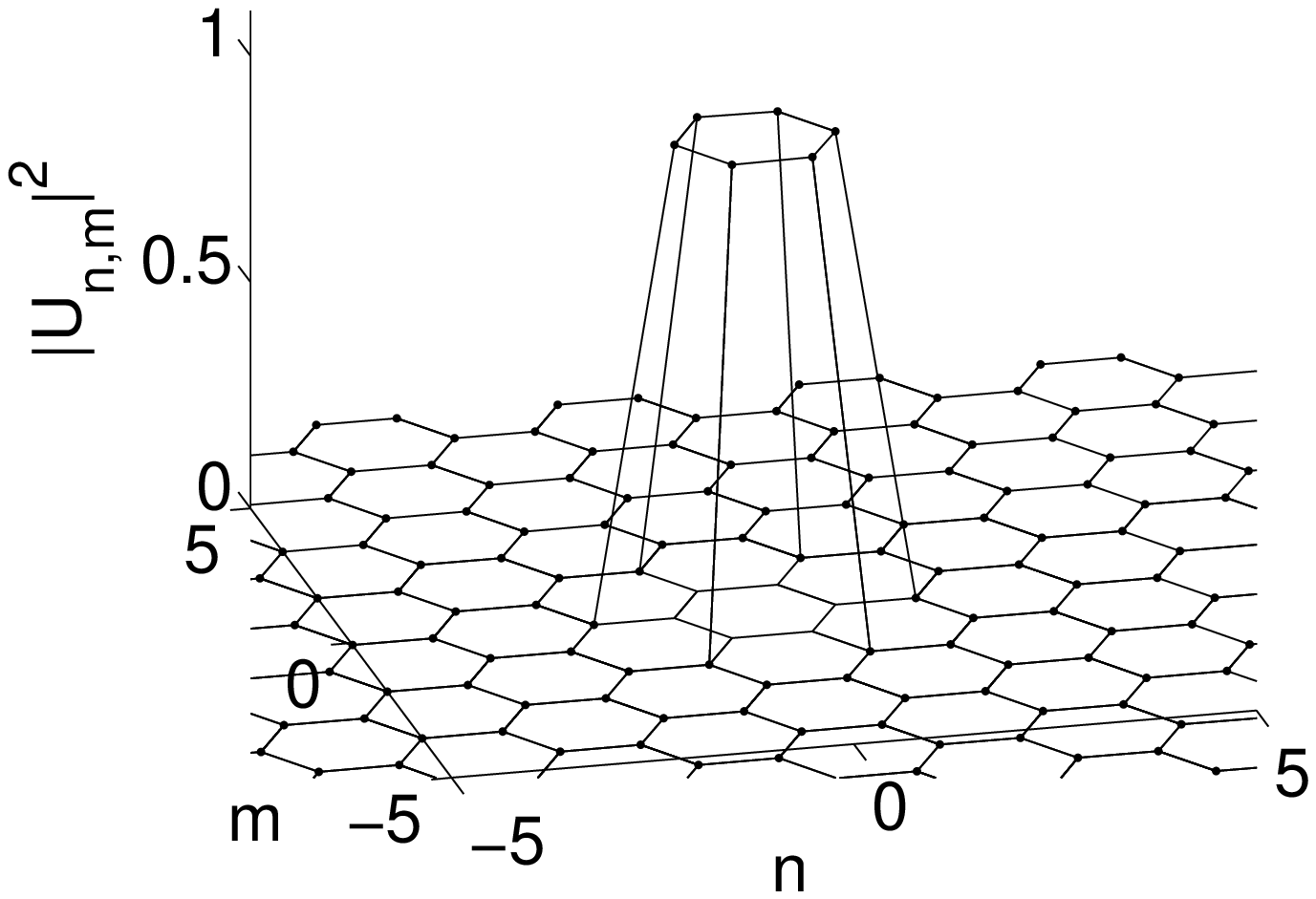}
\includegraphics[width=5cm,height=4cm]{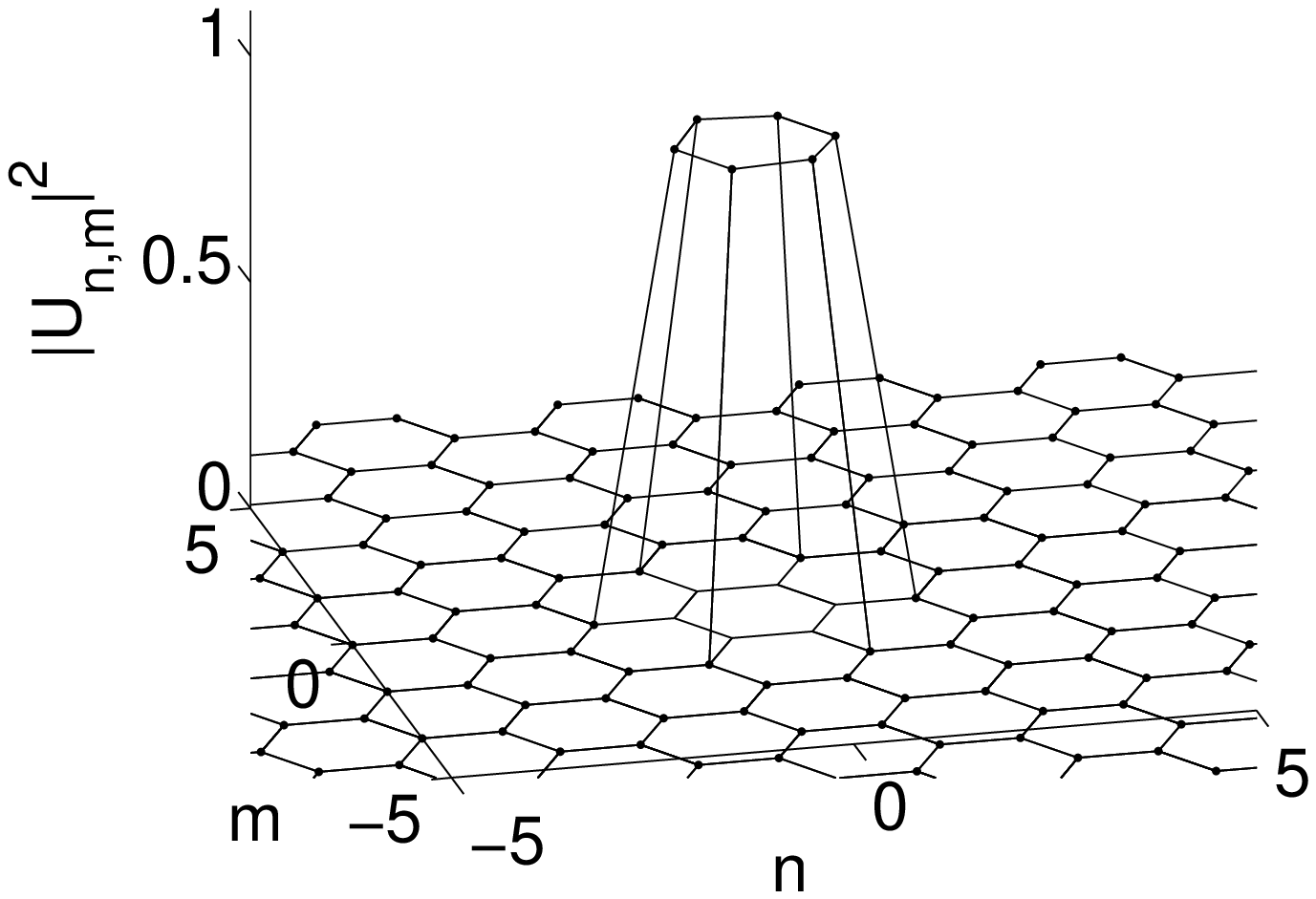}\\
\includegraphics[width=5cm,height=4cm]{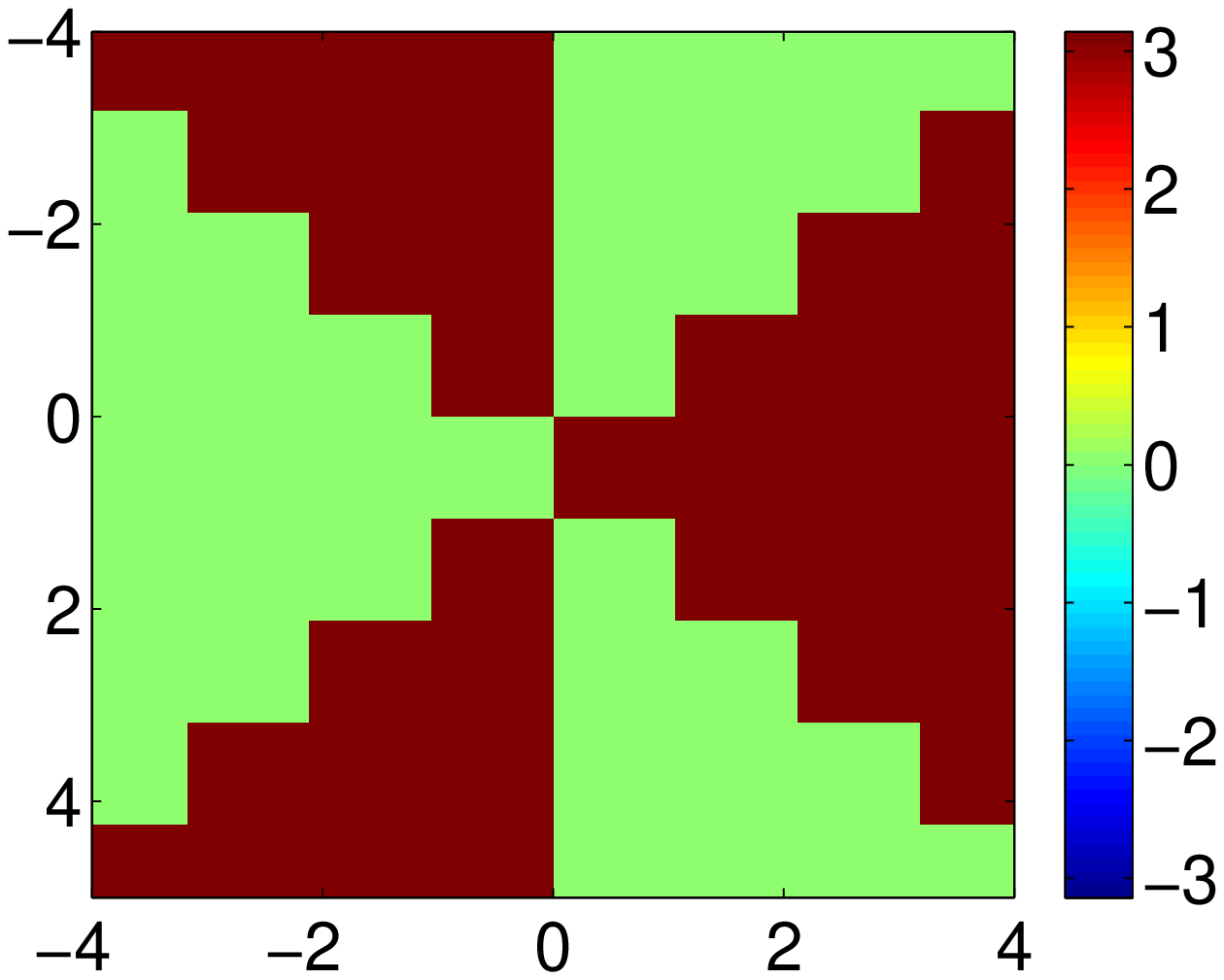}
\includegraphics[width=5cm,height=4cm]{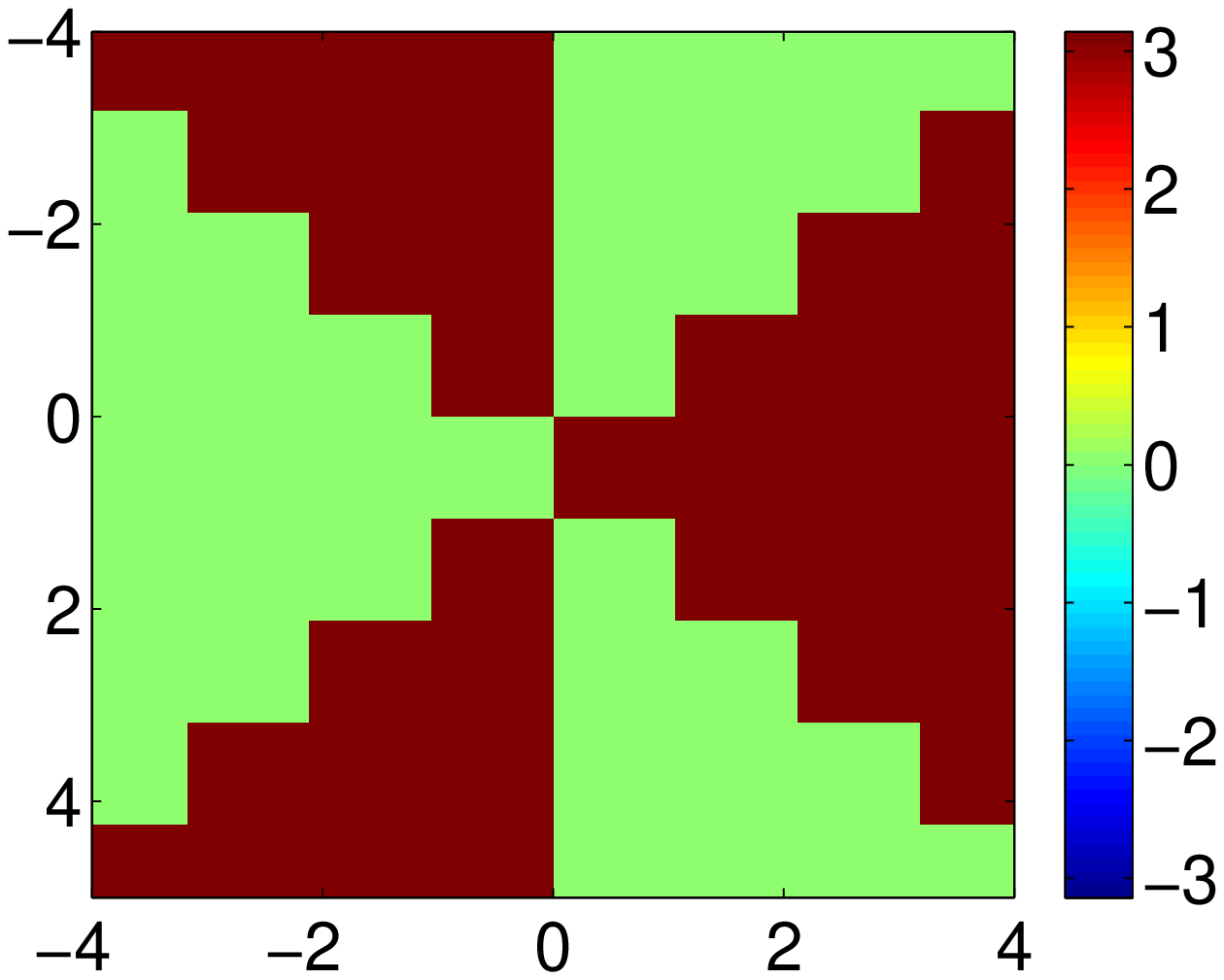}\\
\includegraphics[width=5cm,height=4cm]{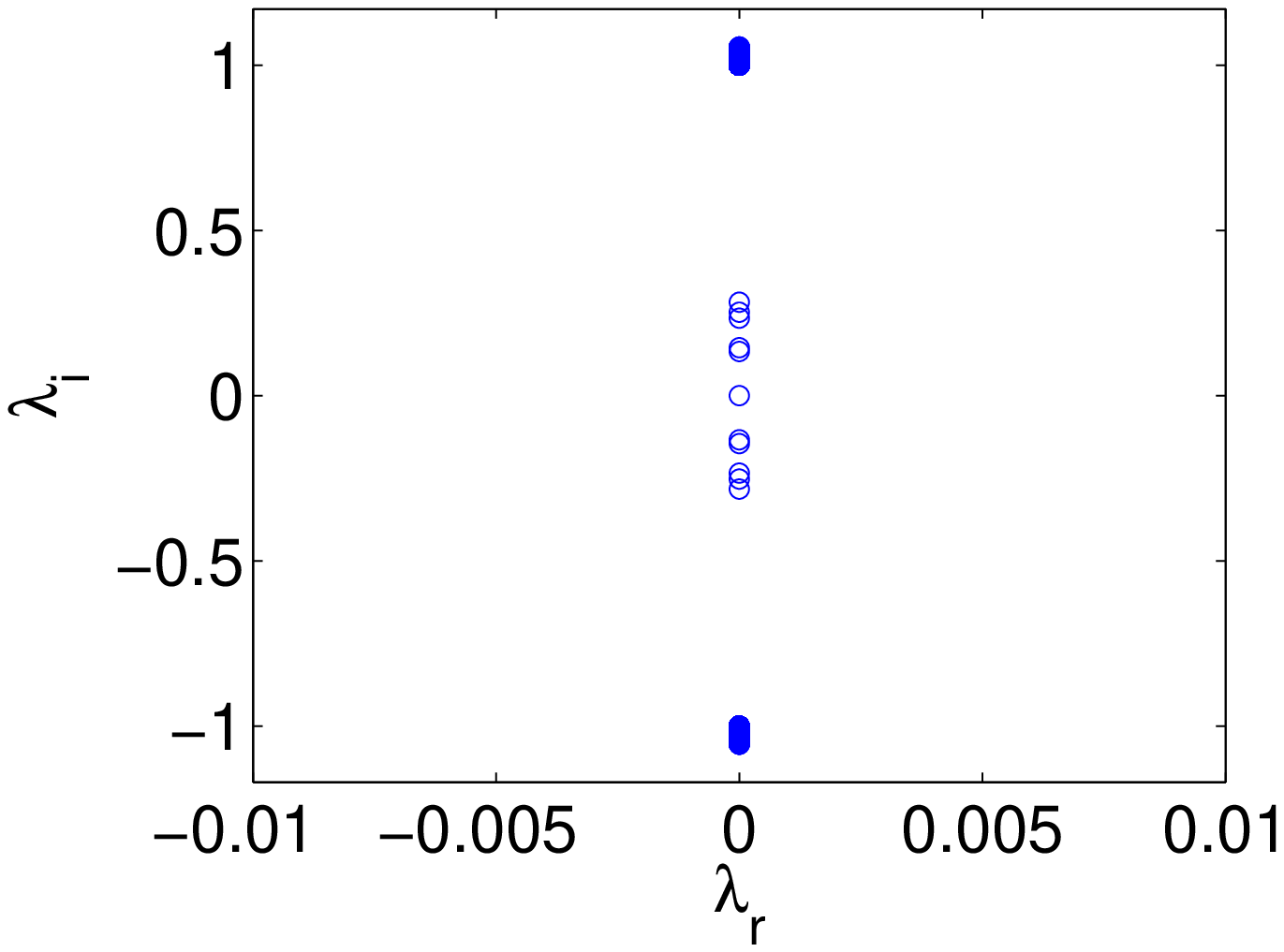}
\includegraphics[width=5cm,height=4cm]{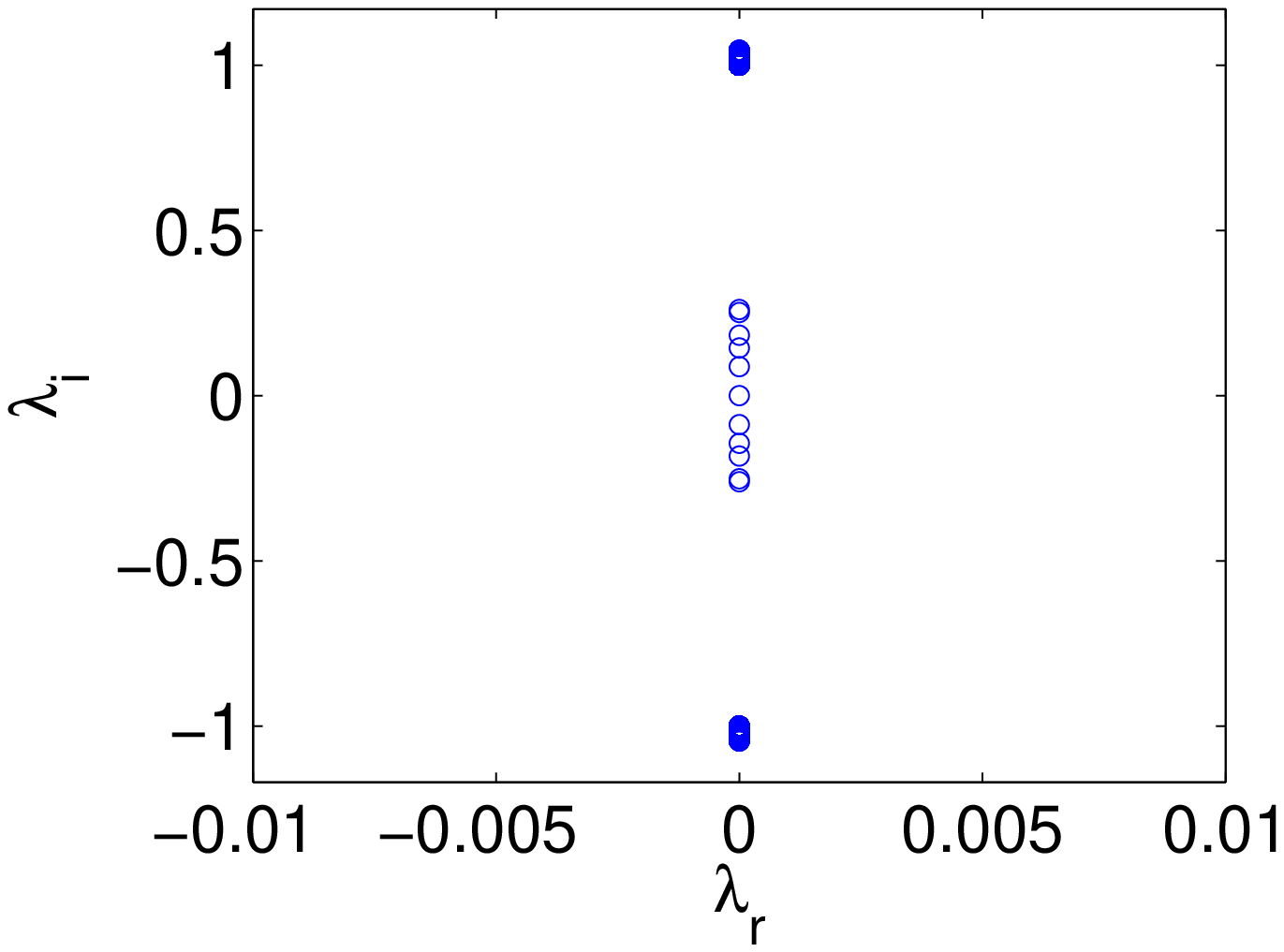}\\
\vspace{-0.4cm}
\caption{Honeycomb six-site out-of-phase configuration, i.e. honeycomb 
$[0, \pi, 0, \pi, 0, \pi]$ configuration ($\varepsilon = 0.01$). The top 
row displays the modulus squared of the field
with $\delta=0.8$ (left panel) and 
$\delta=0.3$ (right panel). 
At $\delta=0$ this 
is equivalent to the two configurations along a line of the form:
$[0, \pi, 0]$ and 
$[\pi, 0, \pi]$. Hence it retains its stability throughout
the interval $0 \leq \delta \leq 1$.}
\label{fig:hon6site_0_pi}
\end{figure}

\begin{figure}[tbh]
\centering
\includegraphics[width=5cm,height=4cm]{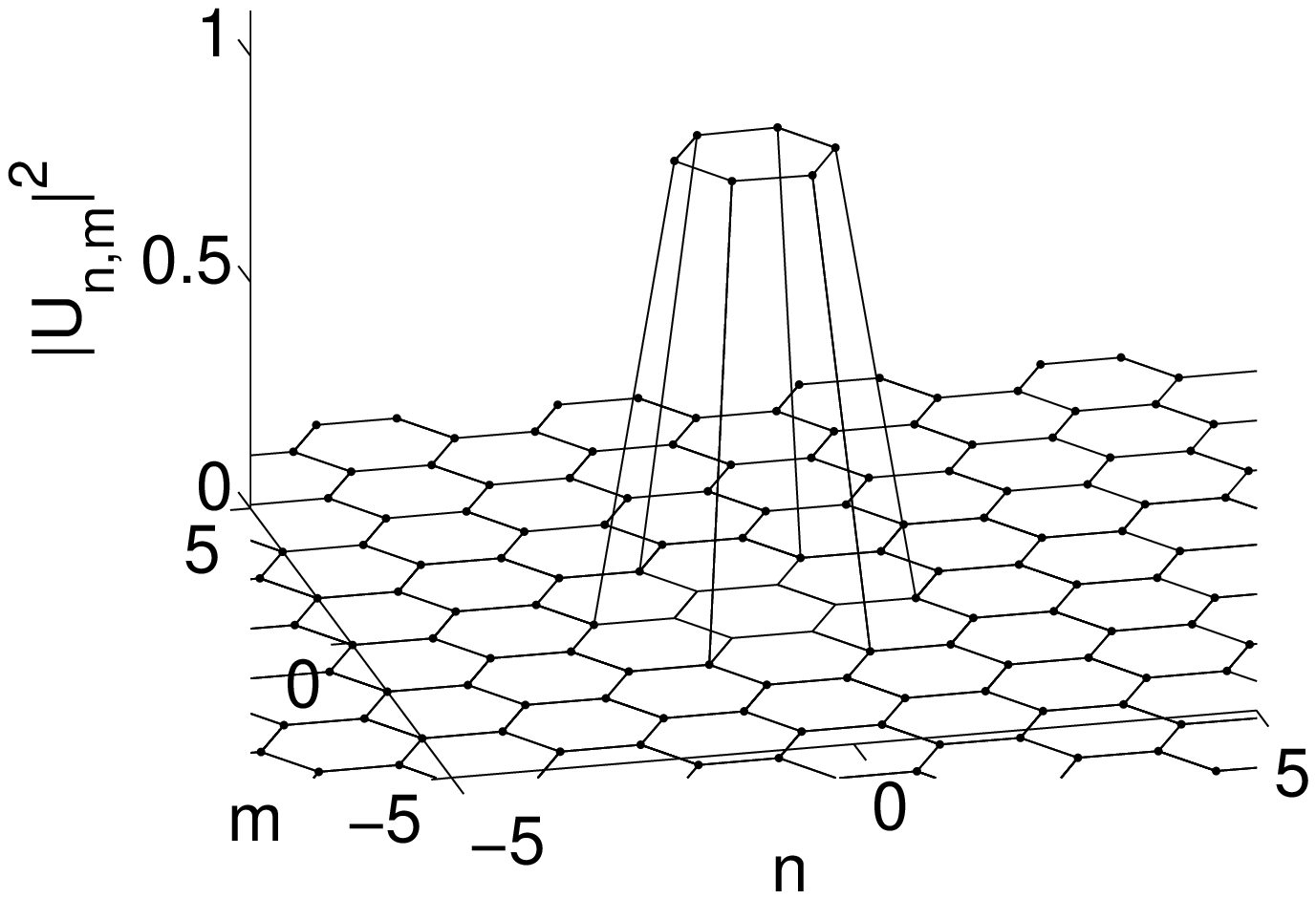}
\includegraphics[width=5cm,height=4cm]{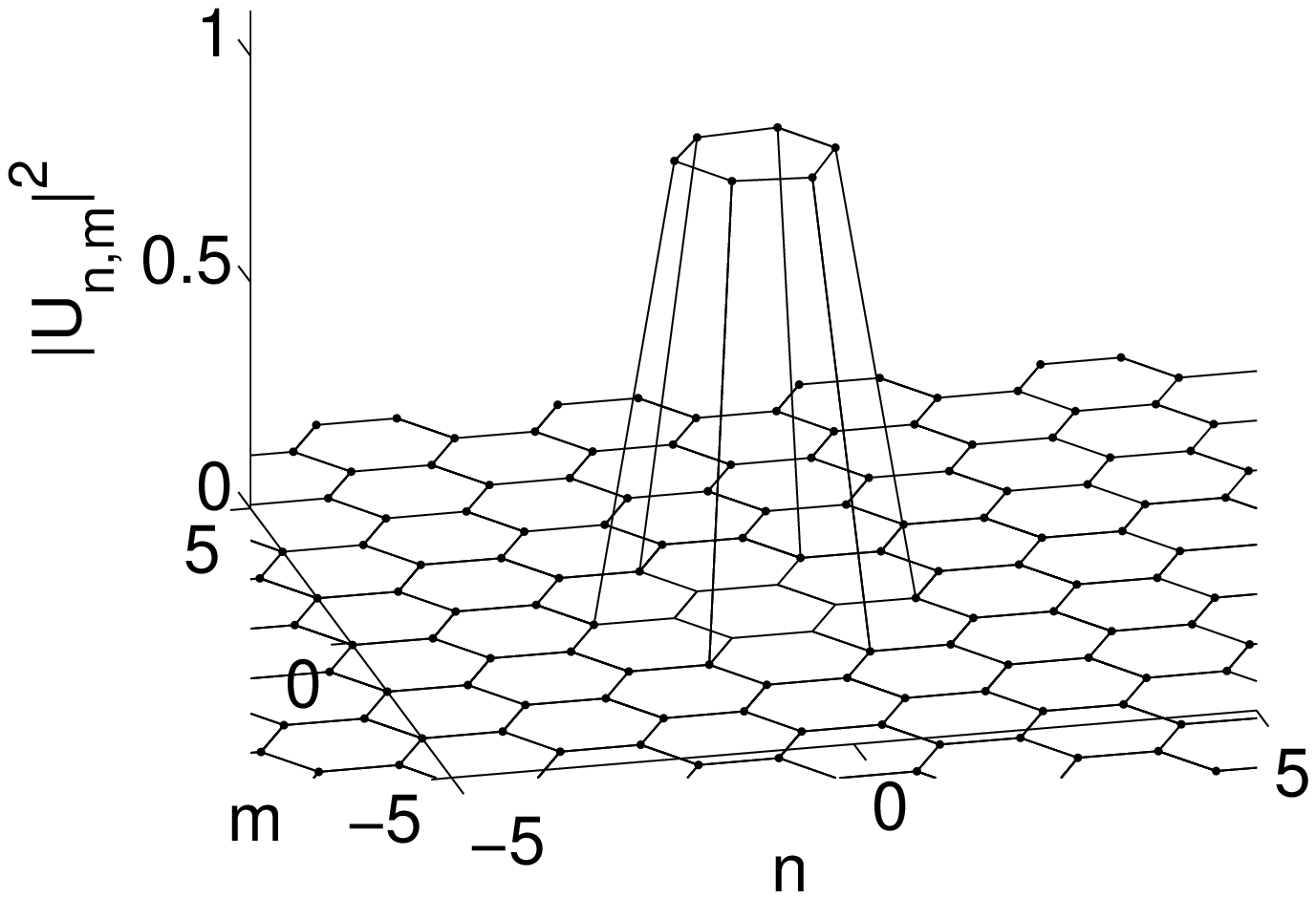}\\
\includegraphics[width=5cm,height=4cm]{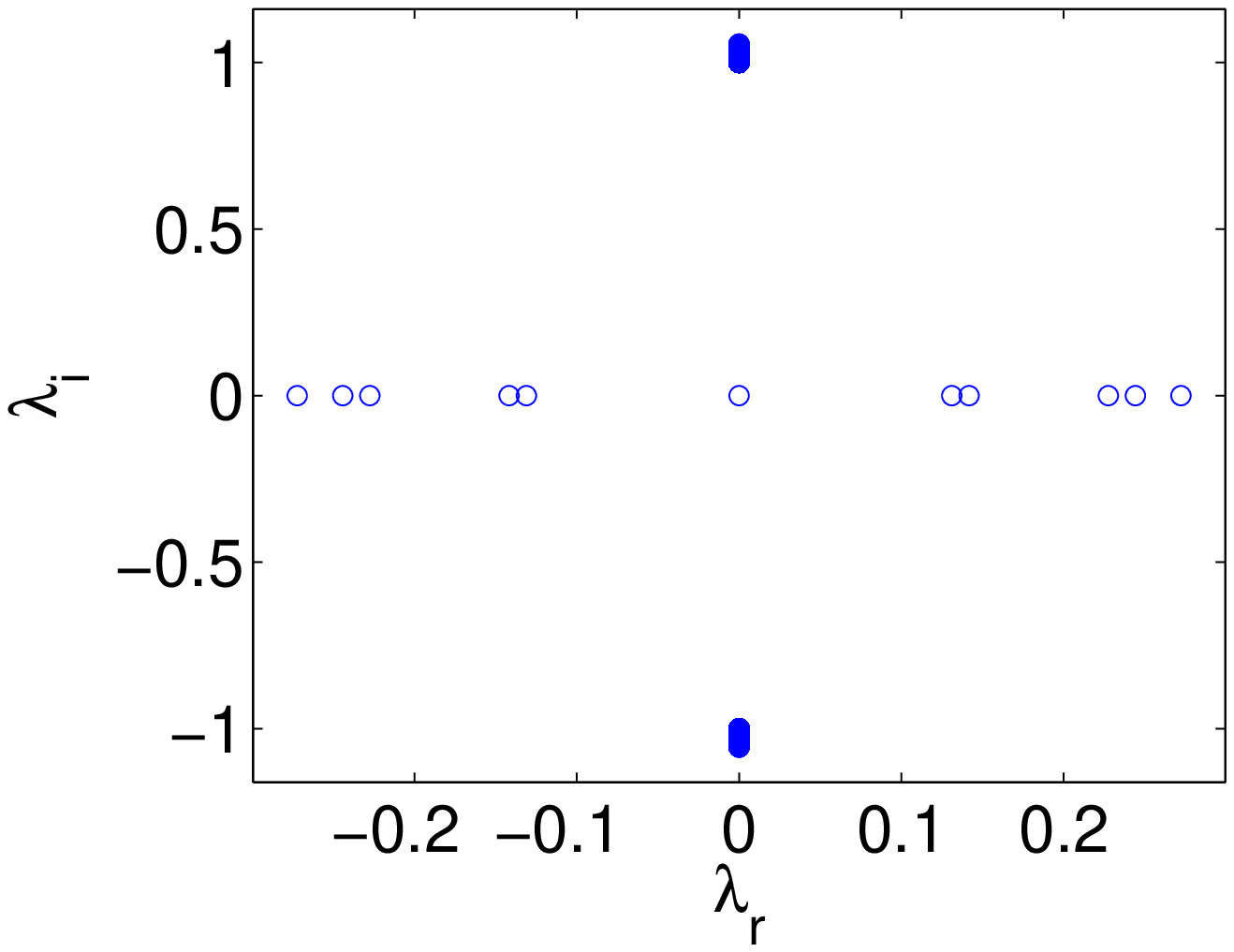}
\includegraphics[width=5cm,height=4cm]{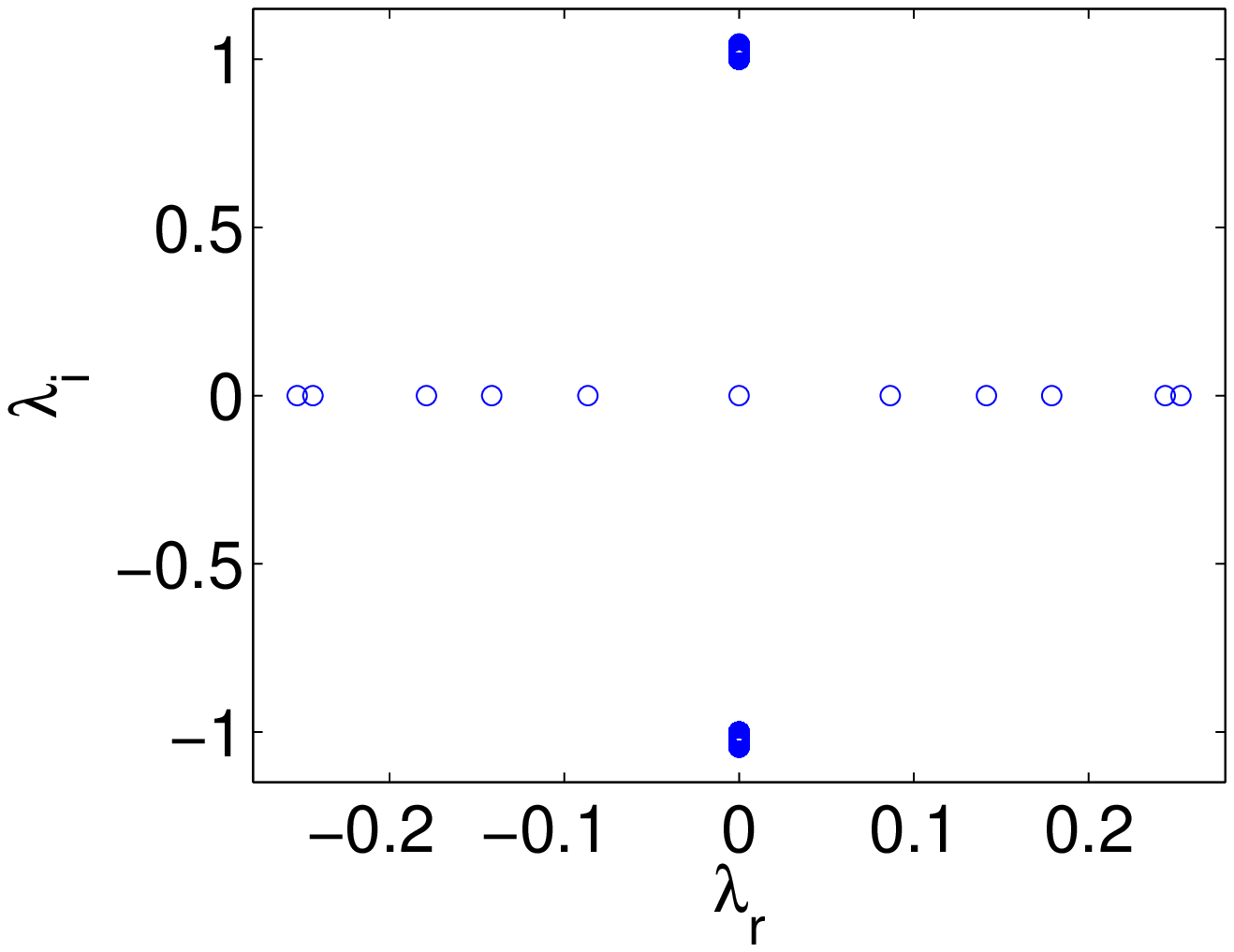}\\
\vspace{-0.4cm}
\caption{Honeycomb six-site in-phase configuration, i.e. honeycomb $[0, 0, 0, 0, 0, 0]$ configuration
($\varepsilon = 0.01$) The top row displays the modulus squared 
of the field with anisotropy 
$0.8$ (left column) and $0.3$ (right column). 
As expected by the adjacency of sites with the same phase, the 
anisotropy cannot prevent this configuration from being highly unstable.}
\label{fig:hon6site_0_0}
\end{figure}

In addition to these vortex configurations, 
in Figs.~\ref{fig:hon6site_0_pi} and~\ref{fig:hon6site_0_0},
we also examined the states with phase configurations:
$[0, \pi, 0, \pi, 0, \pi]$ and $[0, 0, 0, 0, 0, 0]$, respectively.
The former, as expected (for this small $\varepsilon$ and
given its alternating phase structure) is found to be linearly
stable for all the considered values of $\delta$, while
the latter is found to be highly unstable bearing 5 distinct
real eigenvalue pairs (as anticipated due to the presence
of sites of the same phase adjacent to each other).

\section{Dynamics}

In this section we numerically examine the nonlinear dynamics of the 
unstable solutions discussed in the previous sections. Each 
unstable solution is perturbed slightly in the direction of 
the eigenvector corresponding to the most unstable eigenvalue,
in order to seed the relevant instabilities. 
A fourth order (explicit) Runge-Kutta algorithm (RK4) has
been used in order to obtain the relevant dynamical evolution results.
It is observed that the coupling significantly
controls the nature of the dynamical evolution. This 
is natural since a larger coupling allows nearest 
neighbors to interact more strongly. In the 
hexagonal lattice, we show the 
evolution at a fixed background coupling of $\varepsilon = 0.01$ 
and also $\varepsilon = 0.2$. In all cases, the smaller coupling 
leads to a robust (multi-site) 
breather form, while the larger coupling produces a single 
robust site. For the honeycomb lattice case, a coupling of 
$\varepsilon = 0.2$ is used. All the unstable honeycomb 
six-site solutions evolve into multi-site breathers (where the
number of sites participating with a large norm in the final
configuration varies from case to case; see below). In either 
lattice we hold the anisotropy fixed at $\delta = 0.8$ for 
instabilities above the critical threshold of $\delta = 0.5$, 
and to $\delta = 0.2$ if the instability occurs below $\delta = 0.5$. 
The one exception is for the charge 2 vortex where we present the 
dynamics at $\delta = 0.6$ (instead of $\delta = 0.8$), i.e. 
after the onset of instability but before the critical 
transformation point of that state.

\subsection{Hexagonal Lattice}

The solutions found to be unstable in the hexagonal geometry 
are  (1) $[0, \pi, 0]$ with 
the anisotropy between the $0$ phase sites when $\delta > 0.5$
(due to the bifurcation of the vortex state),
(2) $[0, \pi, 0]$ with the anisotropy between the site with 
phase $\pi$ and one of the sites with phase $0$, (3) $[0, 0, 0]$.

\begin{figure}[tbh]
\centering
\includegraphics[width=5cm,height=4cm]{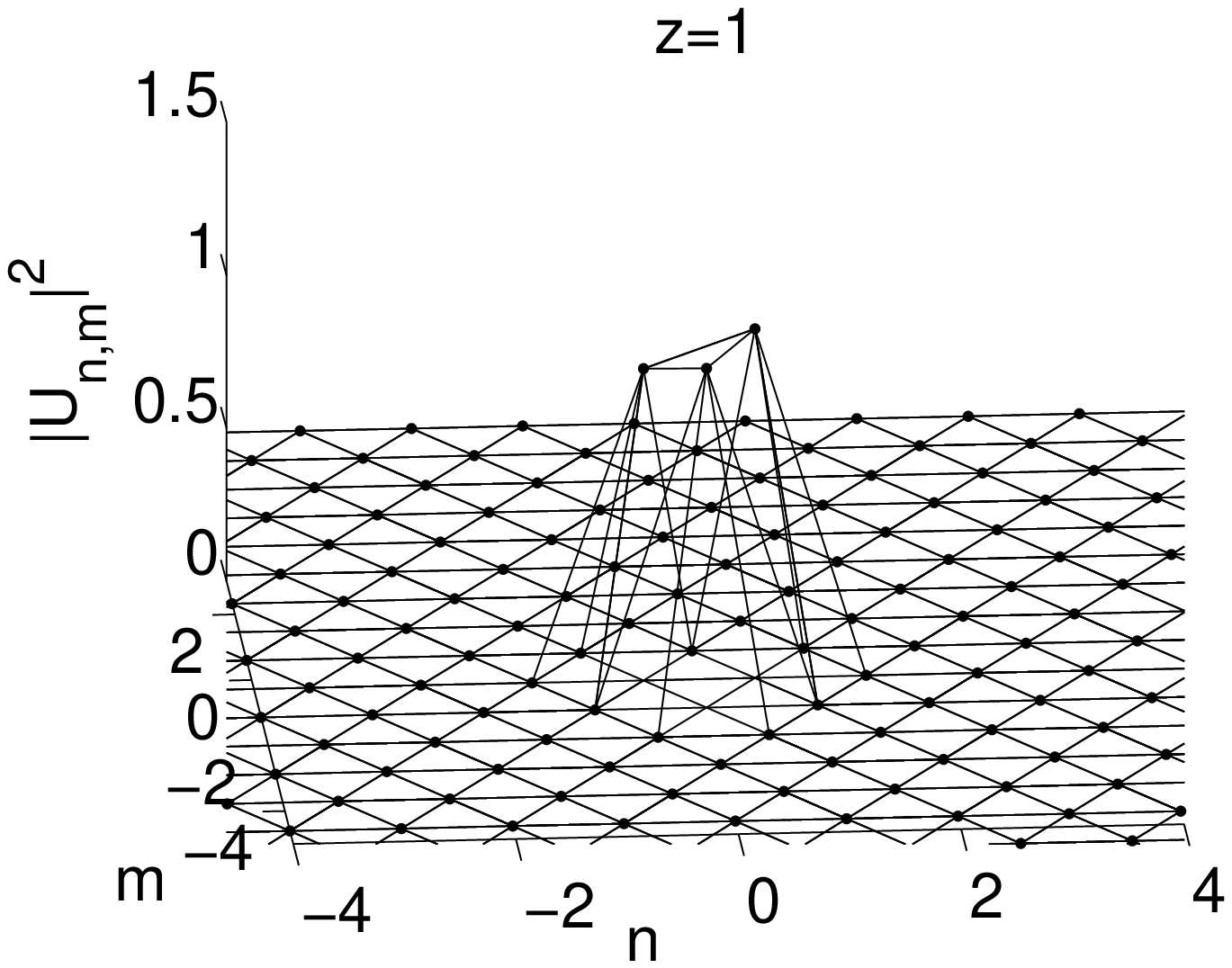}
\includegraphics[width=5cm,height=4cm]{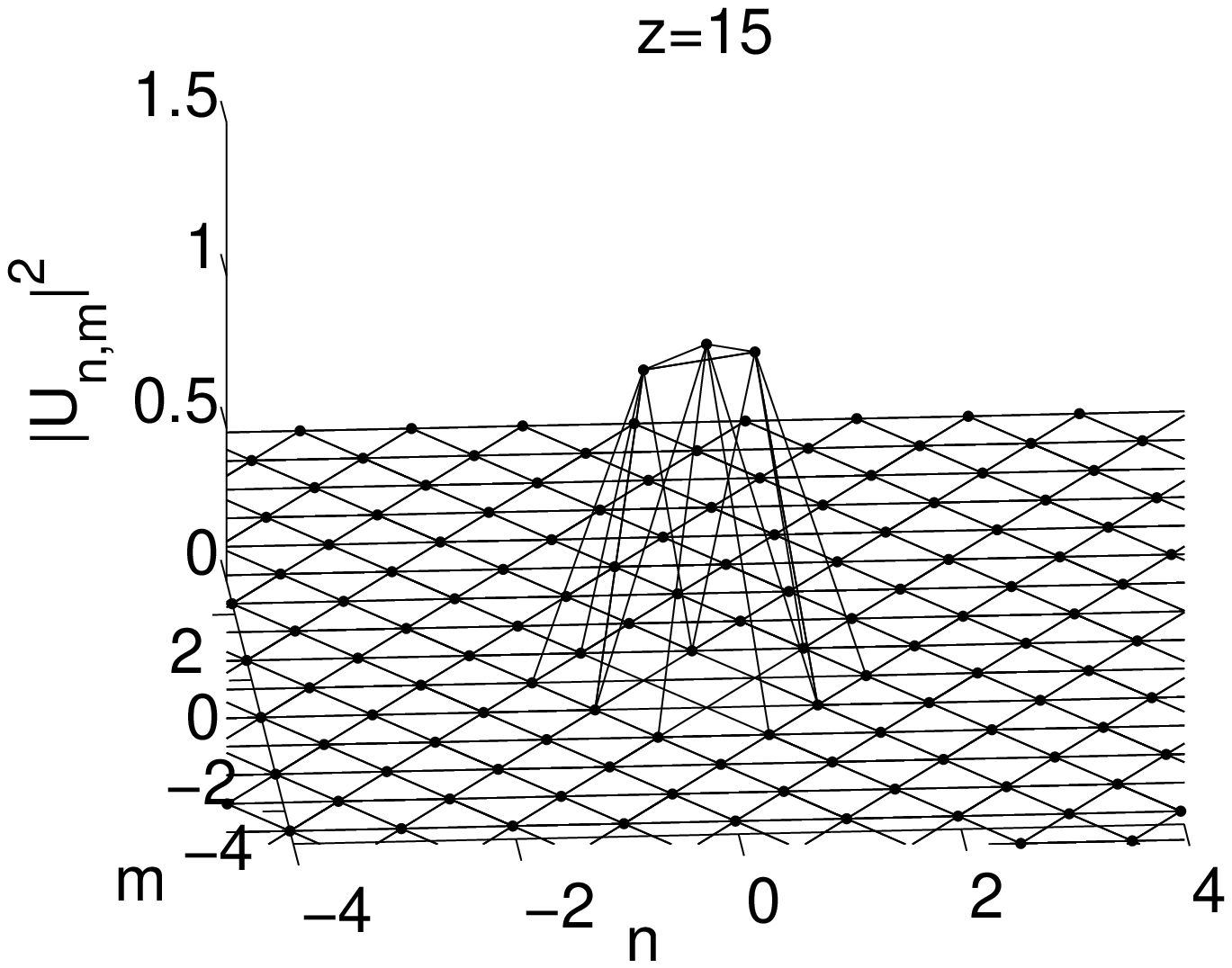}\\
\includegraphics[width=5cm,height=4cm]{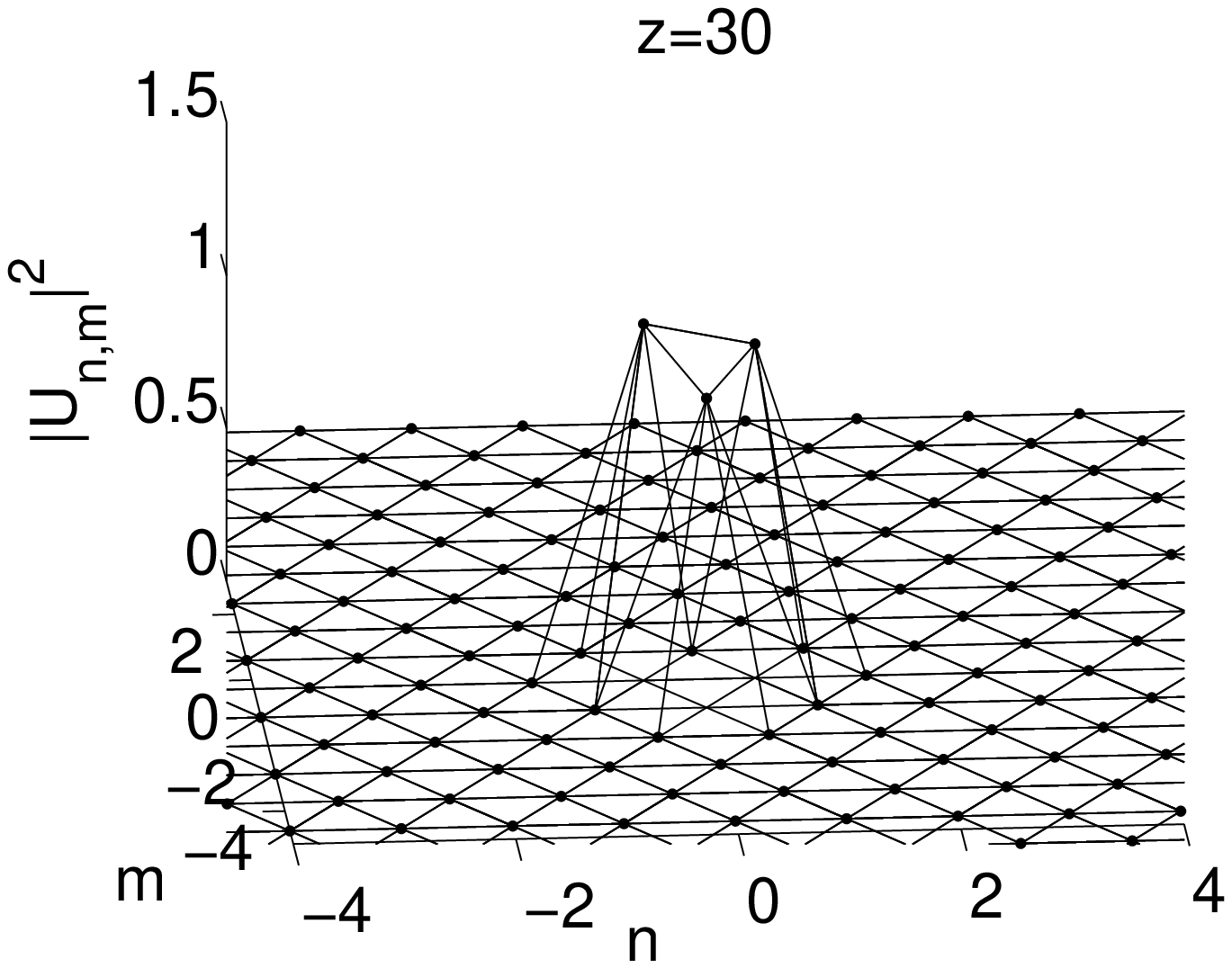}
\includegraphics[width=5cm,height=4cm]{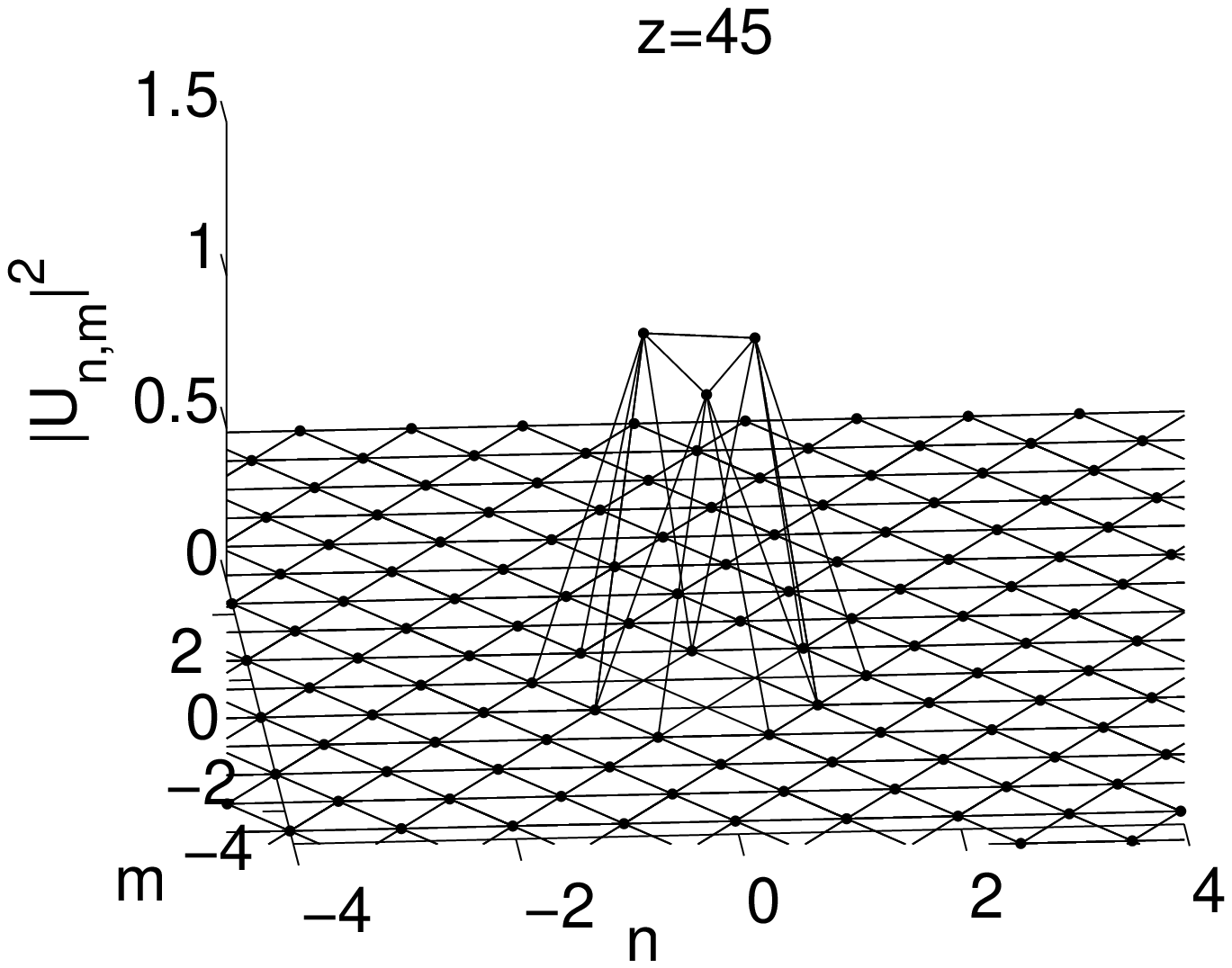}\\
\includegraphics[width=5cm,height=4cm]{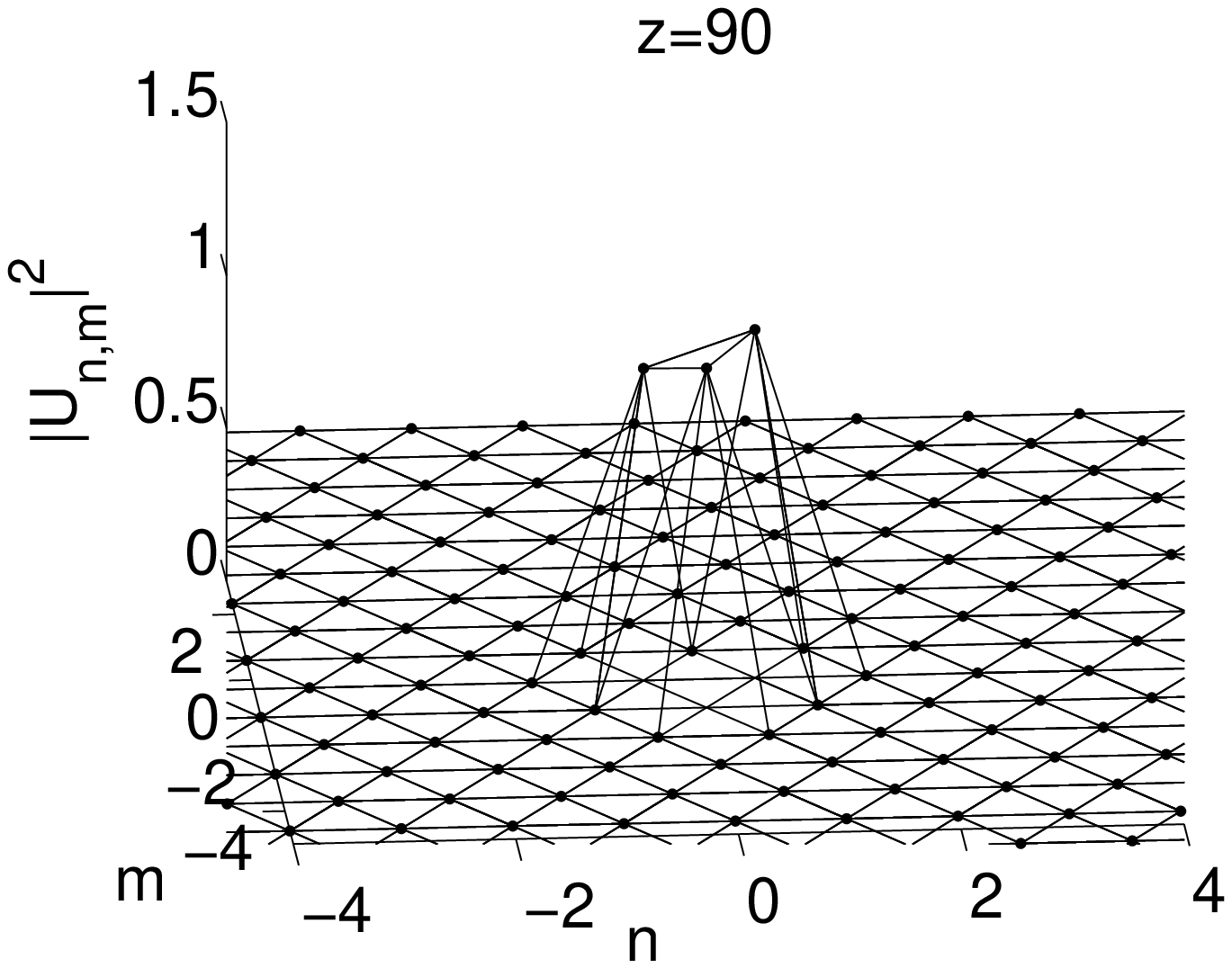}
\includegraphics[width=5cm,height=4cm]{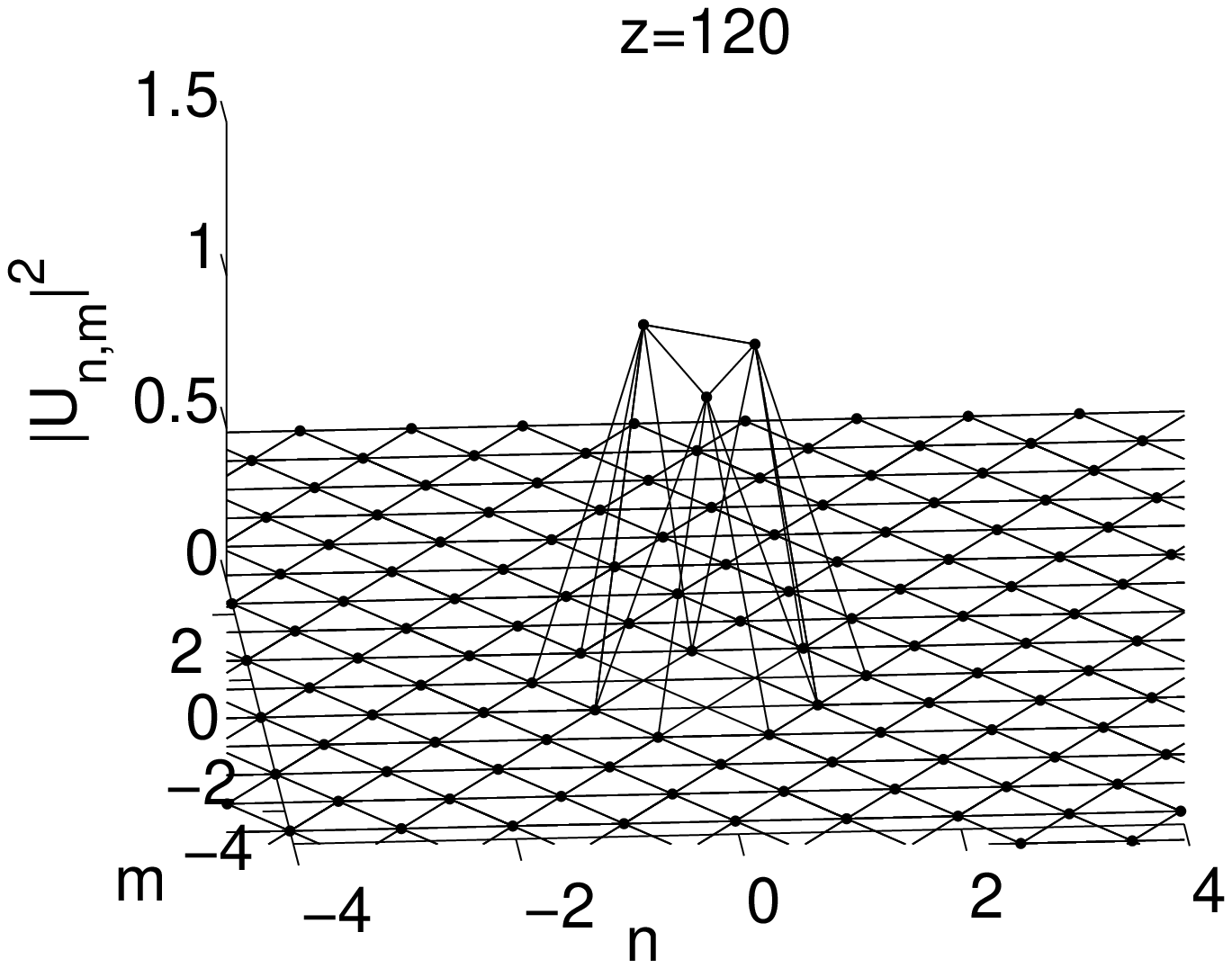}\\
\includegraphics[width=5cm,height=4cm]{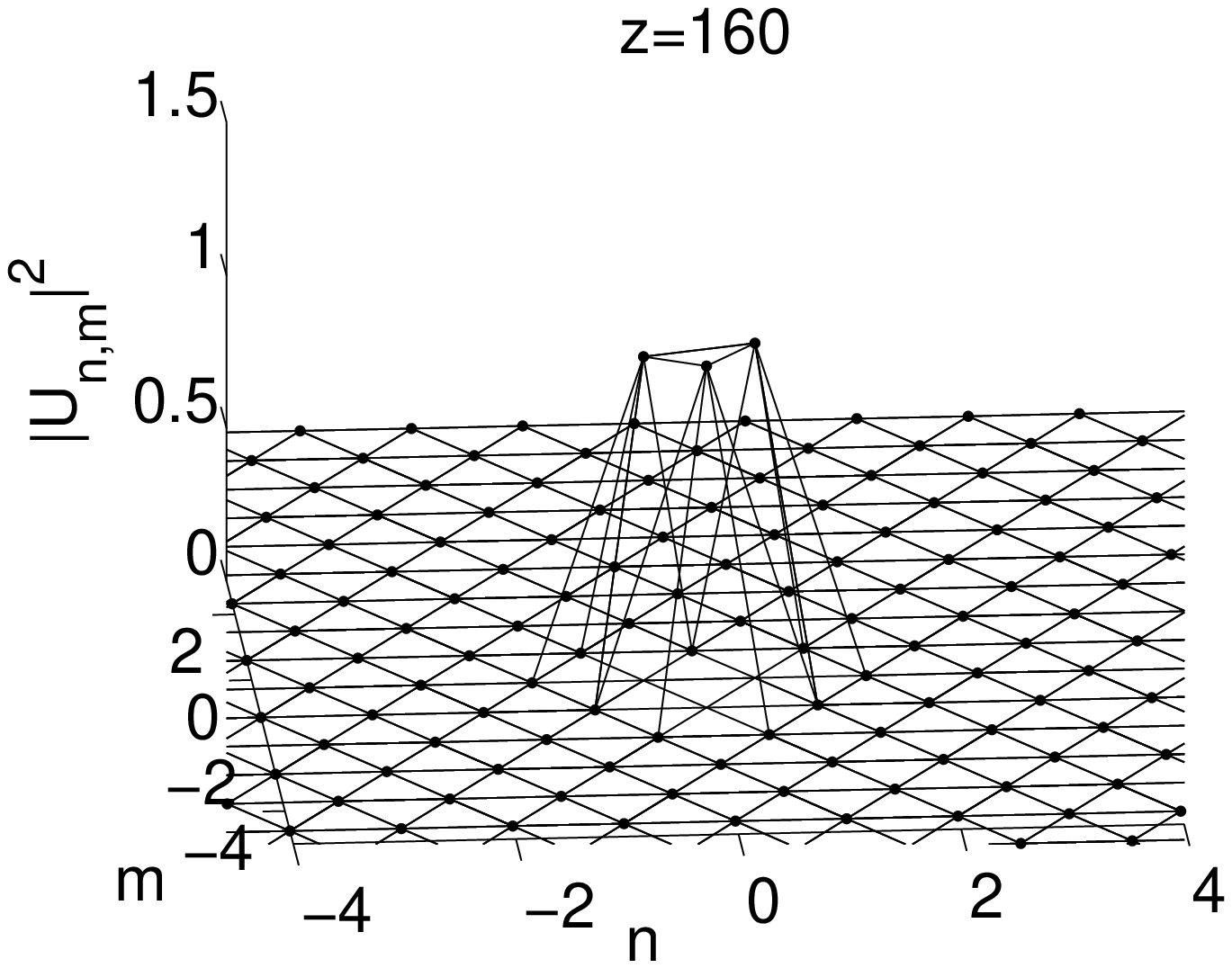}
\includegraphics[width=5cm,height=4cm]{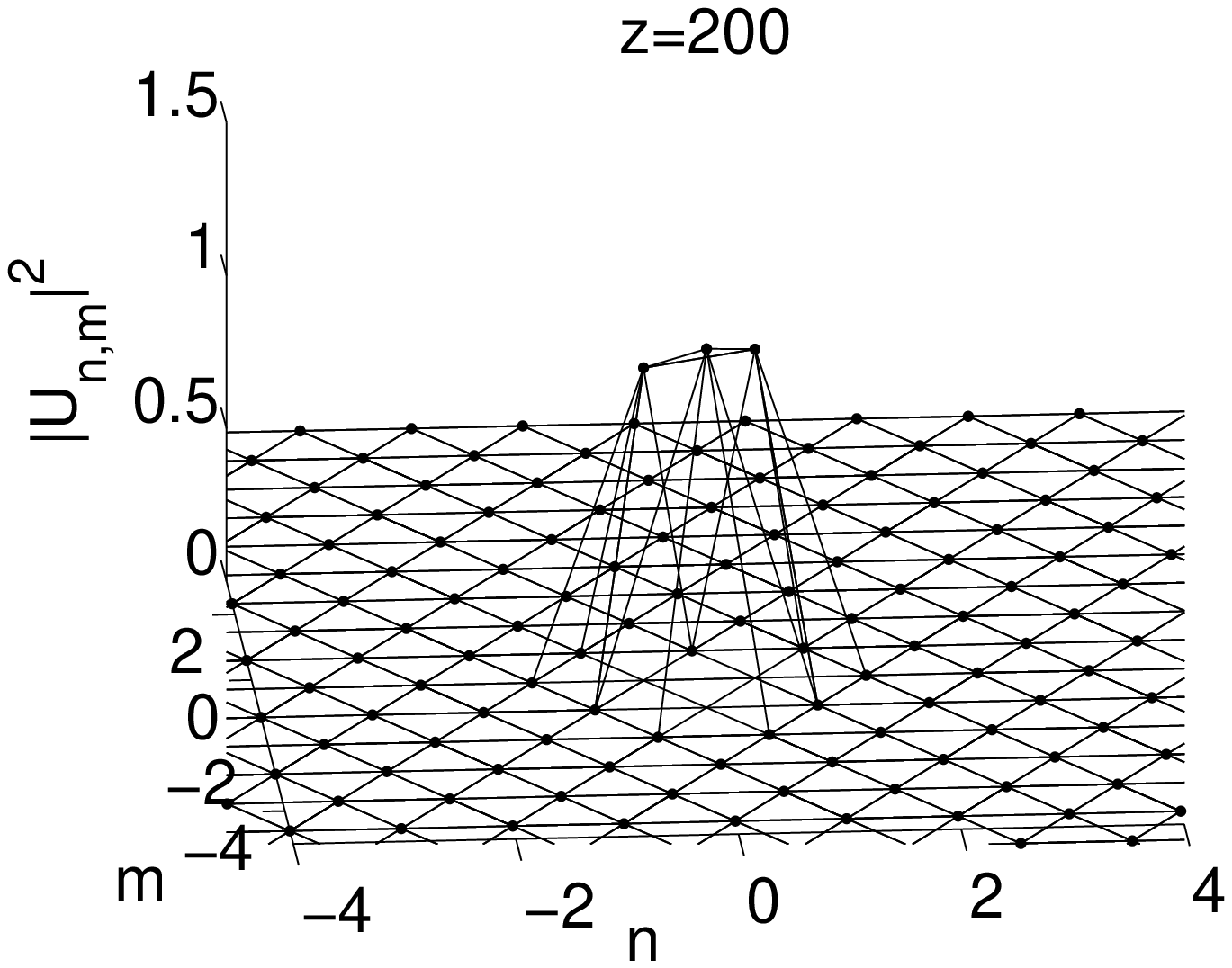}
\vspace{-0.4cm}
\caption{RK$4$ results from the hexagonal three-site $[0, \pi, 0]$ with anisotropy between the sites with phase $0$ at $\delta = 0.80$, $C=0.01$ at $z=1, 15, 30, 45, 90, 120, 160, 200$. The formation of a breathing pattern is clearly
observed.}
\label{fig:Hex3site_0-pi-0_alt_an0_8_C0_01_dyn}
\end{figure}

\begin{figure}[tbh]
\centering
\includegraphics[width=5cm,height=4cm]{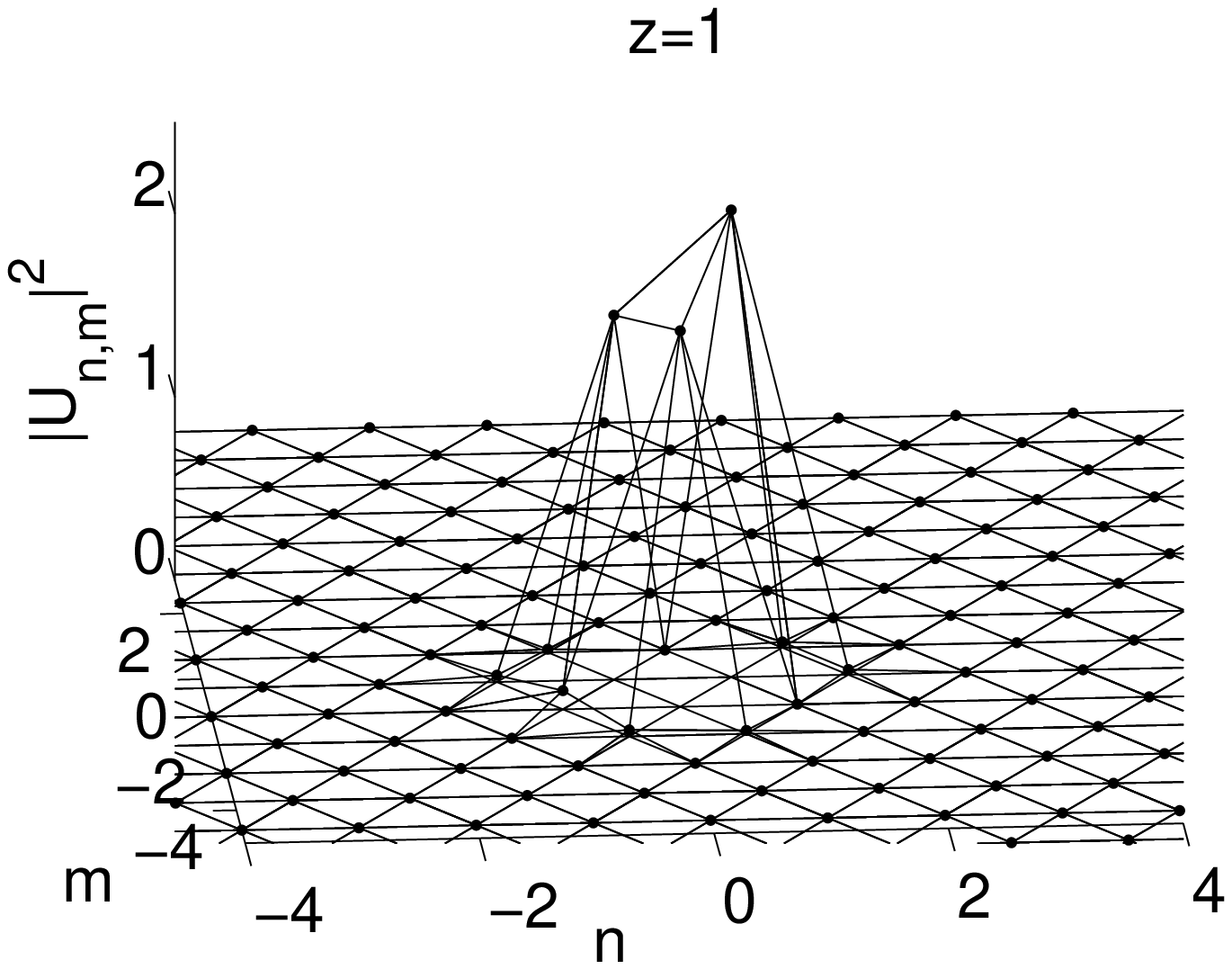}
\includegraphics[width=5cm,height=4cm]{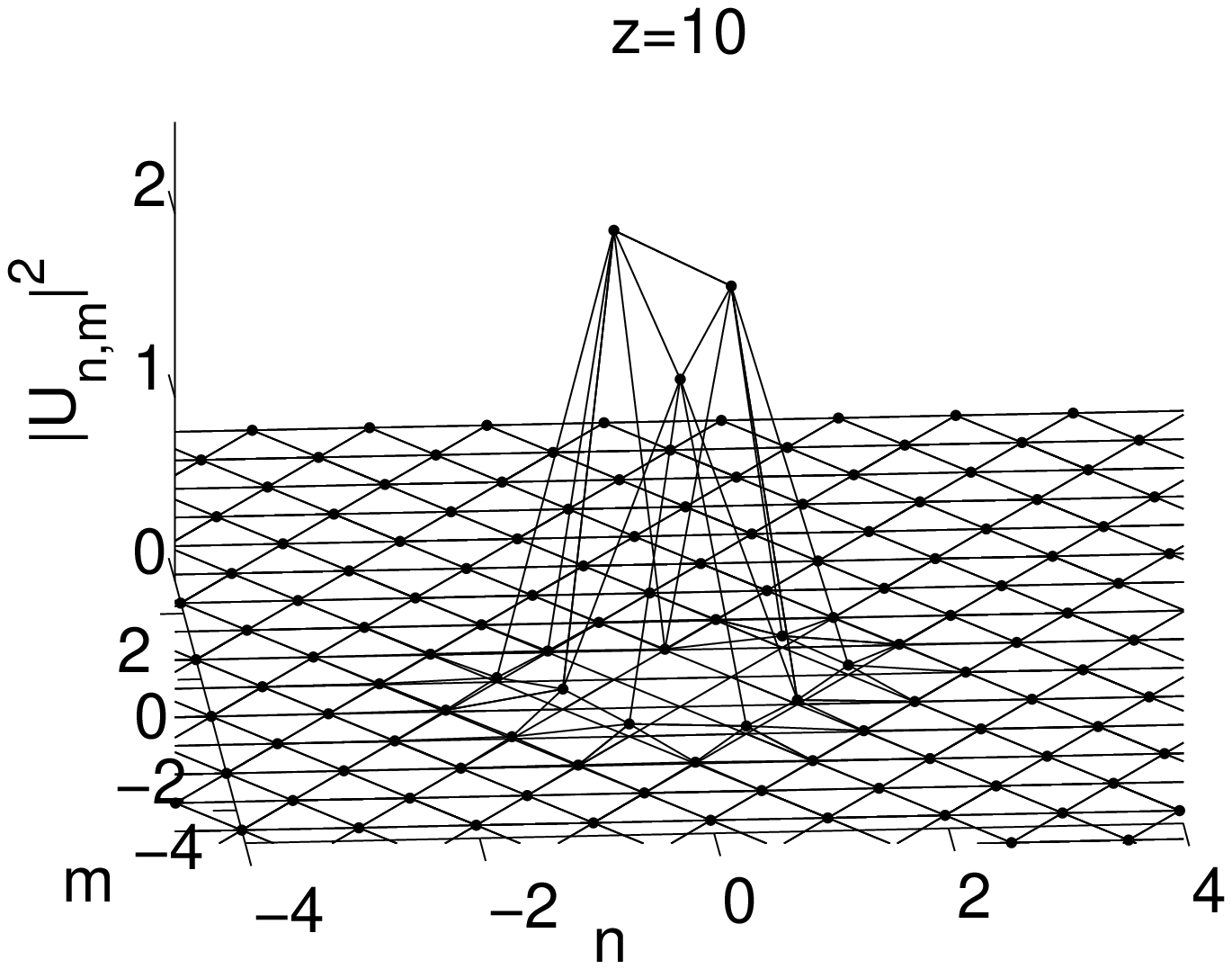}\\
\includegraphics[width=5cm,height=4cm]{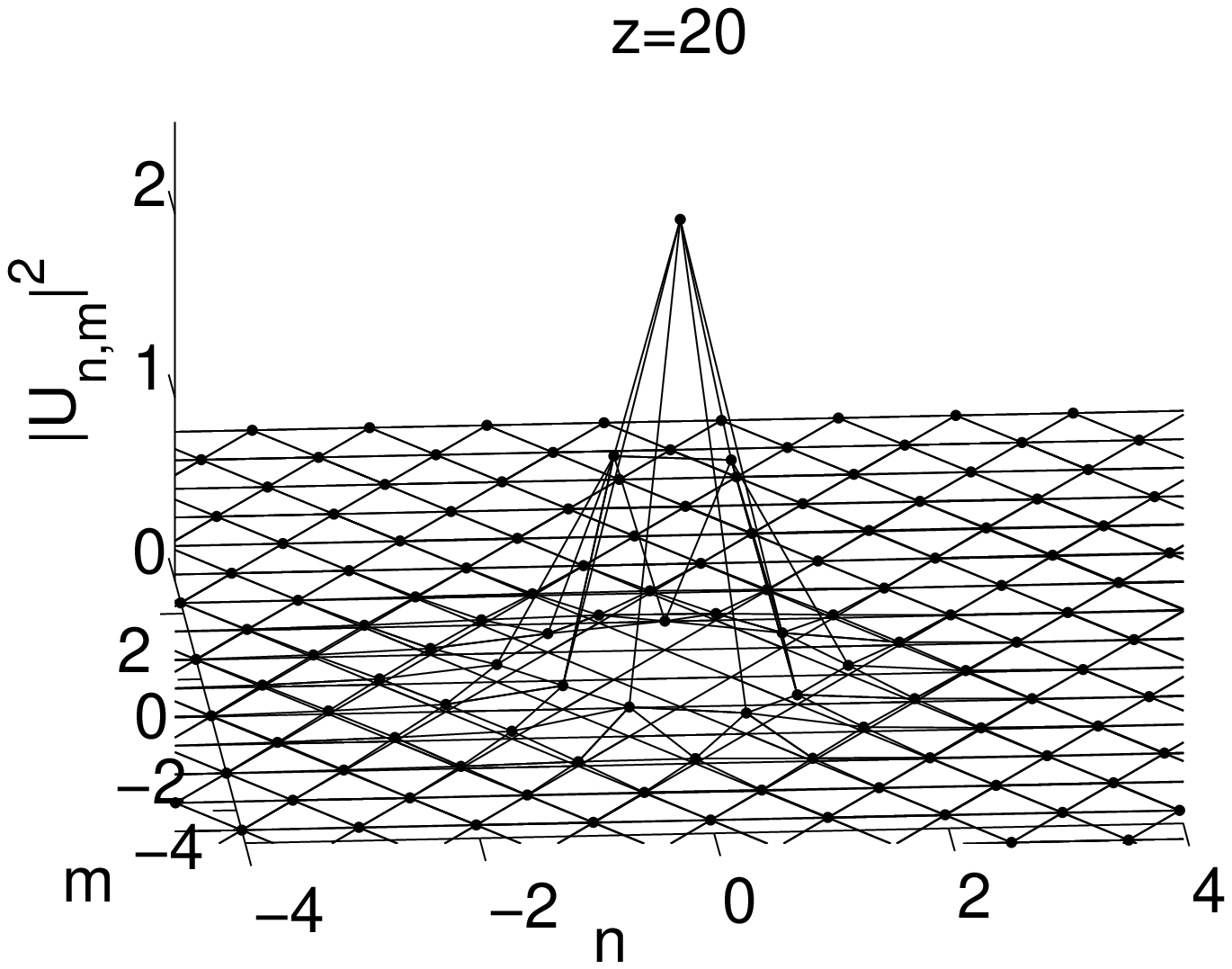}
\includegraphics[width=5cm,height=4cm]{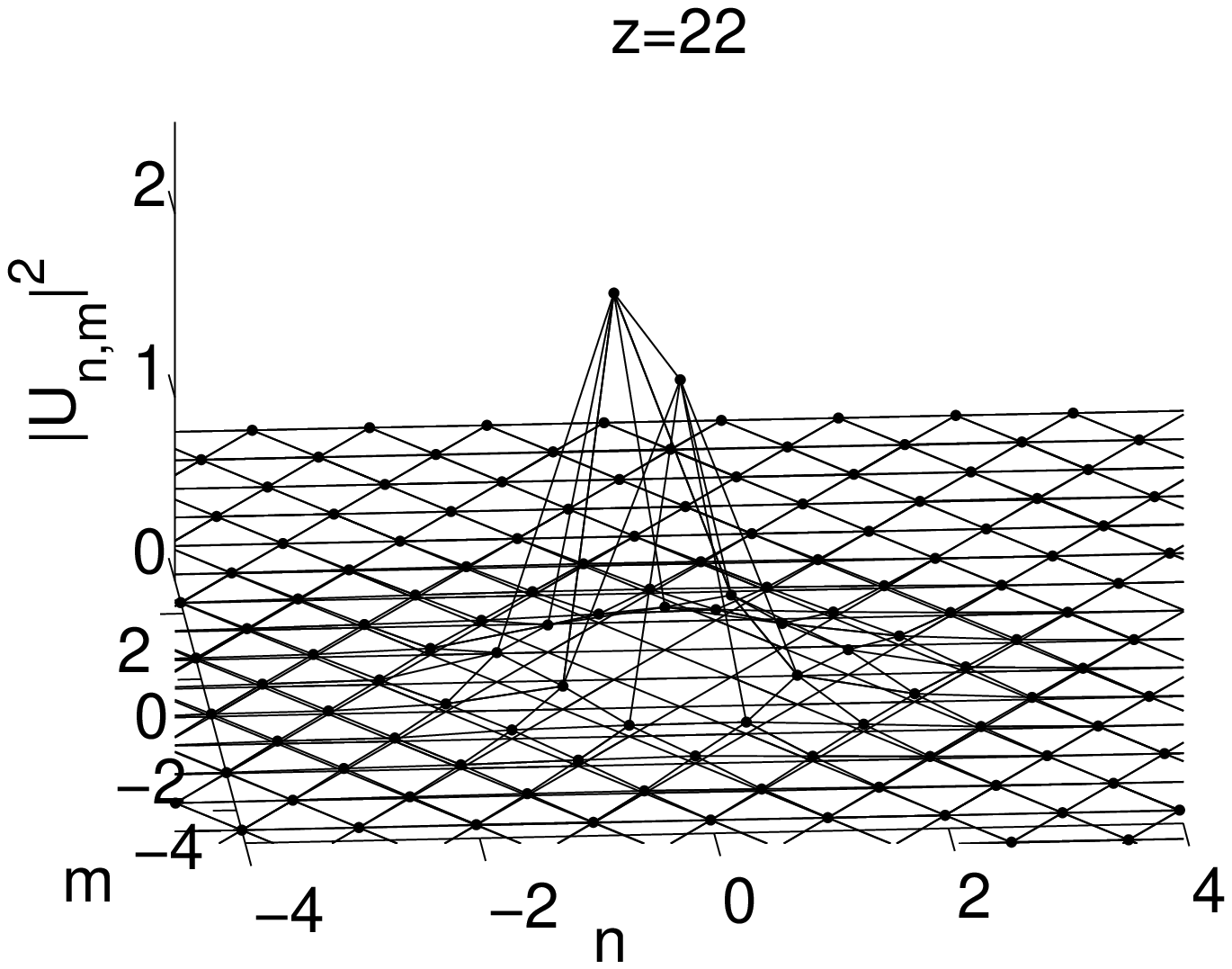}\\
\includegraphics[width=5cm,height=4cm]{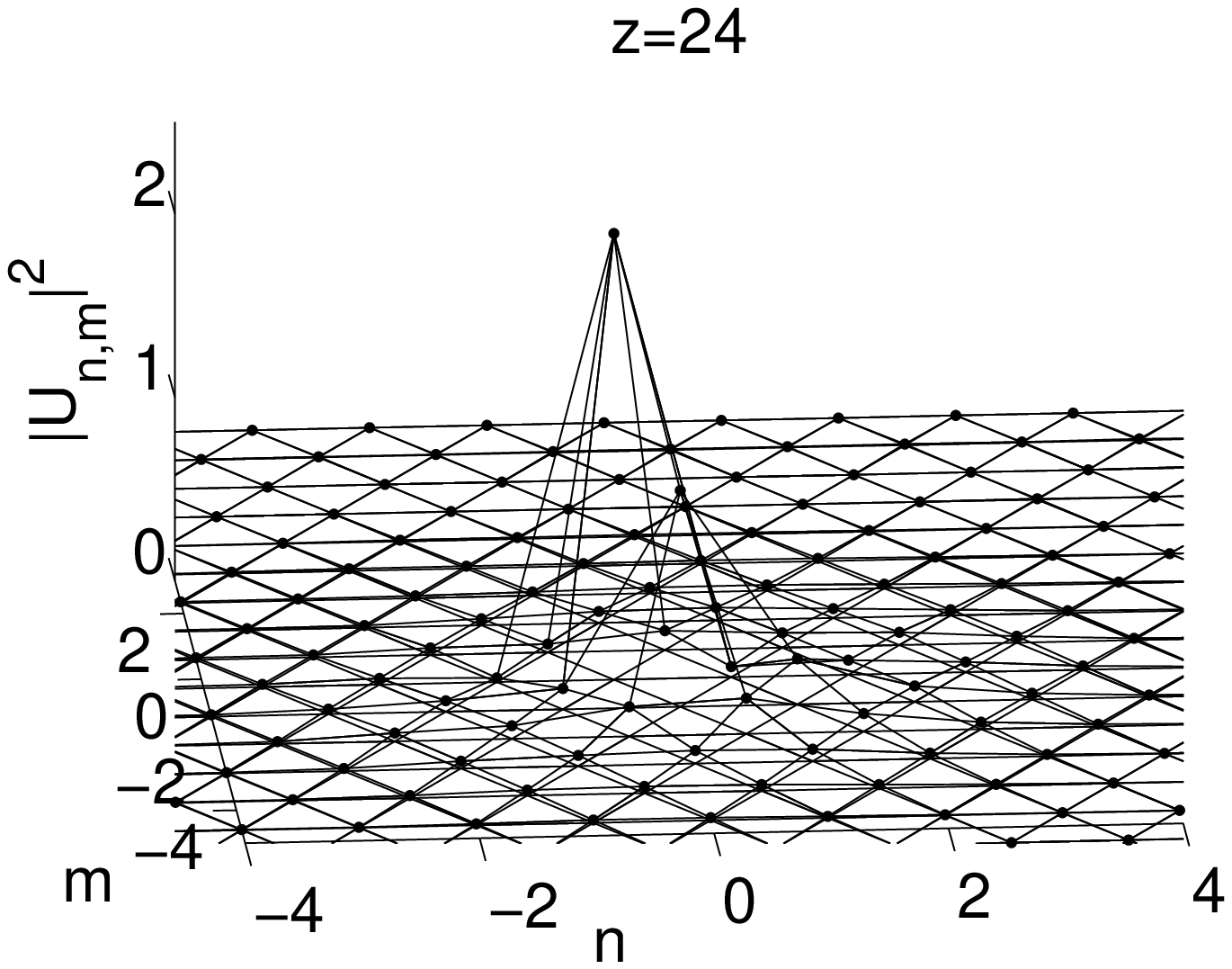}
\includegraphics[width=5cm,height=4cm]{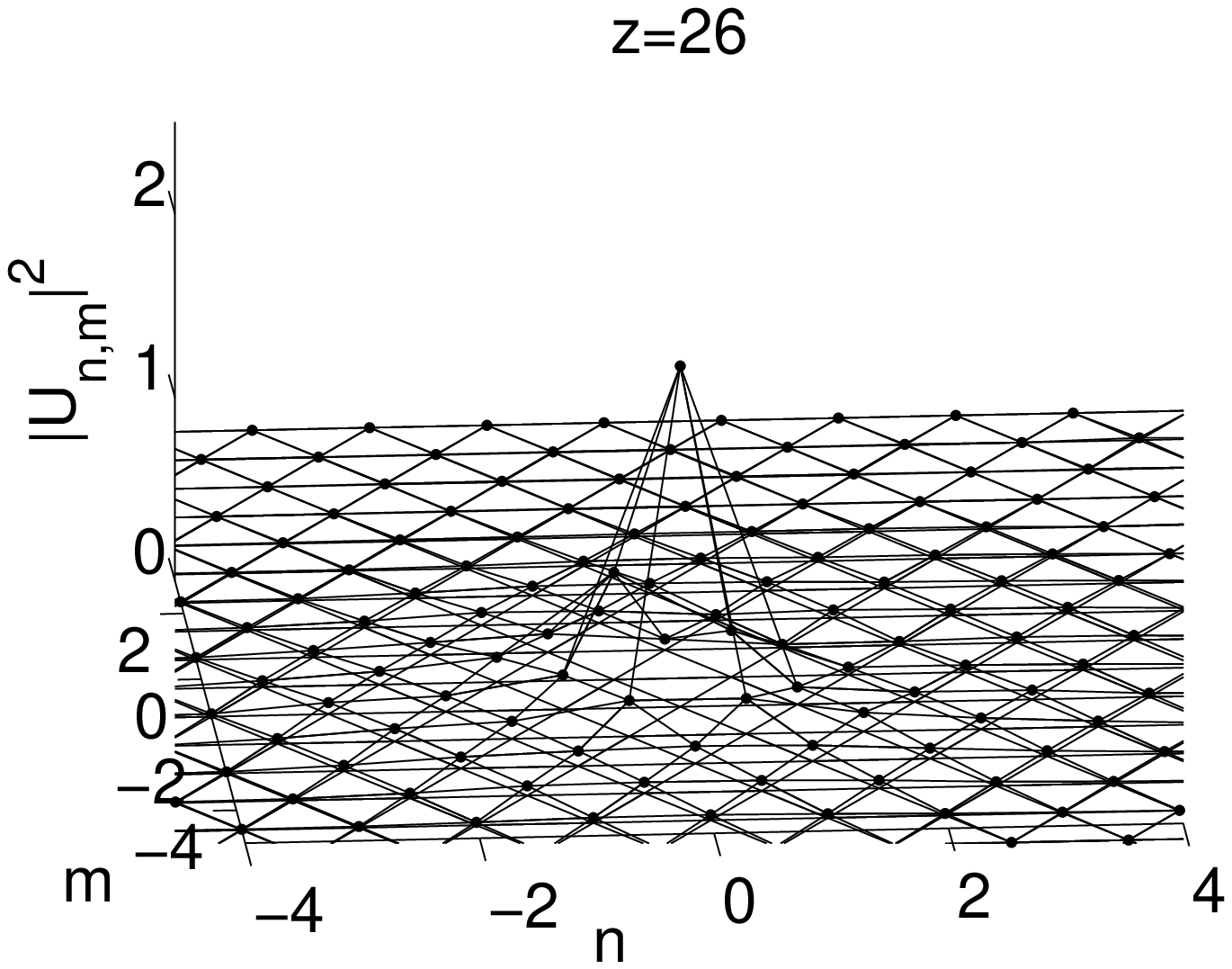}\\
\includegraphics[width=5cm,height=4cm]{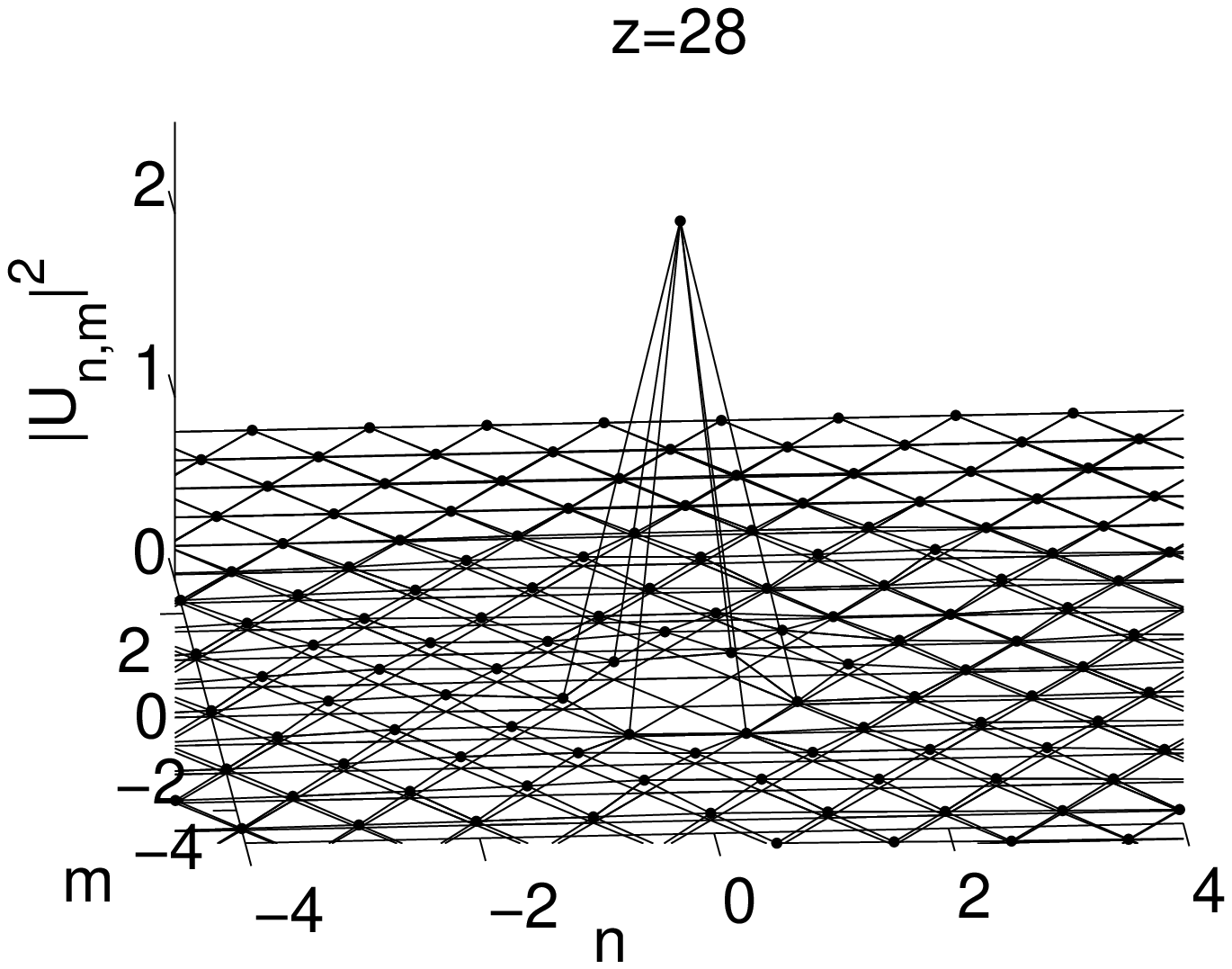}
\includegraphics[width=5cm,height=4cm]{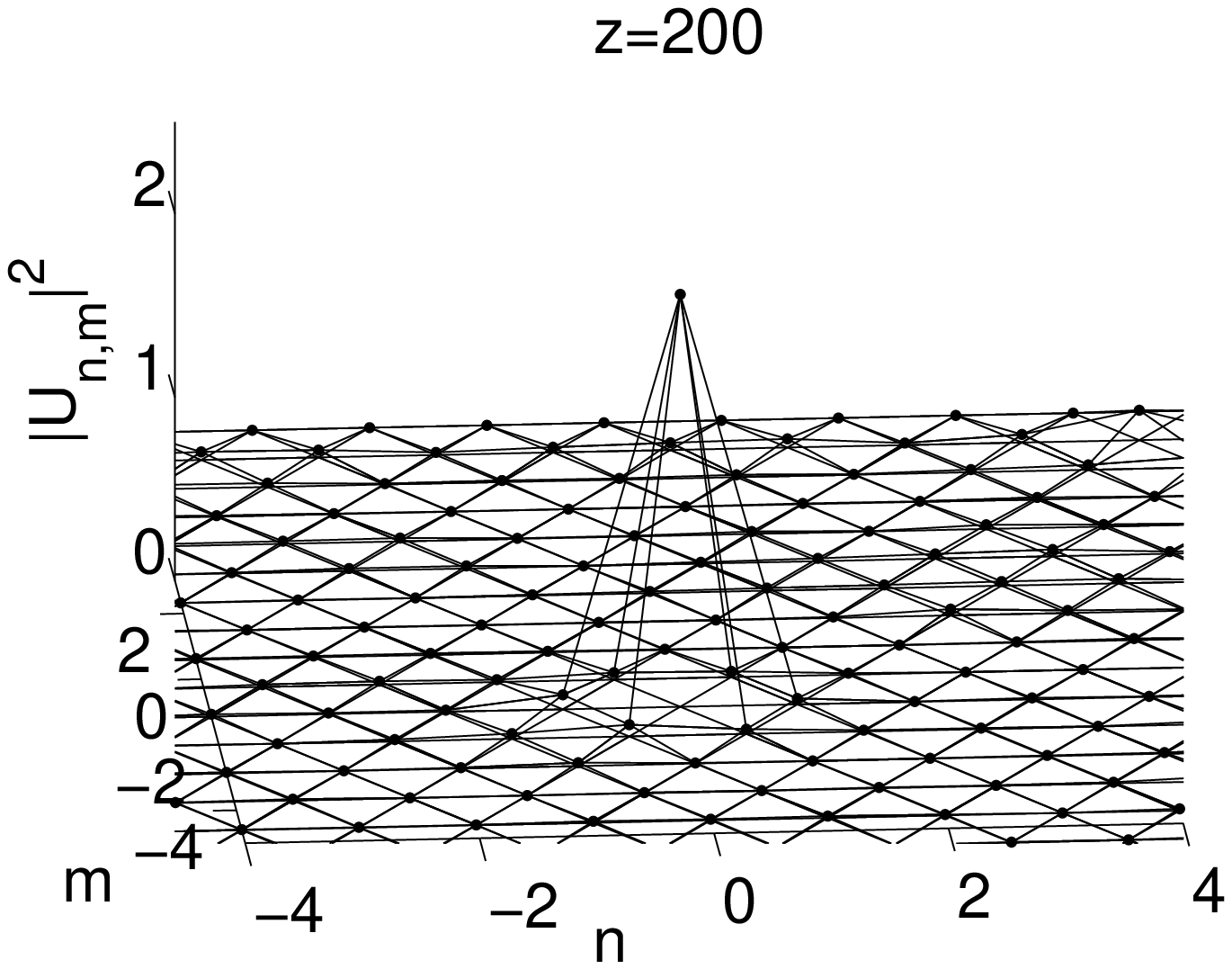}
\vspace{-0.4cm}
\caption{RK$4$ results from the hexagonal three-site $[0, \pi, 0]$ with anisotropy between the sites with phase $0$ at $\delta = 0.80$, 
$\varepsilon=0.2$ at $z=1, 10, 20, 22, 24, 26, 28, 200$. The emergence of a localized state centered on
a single site is clearly evident.}
\label{fig:Hex3site_0-pi-0_alt_an0_8_C0_2_dyn}
\end{figure}

(1) Figures \ref{fig:Hex3site_0-pi-0_alt_an0_8_C0_01_dyn} and 
\ref{fig:Hex3site_0-pi-0_alt_an0_8_C0_2_dyn} exhibit the dynamics
in the anisotropic hexagonal lattice for $[0, \pi, 0]$ when the 
anisotropy is prescribed to be between the sites with phase $0$ with
$\varepsilon = 0.01$ and $\varepsilon = 0.2$ respectively.
In both instances, we use $\delta = 0.8$. In the case of the weaker 
coupling, the propagating solution shows oscillatory (i.e., breathing)
behavior. 
In fact, we observe similar features for the weak coupling
case of $\varepsilon = 0.01$ for all the 
unstable solutions studied here. For the larger coupling of 
$\varepsilon = 0.2$ in ~Fig.~\ref{fig:Hex3site_0-pi-0_alt_an0_8_C0_2_dyn} 
we see a clear destruction of the original waveform and an
emergence of a single surviving site that persists. 
It may be interesting in future work to explore the transition 
regime between weak and strong coupling and the associated
implications for the nature of the resulting states.

(2) The $[0, \pi, 0]$ solution with anisotropy between the site 
with phase $\pi$ and one with $0$ is unstable 
for all values of $\delta$, as discussed previously. As in the previous 
example, for $\varepsilon = 0.01$ the dynamics shows an oscillatory
movement during propagation.
However, 
when the coupling is set to $\varepsilon = 0.2$, destruction 
of the wave is observed with, as before, a single site persisting 
for long times. Given the similarity of these findings to those of 
Figs.~\ref{fig:Hex3site_0-pi-0_alt_an0_8_C0_01_dyn} and 
\ref{fig:Hex3site_0-pi-0_alt_an0_8_C0_2_dyn} we omit them here.

\begin{figure}[tbh]
\centering
\includegraphics[width=5cm,height=4cm]{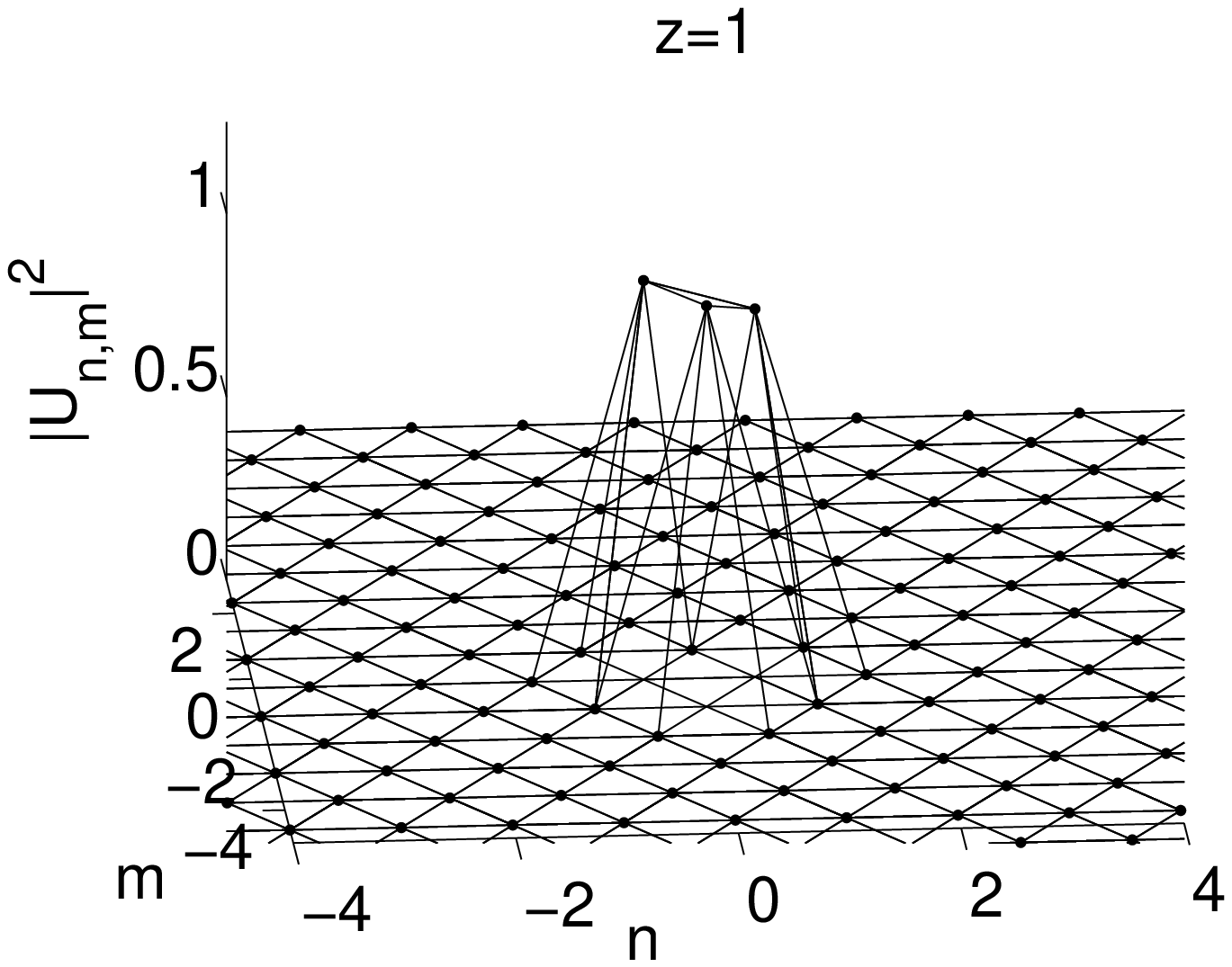}
\includegraphics[width=5cm,height=4cm]{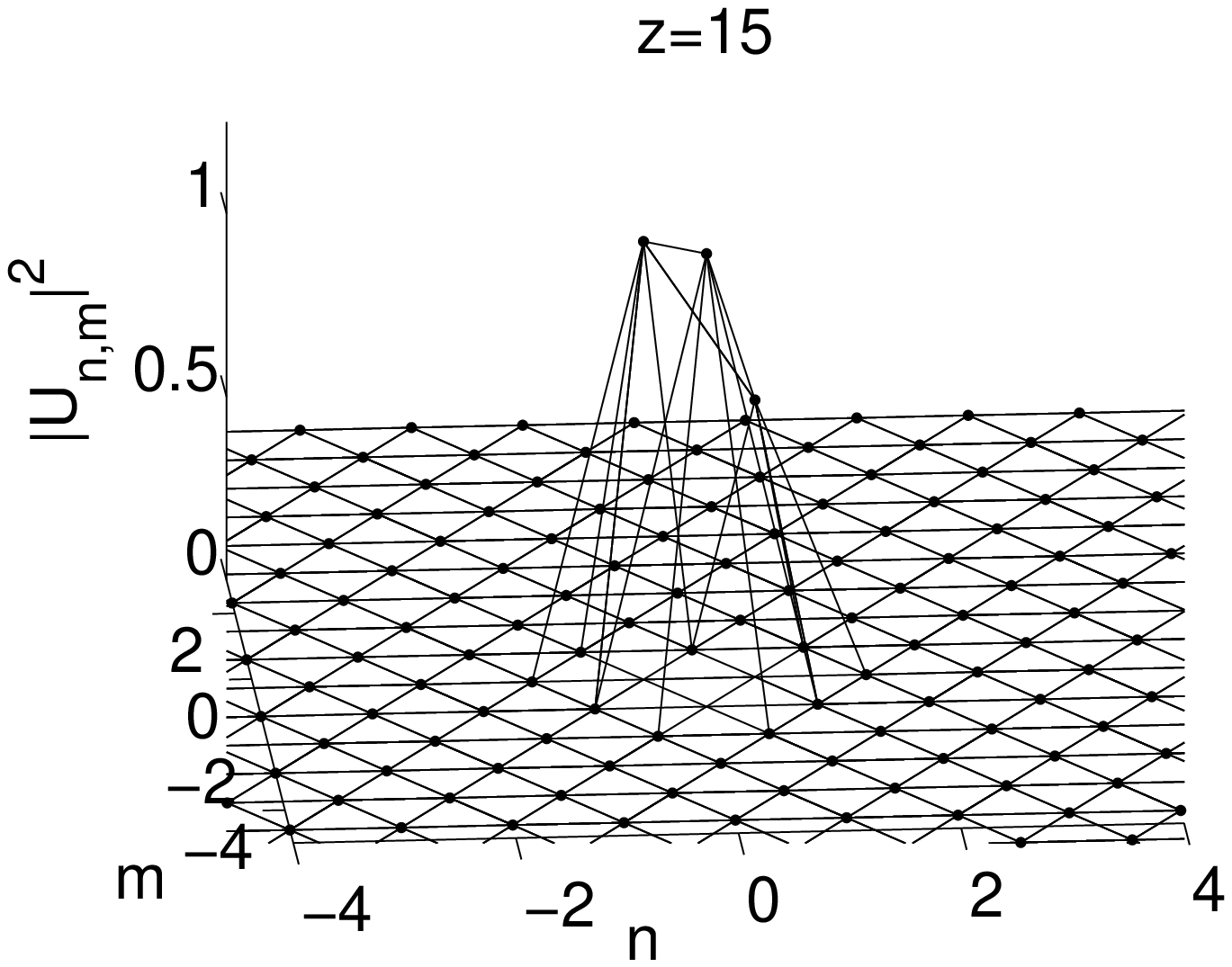}\\
\includegraphics[width=5cm,height=4cm]{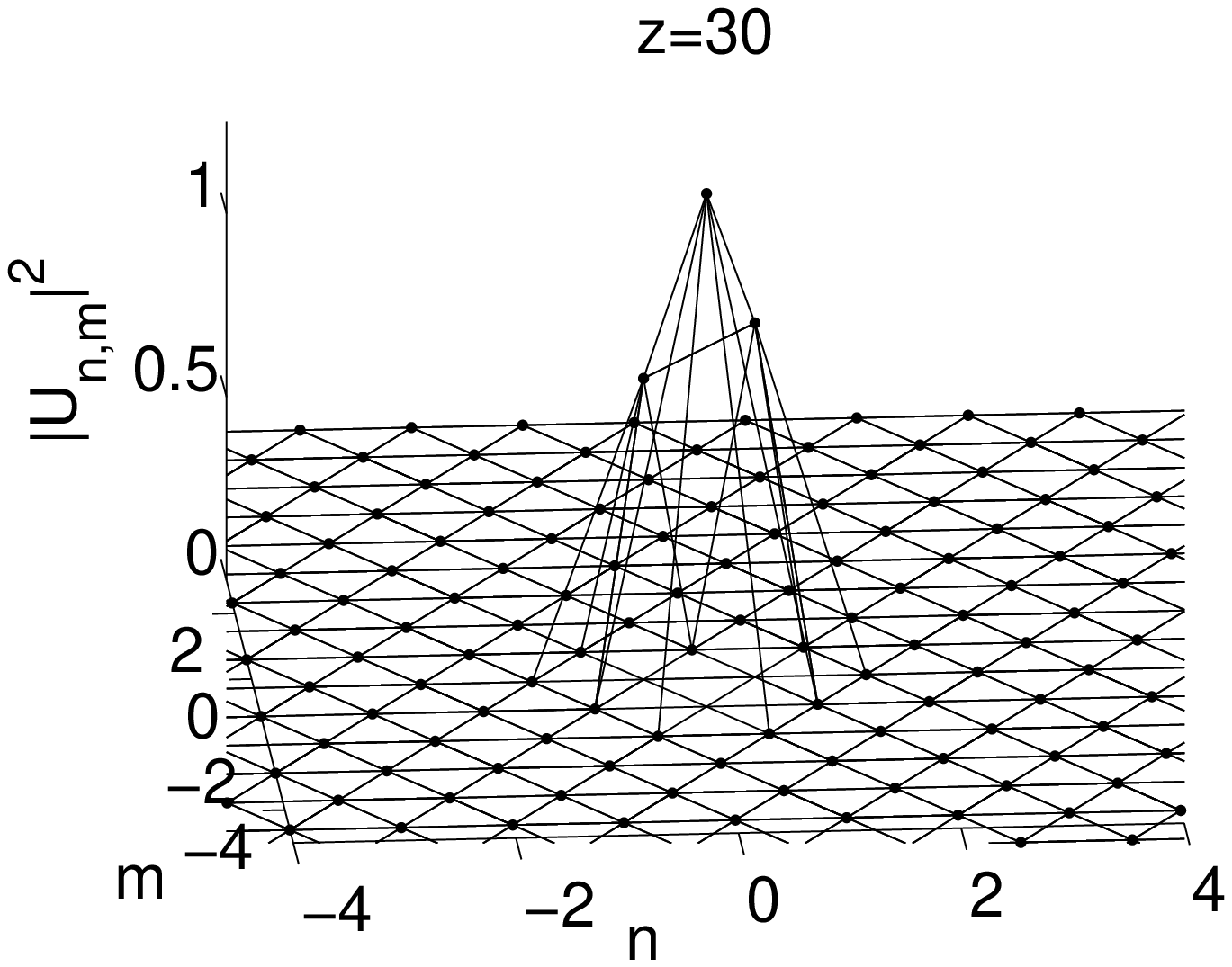}
\includegraphics[width=5cm,height=4cm]{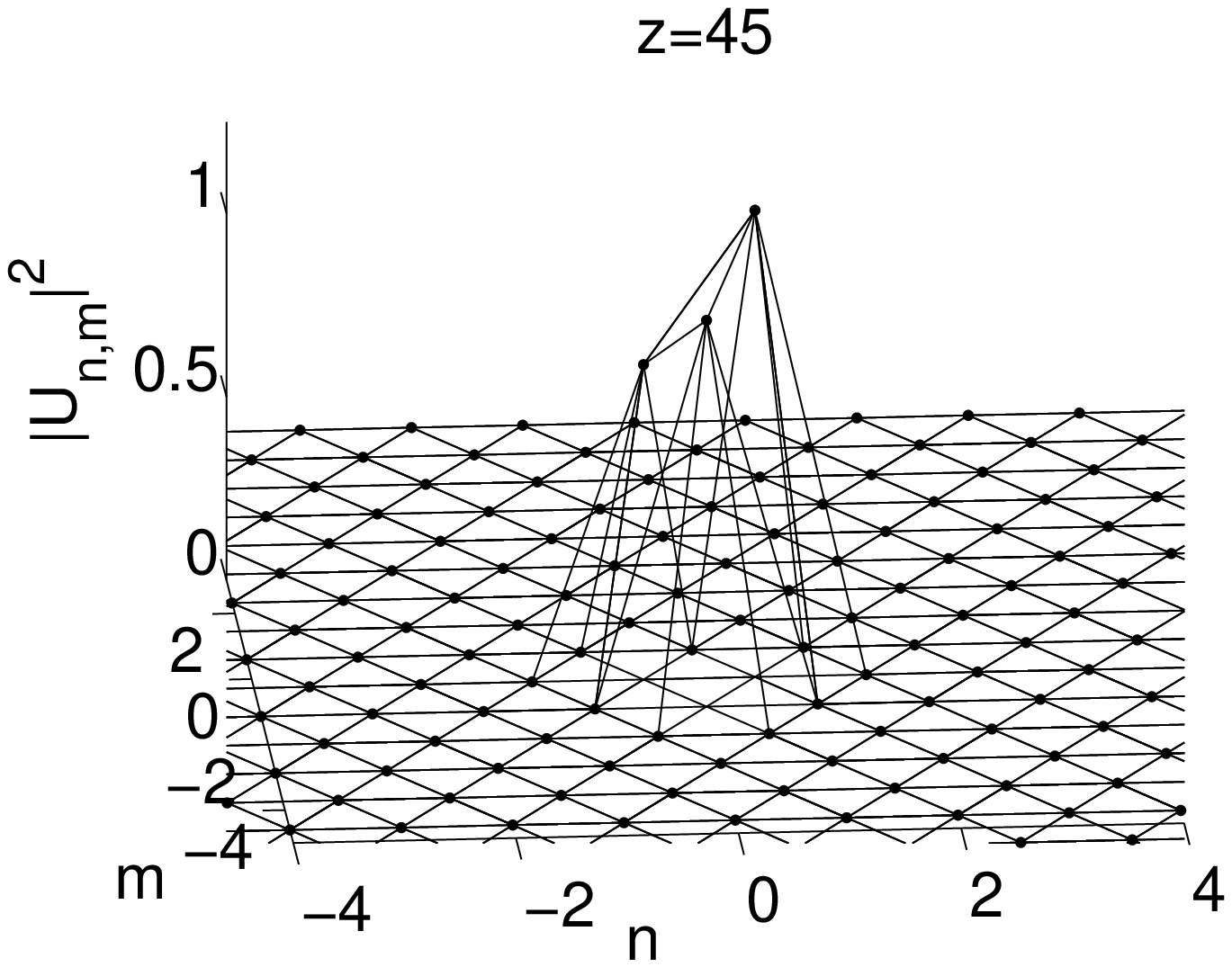}\\
\includegraphics[width=5cm,height=4cm]{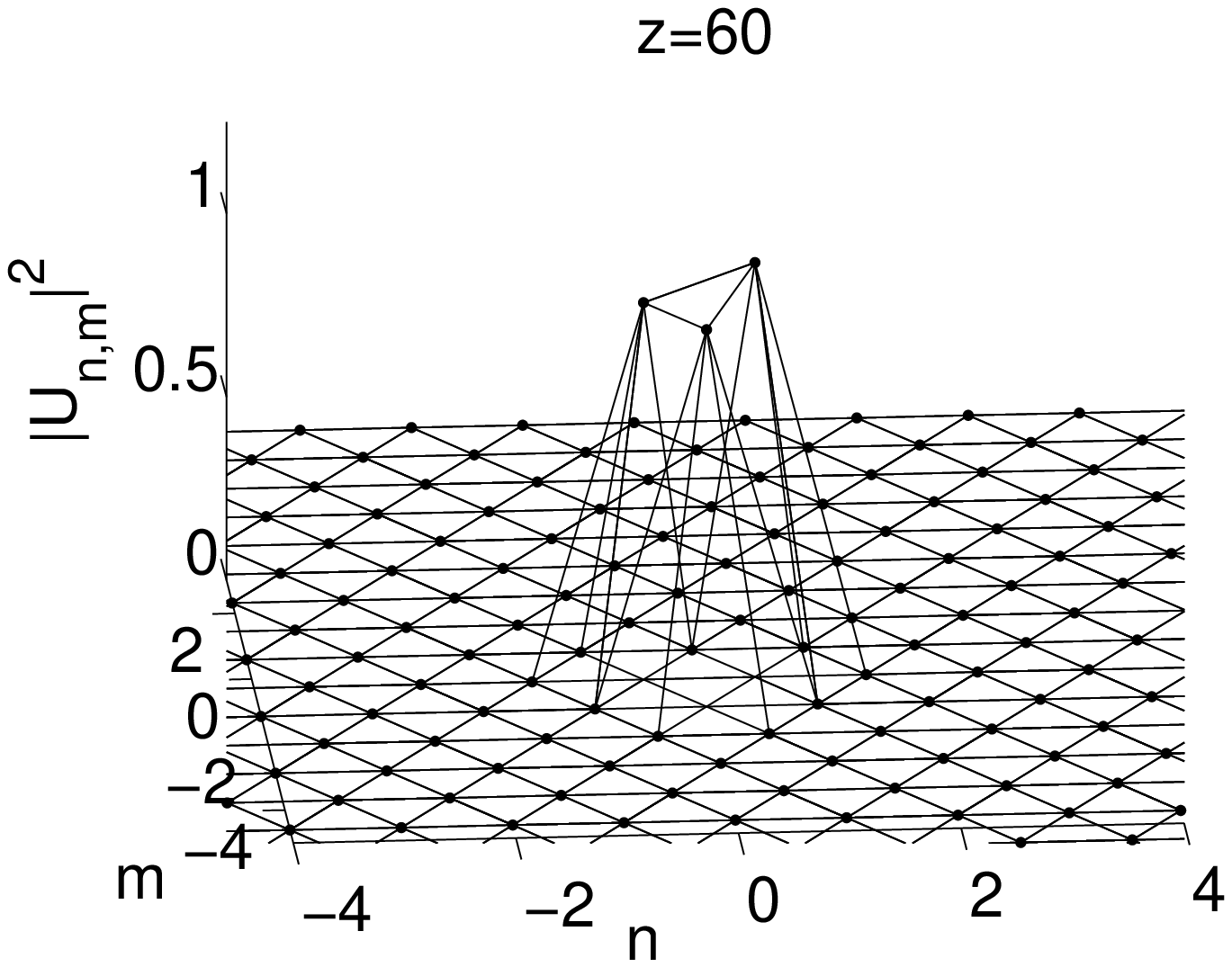}
\includegraphics[width=5cm,height=4cm]{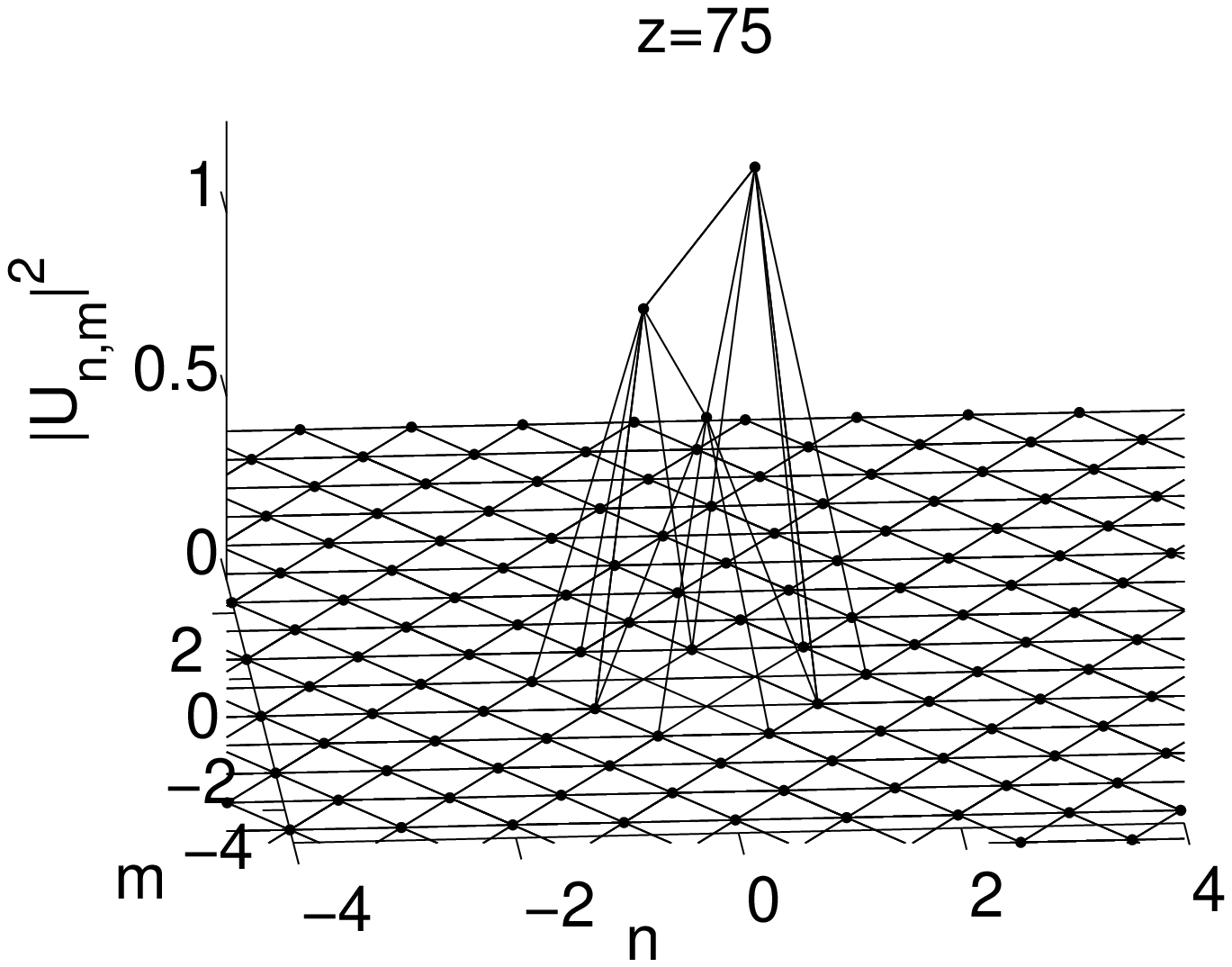}\\
\includegraphics[width=5cm,height=4cm]{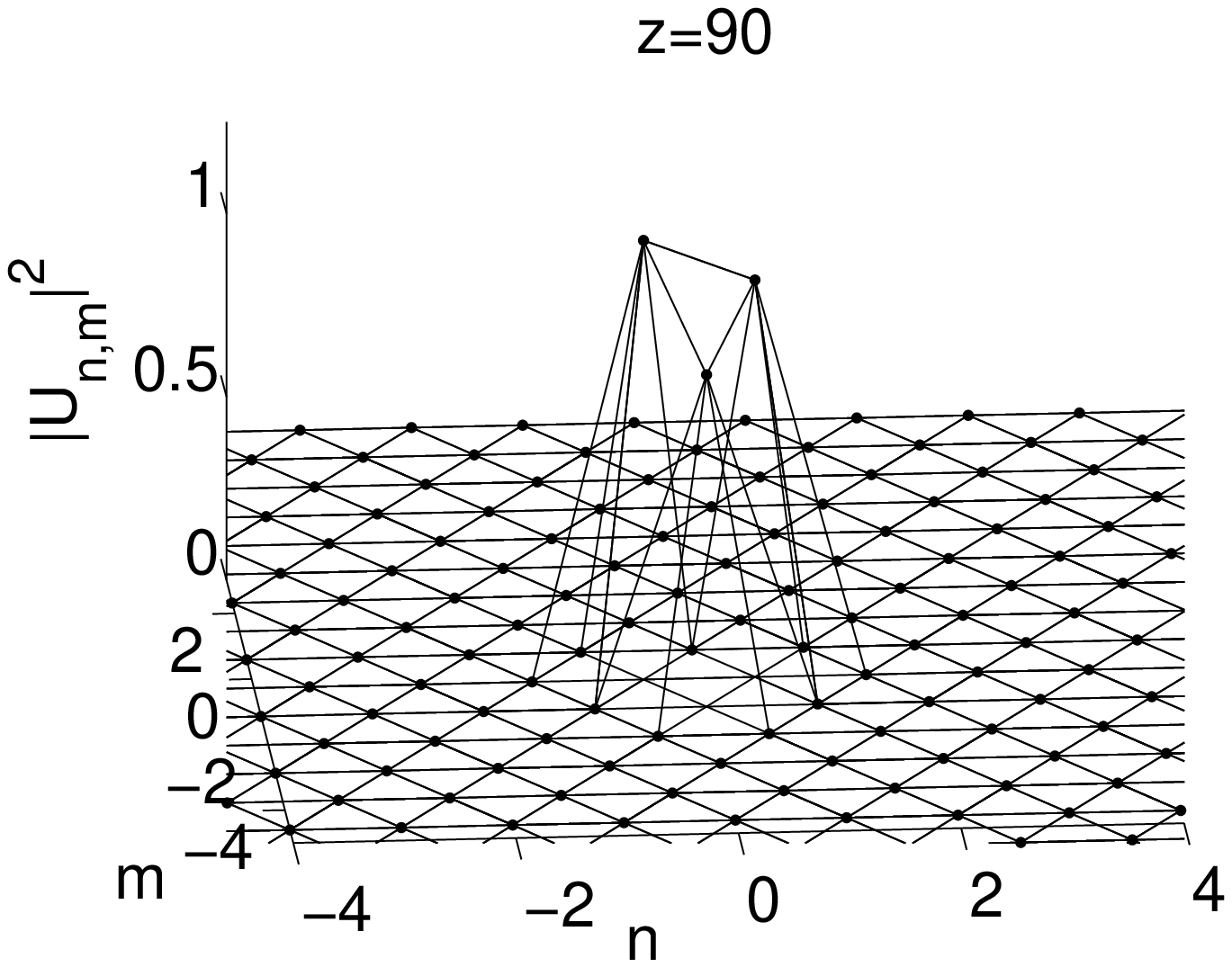}
\includegraphics[width=5cm,height=4cm]{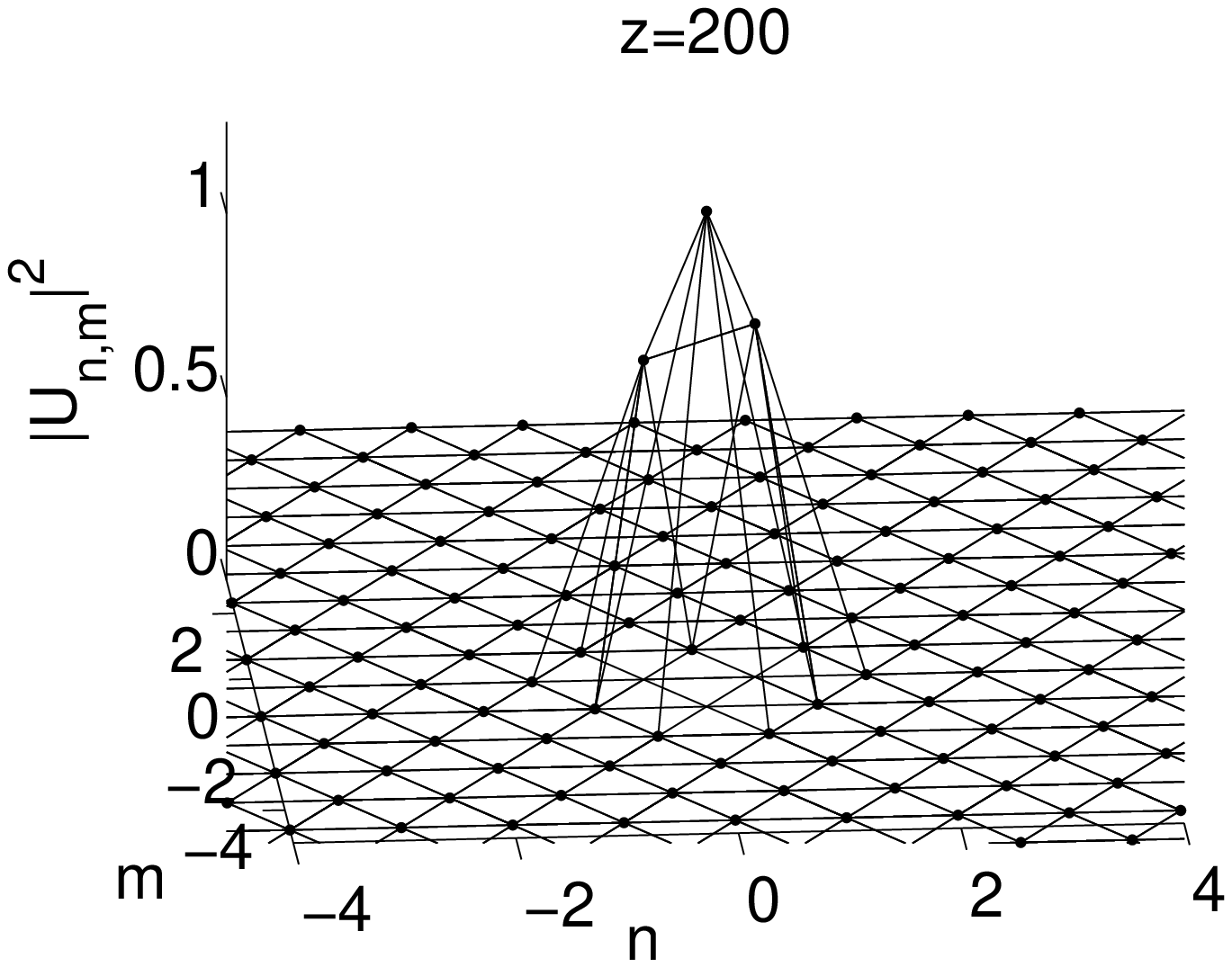}
\vspace{-0.4cm}
\caption{RK$4$ results from the hexagonal three-site $[0, 0, 0]$ at $\delta = 0.80$, $\varepsilon=0.01$ at $z=1, 15, 30, 45, 60, 75, 90, 200$.}
\label{fig:Hex3site_0-0-0_an0_8_C0_01_dyn}
\end{figure}

\begin{figure}[tbh]
\centering
\includegraphics[width=5cm,height=4cm]{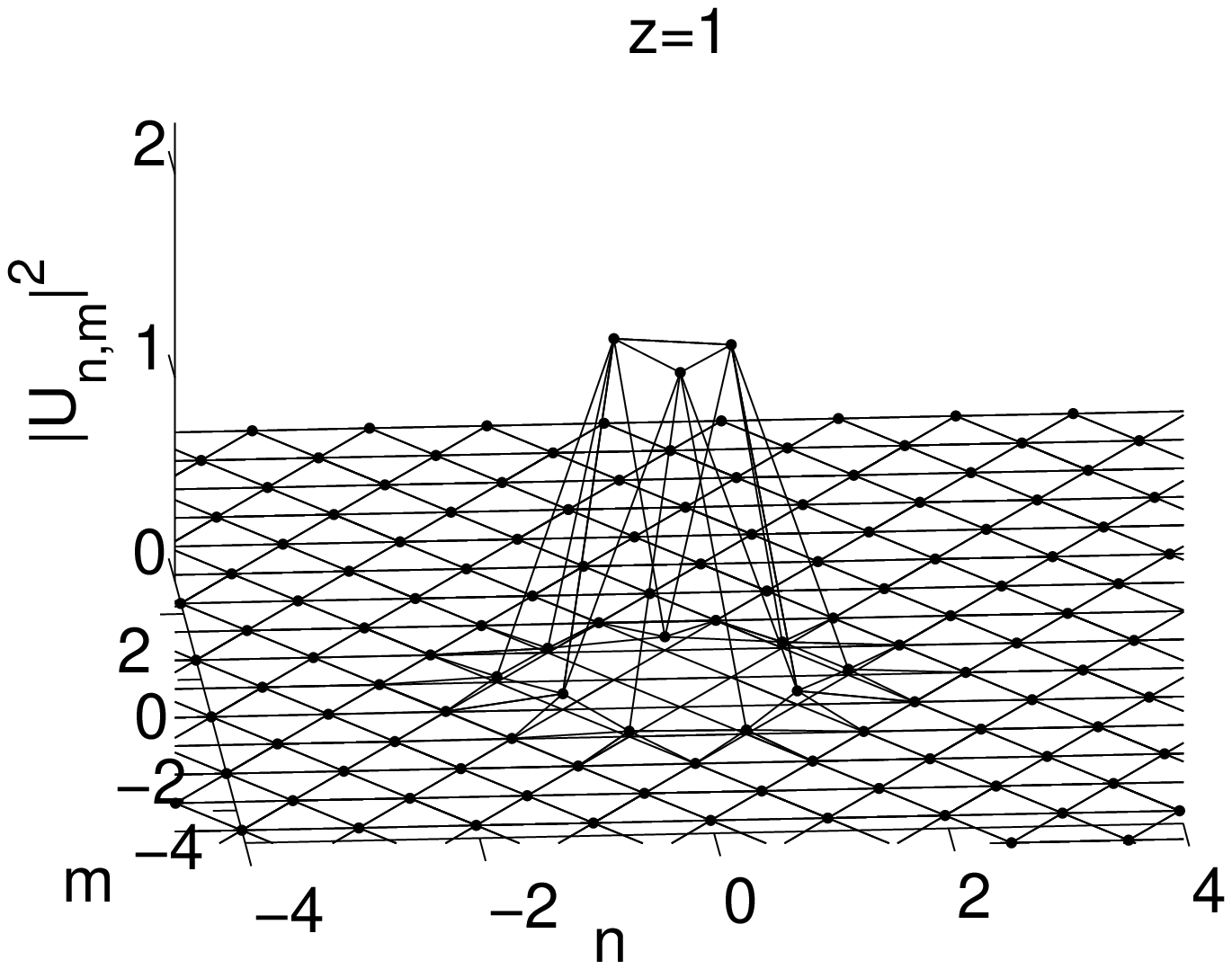}
\includegraphics[width=5cm,height=4cm]{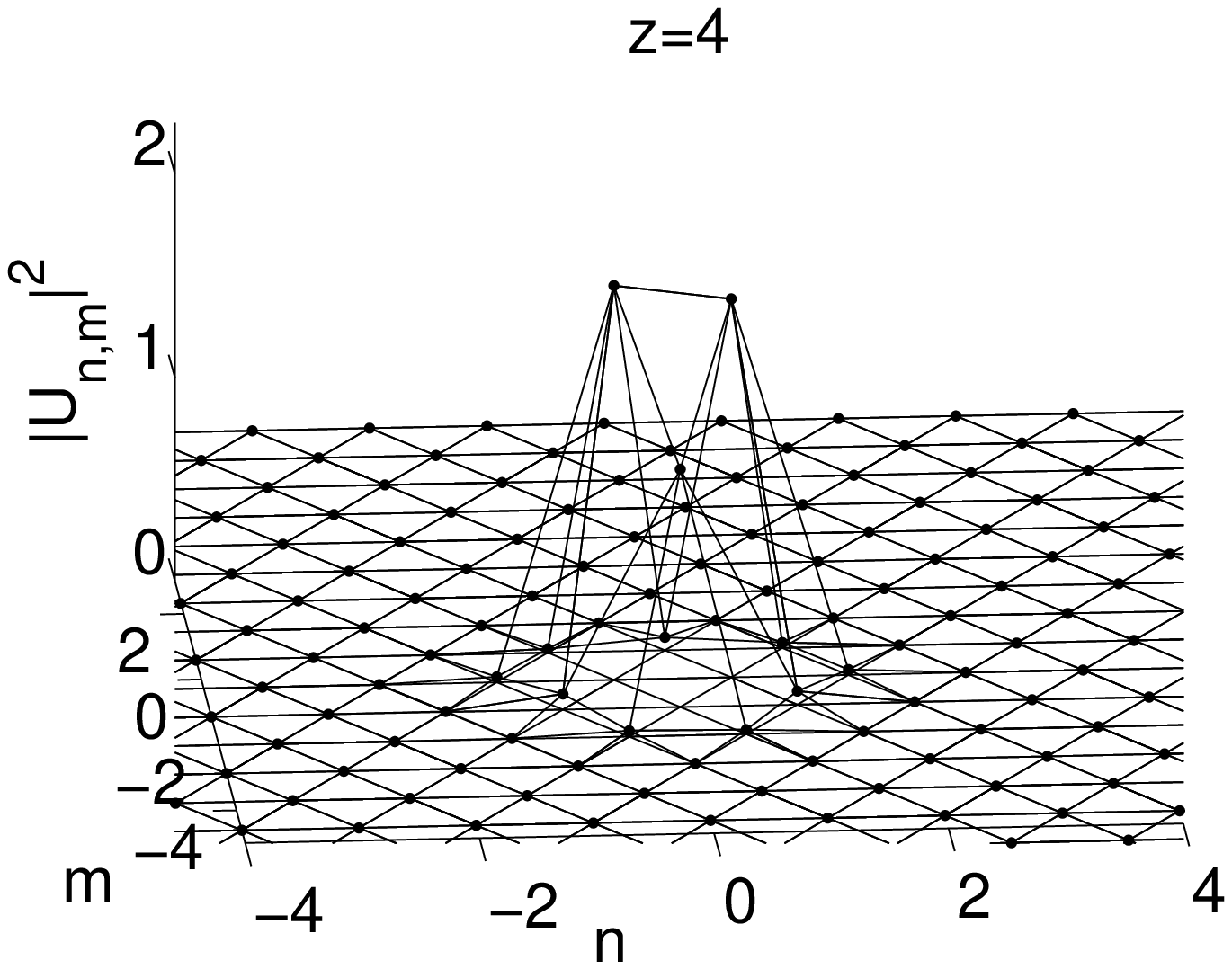}\\
\includegraphics[width=5cm,height=4cm]{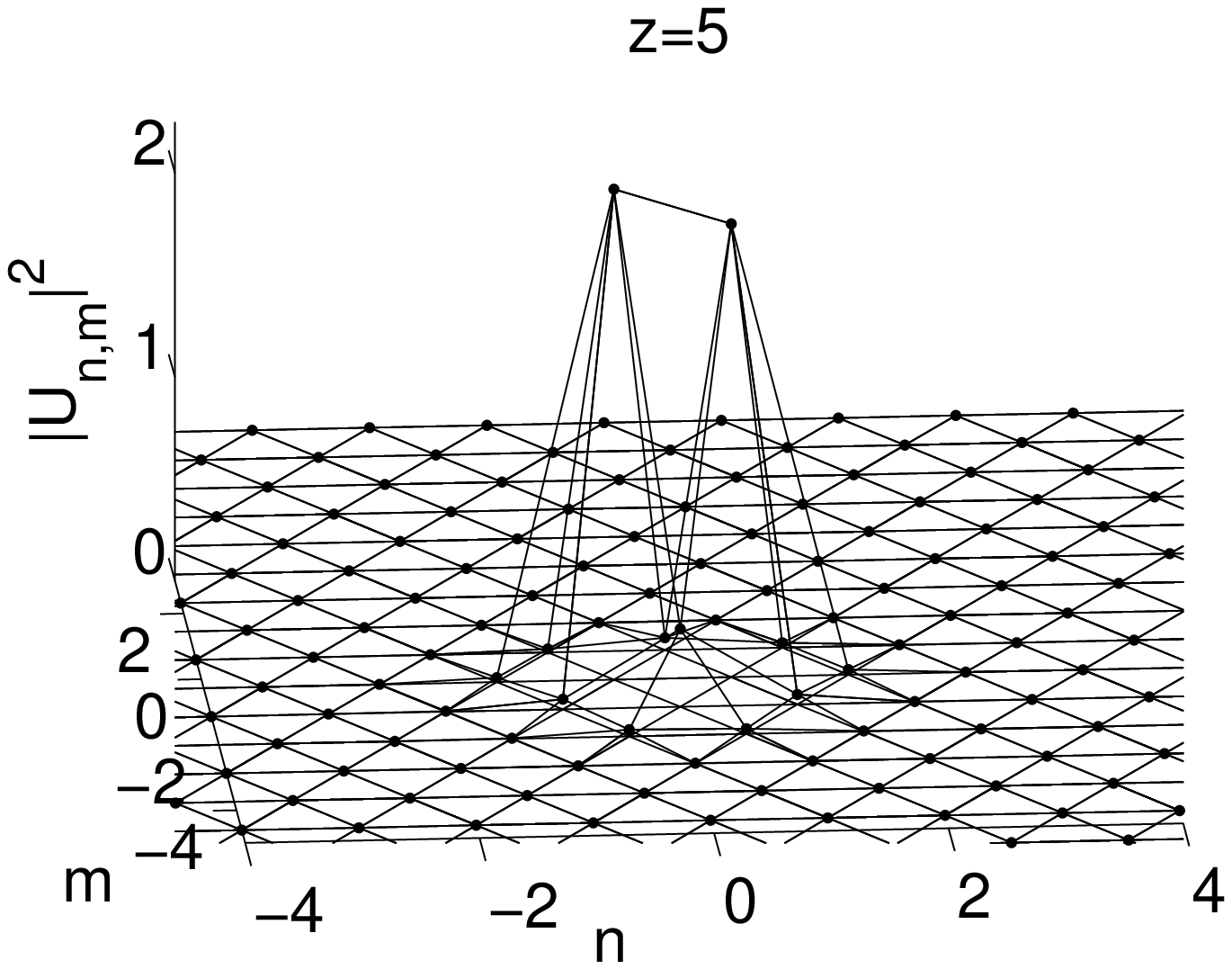}
\includegraphics[width=5cm,height=4cm]{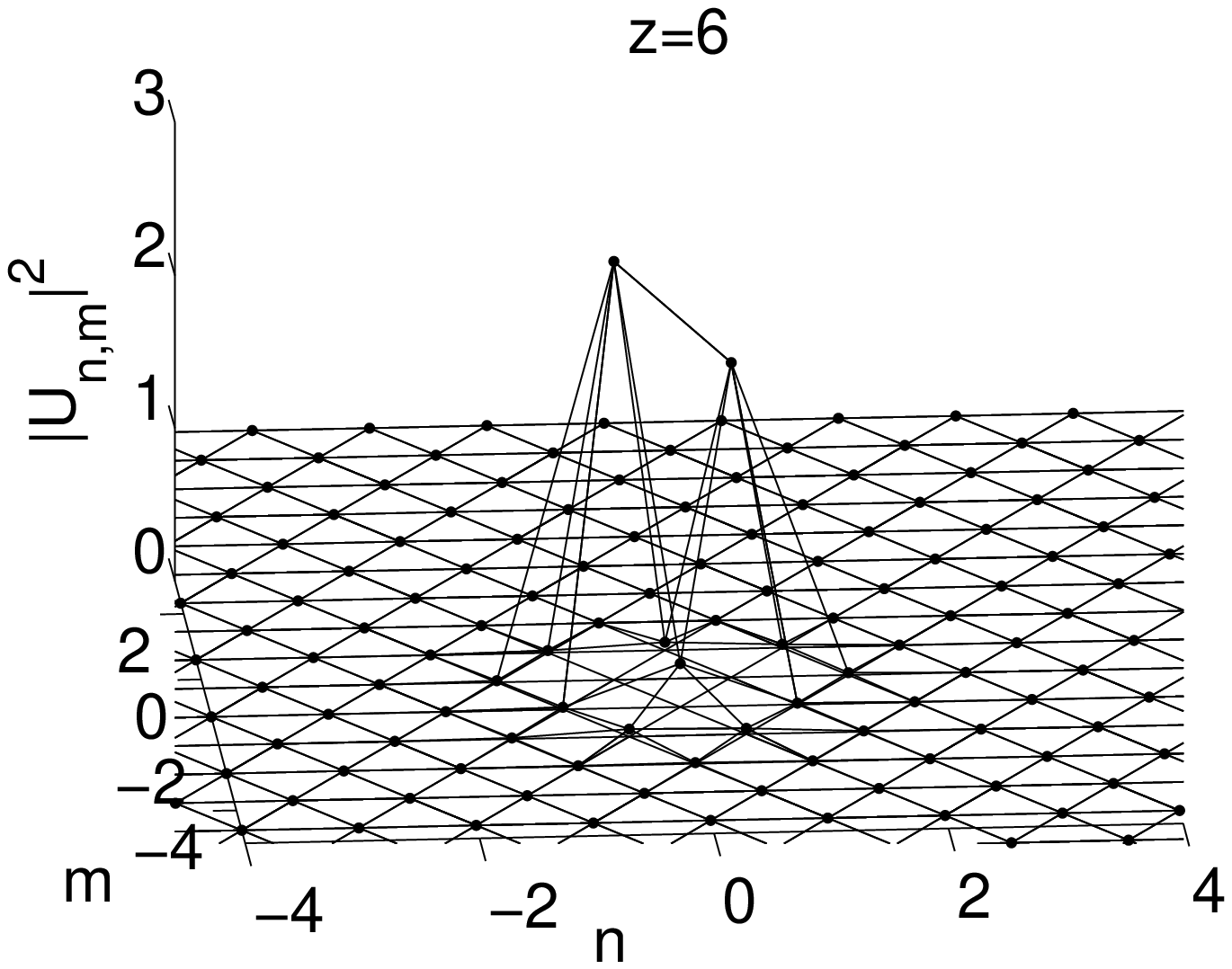}\\
\includegraphics[width=5cm,height=4cm]{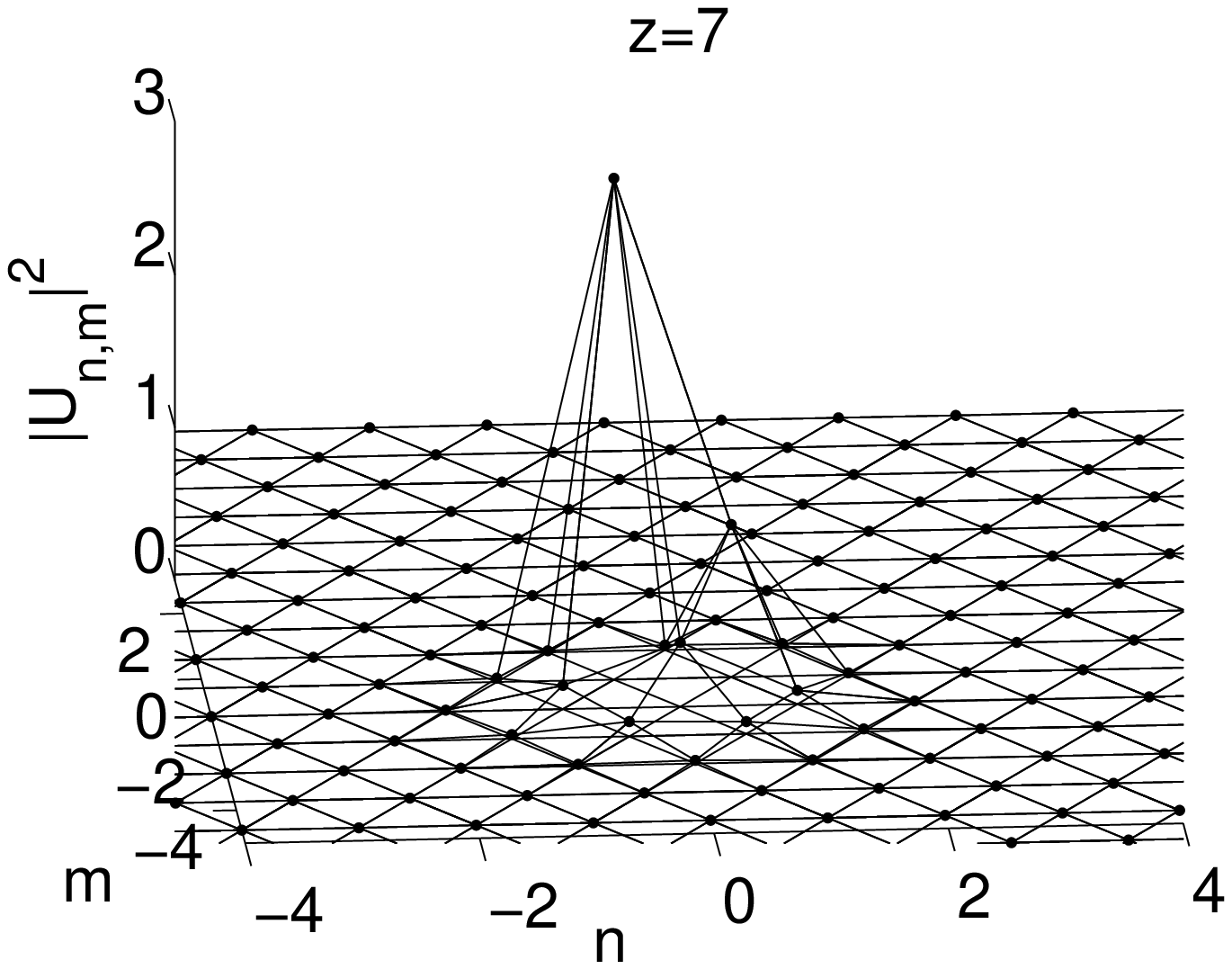}
\includegraphics[width=5cm,height=4cm]{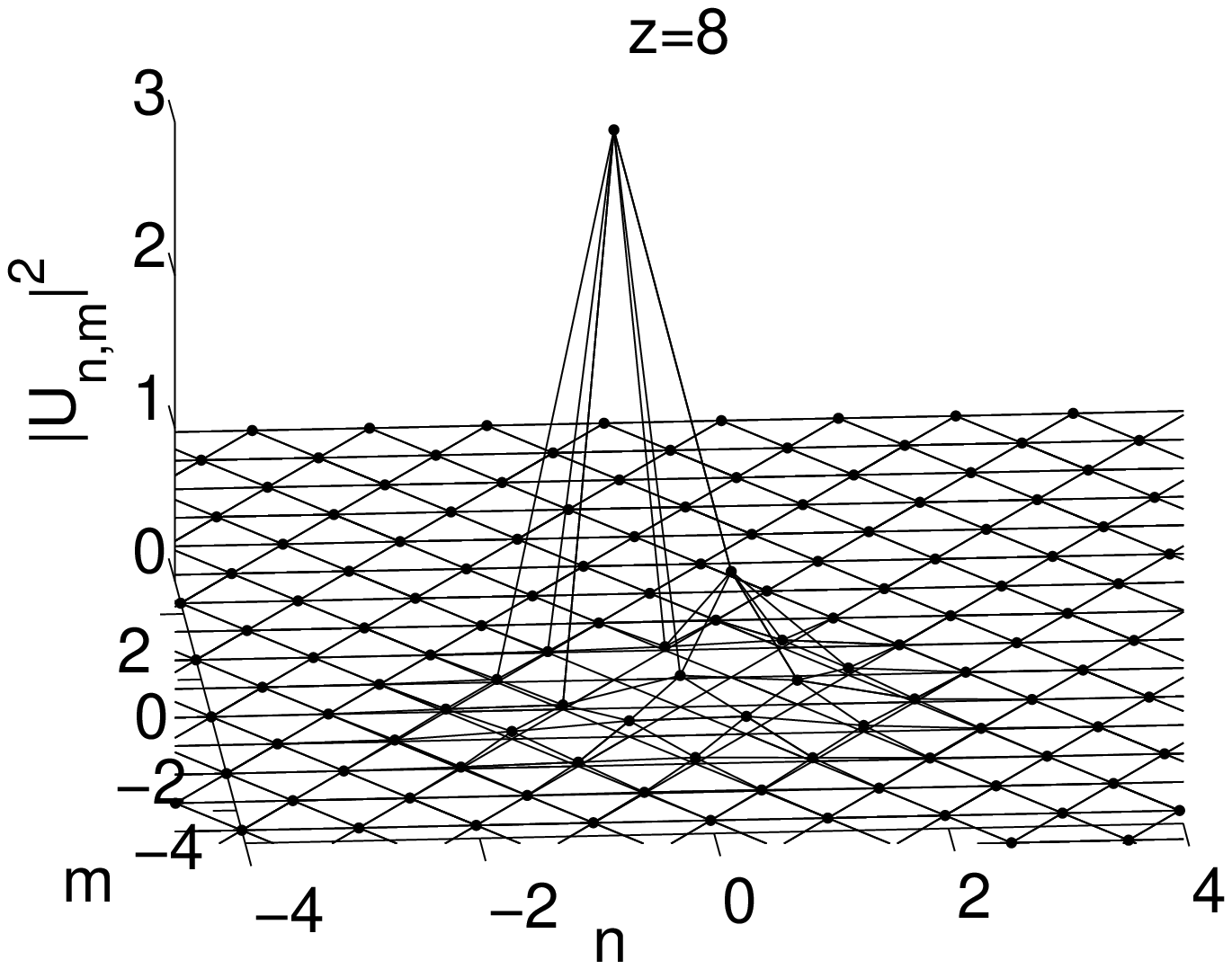}\\
\includegraphics[width=5cm,height=4cm]{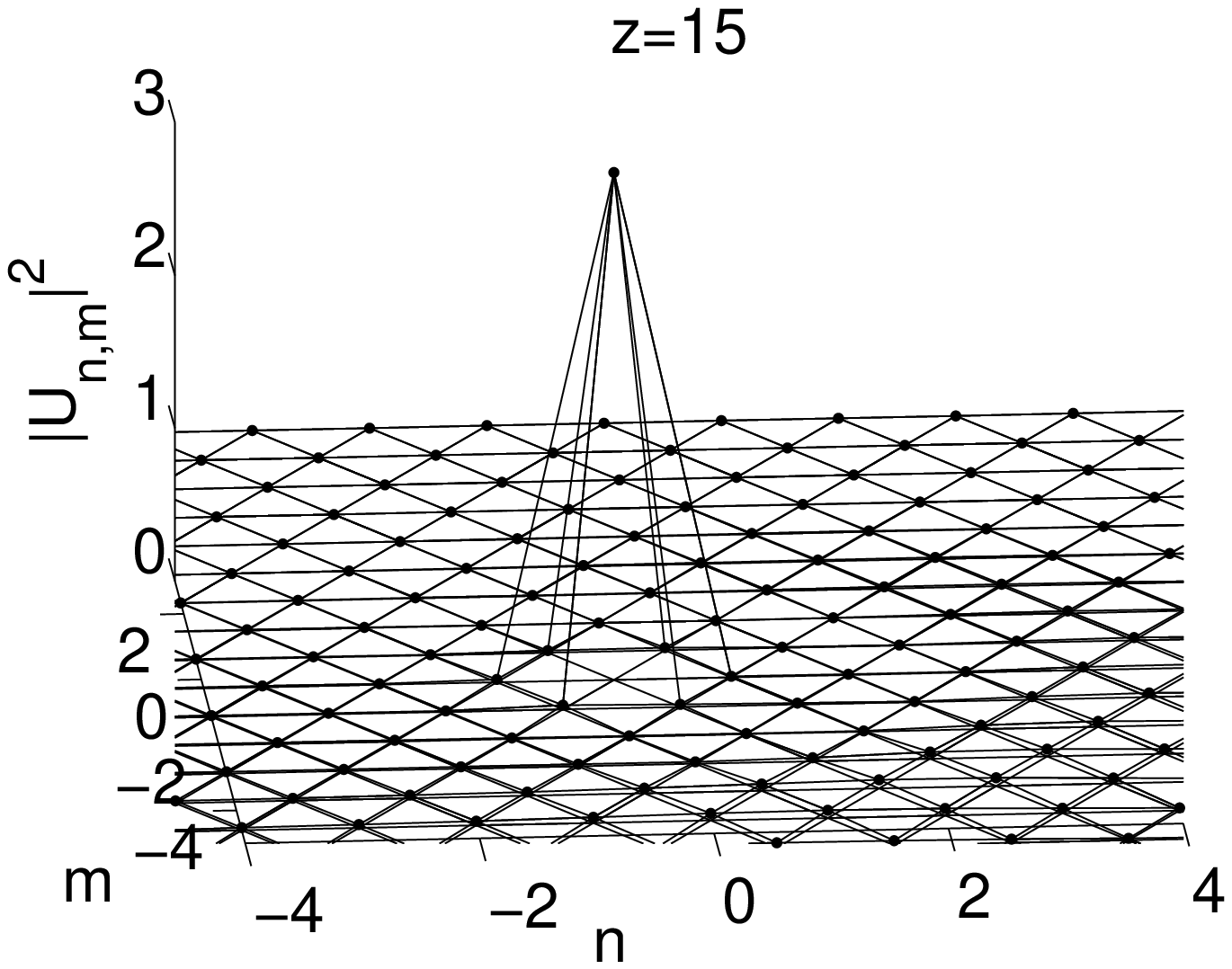}
\includegraphics[width=5cm,height=4cm]{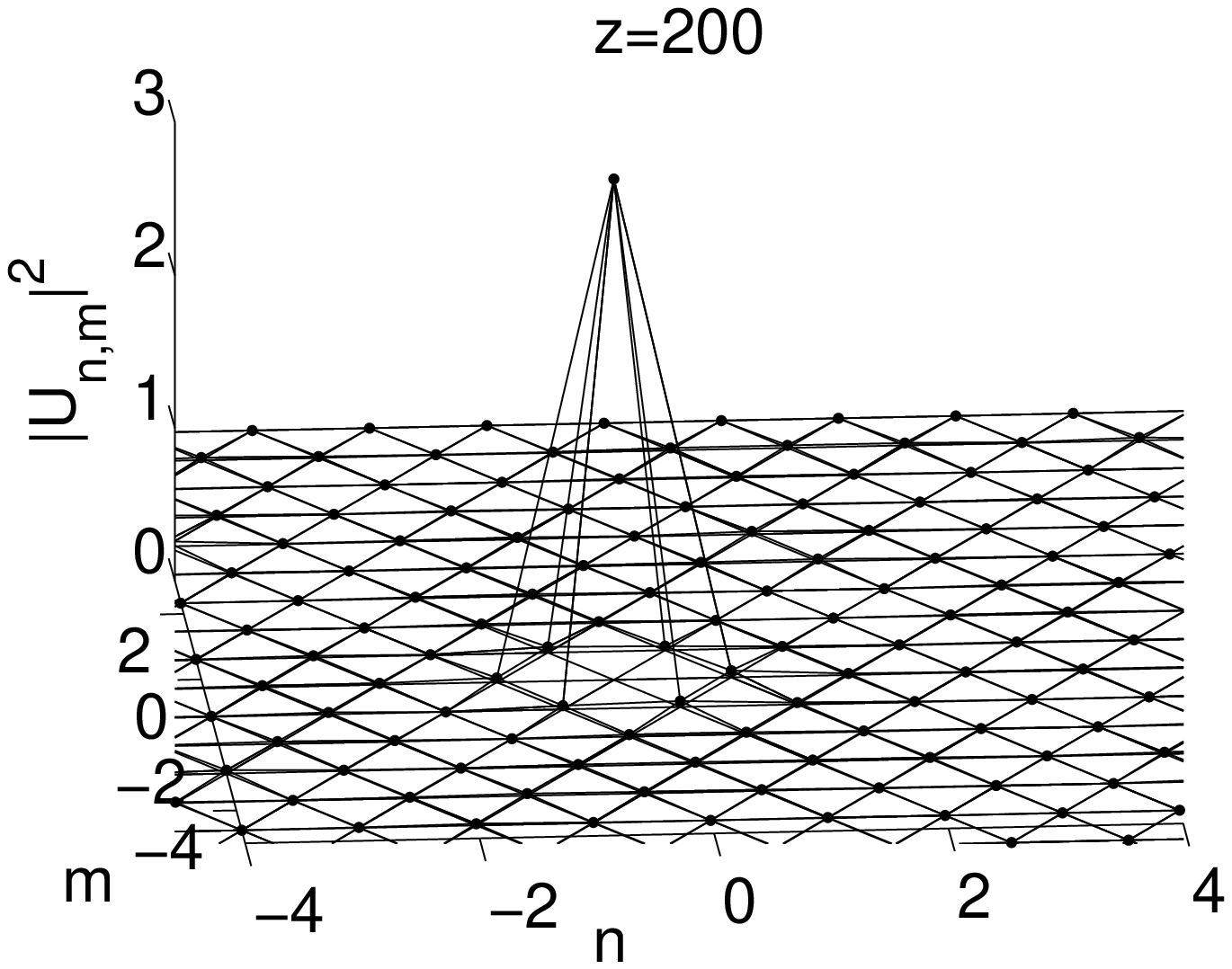}
\vspace{-0.4cm}
\caption{RK$4$ results from the hexagonal three-site $[0, 0, 0]$ at $\delta = 0.80$, $\varepsilon=0.2$ at $z=1, 4, 5, 6, 7, 8, 15, 200$.}
\label{fig:Hex3site_0-0-0_an0_8_C0_2_dyn}
\end{figure}

(3) Finally the evolution of the form $[0, 0, 0]$ is seen in 
Figs.~ \ref{fig:Hex3site_0-0-0_an0_8_C0_01_dyn} 
and~\ref{fig:Hex3site_0-0-0_an0_8_C0_2_dyn}.
The solution is unstable  
for all $0 \leq \delta \leq 1$. We see essentially the same 
qualitative behavior as for the previous two cases, as regards the asymptotic
fate of the unstable waveforms. We have also checked that this
phenomenology arises for different values of $\delta$, such as 
$\delta=0.2$ (results not shown here).

\subsection{Honeycomb Lattice}

Finally, we now turn to case examples of the dynamical
evolution on the honeycomb lattice. 
The unstable solutions that we study in this case
are: (1) the charge one vortex, (2) the 
charge-2 vortex for
$\delta < 0.716$ (below $0.5$, recall that this states morphs
into an unstable $[0, 0 + \pi, 0, 0, 0 + \pi, 0]$ state), 
(3) the in-phase solution $[0, 0, 0, 0, 0, 0]$.
Given that the results for weak coupling are similarly
(breathing) as in the previous subsection, we focus on the
case of the larger coupling $\varepsilon=0.2$.

\begin{figure}[tbh]
\centering
\includegraphics[width=5cm,height=4cm]{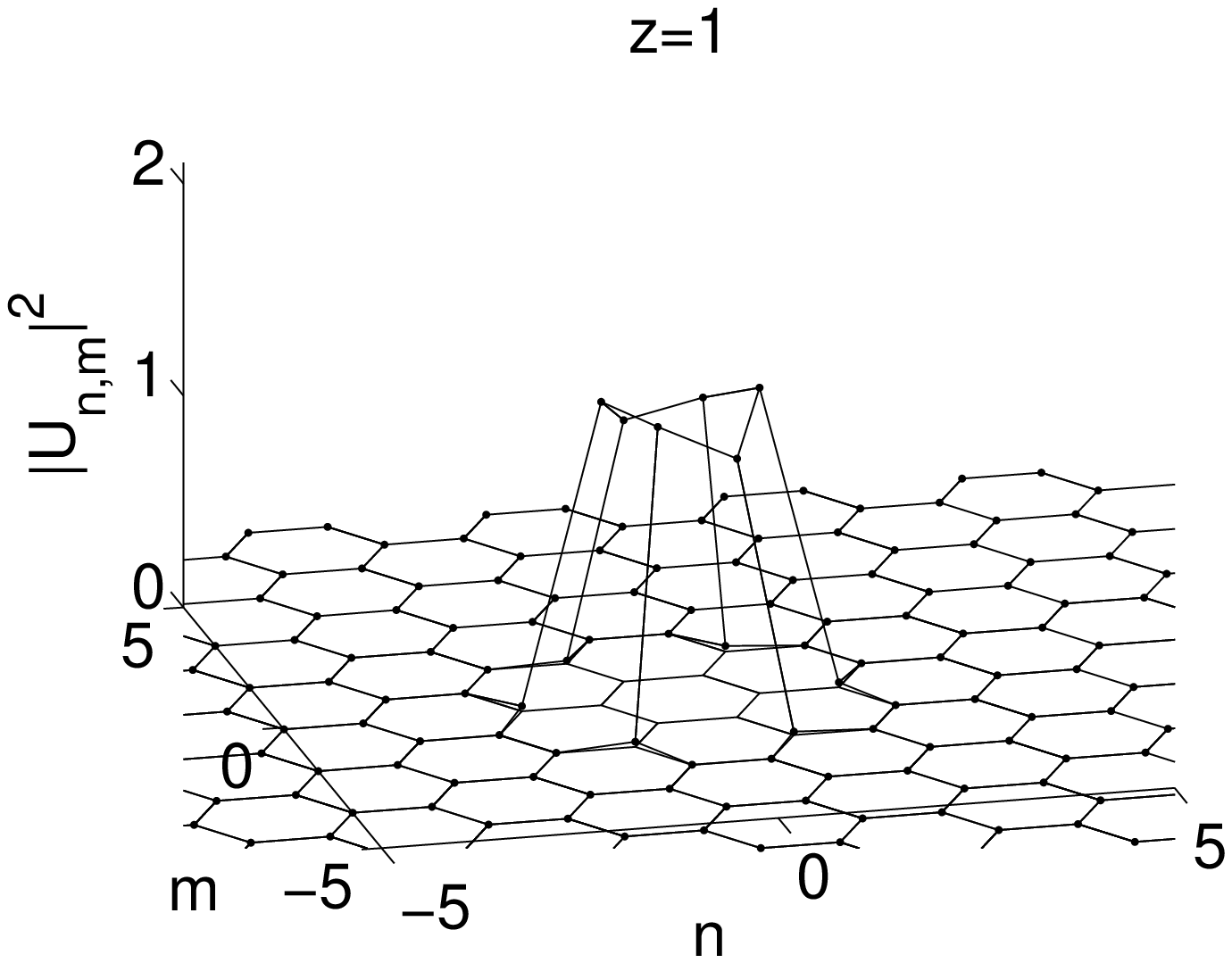}
\includegraphics[width=5cm,height=4cm]{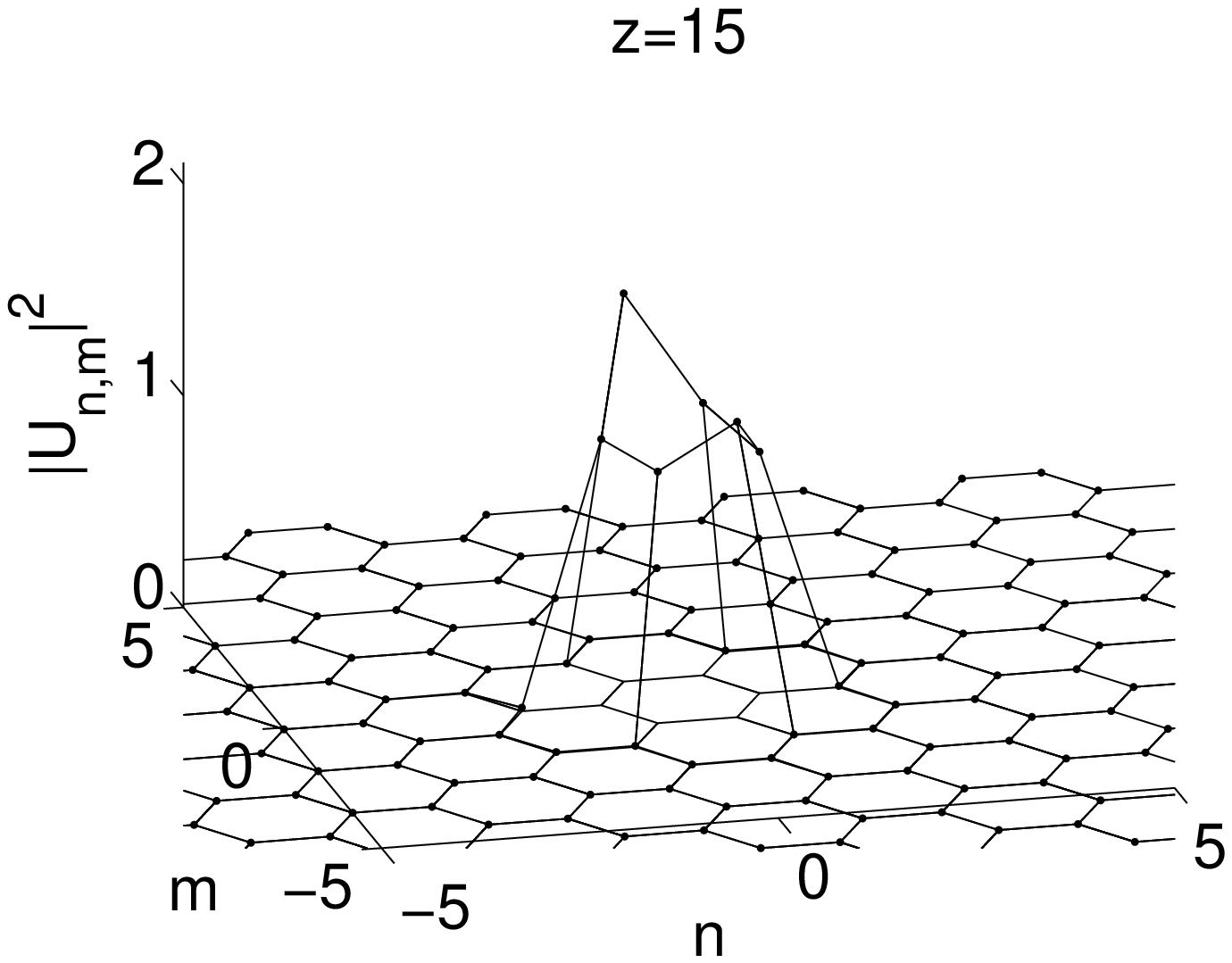}\\
\includegraphics[width=5cm,height=4cm]{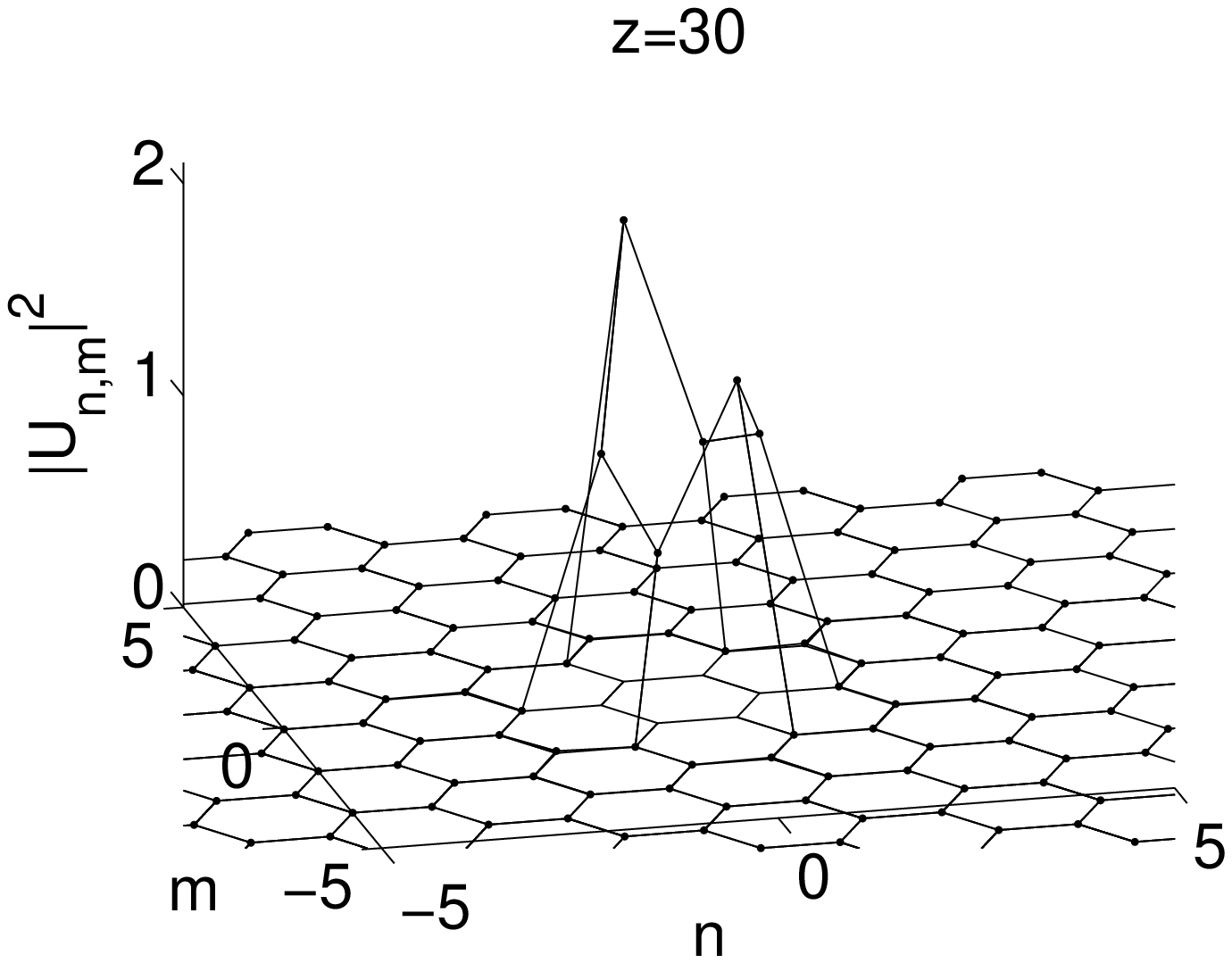}
\includegraphics[width=5cm,height=4cm]{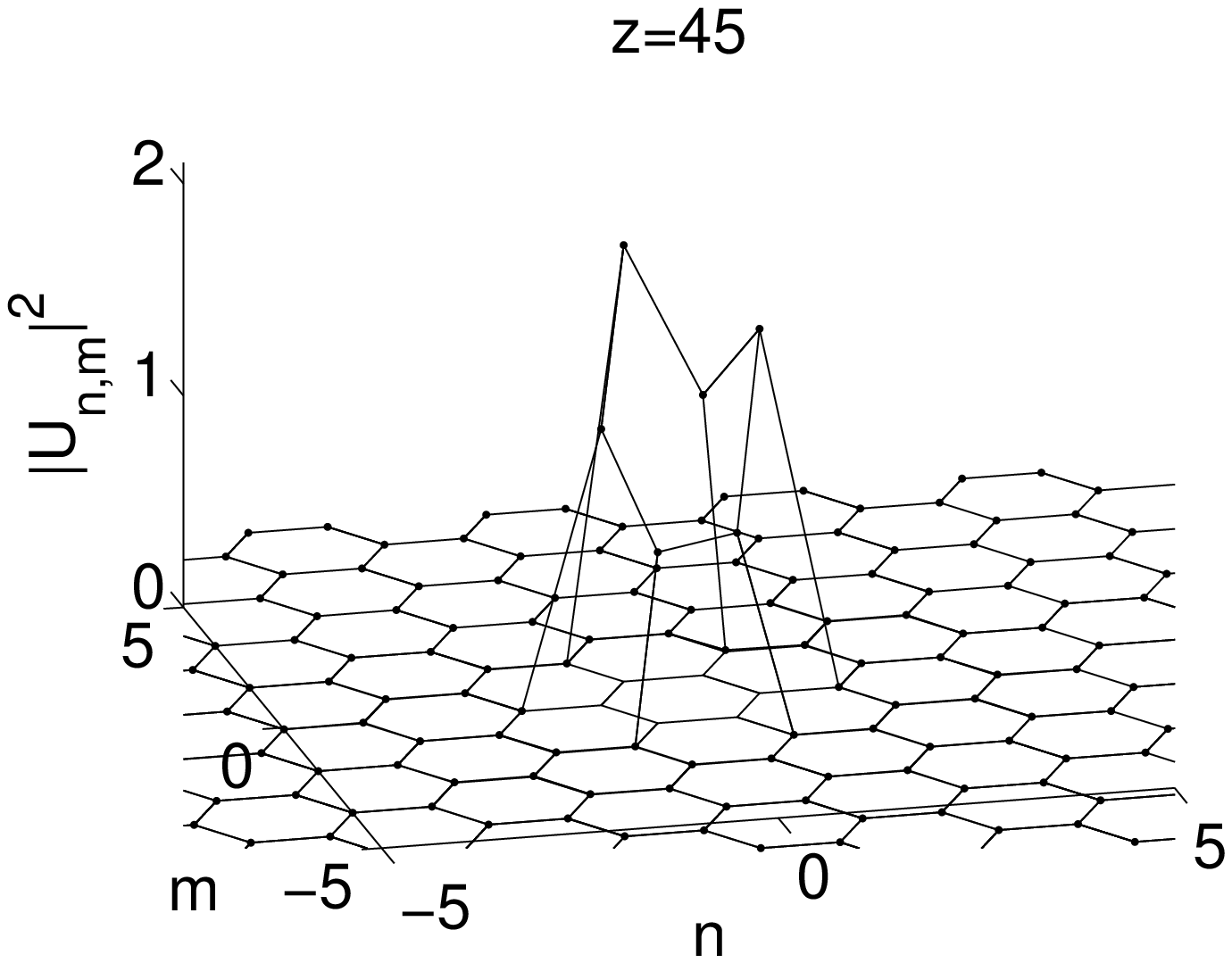}\\
\includegraphics[width=5cm,height=4cm]{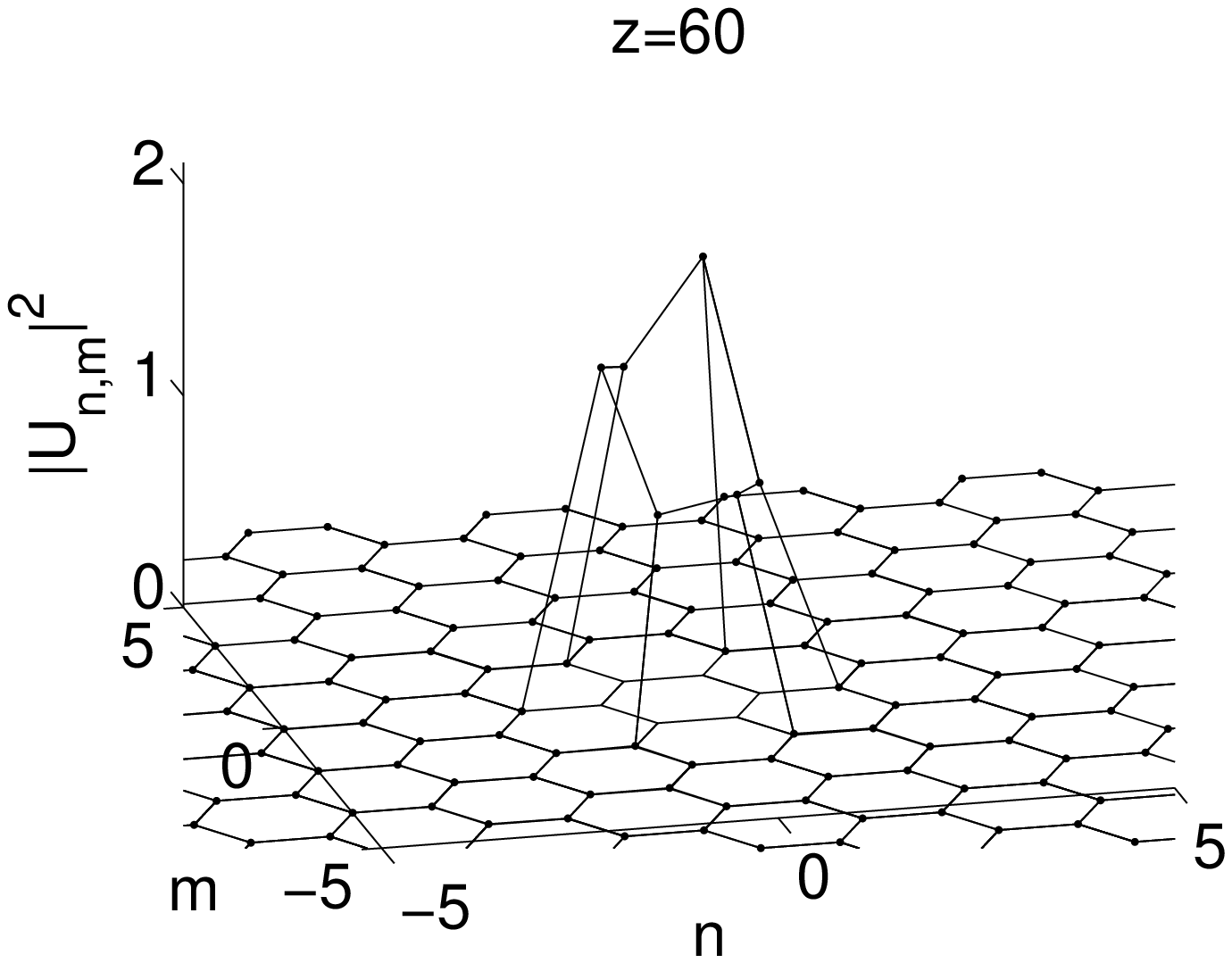}
\includegraphics[width=5cm,height=4cm]{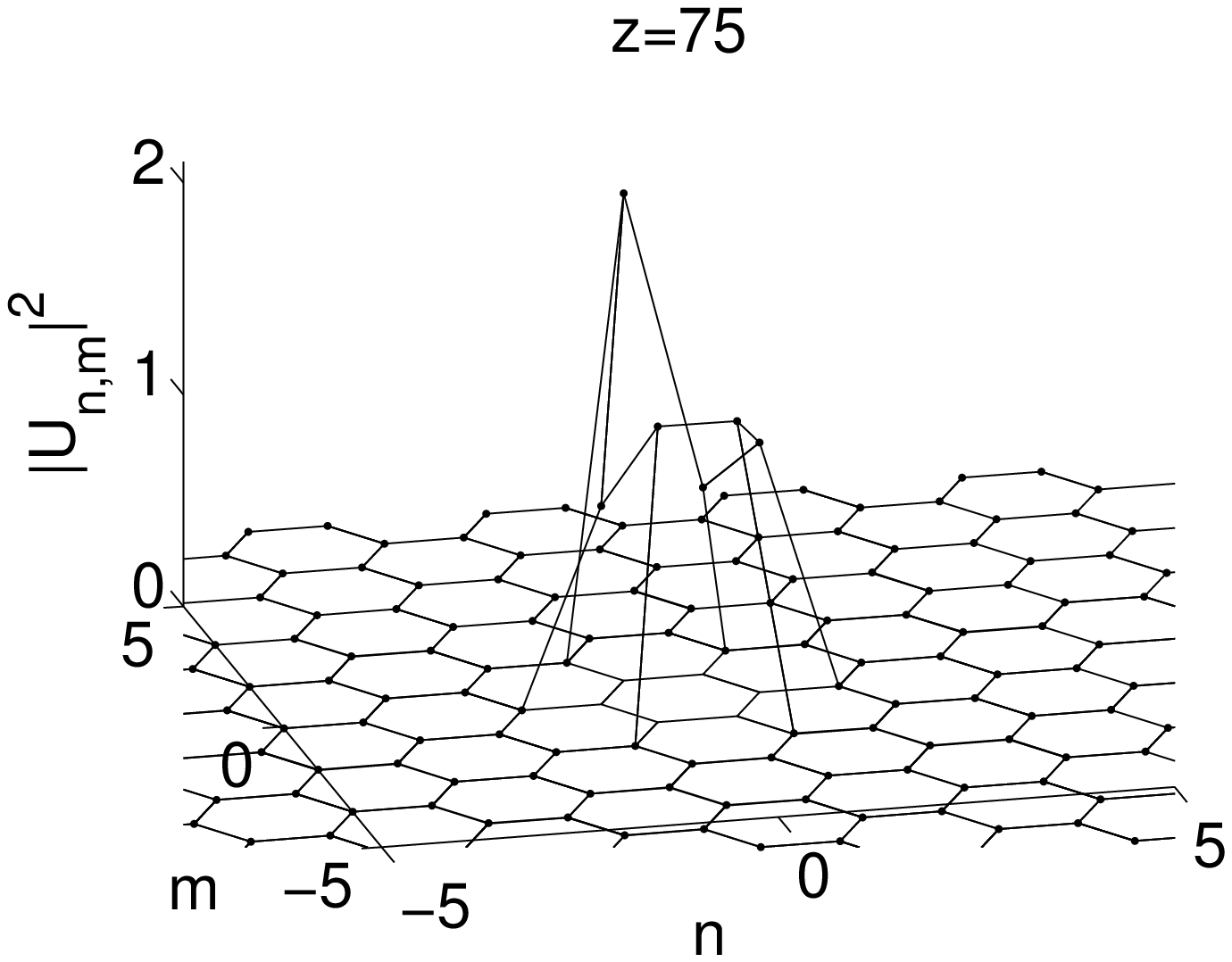}\\
\includegraphics[width=5cm,height=4cm]{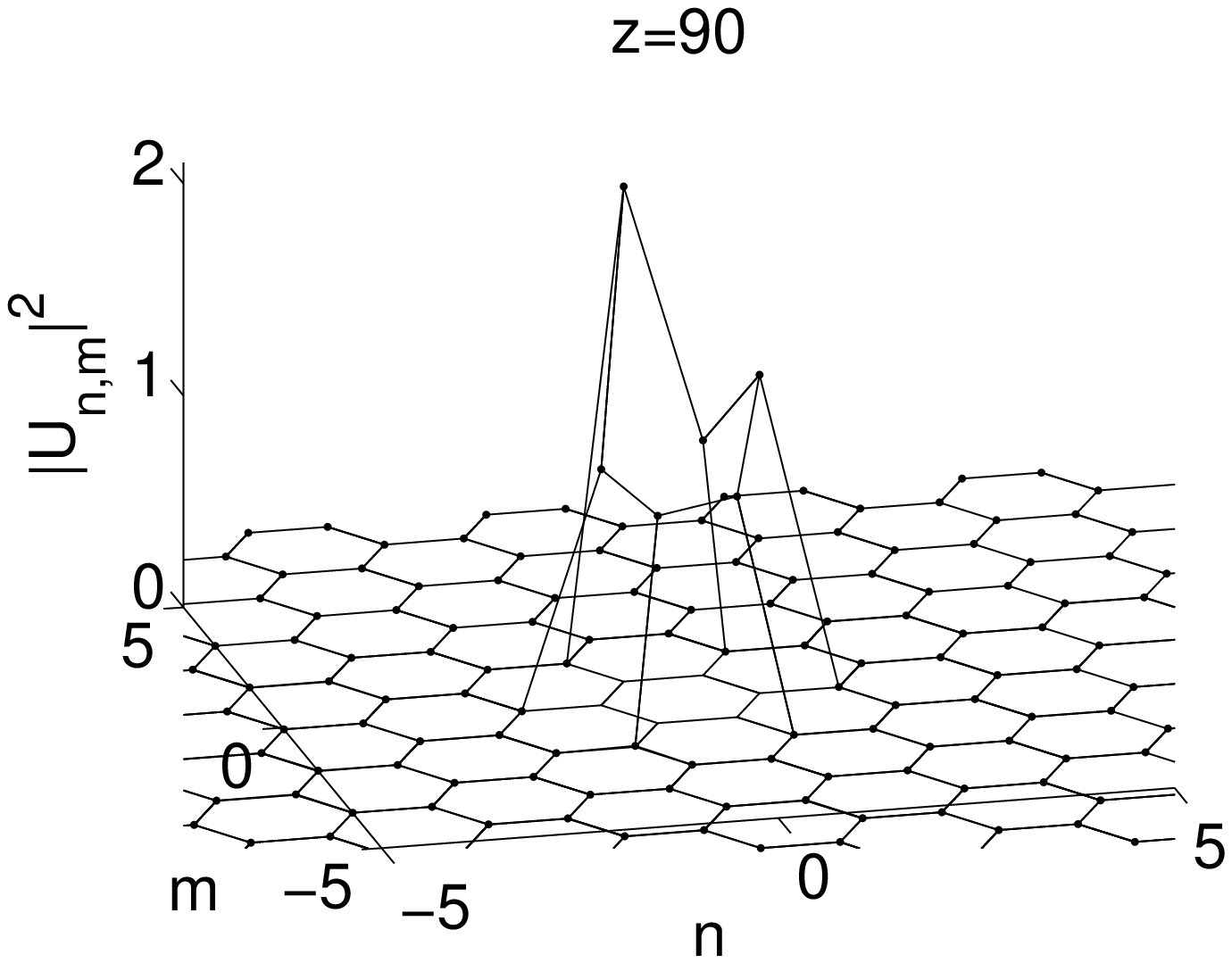}
\includegraphics[width=5cm,height=4cm]{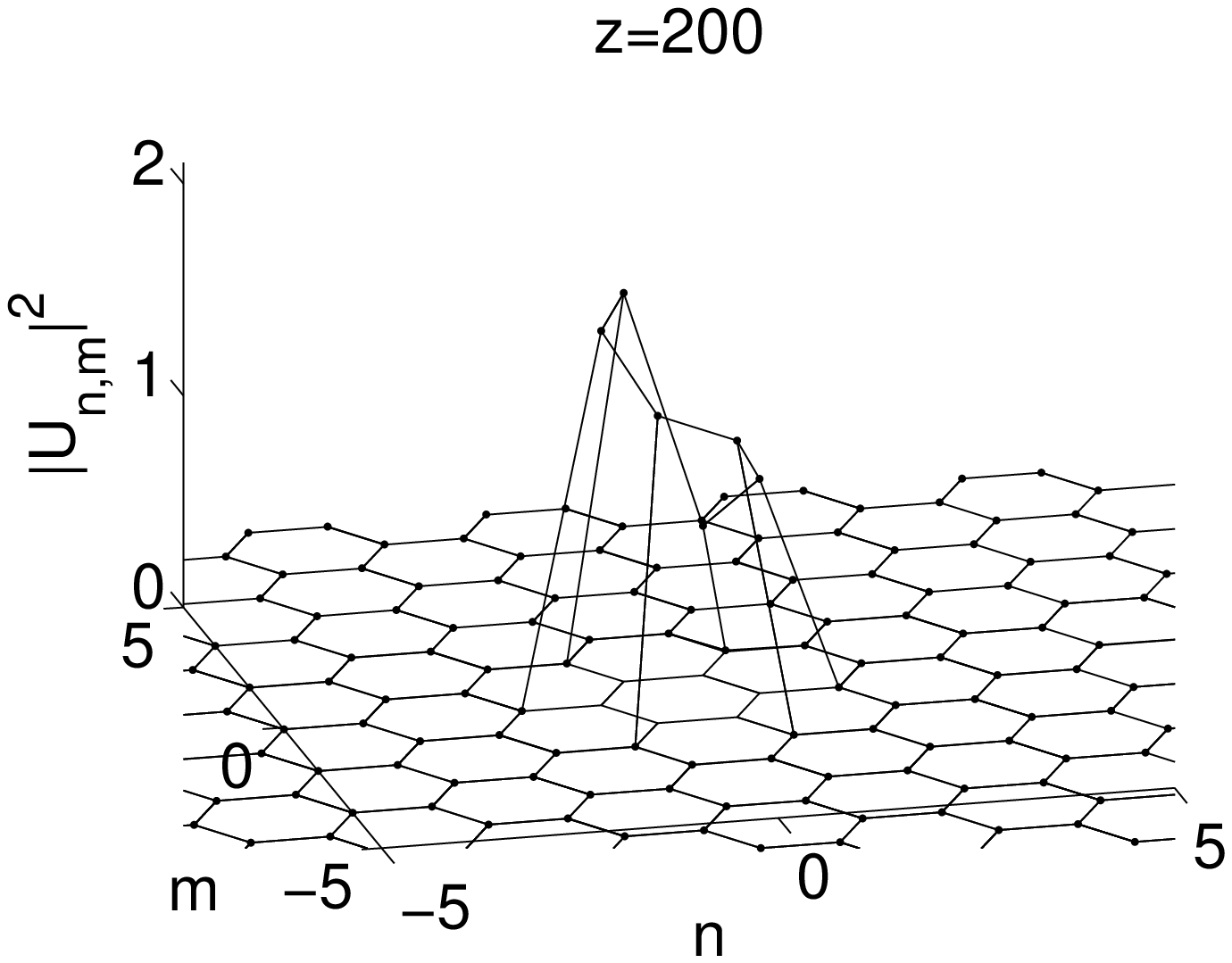}
\vspace{-0.4cm}
\caption{RK$4$ results from the six site honeycomb charge $1$ vortex at $\delta = 0.80$, $\varepsilon=0.2$ at $z=1, 15, 30, 45, 60, 90, 200$. A multi-site
breather emerges with an associated repartitioning of the norm of the
solution.}
\label{fig:Hon6site_C1_an0_8_C0_2_dyn}
\end{figure}

\begin{figure}[tbh]
\centering
\includegraphics[width=5cm,height=4cm]{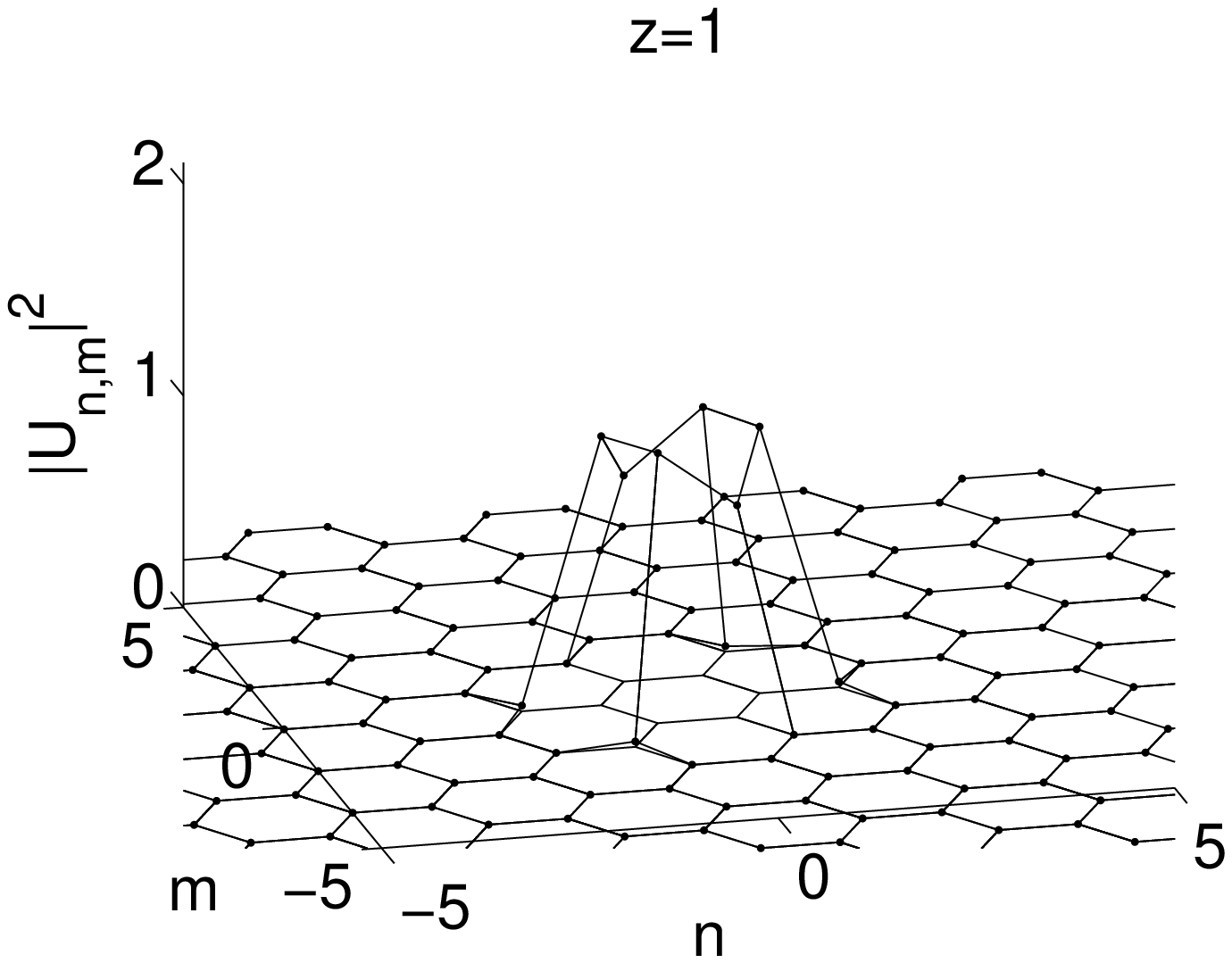}
\includegraphics[width=5cm,height=4cm]{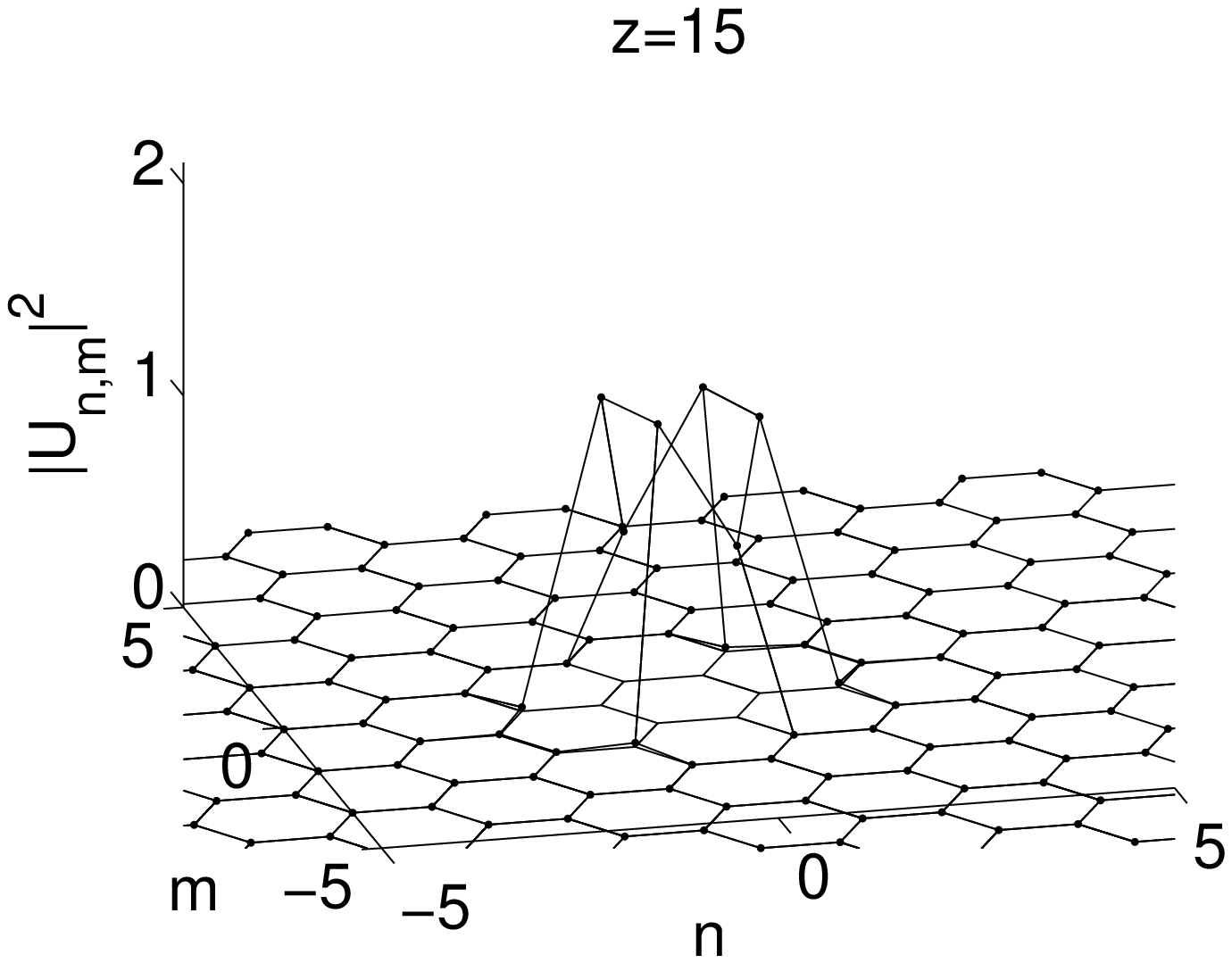}\\
\includegraphics[width=5cm,height=4cm]{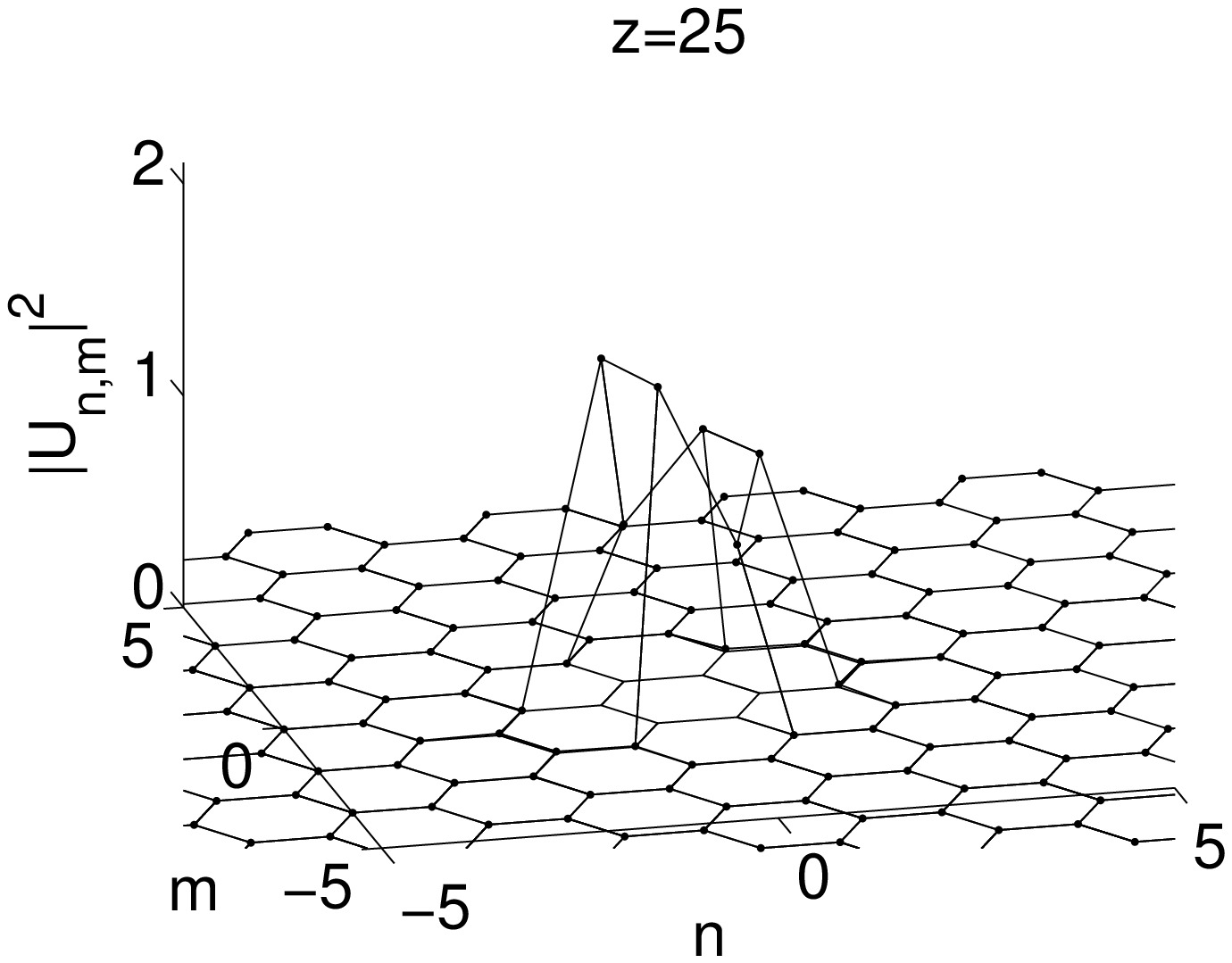}
\includegraphics[width=5cm,height=4cm]{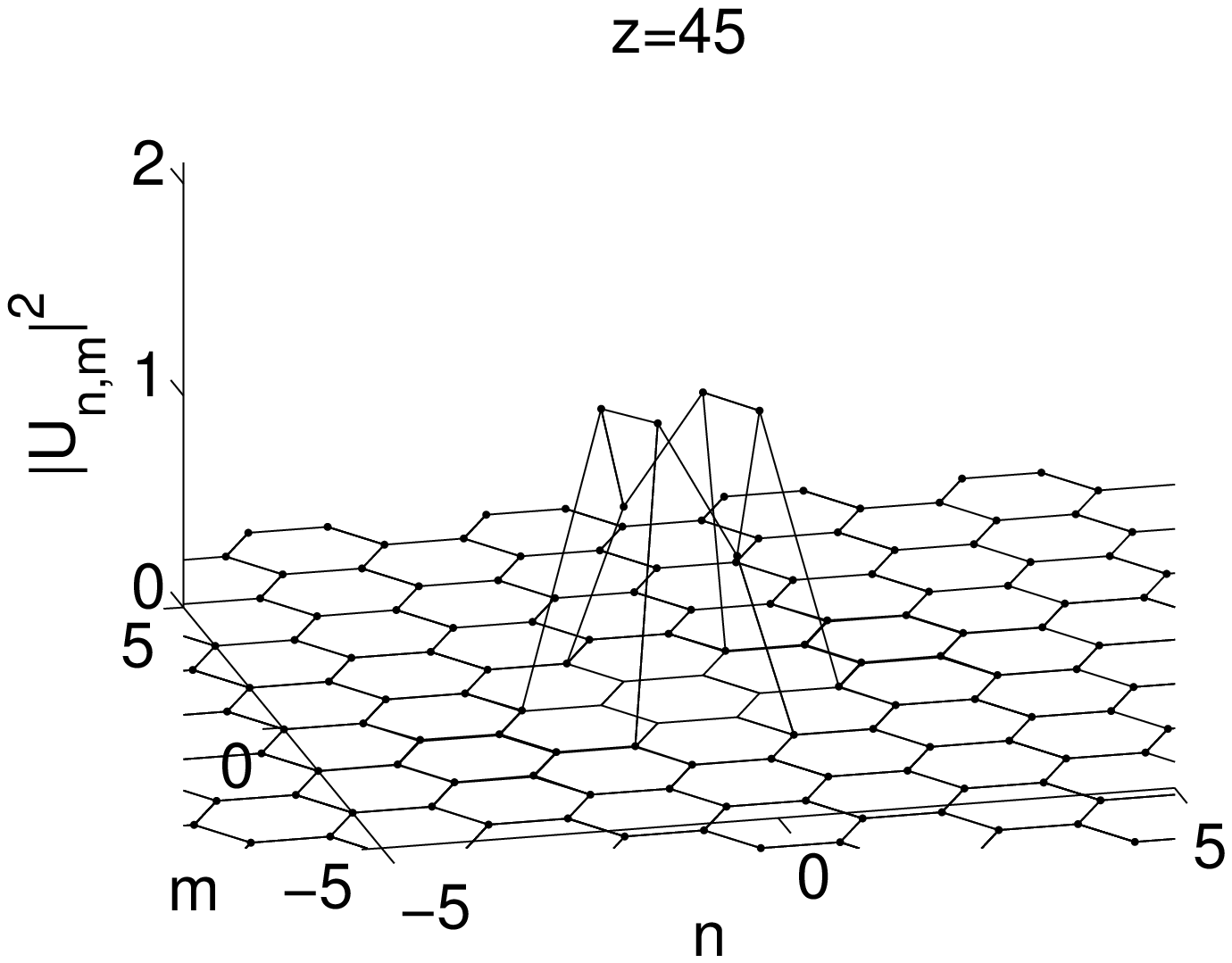}\\
\includegraphics[width=5cm,height=4cm]{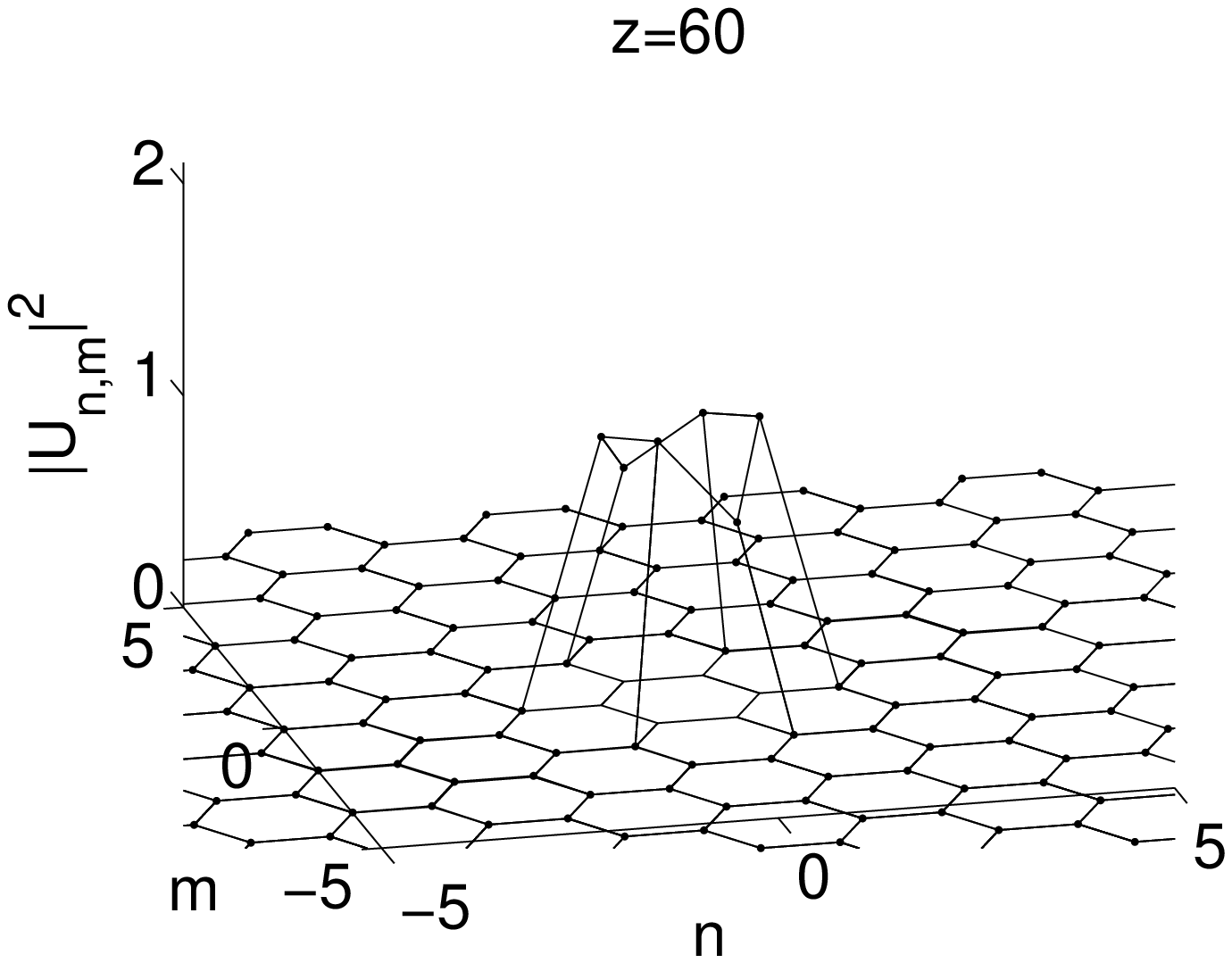}
\includegraphics[width=5cm,height=4cm]{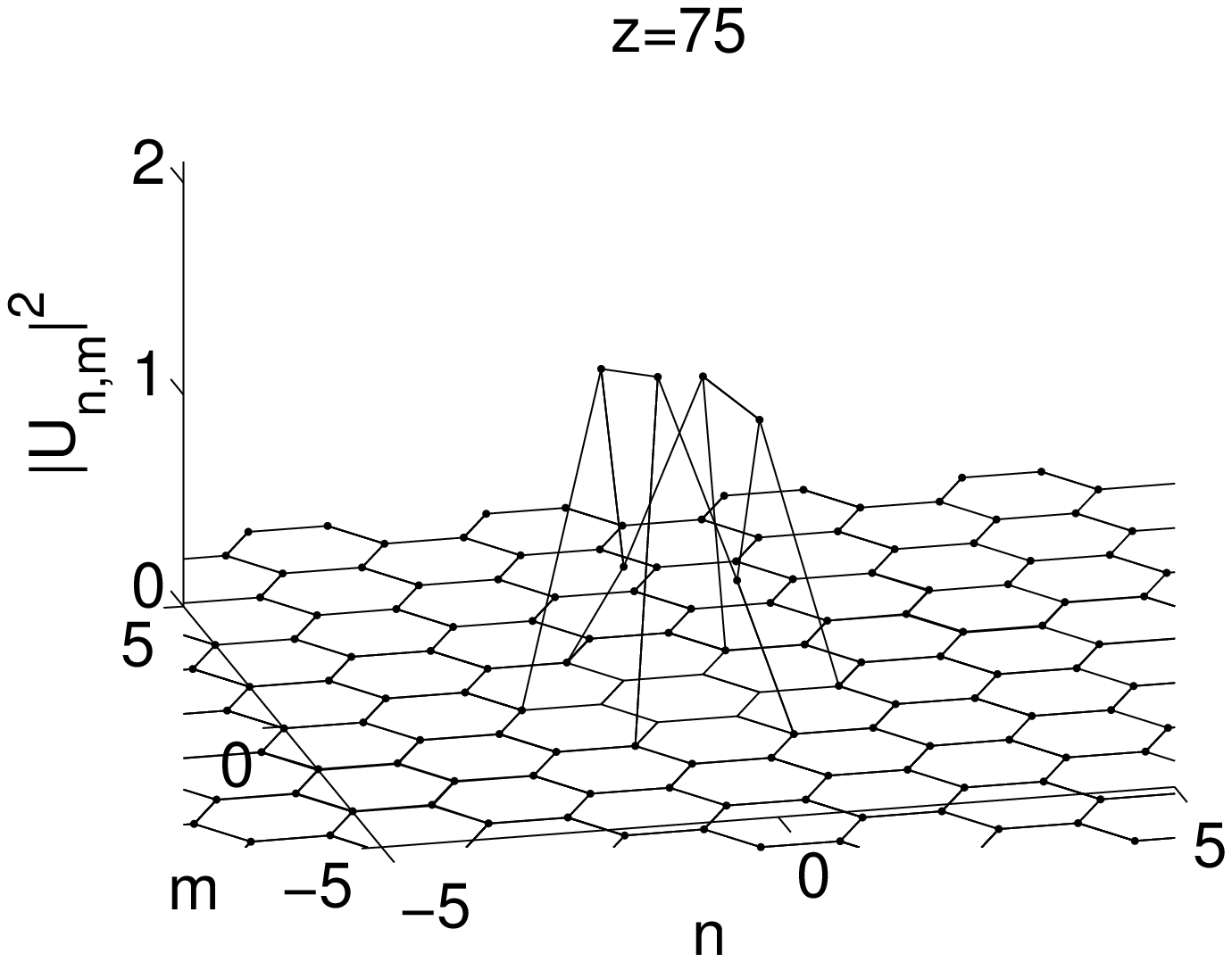}\\
\includegraphics[width=5cm,height=4cm]{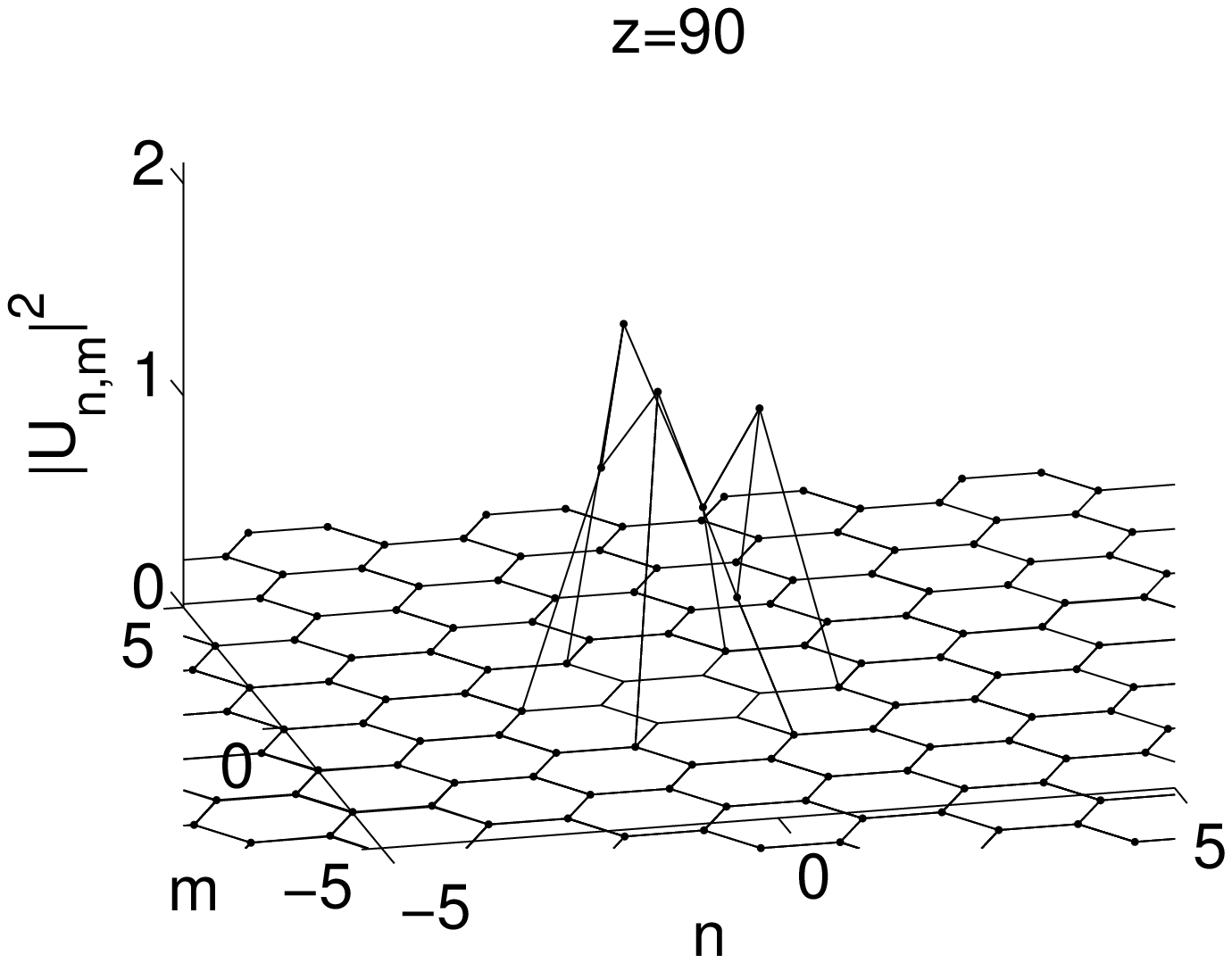}
\includegraphics[width=5cm,height=4cm]{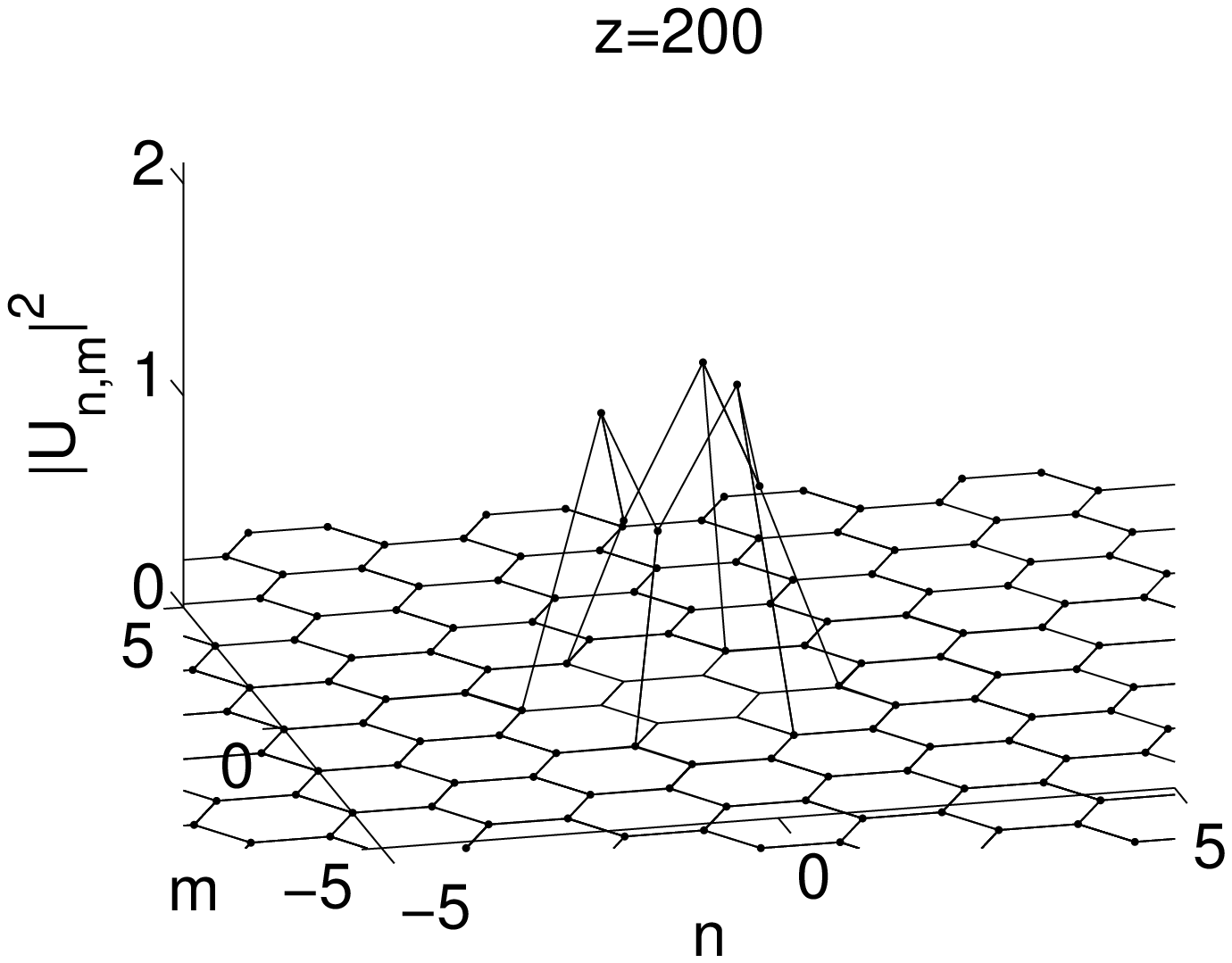}
\vspace{-0.4cm}
\caption{Similar to the above case, i.e., 
RK$4$ results from the six site honeycomb charge $1$ vortex at $\delta = 0.20$, $C=0.2$ at $z=1, 15, 25, 45, 60, 90, 200$.}
\label{fig:Hon6site_C1_an0_2_C0_2_dyn}
\end{figure}

(1) The charge $1$ vortex is unstable for the full range of the 
anisotropy considered herein. 
In Fig.~\ref{fig:Hon6site_C1_an0_8_C0_2_dyn}, we explore its
unstable dynamics for a
coupling strength of $\varepsilon = 0.2$ and $\delta=0.8$; similar
results have also been found for other values of $\delta$, as
e.g. in Fig.~\ref{fig:Hon6site_C1_an0_2_C0_2_dyn} for $\delta=0.2$.
The dynamics yields a multi-site excitation,
but with a repartitioning of the relevant intensity so that
some sites are dominant in amplitude in comparison to others.

\begin{figure}[tbh]
\centering
\includegraphics[width=5cm,height=4cm]{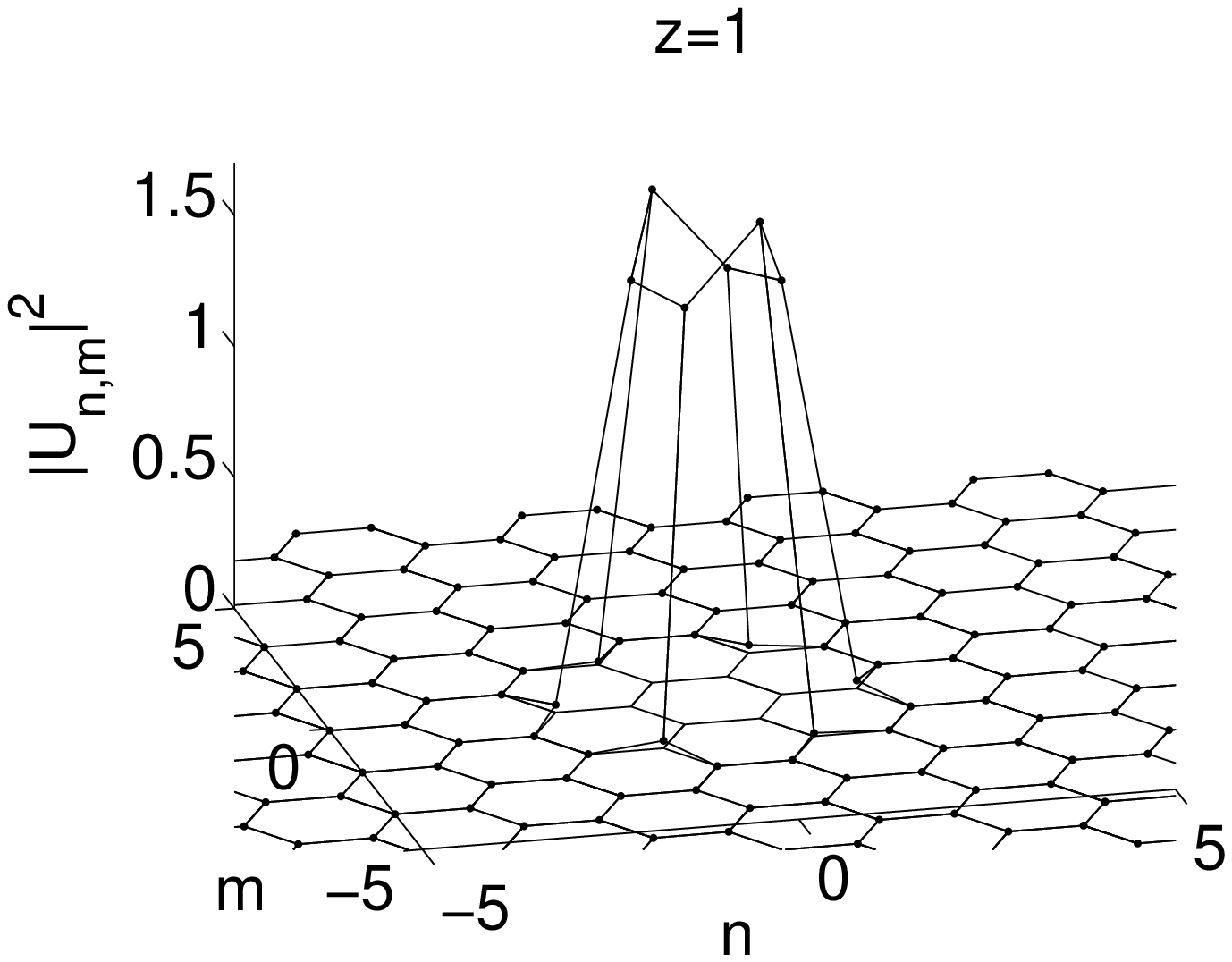}
\includegraphics[width=5cm,height=4cm]{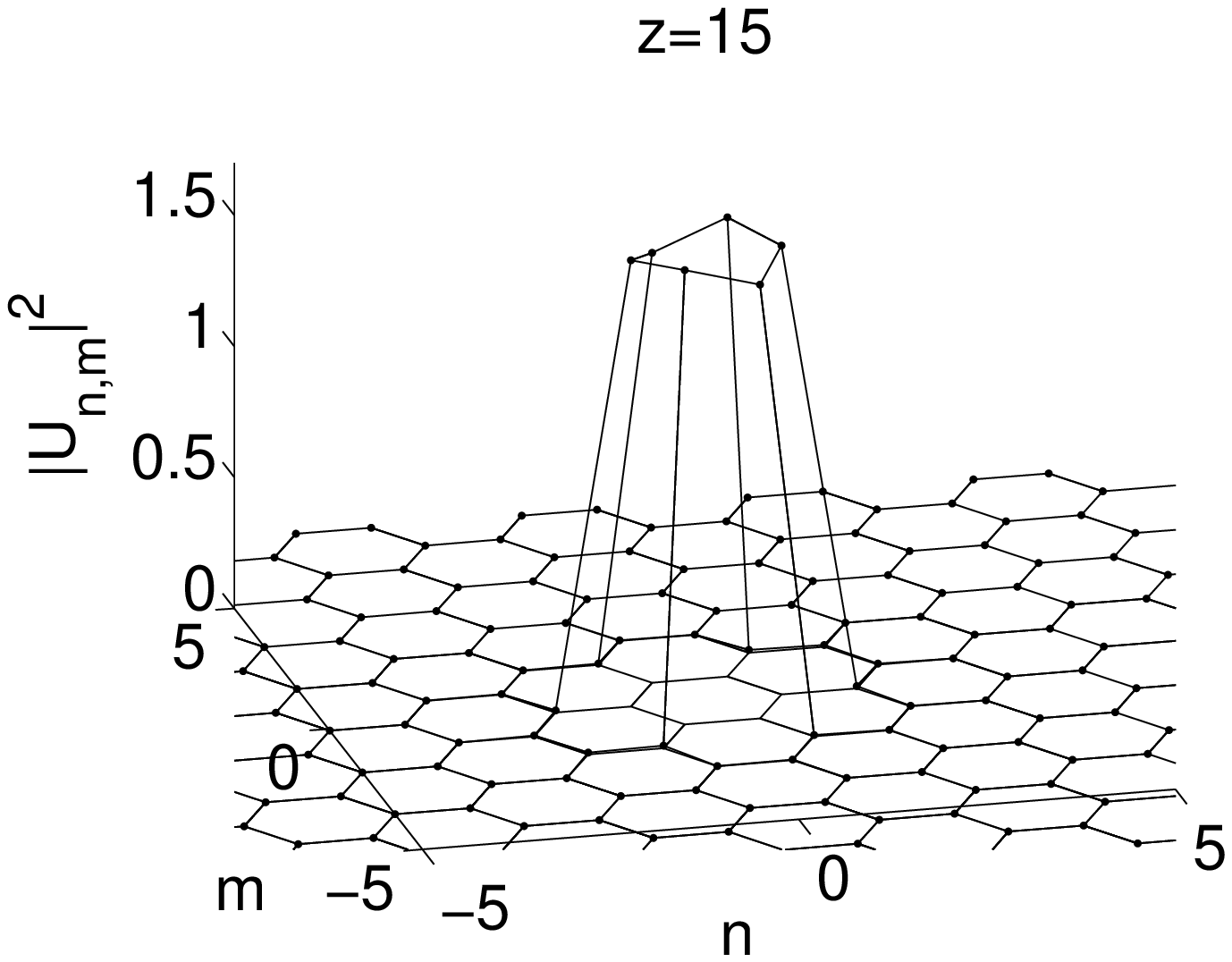}\\
\includegraphics[width=5cm,height=4cm]{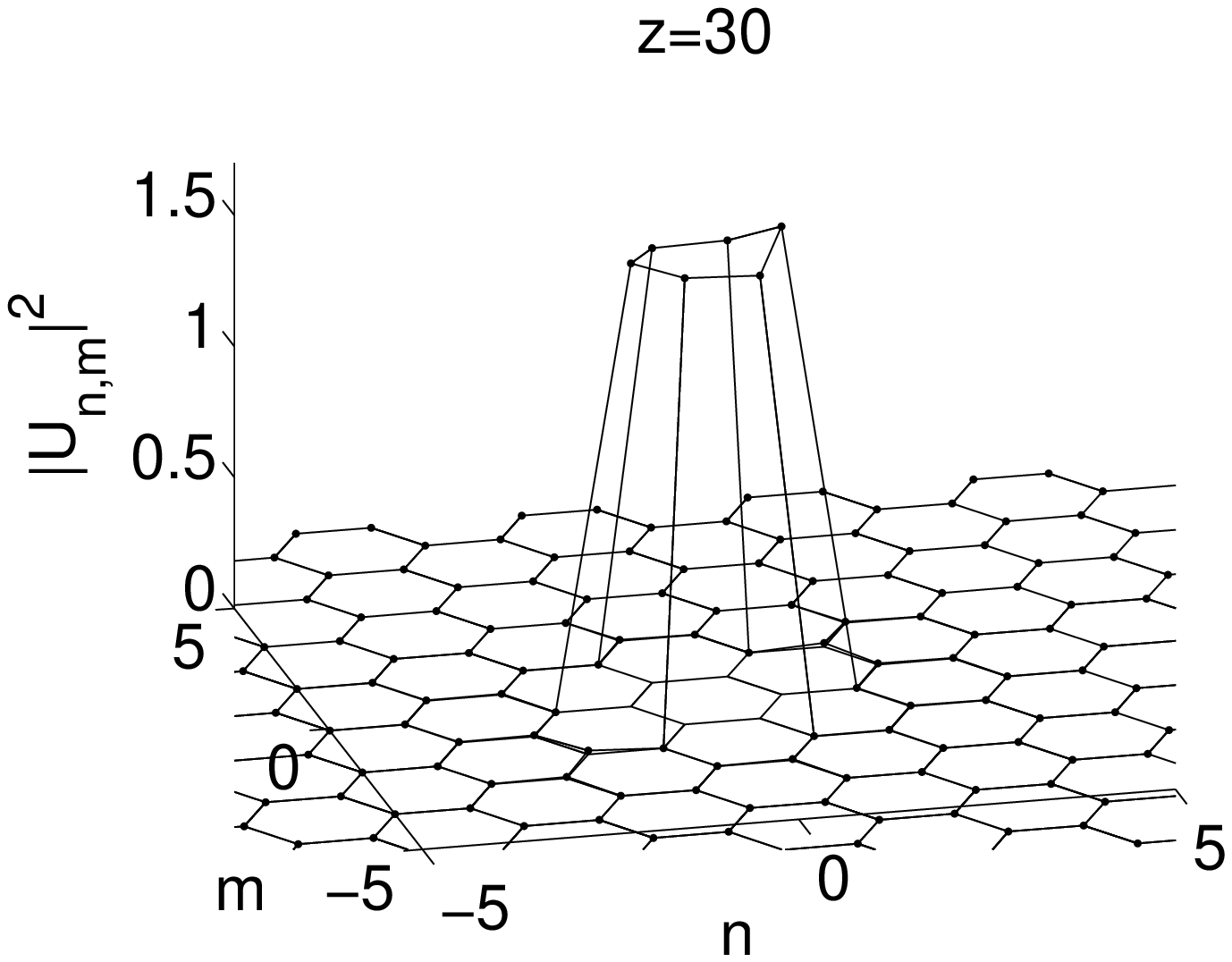}
\includegraphics[width=5cm,height=4cm]{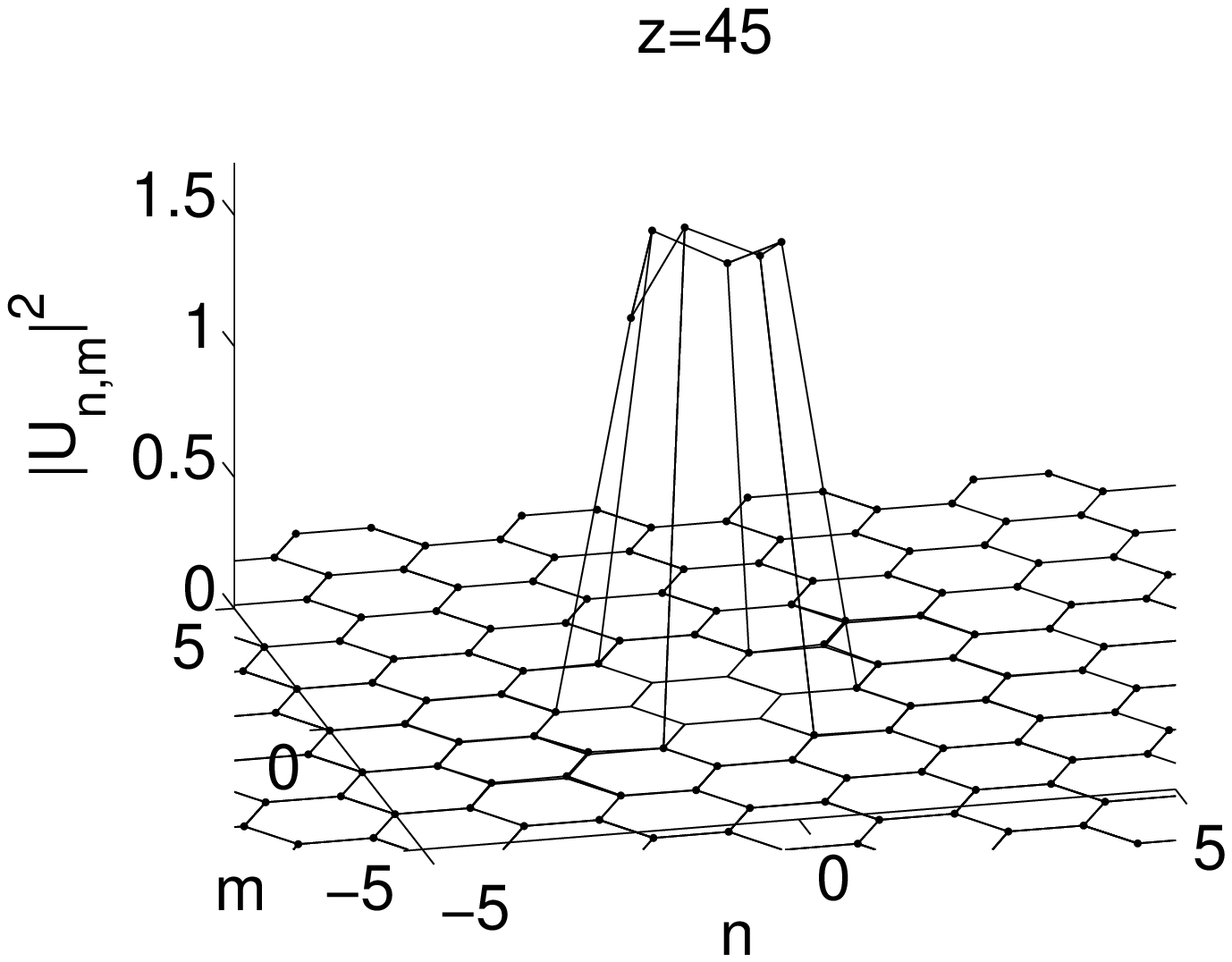}\\
\includegraphics[width=5cm,height=4cm]{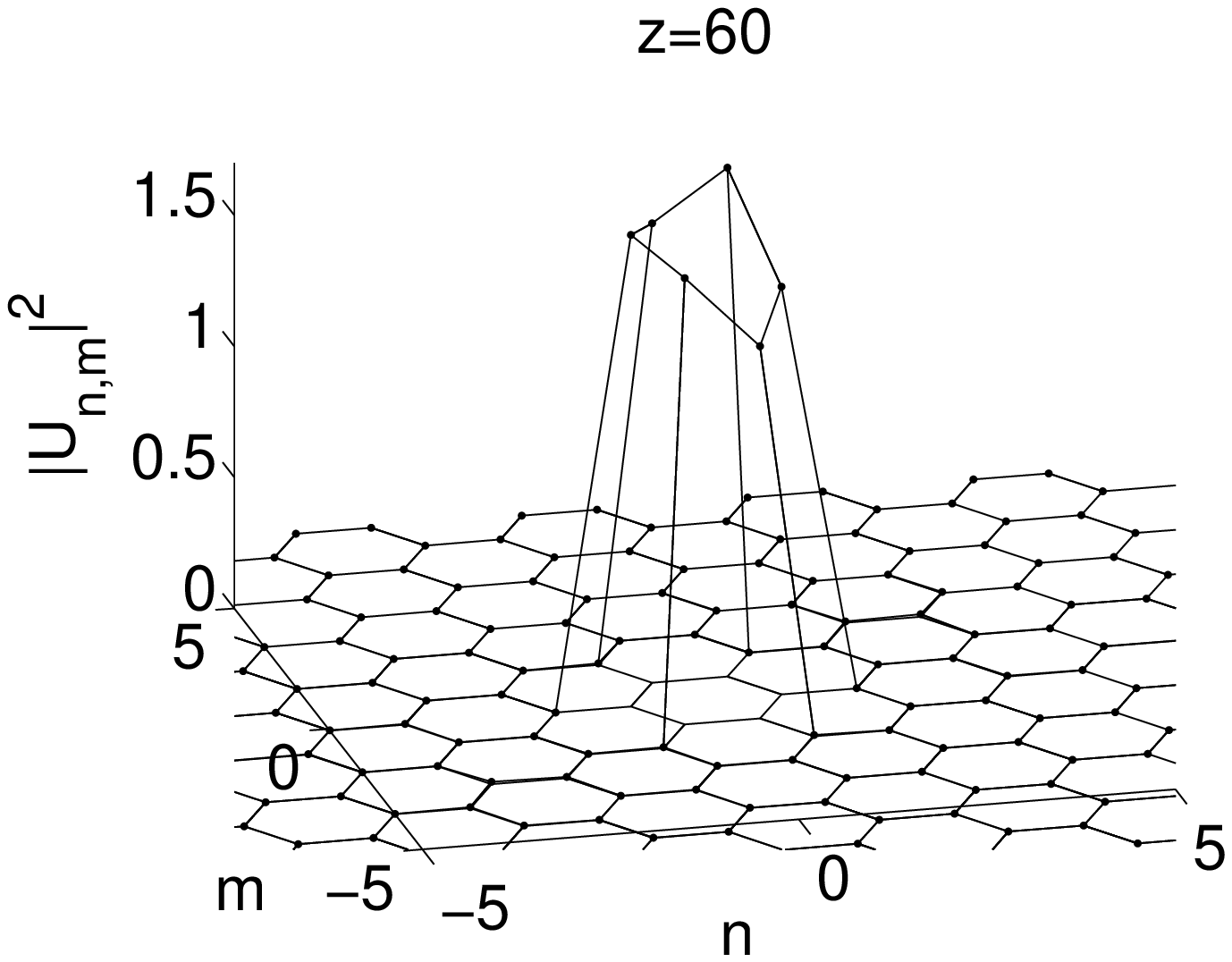}
\includegraphics[width=5cm,height=4cm]{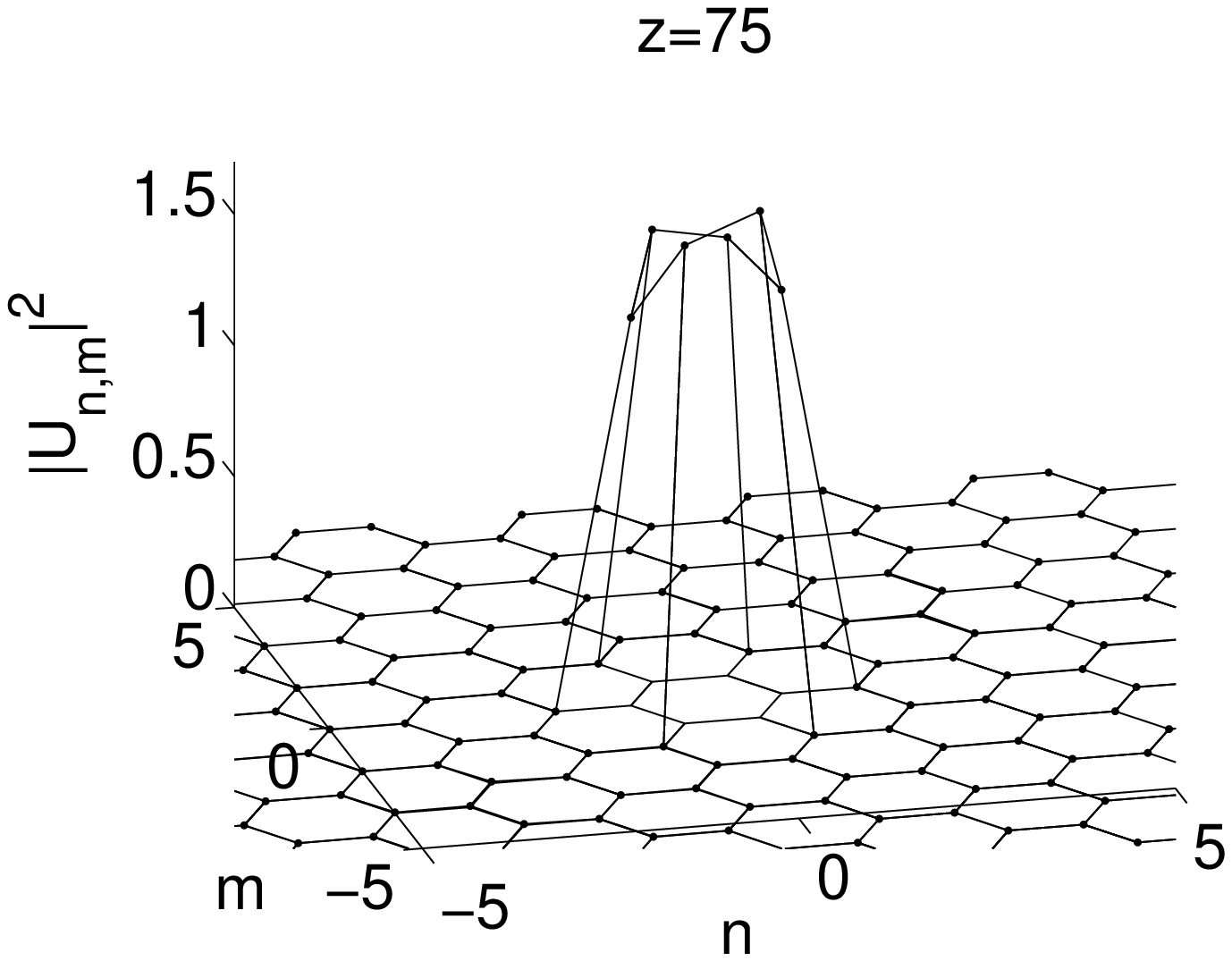}\\
\includegraphics[width=5cm,height=4cm]{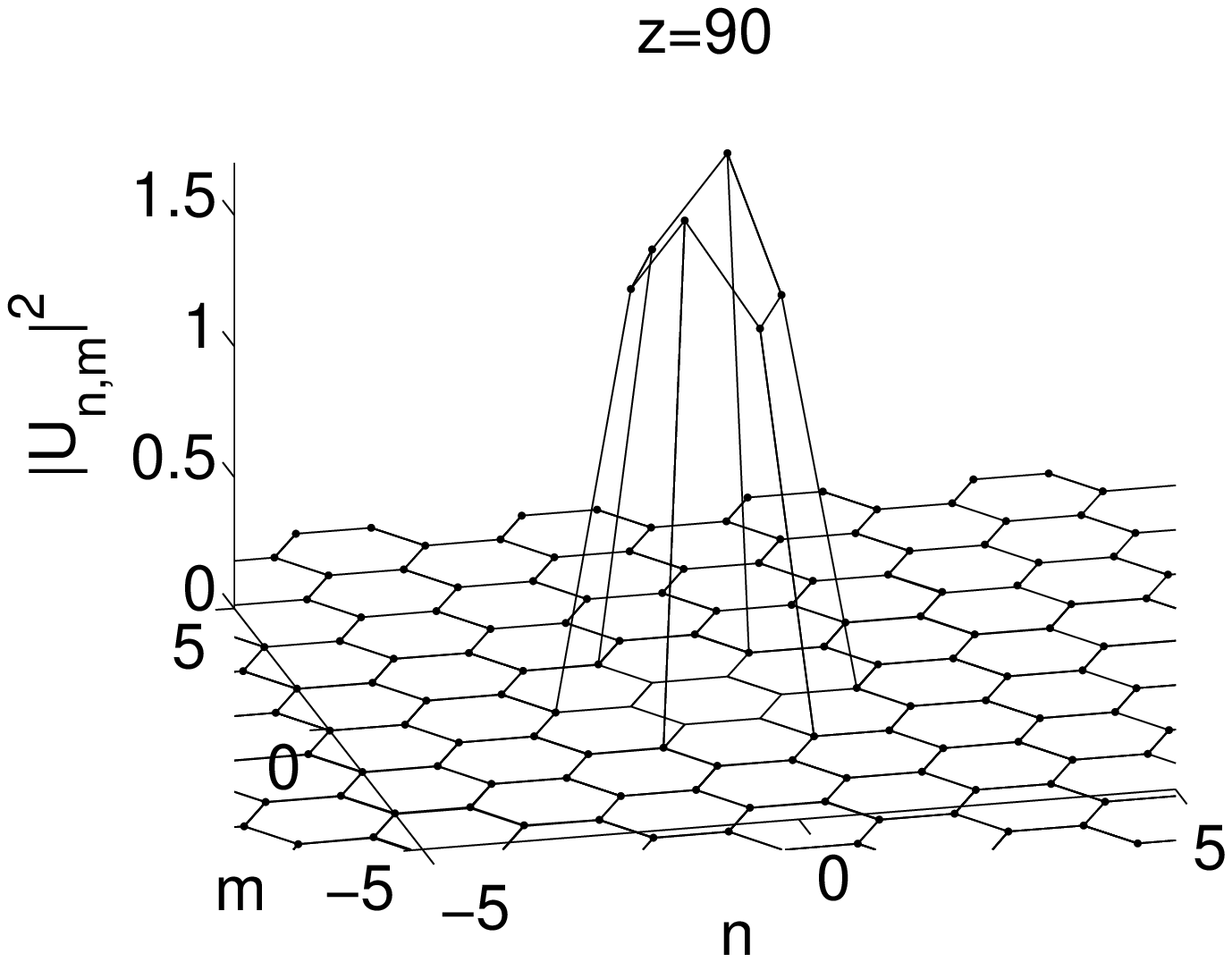}
\includegraphics[width=5cm,height=4cm]{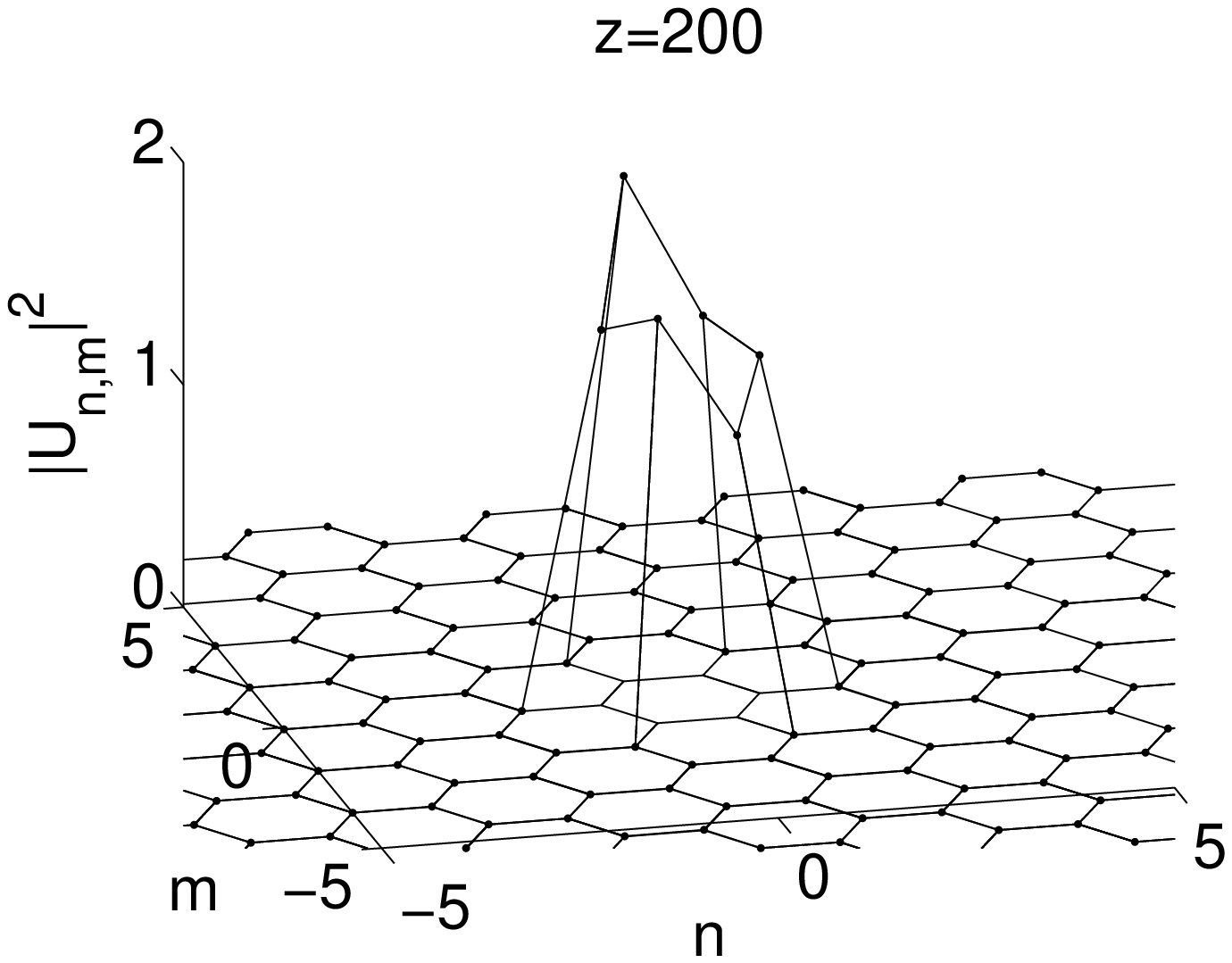}
\vspace{-0.4cm}
\caption{RK$4$ results from the six site charge-2 vortex at $\delta = 0.60$, $\varepsilon=0.2$ at $z=1, 15, 25, 45, 60, 90, 200$. A complex multi-site
breathing pattern results from the instability.}
\label{fig:Hon6site_C2_an0_6_C0_2_dyn}
\end{figure}

\begin{figure}[tbh]
\centering
\includegraphics[width=5cm,height=4cm]{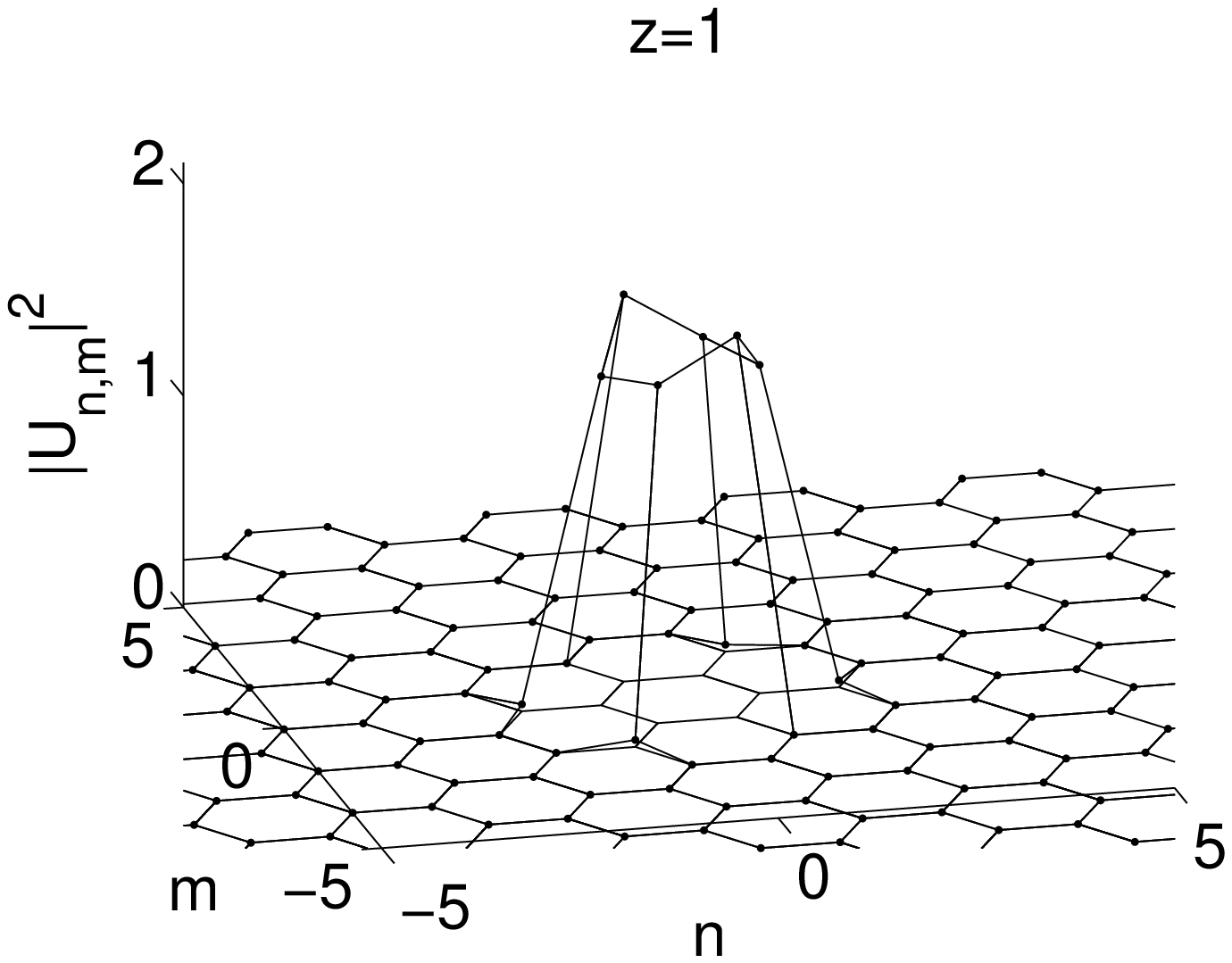}
\includegraphics[width=5cm,height=4cm]{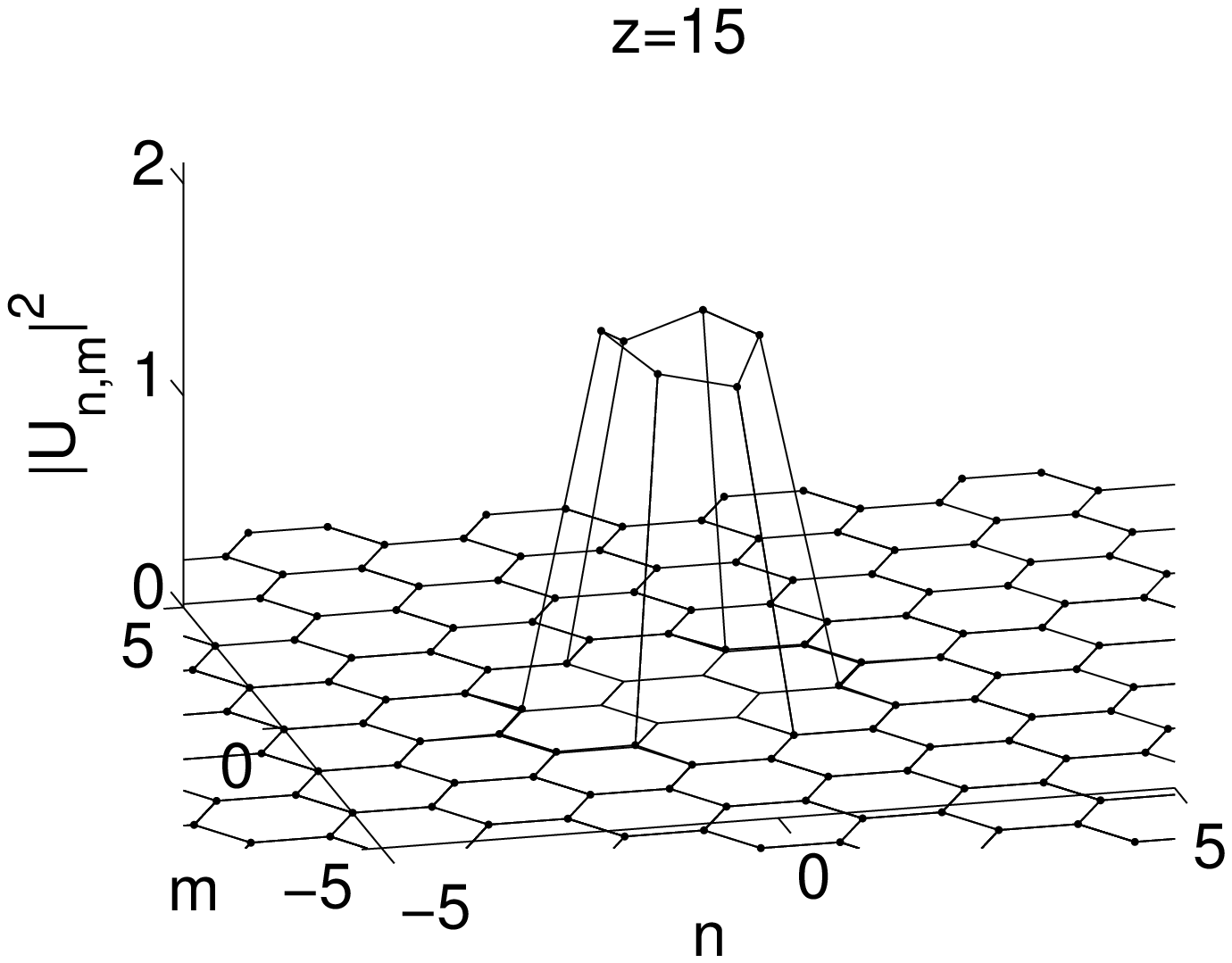}\\
\includegraphics[width=5cm,height=4cm]{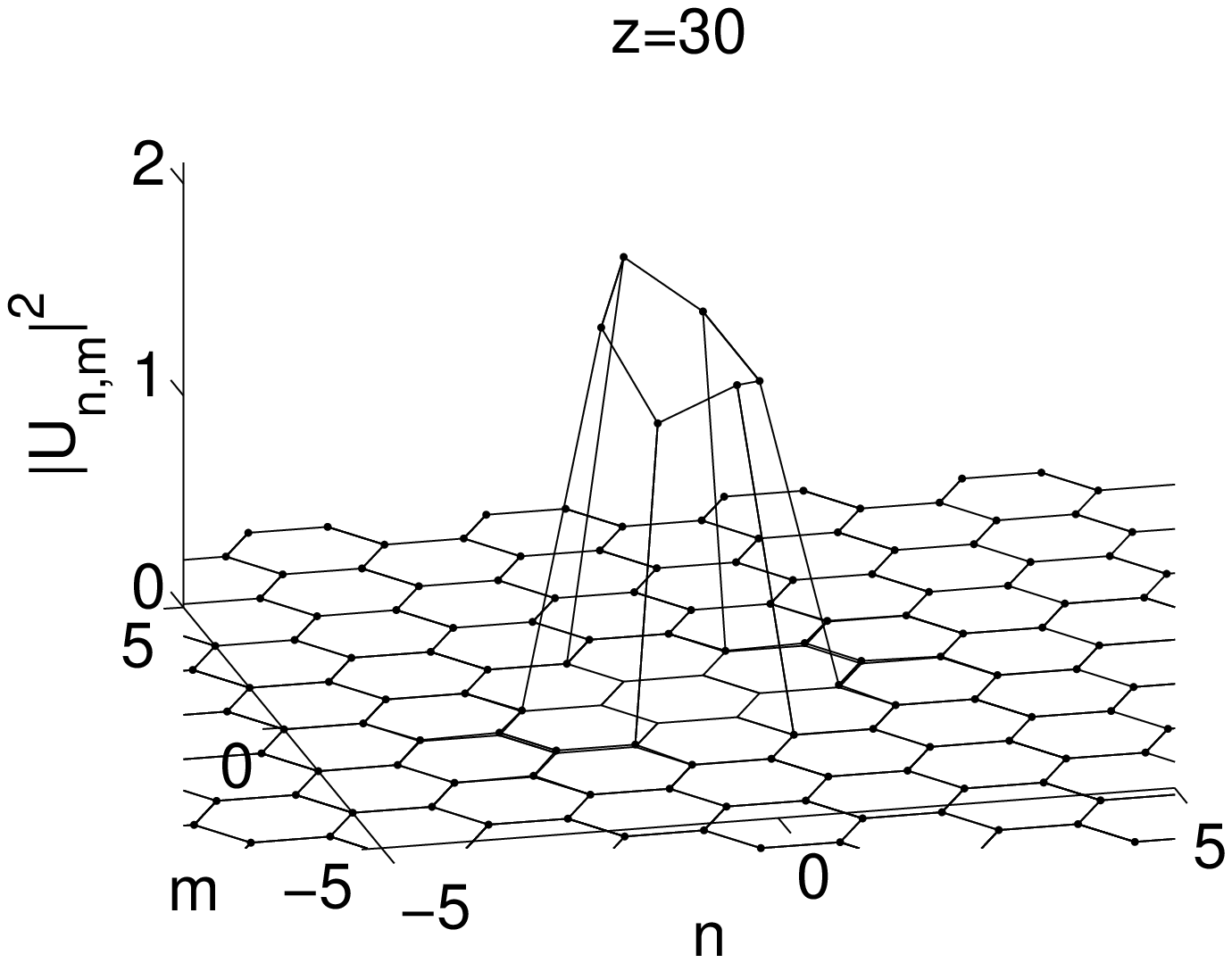}
\includegraphics[width=5cm,height=4cm]{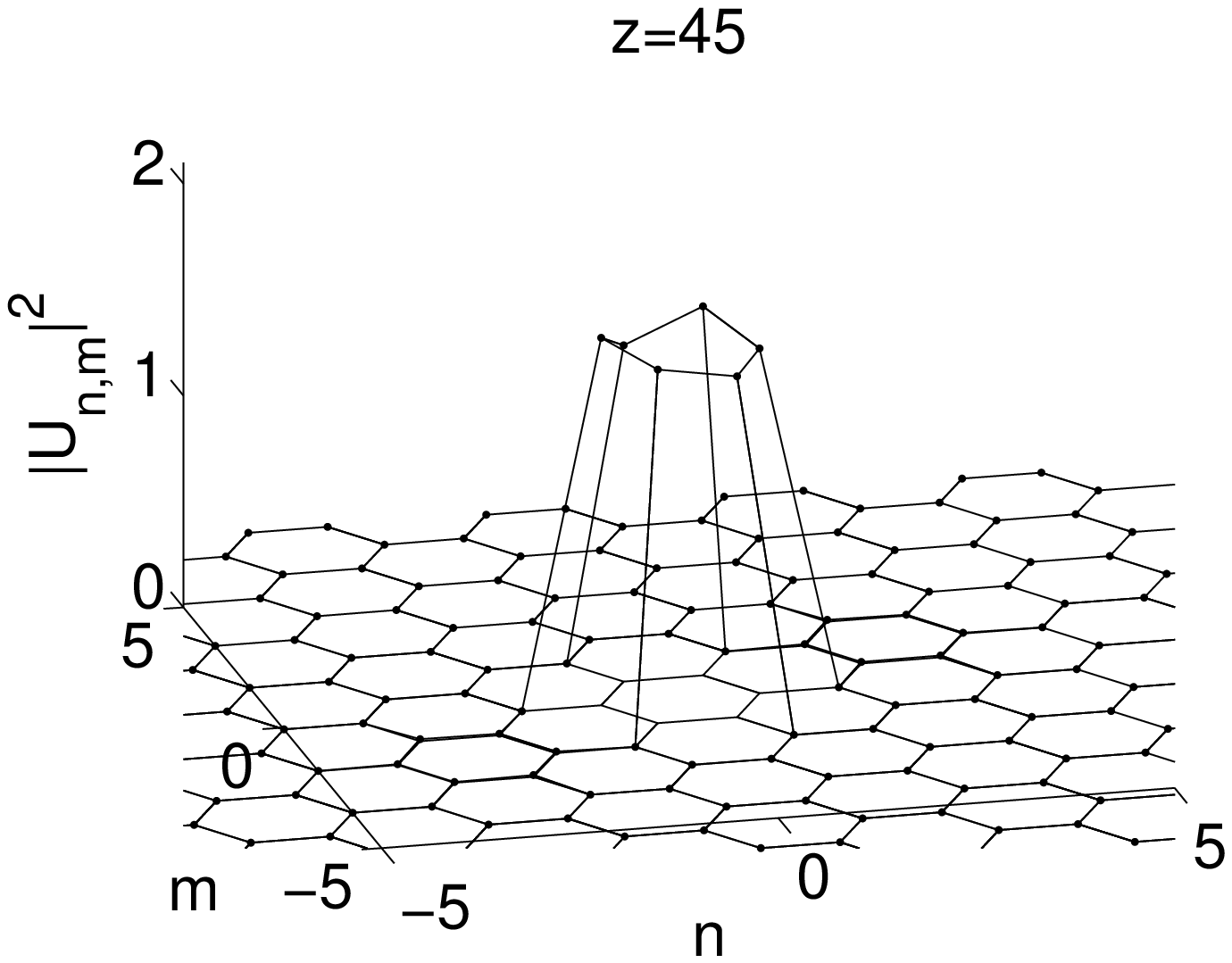}\\
\includegraphics[width=5cm,height=4cm]{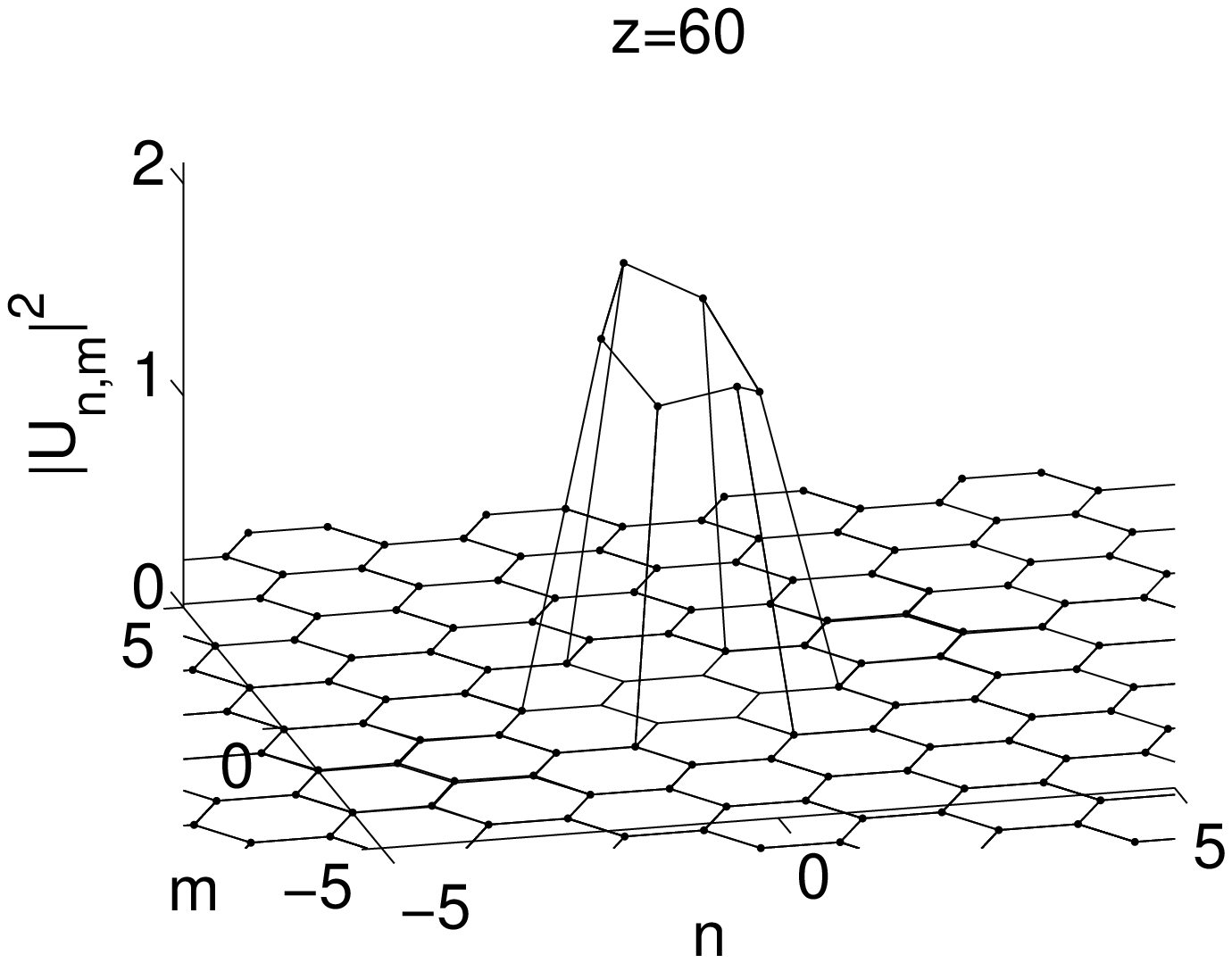}
\includegraphics[width=5cm,height=4cm]{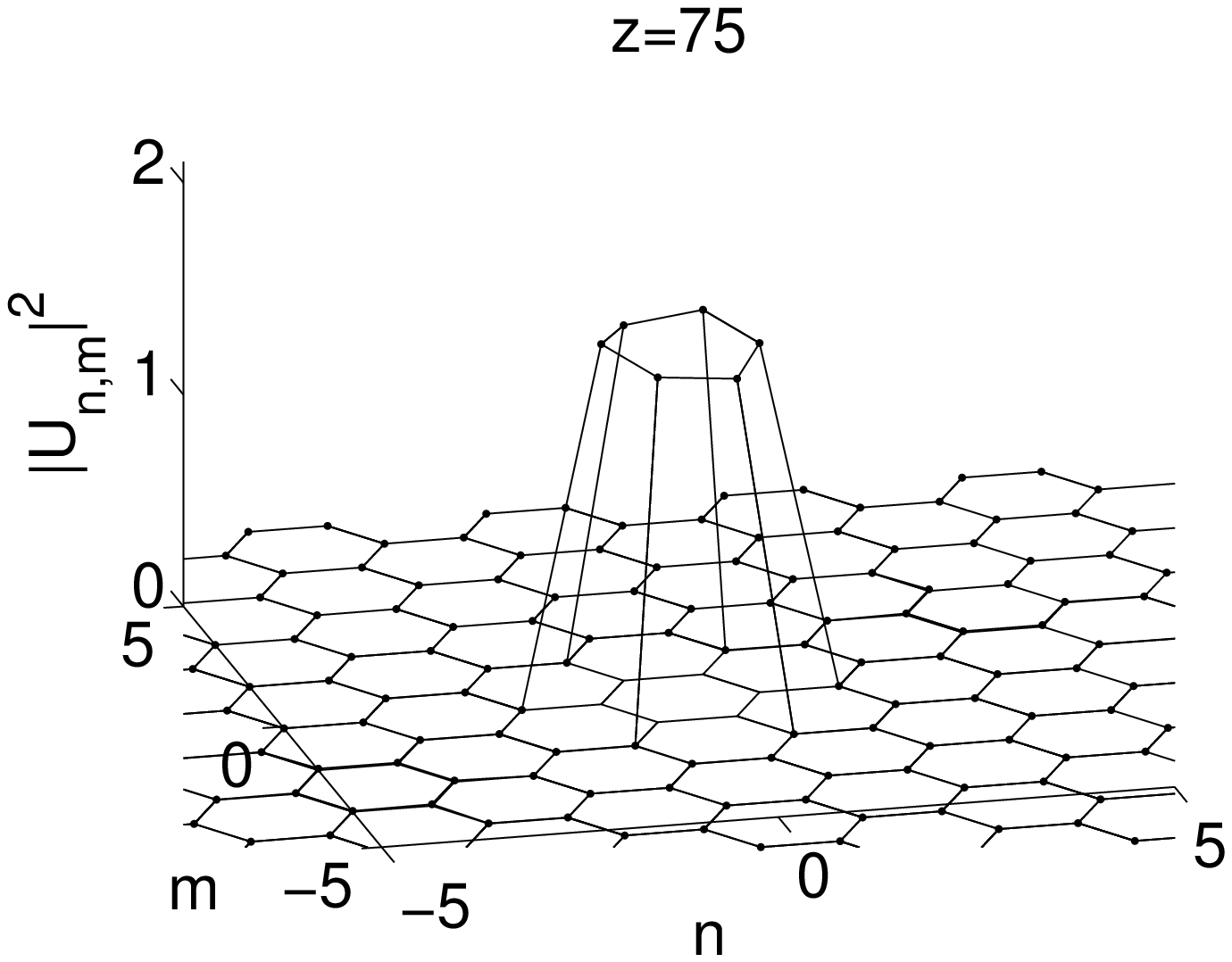}\\
\includegraphics[width=5cm,height=4cm]{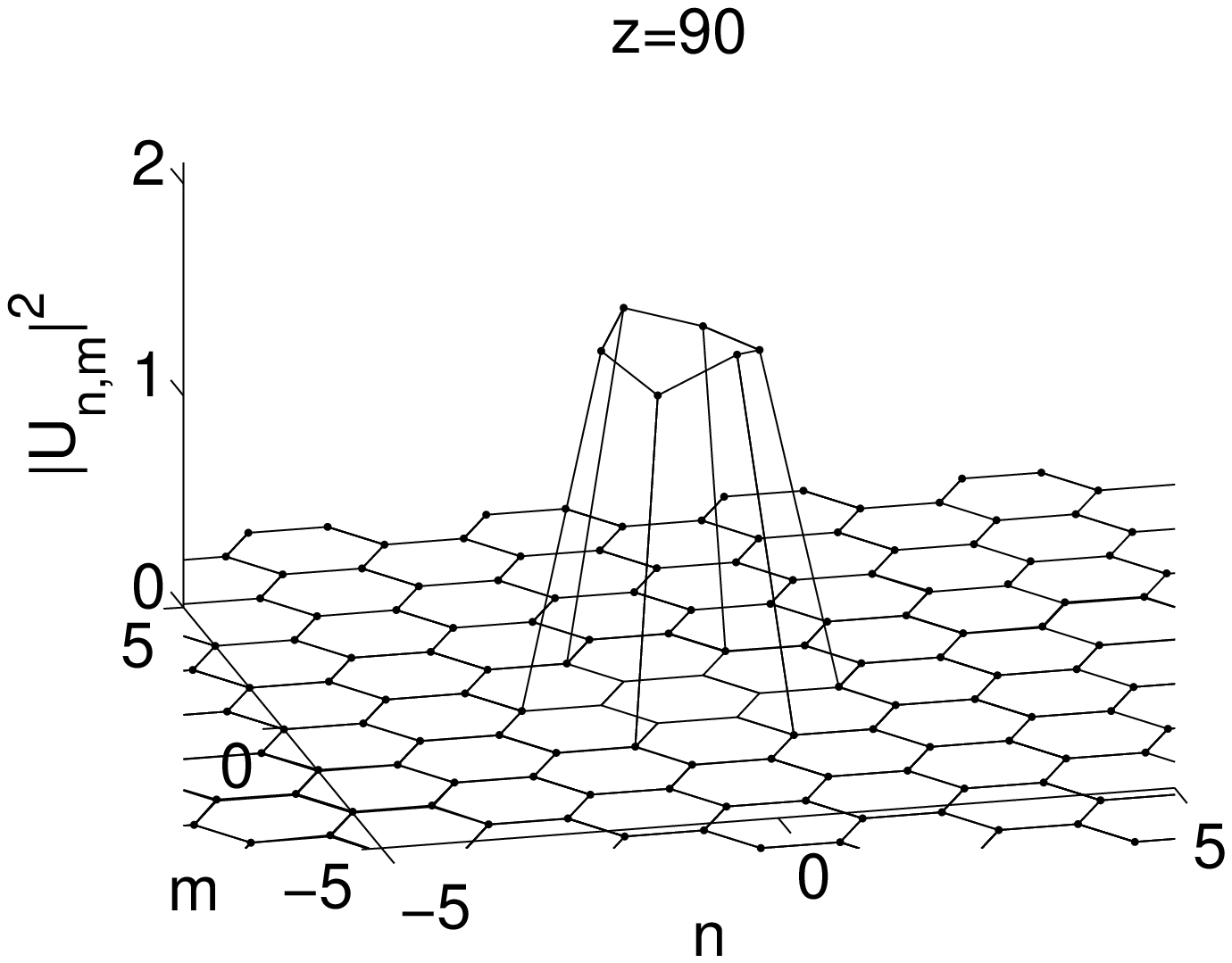}
\includegraphics[width=5cm,height=4cm]{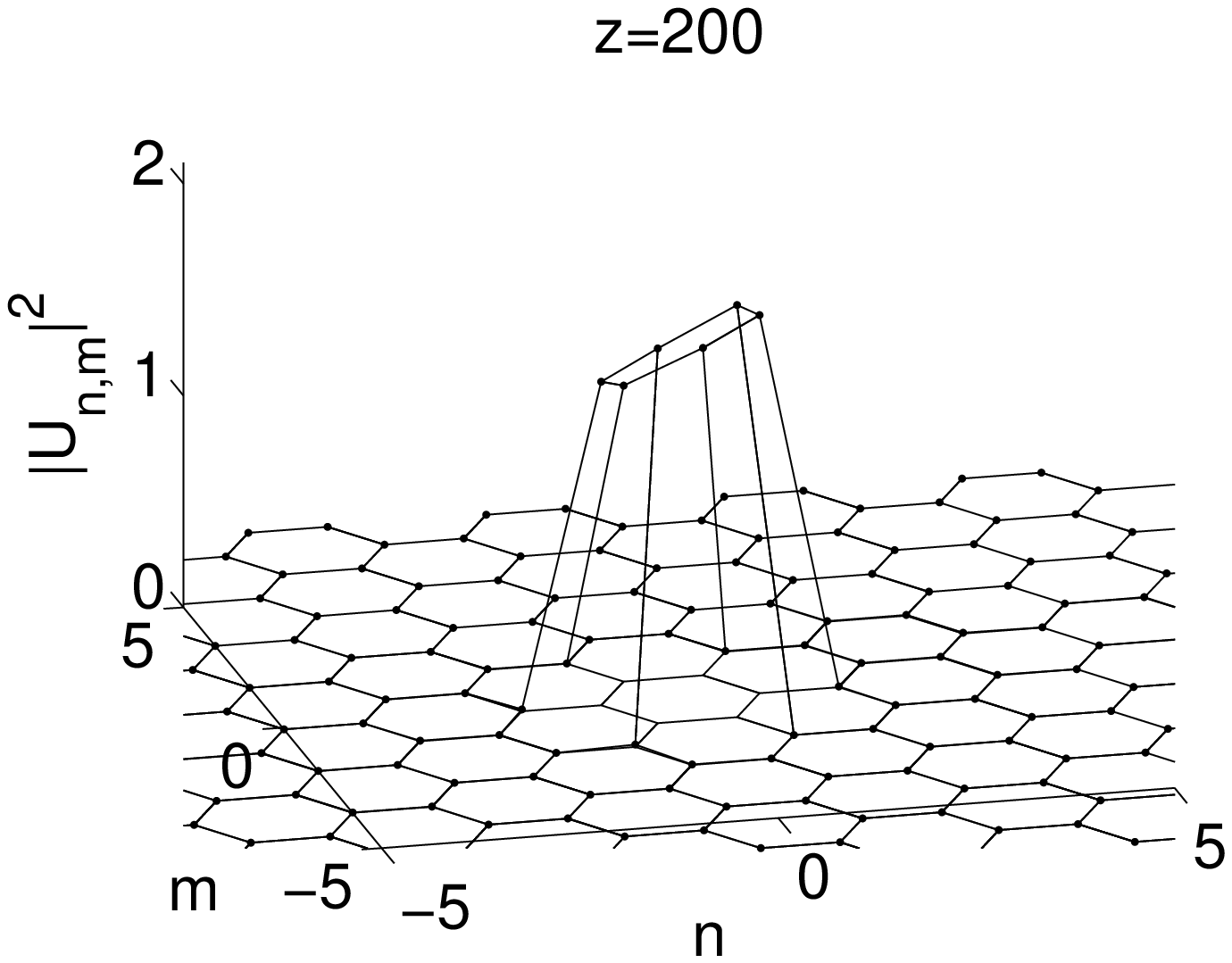}
\vspace{-0.4cm}
\caption{RK$4$ results from the six site waveform $[0, 0 + \pi, 0, 0, 0 + \pi, 0]$ (resulting
from the bifurcation of a charge-2 vortex) at $\delta = 0.20$, $\varepsilon=0.2$ at $z=1, 15, 25, 45, 60, 90, 200$.}
\label{fig:Hon6site_C2_an0_2_C0_2_dyn}
\end{figure}

(2) In Fig.~\ref{fig:Hon6site_C2_an0_6_C0_2_dyn} we see the
dynamics of the charge 2 vortex after the onset of instability,
at $\delta = 0.6$. The result is a six site breathing structure with
a complex norm redistribution.
In Fig.~\ref{fig:Hon6site_C2_an0_2_C0_2_dyn}, we show the
dynamics of the unstable solution of the form
$[0, 0 + \pi, 0, 0, 0 + \pi, 0]$, resulting from the pitchfork
bifurcation of the charge-2 vortex for the case of $\delta < 0.5$.
Specifically, in this example for $\delta=0.2$, a breathing 
six-site excitation appears to persist.

\begin{figure}[tbh]
\centering
\includegraphics[width=5cm,height=4cm]{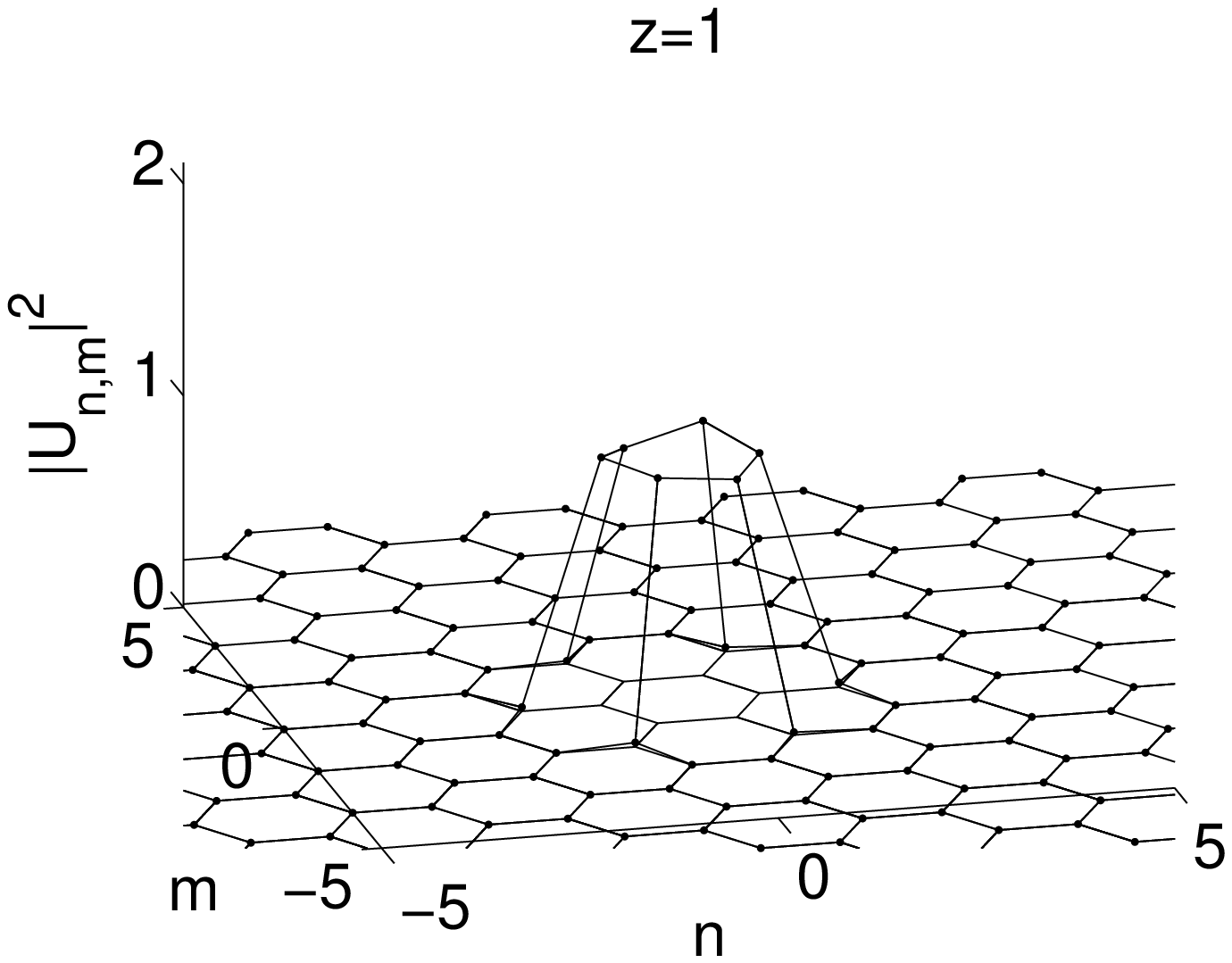}
\includegraphics[width=5cm,height=4cm]{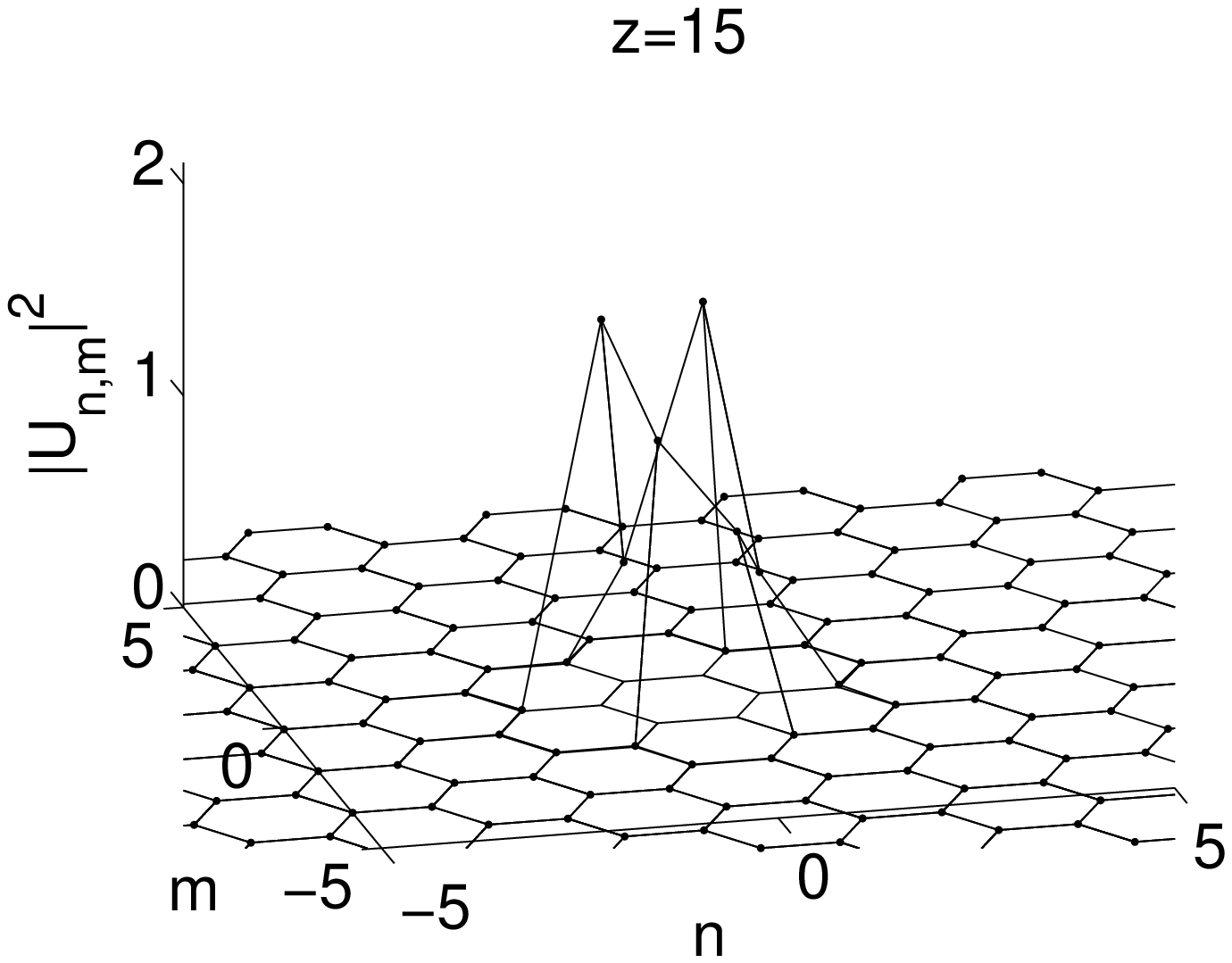}\\
\includegraphics[width=5cm,height=4cm]{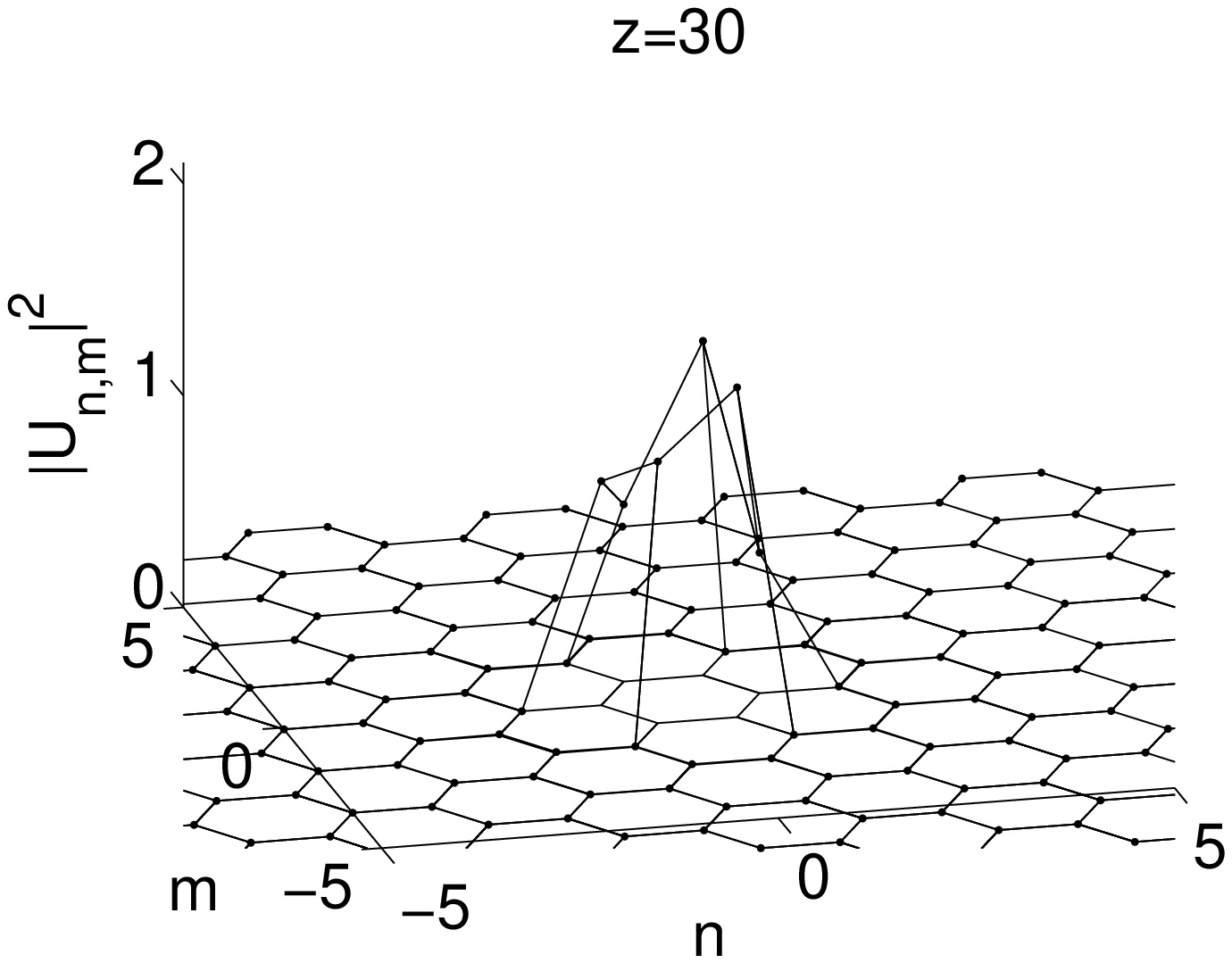}
\includegraphics[width=5cm,height=4cm]{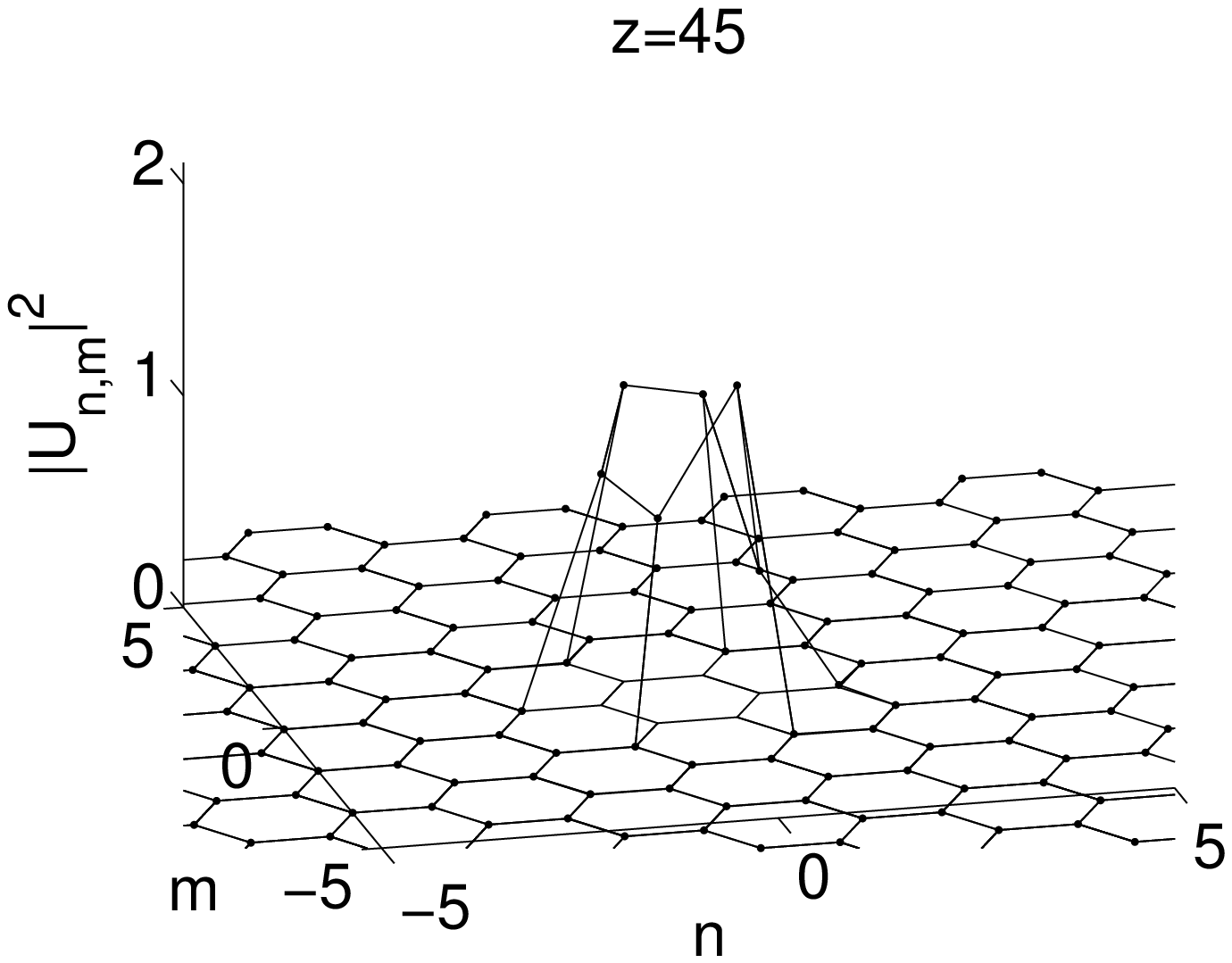}\\
\includegraphics[width=5cm,height=4cm]{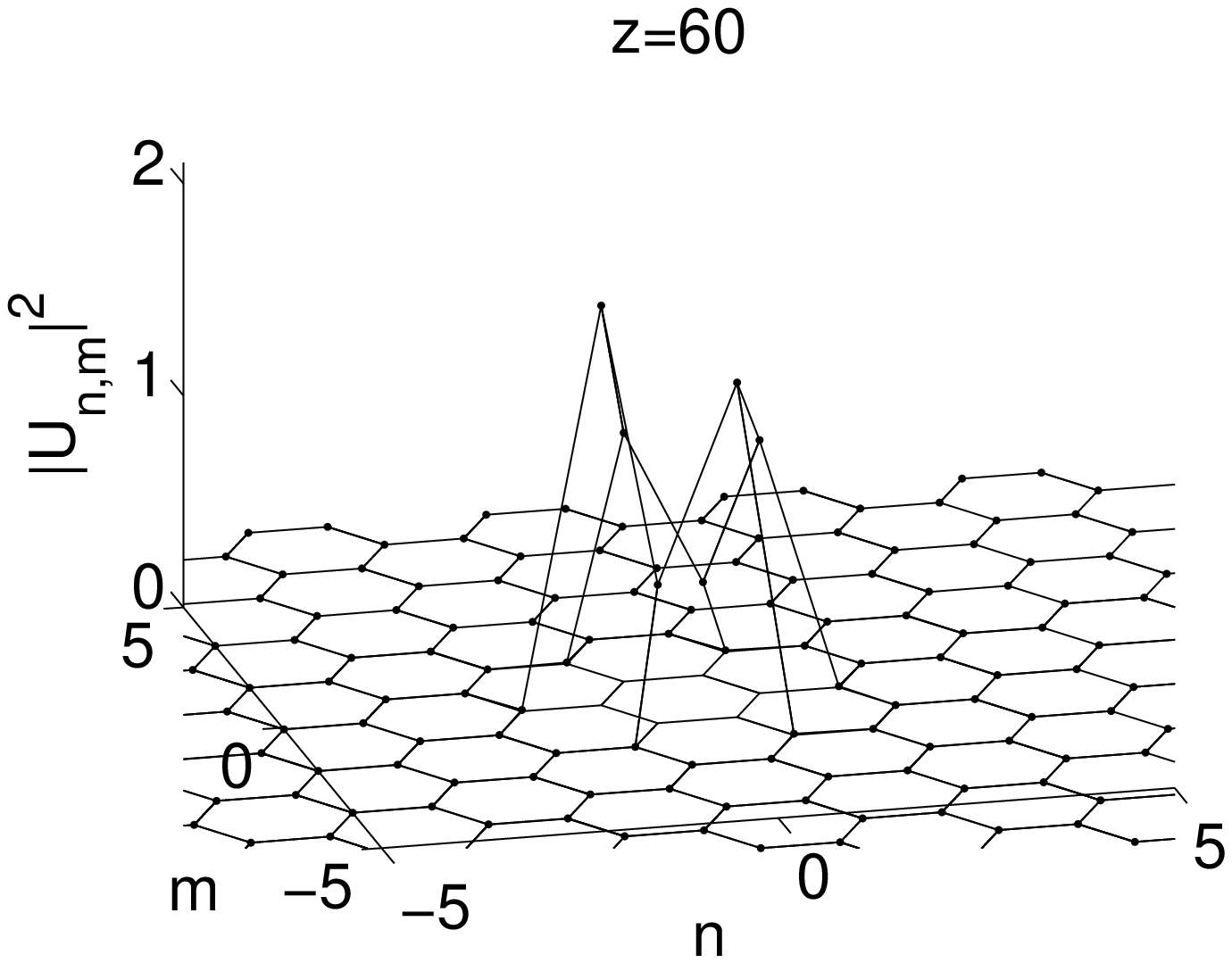}
\includegraphics[width=5cm,height=4cm]{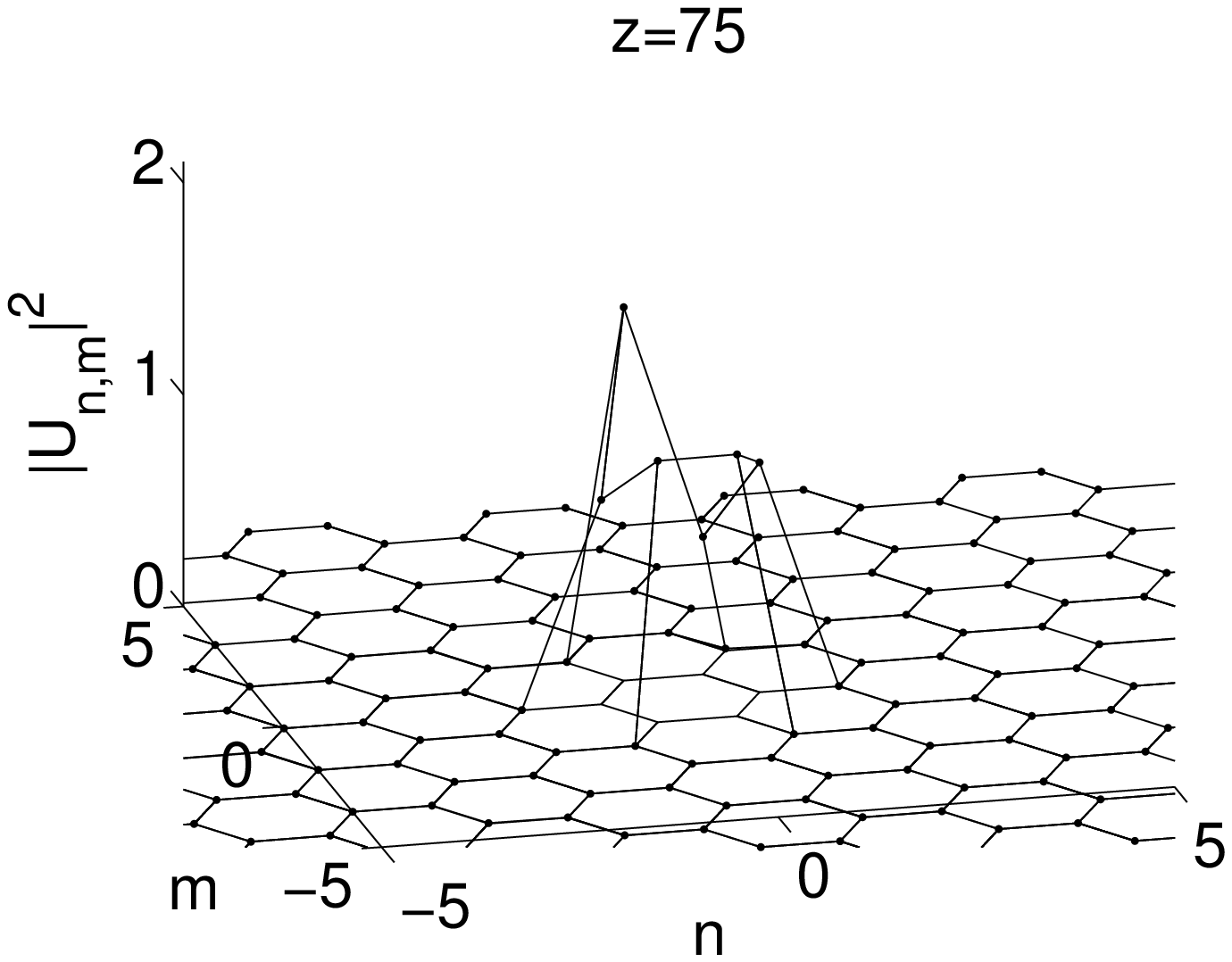}\\
\includegraphics[width=5cm,height=4cm]{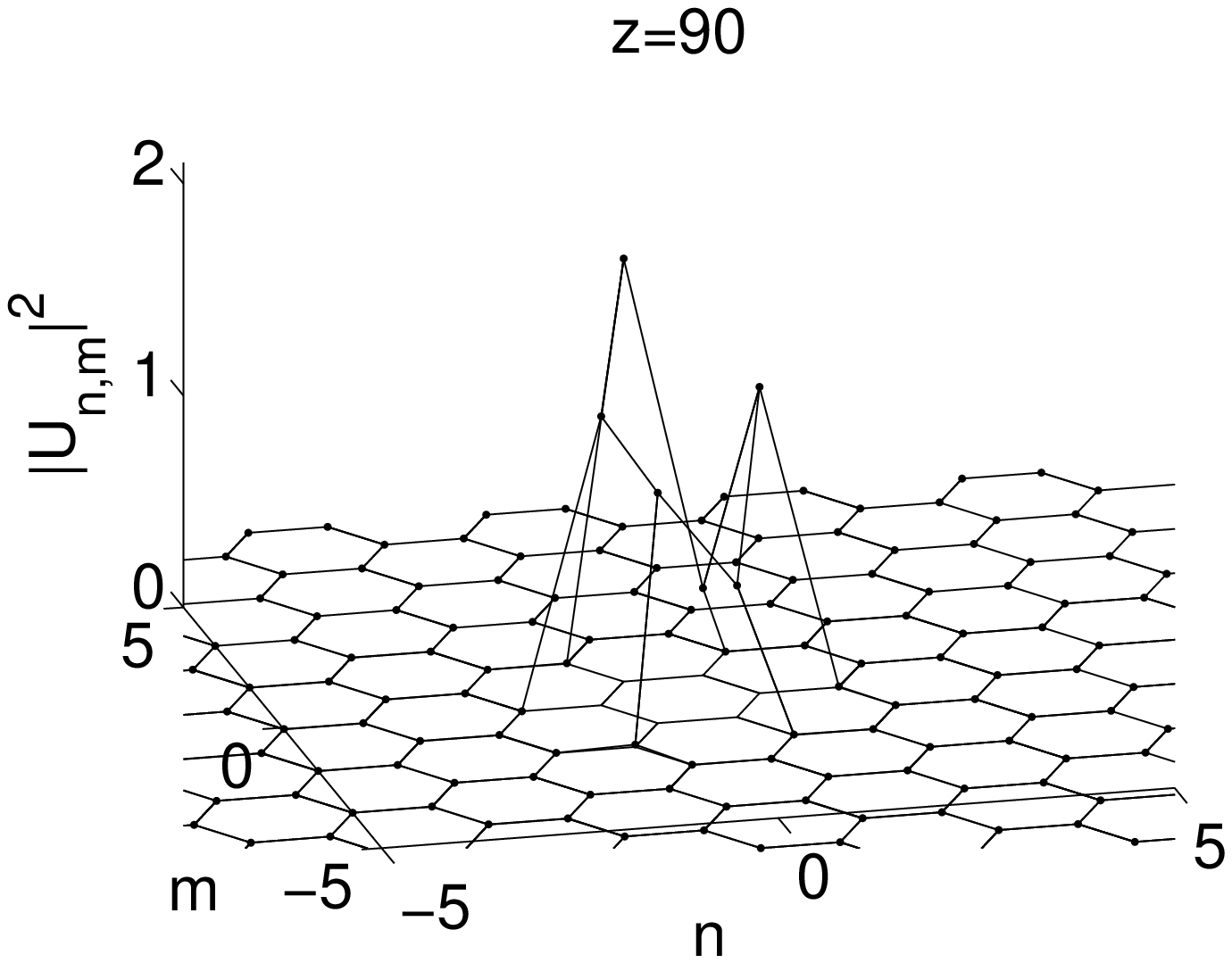}
\includegraphics[width=5cm,height=4cm]{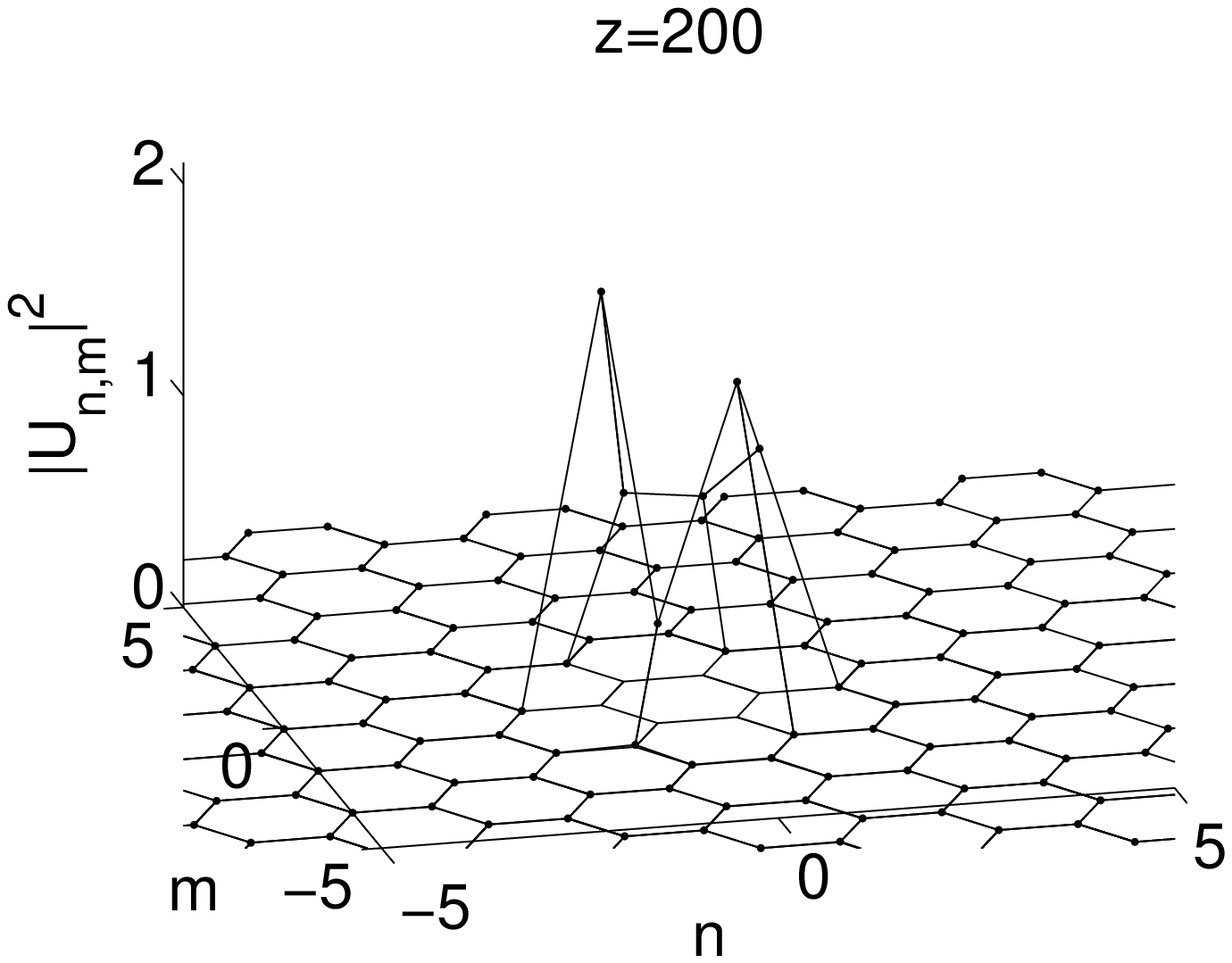}
\vspace{-0.4cm}
\caption{RK$4$ results from the honeycomb six-site $[0, 0, 0, 0, 0, 0]$ configuration at $\delta = 0.80$, $C=0.2$ at $z=1, 15, 25, 45, 60, 90, 200$. A state
with fewer dominant sites appears to emerge from the instability dynamics.}
\label{fig:Hon6site_0-0_an0_8_C0_2_dyn}
\end{figure}

(3) Finally, the dynamical evolution of the
unstable state $[0, 0, 0, 0, 0, 0]$ for $\delta=0.8$ in
Fig.~\ref{fig:Hon6site_0-0_an0_8_C0_2_dyn} (but also for
other values of $\delta$) illustrates the dynamical tendency
of this state towards configurations with fewer --arguably two,
at the final evolution snapshot shown-- dominant
(in amplitude) sites.

\section{Conclusions and Future Challenges}

In summary, in the present work, we have explored the existence,
stability and dynamics of localized states (focusing on multi-site
solitonic and vortex states) in hexagonal and honeycomb lattices.
We considered the prototypical unit cells in each case, namely
a 3-site one in the hexagonal case and a 6-site one
in the honeycomb case. Analytical considerations in the vicinity
of the anti-continuum limit permitted us to identify the states
in the presence of anisotropy in an approximate analytical form
and gave us the ability to consider the linear (spectral) stability
eigenvalues and obtain approximate analytical expressions for them.
These results allowed us to elucidate the pitchfork bifurcations
that lead to the disappearance of states such as vortices 
(and the destabilization of other solitonic states) 
as the path from more
effectively one-dimensional to effectively two-dimensional configurations
is traversed. These existence and stability findings were also found
to be in good agreement with detailed numerical continuations
(over the anisotropy parameter), at least for small values of 
the coupling. Finally, the dynamics of the relevant structures
were examined, allowing us to identify some gross features, including
the breathing nature of the instability for very weak $\varepsilon$
and the potential for stronger localization (typically to a smaller number
of sites) ensuing as a result of instability for stronger $\varepsilon$.

A significant number of possibilities emerge from
the present work for future explorations. On the one hand,
it would be relevant and interesting to examine in more detail
the dynamical evolution scenarios of the model, and to provide a
more systematic characterization of the propagation outcomes
for cases of both weaker and stronger coupling. On the other hand,
extending similar studies to the case of Kagom{\'e} lattices
and their flat bands, identifying the spectral properties
not only of the solitons/vortices~\cite{kag1} but also of the 
compactly supported structures~\cite{kag2} identified therein
would be a timely theme. Finally, extending such 
considerations to three-dimensional lattices of different types
would also pose significant new challenges and can be expected to
feature intriguing bifurcation phenomena and states of interest.
Efforts along these directions are presently underway and will
be reported in future publications.

\clearpage

\clearpage

\clearpage

\clearpage

\clearpage

\clearpage

\clearpage

\clearpage

\clearpage

\clearpage

\end{document}